\documentclass[11pt,a4paper]{report}
\usepackage{amsbsy}
\usepackage{bbm}
\usepackage{bm}
\usepackage{epsfig}
\usepackage{epstopdf}
\usepackage{dsfont}
\usepackage{soul}
\usepackage{color}
\definecolor{darkgreen}{RGB}{50,190,50}
\definecolor{darkblue}{RGB}{0,0,190}
\definecolor{darkred}{RGB}{238,0,0}
\usepackage{makeidx}
\makeindex
\usepackage{idxlayout}
\usepackage{ragged2e}

\usepackage{amsmath, amssymb, amsthm, amsbsy}
\usepackage{graphicx,color}
\usepackage[left=1.25in, right=1.25in, top=1in, bottom=1in, includefoot, headheight=13.6pt]{geometry}
\usepackage[square, comma, numbers, sort&compress]{natbib}
\usepackage[T1]{fontenc}

\usepackage{palatino}
\usepackage[small, bf, margin=15pt, tableposition=top]{caption}
\numberwithin{equation}{chapter}

\usepackage{fancyheadings}
\makeatletter
\renewcommand{\chaptermark}[1]{\markboth{\textsc{\@chapapp}\ \thechapter:\ #1}{}}
\makeatother
\fancypagestyle{plain}{%
  \fancyhf{}%
  \fancyfoot[C]{\thepage}%
}
\fancypagestyle{CV}{%
\fancyhead[RO,RE]{\nouppercase{\textsc{Curriculum Vitae}}}}

\usepackage{sectsty}
\chapterfont{\large\sc\centering}
\chaptertitlefont{\centering}
\subsubsectionfont{\centering}

\usepackage{epstopdf}
\usepackage[bookmarks=true, colorlinks]{hyperref}
\usepackage[figure,table]{hypcap}
\hypersetup{
	bookmarksnumbered,
	pdfstartview={FitH},
	citecolor={darkgreen},
    linkcolor={darkred},
    linktoc={page},
	urlcolor={darkblue},
	pdfpagemode={UseOutlines}}
\usepackage{tocloft}
\makeatletter
\newcommand\org@hypertarget{}
\let\org@hypertarget\hypertarget
\renewcommand\hypertarget[2]{%
  \Hy@raisedlink{\org@hypertarget{#1}{}}#2%
  }
\makeatother
\newcommand{\tr}{\textnormal{Tr}}

\newcommand{\bra}[1]{\ensuremath{\left\langle\right. #1 \left.\right|}}
\newcommand{\ket}[1]{\ensuremath{\left|\right. #1 \left.\right\rangle}}
\newcommand{\brahat}[1]{\ensuremath{\left\langle\right. \widehat{#1} \left.\right|}}
\newcommand{\kethat}[1]{\ensuremath{\left|\right. \widehat{#1} \left.\right\rangle}}
\newcommand{\brahatk}[2]{\ensuremath{\left\langle\right. \widehat{#1}_{#2}\! \left.\right|}}
\newcommand{\kethatk}[2]{\ensuremath{\left|\right. \widehat{#1}_{#2} \!\left.\right\rangle}}

\newcommand{\fbra}[1]{\ensuremath{\left\langle\!\left\langle\right.\right.\! #1 \!\left.\left.\right|\hspace*{-0.75pt}\right|}}
\newcommand{\fket}[1]{\ensuremath{\left|\hspace*{-0.75pt}\left|\right.\right.\! #1 \!\left.\left.\right\rangle\!\right\rangle}}
\newcommand{\fbrap}[1]{\ensuremath{{}^{\protect\raisebox{0.5pt}{\tiny{$+$}}\hspace*{-4.5pt}}
\left\langle\!\left\langle\right.\right.\! #1 \!\left.\left.\right|\hspace*{-0.75pt}\right|}}
\newcommand{\fbram}[1]{\ensuremath{{}^{\protect\raisebox{0.5pt}{\tiny{$-$}}\hspace*{-4.5pt}}
\left\langle\!\left\langle\right.\right.\! #1 \!\left.\left.\right|\hspace*{-0.75pt}\right|}}
\newcommand{\fketp}[1]{\ensuremath{\left|\hspace*{-0.75pt}\left|\right.\right.\! #1 \!\left.\left.\right\rangle\!\right\rangle
^{\hspace*{-2pt}\protect\raisebox{0.0pt}{\tiny{$+$}}}}}
\newcommand{\fketm}[1]{\ensuremath{\left|\hspace*{-0.75pt}\left|\right.\right.\! #1 \!\left.\left.\right\rangle\!\right\rangle
^{\hspace*{-2pt}\protect\raisebox{0.0pt}{\tiny{$-$}}}}}
\newcommand{\fbrahat}[1]{\ensuremath{\left\langle\!\left\langle\right.\right.\! \widehat{#1} \!\left.\left.\right|\hspace*{-0.75pt}\right|}}
\newcommand{\fkethat}[1]{\ensuremath{\left|\hspace*{-0.75pt}\left|\right.\right.\! \widehat{#1} \!\left.\left.\right\rangle\!\right\rangle}}

\newcommand{\fbrahatkp}[2]{\ensuremath{{}^{\protect\raisebox{0.5pt}{\tiny{$+$}}\hspace*{-4.5pt}}
\left\langle\!\left\langle\right.\right.\! \widehat{#1}_{#2} \!\left.\left.\right|\hspace*{-0.75pt}\right|}}
\newcommand{\fbrahatkm}[2]{\ensuremath{{}^{\protect\raisebox{0.5pt}{\tiny{$-$}}\hspace*{-4.5pt}}
\left\langle\!\left\langle\right.\right.\! \widehat{#1}_{#2} \!\left.\left.\right|\hspace*{-0.75pt}\right|}}

\newcommand{\fbrahatkpup}[2]{\ensuremath{{}^{\protect\raisebox{2.0pt}{\tiny{$+$}}\hspace*{-4.5pt}}
\left\langle\!\left\langle\right.\right.\! \widehat{#1}_{#2} \!\left.\left.\right|\hspace*{-0.75pt}\right|}}
\newcommand{\fbrahatkmup}[2]{\ensuremath{{}^{\protect\raisebox{2.0pt}{\tiny{$-$}}\hspace*{-4.5pt}}
\left\langle\!\left\langle\right.\right.\! \widehat{#1}_{#2} \!\left.\left.\right|\hspace*{-0.75pt}\right|}}

\newcommand{\fkethatkp}[2]{\ensuremath{\left|\hspace*{-0.75pt}\left|\right.\right.\! \widehat{#1}_{#2} \!\left.\left.\right\rangle\!\right\rangle^{\hspace*{-2pt}\protect\raisebox{0.0pt}{\tiny{$+$}}}}}
\newcommand{\fkethatkm}[2]{\ensuremath{\left|\hspace*{-0.75pt}\left|\right.\right.\! \widehat{#1}_{#2} \!\left.\left.\right\rangle\!\right\rangle^{\hspace*{-2pt}\protect\raisebox{0.0pt}{\tiny{$-$}}}}}

\newcommand{\comm}[2]{\ensuremath{\left[\right.\! #1 \,, #2 \!\left.\right]}}
\newcommand{\anticomm}[2]{\ensuremath{\left\{\right.\! #1 \,, #2 \!\left.\right\}}}

\newcommand{\scpr}[2]{\ensuremath{\left\langle\right. #1 \,\left|\right. #2 \left.\right\rangle}}
\newcommand{\fscpr}[2]{\ensuremath{\left\langle\!\hspace*{-0.7pt}\left\langle\right.\right.\! #1 \!\left.\left.\right|\!\right| #2 \!\left.\left.\right\rangle\!\hspace*{-0.7pt}\right\rangle}}
\newcommand{\expval}[1]{\ensuremath{\left\langle\right. #1 \left.\right\rangle}}

\newcommand{\pr}{^{\prime}}
\newcommand{\prpr}{^{\prime\hspace*{-0.5pt}\prime}}
\newcommand{\chii}[1]{\ensuremath{\chi_{\protect\raisebox{-2.0pt}{\scriptsize{$ #1 $}}}}}
\newcommand{\lambdai}[1]{\ensuremath{\lambda_{\hspace*{0.5pt} #1 }}}
\newcommand{\pii}[1]{\ensuremath{p_{\hspace*{0.2pt}\protect\raisebox{-1.0pt}{\scriptsize{$ #1 $}}}}}
\newcommand{\rhoA}[1]{\ensuremath{\rho_{\hspace*{-1.0pt}\protect\raisebox{-1.0pt}{\tiny{$ #1 $}}}}}
\newcommand{\idN}[1]{\ensuremath{\mathds{1}_{\hspace*{-1.0pt}\protect\raisebox{-1.0pt}{\scriptsize{$ #1 $}}}}}
\newcommand{\idNtiny}[1]{\ensuremath{\mathds{1}_{\hspace*{-1.0pt}\protect\raisebox{-1.0pt}{\tiny{$ #1 $}}}}}

\newcommand{\OpNtiny}[1]{\ensuremath{\mathcal{O}_{\hspace*{-1.0pt}\protect\raisebox{-1.0pt}{\tiny{$ #1 $}}}}}
\newcommand{\sub}[3]{\ensuremath{_{\hspace{#1 pt}\protect\raisebox{#2 pt}{\scriptsize{$ #3$}}}}}
\newcommand{\subtiny}[3]{\ensuremath{_{\hspace{#1 pt}\protect\raisebox{#2 pt}{\tiny{$ #3$}}}}}
\newcommand{\oversmile}[1]{\ensuremath{\overset{\protect\raisebox{0.0pt}{\tiny{$\smile$}}}{\rule{0pt}{6.0pt}\smash{#1}}}}

\newcommand{\phii}[1]{\ensuremath{\phi_{\hspace*{0.5pt}\protect\raisebox{-1.0pt}{\scriptsize{$ #1 $}}}}}
\newcommand{\phiitilde}[1]{\ensuremath{\tilde{\phi}_{\hspace*{0.5pt}\protect\raisebox{-1.0pt}{\scriptsize{$ #1 $}}}}}
\newcommand{\phiihat}[1]{\ensuremath{\widehat{\phi}_{\hspace*{0.5pt}\protect\raisebox{-1.0pt}{\scriptsize{$ #1 $}}}}}
\newcommand{\psii}[1]{\ensuremath{\psi_{\hspace*{0.5pt}\protect\raisebox{-1.0pt}{\scriptsize{$ #1 $}}}}}

\newcommand{\psiihat}[1]{\ensuremath{\widehat{\psi}_{\hspace*{0.5pt}\protect\raisebox{-1.0pt}{\scriptsize{$ #1 $}}}}}

\newcommand{\an}[1]{\ensuremath{a_{\hspace*{0.2pt}\protect\raisebox{-1.0pt}{\scriptsize{$ #1 $}}}}}
\newcommand{\adn}[1]{\ensuremath{a^{\dagger}_{\hspace*{0.2pt}\protect\raisebox{-0.5pt}{\scriptsize{$ #1 $}}}}}
\newcommand{\antilde}[1]{\ensuremath{\tilde{a}_{\hspace*{0.2pt}\protect\raisebox{-1.0pt}{\scriptsize{$ #1 $}}}}}
\newcommand{\adntilde}[1]{\ensuremath{\tilde{a}^{\dagger}_{\hspace*{0.2pt}\protect\raisebox{-0.5pt}{\scriptsize{$ #1 $}}}}}
\newcommand{\anhat}[1]{\ensuremath{\widehat{a}_{\hspace*{0.2pt}\protect\raisebox{-1.0pt}{\scriptsize{$ #1 $}}}}}
\newcommand{\adnhat}[1]{\ensuremath{\widehat{a}^{\dagger}_{\hspace*{0.2pt}\protect\raisebox{-0.5pt}{\scriptsize{$ #1 $}}}}}
\newcommand{\bn}[1]{\ensuremath{b_{#1}}}
\newcommand{\bdn}[1]{\ensuremath{b^{\dagger}_{#1}}}
\newcommand{\cn}[1]{\ensuremath{c_{#1}}}
\newcommand{\cdn}[1]{\ensuremath{c^{\dagger}_{#1}}}
\newcommand{\bnhat}[1]{\ensuremath{\widehat{b}_{#1}}}
\newcommand{\bdnhat}[1]{\ensuremath{\widehat{b}^{\hspace*{1.5pt}\dagger}_{#1}}}
\newcommand{\cnhat}[1]{\ensuremath{\widehat{c}_{#1}}}
\newcommand{\cdnhat}[1]{\ensuremath{\widehat{c}^{\hspace*{1.5pt}\dagger}_{#1}}}

\newcommand{\GammaStk}[1]{\ensuremath{\Gamma_{\protect\raisebox{0pt}{\tiny{$\operatorname{St-}\hspace*{-0.6pt}#1$}}}}}
\newcommand{\Gammak}[1]{\ensuremath{\Gamma_{\protect\raisebox{0pt}{\tiny{$#1$}}}}}
\newcommand{\Gammahatk}[1]{\ensuremath{\widehat{\Gamma}_{\protect\raisebox{0pt}{\tiny{$#1$}}}}}

\newcommand{\Gammahatkh}[2]{\ensuremath{\widehat{\Gamma}_{\protect\raisebox{0pt}{\tiny{$#1$}}}
^{\hspace*{0.5pt}\protect\raisebox{0.0pt}{\tiny{$(#2)$}}}}}
\newcommand{\Cmn}[1]{\ensuremath{C_{\protect\raisebox{0pt}{\tiny{$#1$}}}}}
\newcommand{\Chatmn}[1]{\ensuremath{\widehat{C}_{\protect\raisebox{0pt}{\tiny{$#1$}}}}}
\newcommand{\Chatmnh}[2]{\ensuremath{\widehat{C}_{\protect\raisebox{0pt}{\tiny{$#1$}}}
^{\hspace*{0.5pt}\protect\raisebox{0.0pt}{\tiny{$(#2)$}}}}}
\newcommand{\num}[1]{\ensuremath{\overset{\protect\raisebox{0.0pt}{\hspace*{0.5pt}\tiny{$\smile$}}}{\rule{0pt}{3.0pt}\smash{\nu}}_{\!\small{#1}}}}
\newcommand{\numh}[2]{\ensuremath{\overset{\protect\raisebox{0.0pt}{\hspace*{0.5pt}\tiny{$\smile$}}}{\rule{0pt}{3.0pt}\smash{\nu}}
_{\!\small{#1}}^{\hspace*{0.5pt}\protect\raisebox{0.0pt}{\tiny{$(#2)$}}}}}
\newcommand{\GammaTMShath}[1]{\ensuremath{\widehat{\Gamma}_{\protect\raisebox{0pt}{\tiny{$\mathrm{TMS}$}}}
^{\hspace*{0.5pt}\protect\raisebox{0.0pt}{\tiny{$(#1)$}}}}}

\newcommand{\xLR}[1]{\ensuremath{x_{\hspace*{-0.5pt}\protect\raisebox{-0.5pt}{\tiny{$#1$}}}}}
\newcommand{\xiLR}[1]{\ensuremath{\xi_{\hspace*{-0.5pt}\protect\raisebox{-0.5pt}{\tiny{$#1$}}}}}
\newcommand{\chiLR}[1]{\ensuremath{\chi_{\hspace*{-0.5pt}\protect\raisebox{-1.5pt}{\tiny{$#1$}}}}}
\newcommand{\tildexiLR}[1]{\ensuremath{\tilde{\xi}_{\hspace*{-0.5pt}\protect\raisebox{-0.5pt}{\tiny{$#1$}}}}}

\newcommand{\MRalpha}[1]{{{}_{\text{o}}\alpha_{\hspace*{0.2pt}\protect\raisebox{-1.0pt}{\scriptsize{$ #1 $}}}}}
\newcommand{\MRbeta}[1]{\ensuremath{{}_{\text{o}}\beta_{\hspace*{0.2pt}\protect\raisebox{-1.0pt}{\scriptsize{$ #1 $}}}}}
\newcommand{\MRalphah}[1]{\ensuremath{{}_{\text{o}}\alpha^{\hspace*{0.5pt}\protect\raisebox{0.0pt}{\tiny{$(#1)$}}}}}
\newcommand{\MRbetah}[1]{{{}_{\text{o}}\beta^{\hspace*{0.5pt}\protect\raisebox{0.0pt}{\tiny{$(#1)$}}}}}
\newcommand{\MRalphahmn}[2]{\ensuremath{{}_{\text{o}}\alpha^{\hspace*{0.5pt}\protect\raisebox{0.0pt}{\tiny{$(#1)$}}}
_{\hspace*{0.2pt}\protect\raisebox{-1.0pt}{\scriptsize{$ #2 $}}}}}
\newcommand{\MRbetahmn}[2]{{{}_{\text{o}}\beta^{\hspace*{0.5pt}\protect\raisebox{0.0pt}{\tiny{$(#1)$}}}
_{\hspace*{0.2pt}\protect\raisebox{-1.0pt}{\scriptsize{$ #2 $}}}}}
\newcommand{\MRalphahsymb}[2]{{{}_{\text{o}}\alpha^{\hspace*{0.5pt}\protect\raisebox{0.0pt}{\tiny{$(#1)$}\hspace*{0.2pt}$#2$}}}}
\newcommand{\MRbetahsymb}[2]{{{}_{\text{o}}\beta^{\hspace*{0.5pt}\protect\raisebox{0.0pt}{\tiny{$(#1)$}\hspace*{0.2pt}$#2$}}}}

\newcommand{\MRA}[1]{{{}_{\text{o}}A_{\hspace*{0.2pt}\protect\raisebox{-1.0pt}{\scriptsize{$ #1 $}}}}}
\newcommand{\MRAh}[1]{{{}_{\text{o}}A^{\hspace*{0.5pt}\protect\raisebox{0.0pt}{\tiny{$(#1)$}}}}}
\newcommand{\MRAhmn}[2]{{{}_{\text{o}}A^{\hspace*{0.5pt}\protect\raisebox{0.0pt}{\tiny{$(#1)$}}}
_{\hspace*{0.2pt}\protect\raisebox{-1.0pt}{\scriptsize{$ #2 $}}}}}
\newcommand{\MRAstar}[1]{{{}_{\text{o}}A^{*}_{\hspace*{0.2pt}\protect\raisebox{-1.0pt}{\scriptsize{$ #1 $}}}}}

\newcommand{\BBBA}[1]{{{}_{\text{B}}\mathcal{A}_{\hspace*{0.2pt}\protect\raisebox{-1.0pt}{\scriptsize{$ #1 $}}}}}
\newcommand{\BBBalpha}[1]{{}_{\text{B}}\alpha_{\hspace*{0.2pt}\protect\raisebox{-1.0pt}{\scriptsize{$ #1 $}}}}
\newcommand{\BBBbeta}[1]{{{}_{\text{B}}\beta_{\hspace*{0.2pt}\protect\raisebox{-1.0pt}{\scriptsize{$ #1 $}}}}}
\newcommand{\BBBalphah}[1]{\ensuremath{{}_{\text{B}}\alpha^{\hspace*{0.5pt}\protect\raisebox{0.0pt}{\tiny{$(#1)$}}}}}
\newcommand{\BBBbetah}[1]{{{}_{\text{B}}\beta^{\hspace*{0.5pt}\protect\raisebox{0.0pt}{\tiny{$(#1)$}}}}}
\newcommand{\BBBalphahmn}[2]{\ensuremath{{}_{\text{B}}\alpha^{\hspace*{0.5pt}\protect\raisebox{0.0pt}{\tiny{$(#1)$}}}
_{\hspace*{0.2pt}\protect\raisebox{-1.0pt}{\scriptsize{$ #2 $}}}}}
\newcommand{\BBBbetahmn}[2]{{{}_{\text{B}}\beta^{\hspace*{0.5pt}\protect\raisebox{0.0pt}{\tiny{$(#1)$}}}
_{\hspace*{0.2pt}\protect\raisebox{-1.0pt}{\scriptsize{$ #2 $}}}}}
\newcommand{\BBBAhmn}[2]{{{}_{\text{B}}A^{\hspace*{0.5pt}\protect\raisebox{0.0pt}{\tiny{$(#1)$}}}
_{\hspace*{0.2pt}\protect\raisebox{-1.0pt}{\scriptsize{$ #2 $}}}}}

\newcommand{\ACalphahmn}[2]{\ensuremath{{}_{\alpha\text{c}}\alpha^{\hspace*{0.5pt}\protect\raisebox{0.0pt}{\tiny{$(#1)$}}}
_{\hspace*{0.2pt}\protect\raisebox{-1.0pt}{\scriptsize{$ #2 $}}}}}
\newcommand{\ACbetahmn}[2]{{{}_{\alpha\text{c}}\beta^{\hspace*{0.5pt}\protect\raisebox{0.0pt}{\tiny{$(#1)$}}}
_{\hspace*{0.2pt}\protect\raisebox{-1.0pt}{\scriptsize{$ #2 $}}}}}
\newcommand{\ACAhmn}[2]{\ensuremath{{}_{\alpha\text{c}}A^{\hspace*{0.5pt}\protect\raisebox{0.0pt}{\tiny{$(#1)$}}}
_{\hspace*{0.2pt}\protect\raisebox{-1.0pt}{\scriptsize{$ #2 $}}}}}

\newcommand{\GBh}[1]{\ensuremath{\mathcal{B}^{\hspace*{0.5pt}\protect\raisebox{0.0pt}{\tiny{$(#1)$}}}}}
\newcommand{\GBhmn}[2]{\ensuremath{\mathcal{B}^{\hspace*{0.5pt}\protect\raisebox{0.0pt}{\tiny{$(#1)$}}}
_{\hspace*{0.2pt}\protect\raisebox{-1.0pt}{\scriptsize{$ #2 $}}}}}
\newcommand{\alphamn}[1]{\ensuremath{\alpha_{\hspace*{0.2pt}\protect\raisebox{-1.0pt}{\scriptsize{$ #1 $}}}}}
\newcommand{\betamn}[1]{\ensuremath{\beta_{\hspace*{0.2pt}\protect\raisebox{-1.0pt}{\scriptsize{$ #1 $}}}}}
\newcommand{\alphamnstar}[1]{\ensuremath{\alpha^{\hspace*{0.5pt}\protect\raisebox{0.0pt}{$*$}}_{\hspace*{0.2pt}\protect\raisebox{-1.0pt}{\scriptsize{$ #1 $}}}}}
\newcommand{\betamnstar}[1]{\ensuremath{\beta^{\hspace*{0.5pt}\protect\raisebox{0.0pt}{$*$}}
_{\hspace*{0.2pt}\protect\raisebox{-1.0pt}{\scriptsize{$ #1 $}}}}}
\newcommand{\Amn}[1]{\ensuremath{A_{\hspace*{0.2pt}\protect\raisebox{-1.0pt}{\scriptsize{$ #1 $}}}}}
\newcommand{\Amnstar}[1]{\ensuremath{A^{*}_{\hspace*{0.2pt}\protect\raisebox{-1.0pt}{\scriptsize{$ #1 $}}}}}
\newcommand{\alphahmn}[2]{\ensuremath{\alpha^{\hspace*{0.5pt}\protect\raisebox{0.0pt}{\tiny{$(#1)$}}}_{\hspace*{0.2pt}\protect\raisebox{-1.0pt}{\scriptsize{$ #2 $}}}}}
\newcommand{\betahmn}[2]{\ensuremath{\beta^{\hspace*{0.5pt}\protect\raisebox{0.0pt}{\tiny{$(#1)$}}}_{\hspace*{0.2pt}\protect\raisebox{-1.0pt}{\scriptsize{$ #2 $}}}}}
\newcommand{\alphahmnstar}[2]{\ensuremath{\alpha^{\hspace*{0.5pt}\protect\raisebox{0.0pt}{\tiny{$(#1)$}$*$}}_{\hspace*{0.2pt}\protect\raisebox{-1.0pt}{\scriptsize{$ #2 $}}}}}
\newcommand{\betahmnstar}[2]{\ensuremath{\beta^{\hspace*{0.5pt}\protect\raisebox{0.0pt}{\tiny{$(#1)$}$*$}}_{\hspace*{0.2pt}\protect\raisebox{-1.0pt}{\scriptsize{$ #2 $}}}}}
\newcommand{\Ahmn}[2]{\ensuremath{A^{\hspace*{0.5pt}\protect\raisebox{0.0pt}{\tiny{$(#1)$}}}_{\hspace*{0.2pt}\protect\raisebox{-1.0pt}{\scriptsize{$ #2 $}}}}}
\newcommand{\Ahmnstar}[2]{\ensuremath{A^{\hspace*{0.5pt}\protect\raisebox{0.0pt}{\tiny{$(#1)$}$*$}}_{\hspace*{0.2pt}\protect\raisebox{-1.0pt}{\scriptsize{$ #2 $}}}}}

\newcommand{\Vmn}[1]{\ensuremath{V_{\hspace*{0.2pt}\protect\raisebox{-1.0pt}{\scriptsize{$ #1 $}}}}}
\newcommand{\Vmnstar}[1]{\ensuremath{V^{\hspace*{0.5pt}\protect\raisebox{0.0pt}{$*$}}_{\hspace*{0.2pt}\protect\raisebox{-1.0pt}{\scriptsize{$ #1 $}}}}}
\newcommand{\Vhmn}[2]{\ensuremath{V^{\hspace*{0.5pt}\protect\raisebox{0.0pt}{\tiny{$(#1)$}}}_{\hspace*{0.2pt}\protect\raisebox{-1.0pt}{\scriptsize{$ #2 $}}}}}
\newcommand{\Vhmnstar}[2]{\ensuremath{V^{\hspace*{0.5pt}\protect\raisebox{0.0pt}{\tiny{$(#1)$}$*$}}_{\hspace*{0.2pt}\protect\raisebox{-1.0pt}{\scriptsize{$ #2 $}}}}}
\newcommand{\fVmn}[1]{\ensuremath{\mathcal{V}_{\hspace*{0.2pt}\protect\raisebox{-1.0pt}{\scriptsize{$ #1 $}}}}}

\newcommand{\fVhmn}[2]{\ensuremath{\mathcal{V}^{\hspace*{0.5pt}\protect\raisebox{0.0pt}{\tiny{$(#1)$}}}_{\hspace*{0.2pt}\protect\raisebox{-1.0pt}{\scriptsize{$ #2 $}}}}}
\newcommand{\fVhmnstar}[2]{\ensuremath{\mathcal{V}^{\hspace*{0.5pt}\protect\raisebox{0.0pt}{\tiny{$(#1)$}$*$}}_{\hspace*{0.2pt}\protect\raisebox{-1.0pt}{\scriptsize{$ #2 $}}}}}

\newcommand{\Salphamn}[1]{\ensuremath{{}_{\text{s}}\alpha_{\hspace*{0.2pt}\protect\raisebox{-1.0pt}{\scriptsize{$ #1 $}}}}}
\newcommand{\Sbetamn}[1]{{{}_{\text{s}}\beta_{\hspace*{0.2pt}\protect\raisebox{-1.0pt}{\scriptsize{$ #1 $}}}}}
\newcommand{\SAmn}[1]{\ensuremath{{}_{\text{s}}A_{\hspace*{0.2pt}\protect\raisebox{-1.0pt}{\scriptsize{$ #1 $}}}}}

\newcommand{\Sbetahmn}[2]{{{}_{\text{s}}\beta^{\hspace*{0.5pt}\protect\raisebox{0.0pt}{\tiny{$(#1)$}}}
_{\hspace*{0.2pt}\protect\raisebox{-1.0pt}{\scriptsize{$ #2 $}}}}}

\newcommand{\Mmn}[1]{\ensuremath{\mathcal{M}_{\hspace*{0.2pt}\protect\raisebox{-1.0pt}{\scriptsize{$ #1 $}}}}}
\newcommand{\Mmntrans}[1]{\ensuremath{\mathcal{M}^{T}_{\hspace*{0.2pt}\protect\raisebox{-1.0pt}{\scriptsize{$ #1 $}}}}}
\newcommand{\Mhmn}[2]{\ensuremath{\mathcal{M}^{\hspace*{0.5pt}\protect\raisebox{0.0pt}{\tiny{$(#1)$}}}_{\hspace*{0.2pt}\protect\raisebox{-1.0pt}{\scriptsize{$ #2 $}}}}}

\newcommand{\Gh}[1]{G^{\hspace*{0.5pt}\protect\raisebox{0.0pt}{\tiny{$(#1)$}}}}
\newcommand{\Ghstar}[1]{G^{\hspace*{0.5pt}\protect\raisebox{0.0pt}{\tiny{$(#1)$}$*$}}}
\newcommand{\Ghn}[2]{G^{\hspace*{0.5pt}\protect\raisebox{0.0pt}{\tiny{$(#1)$}}}_{#2}}
\newcommand{\Ghnstar}[2]{G^{\hspace*{0.5pt}\protect\raisebox{0.0pt}{\tiny{$(#1)$}$*$}}_{#2}}
\newcommand{\vLR}[1]{\ensuremath{\lambda_{\hspace*{-0.5pt}\protect\raisebox{-0.5pt}{\tiny{$#1$}}}}}

\newcommand{\rhoh}[1]{\ensuremath{\rho^{\hspace*{0.5pt}\protect\raisebox{0.0pt}{\tiny{$(#1)$}}}}}
\newcommand{\rhohsymb}[2]{\ensuremath{\rho^{\hspace*{0.5pt}\protect\raisebox{0.0pt}{\tiny{$(#1)$}$#2$}}}}
\newcommand{\lambdah}[1]{\ensuremath{\lambda^{\hspace*{0.0pt}\protect\raisebox{0.0pt}{\tiny{$(#1)$}}}}}
\newcommand{\Uh}[1]{\ensuremath{U^{\hspace*{0.5pt}\protect\raisebox{0.0pt}{\tiny{$(#1)$}}}}}
\newcommand{\vhsymb}[2]{\ensuremath{\underline{v}^{\hspace*{0.5pt}\protect\raisebox{0.0pt}{\tiny{$(#1)$} $#2$ }}}}
\newcommand{\frhovackkprkprpr}{\ensuremath{{}_{\mathrm{vac}}\varrho_{\raisebox{-1pt}{\tiny{$\kappa\kappa\pr\hspace*{-1pt}\kappa\prpr$}}}}}

\newcommand{\Negh}[1]{\ensuremath{\mathcal{N}^{\hspace*{0.5pt}\protect\raisebox{0.0pt}{\tiny{$(#1)$}}}}}
\newcommand{\Fidh}[1]{\ensuremath{\mathcal{F}^{\hspace*{0.5pt}\protect\raisebox{0.0pt}{\tiny{$(#1)$}}}}}
\newcommand{\Fidopth}[1]{\ensuremath{\mathcal{F}_{\mathrm{opt}}^{\hspace*{0.5pt}\protect\raisebox{0.0pt}{\tiny{$(#1)$}}}}}

\DeclareMathOperator{\artanh}{artanh}

\DeclareMathOperator{\diag}{diag}
\DeclareMathOperator{\spectr}{spectr}
\newtheorem{theorem}{Theorem}[chapter]

\newtheorem{defi}{Definition}[chapter]
\usepackage{enumerate}
\begin{document}
	
        \title{
            \LARGE{\textbf{CAVITY MODE ENTANGLEMENT IN \\
            RELATIVISTIC QUANTUM INFORMATION}} \\[1.5cm]
		    \LARGE{NICOLAI FRIIS, Mag.~rer.~nat.} \\[1.3cm]
		    \Large{Thesis submitted to the University of Nottingham\\
            for the degree of Doctor of Philosophy} \\ \vspace{1.3cm}
            \Large{DECEMBER 2013}
            }
	    \pdfbookmark[0]{Titlepage}{title}
        \date{}

	    \maketitle
	
    \newpage
    \setcounter{page}{2}\ \\
 	\newpage
    \pdfbookmark[0]{Dedication}{dedication}
    \vspace*{8cm}
    \setcounter{page}{3}
	\begin{center}

%
\begin{tabbing}\hspace*{2cm}\=\hspace*{1.5cm}\=\kill
\>\emph{``To those who do not know mathematics it is difficult to get across a real feeling as}\\
\>\emph{to the beauty, the deepest beauty, of nature ... If you want to learn about nature,}\\
\>\emph{to appreciate nature, it is necessary to understand the language that she speaks in."}\\[2mm]

\>\>\footnotesize{(R.~Feynman, The Character of Physical Law (1965) Ch.~2)}\\[15mm]

\>\emph{``Quantum physics means anything can happen at any time for no reason."}\\[2mm]

\>\>\footnotesize{(H.~J.~Farnsworth)}
\end{tabbing}

 	\end{center}

 	\newpage
    \ \\

    \setcounter{page}{4}
    \newpage
    \pdfbookmark[0]{Abstract}{abstract}
	\chapter*{Abstract}

A central aim of the field of relativistic quantum information (RQI) is the investigation of quantum information tasks and resources taking into account the relativistic aspects of nature. More precisely, it is of fundamental interest to understand how the storage, manipulation, and transmission of information utilizing quantum systems are influenced by the fact that these processes take place in a relativistic spacetime. In particular, many studies in RQI have been focused on the effects of non-uniform motion on entanglement, the main resource of quantum information protocols. Early investigations in this direction were performed in highly idealized settings that prompted questions as to the practical accessibility of these results. To overcome these limitations it is necessary to consider quantum systems that are in principle accessible to localized observers. In this thesis we present such a model, the rigid relativistic cavity, and its extensions, focusing on the effects of motion on entanglement and applications such as quantum teleportation.\\

We study cavities in $(1+1)$ dimensions undergoing non-uniform motion, consisting of segments of uniform acceleration and inertial motion of arbitrary duration that allow the involved velocities to become relativistic. The transitions between segments of different accelerations can be sharp or smooth and higher dimensions can be incorporated. The primary focus lies in the Bogoliubov transformations of the quantum fields, real scalar fields or Dirac fields, confined to the cavities. The Bogoliubov transformations change the particle content and the occupation of the energy levels of the cavity. We show how these effects generate entanglement between the modes of the quantum fields inside a single cavity for various initial states. The entanglement between several cavities, on the other hand, is degraded by the non-uniform motion, influencing the fidelity of tasks such as teleportation. An extensive analysis of both situations and a setup for a possible simulation of these effects in a table-top experiment are presented.

    \newpage
    \pdfbookmark[0]{Acknowledgements}{acknowledgements}
 	\chapter*{Acknowledgements}

My gratitude goes to all the fantastic people that have supported me in the pursuit of my goals, specifically throughout my PhD studies. First of all I would like to thank my family and all those of my friends, which I could not have been there for as much as I would have liked, for their continued encouragement and friendship. In particular, I would like to thank \emph{Verena Hofst\"atter} for her patience and ethical counsel, and a special thanks to those adventurers that came all the way to Nottingham to visit me.\\

I would like to express my thanks to the researchers that have guided me with their experience: \emph{Ivette Fuentes}, for her trust, friendship, and openness, and her joy in doing physics; \emph{Gerardo Adesso}, for the nonchalant honesty of his advice; \emph{Jorma Louko}, for his love to detail and complicated mathematical functions; and to \emph{Reinhold A. Bertlmann}, for his spirit, verve, and his socks. During my time in Nottingham I have been surrounded by people that I count as friends as much as colleagues: I want to thank \emph{Antony R.~Lee} for enriching my vocabulary in what he claims to be English; \emph{David Edward Bruschi} for learning to expect the unexpected; and \emph{Sara Tavares} for bothering to organize a conference with us. In addition I would like to thank the entire team in Nottingham, present or past, which I feel proud to have been a part of: \emph{Mehdi Ahmadi}, \emph{Valentina Baccetti}, \emph{Luis C.~Barbado}, \emph{Jason Doukas}, \emph{Andrzej Dragan}, \emph{Karishma Hathlia}, \emph{Giannis Kogias}, \emph{Bartosz Regula}, \emph{Carlos Sab{\'i}n}, \emph{Kevin Truong}, \emph{Luke Westwood}, and \emph{Angela White}.\\

A special thanks to \emph{Marcus Huber}, for his hospitality, friendship, and unconventionality. I am also specifically grateful to \emph{Markus Arndt}, \emph{Jan Bouda}, \emph{Hans J.~Briegel}, \emph{Jacob Dunningham}, \emph{Davide Girolami}, \emph{Lucia Hackerm\"uller}, \emph{Eli Hawkins}, \emph{Beatrix C.~Hiesmayr}, \emph{Helmuth H\"uffel}, \emph{Juan Le\'{o}n}, \emph{Robert B.~Mann}, \emph{Nicolas Menicucci}, \emph{Daniel K.~L.~Oi}, \emph{Sammy Ragy}, \emph{Timothy C.~Ralph}, \emph{Mohsen Razavi}, \emph{Paul Skrzypczyk}, \emph{Vlatko Vedral}, \emph{Silke Weinfurtner}, and \emph{Andreas Winter} for their kind invitations, hospitality, and financial support during academic visits, and, in particular, for giving me the opportunity to present my work at various seminars, workshops and conferences.\\

I am further grateful to all my co-authors, those named above as well as \emph{Andreas Gabriel}, \emph{G\"oran Johansson}, \emph{Philipp K\"ohler}, \emph{Eduardo Mart\'{i}n-Mart\'{i}nez}, \emph{Enrique Solano}, and \emph{Christoph Spengler} for all the engaging discussions, computations, correspondence, and their time and effort during the publication process.\\

Finally, my sincere thanks go to the funding agencies who have made many of the endeavors of the past years possible: the EPSRC (CAF Grant No.\ EP/G00496X/2 to Ivette Fuentes), the Royal Society, Universitas 21 and the University of Nottingham Graduate School, the London Mathematical Society, the Institute of Mathematics and its Applications small grant scheme, the C$\Lambda$P research network, the research support fund of the Edinburgh Mathematical Society, the Science and Technology Facilities Council, the Institute of Physics (IOP) Mathematical and Theoretical Physics group, and the IOP Gravitational Physics group.

%
%
    \newpage
    \pdfbookmark[0]{Contents}{contents}
	\tableofcontents

	\newpage	
    \setcounter{page}{1}
	\pagenumbering{arabic}
 	

    \chapter*{Introduction}\label{chapter:Introduction}
\addcontentsline{toc}{chapter}{Introduction}

\section*{Author's Declaration}\label{sec:author's declaration}
\addcontentsline{toc}{section}{Author's Declaration}

I hereby declare that this thesis was produced by myself and is based on the results that I obtained, together with my collaborators, during the time of my PhD studies at the University of Nottingham. A summary of my publications until this date is listed below, but all of them can also be found in the reference section. In this thesis I present the results of my publications (\ref{Paper:FriisLeeBruschiLouko2012})-(\ref{Paper:FriisLeeLouko2013}) (Refs.~\cite{FriisLeeBruschiLouko2012,
FriisBruschiLoukoFuentes2012,FriisHuberFuentesBruschi2012,FriisFuentes2013,FriisLeeBruschi2013,
FriisLeeTruongSabinSolanoJohanssonFuentes2013,BruschiFriisFuentesWeinfurtner2013,FriisLeeLouko2013}). I am the lead author in all of these publications apart from~\cite[(\ref{Paper:BruschiFriisFuentesWeinfurtner2013})]{BruschiFriisFuentesWeinfurtner2013}, where authors are listed alphabetically. Accordingly, I have made major contributions to all of these articles in terms of conceptual development, computations, proofs, composition, writing, and illustrations, apart from Ref.~\cite[(\ref{Paper:FriisLeeLouko2013})]{FriisLeeLouko2013}, where my contribution is mostly limited to the sections on the Dirac spinor and authors are also listed alphabetically. In addition, the thesis outlines closely related results obtained by my PhD supervisor Ivette Fuentes, and my colleagues David Edward Bruschi, Andrzej Dragan, Daniele Faccio, Antony~R. Lee, and Jorma Louko in Refs.~\cite{BruschiFuentesLouko2012,BruschiLoukoFaccioFuentes2013,BruschiDraganLeeFuentesLouko2013,
BruschiLoukoFaccio2013} during the time of my PhD.\\

\vspace*{-2mm}
A detailed introduction into the topic of the thesis and an outline of its contents are given after the list of publications.

\section*{List of Publications}\label{sec:list of publications}
\addcontentsline{toc}{section}{List of Publications}

\begin{Large}Journal Publications\end{Large}\\

\begin{enumerate}[{(i)\ }]
\vspace*{-5.7mm}
\item{N.~Friis, R.~A.~Bertlmann, M.~Huber, and B.~C.~Hiesmayr,\ \\[0.5mm]
\emph{Relativistic entanglement of two massive particles},\
(Ref.~\cite{FriisBertlmannHuberHiesmayr2010})\\[0.5mm]
\href{http://dx.doi.org/10.1103/PhysRevA.81.042114}{Phys.\ Rev.\ A \textbf{81}, 
042114 (2010)}\ [\href{http://arxiv.org/abs/0912.4863}{arXiv:0912.4863} [quant-ph]].
\label{Paper:FriisBertlmannHuberHiesmayr2010}}\\
\vspace*{-5.7mm}
\item{M.~Huber, N.~Friis, A.~Gabriel, C.~Spengler, and B.~C.~Hiesmayr,\ \\[0.5mm]
\emph{Lorentz invariance of entanglement classes in multipartite systems},\
(Ref.~\cite{HuberFriisGabrielSpenglerHiesmayr2011})\\[0.5mm]
\href{http://dx.doi.org/10.1209/0295-5075/95/20002}{Europhys.\ Lett.\ \textbf{95}, 
20002 (2011)}\ [\href{http://arxiv.org/abs/1011.3374}{arXiv:1011.3374} [quant-ph]].
\label{Paper:HuberFriisGabrielSpenglerHiesmayr2011}}\\
\vspace*{-5.5mm}
\item{N.~Friis, P.~K{\"o}hler, E.~Mart{\'i}n-Mart{\'i}nez, and R.~A.~Bertlmann,\ \\[0.5mm]
\emph{Residual entanglement of accelerated fermions is not nonlocal},\
(Ref.~\cite{FriisKoehlerMartinMartinezBertlmann2011})\\[0.5mm]
\href{http://dx.doi.org/10.1103/PhysRevA.84.062111}{Phys. Rev. A \textbf{84}, 
062111 (2011)}\ [\href{http://arxiv.org/abs/1107.3235}{arXiv:1107.3235} [quant-ph]].
\label{Paper:FriisKoehlerMartinMartinezBertlmann2011}}\\
\vspace*{-5.5mm}
\item{N.~Friis, A.~R.~Lee, D.~E.~Bruschi, and J.~Louko,\ \\[0.5mm]
\emph{Kinematic entanglement degradation of fermionic cavity modes},\
(Ref.~\cite{FriisLeeBruschiLouko2012})\\[0.5mm]
\href{http://dx.doi.org/10.1103/PhysRevD.85.025012}{Phys.\ Rev.\ D \textbf{85}, 
025012 (2012)}\ [\href{http://arxiv.org/abs/1110.6756}{arXiv:1110.6756} [quant-ph]].
\label{Paper:FriisLeeBruschiLouko2012}}\\
\vspace*{-5.5mm}
\item{N.~Friis, D.~E.~Bruschi, J.~Louko, and I.~Fuentes,\
(Ref.~\cite{FriisBruschiLoukoFuentes2012})\\[0.5mm]
\emph{Motion generates entanglement},\ \\[0.5mm]
\href{http://dx.doi.org/10.1103/PhysRevD.85.081701}{Phys.\ Rev.\ D\ \textbf{85}, 
081701(R) (2012)}\ [\href{http://arxiv.org/abs/1201.0549}{arXiv:1201.0549} [quant-ph]].
\label{Paper:FriisBruschiLoukoFuentes2012}}\\
\vspace*{-5.5mm}
\item{N.~Friis, M.~Huber, I.~Fuentes, and D.~E.~Bruschi,\ \\[0.5mm]
\emph{Quantum gates and multipartite entanglement resonances realized by non-uniform \mbox{cavity}
motion},\ (Ref.~\cite{FriisHuberFuentesBruschi2012})\\[0.5mm]
\href{http://dx.doi.org/10.1103/PhysRevD.86.105003}{Phys.\ Rev.\ D \textbf{86}, 
105003 (2012)}\ [\href{http://arxiv.org/abs/1207.1827}{arXiv:1207.1827} [quant-ph]].
\label{Paper:FriisHuberFuentesBruschi2012}}\\
\vspace*{-5.5mm}
\item{N.~Friis and I.~Fuentes,\ \\[0.5mm]
\emph{Entanglement generation in relativistic quantum fields},\
(Ref.~\cite{FriisFuentes2013})\\[0.5mm]
\href{http://dx.doi.org/10.1080/09500340.2012.712725}{J.\ Mod.\ Opt.\ \textbf{60}, 
22 
(2013)}\ [\href{http://arxiv.org/abs/1204.0617}{arXiv:1204.0617} [quant-ph]] \\[0.7mm]
Invited contribution to the Special Issue: Physics in Quantum Electronics.
\label{Paper:FriisFuentes2013}}\\
\vspace*{-5.5mm}
\item{N.~Friis, A.~R.~Lee, and D.~E.~Bruschi\ \\[0.5mm]
\emph{Fermionic mode entanglement in quantum information},\
(Ref.~\cite{FriisLeeBruschi2013})\\[0.5mm]
\href{http://dx.doi.org/10.1103/PhysRevA.87.022338}{Phys.\ Rev.\ A\ \textbf{87}, 
022338 (2013)}\ [\href{http://arxiv.org/abs/1211.7217}{arXiv:1211.7217} [quant-ph]].
\label{Paper:FriisLeeBruschi2013}}\\
\vspace*{-5.5mm}
\item{N.~Friis, A.~R.~Lee, K.~Truong, C.~Sab{\'i}n, E.~Solano, G.~Johansson, and I.~Fuentes,\ \\[0.5mm]
\emph{Relativistic Quantum Teleportation with Superconducting Circuits},\
(Ref.~\cite{FriisLeeTruongSabinSolanoJohanssonFuentes2013})\\[0.5mm]
\href{http://dx.doi.org/10.1103/PhysRevLett.110.113602}{Phys.\ Rev.\ Lett.\ \textbf{110}, 
113602 (2013)}\ [\href{http://arxiv.org/abs/1211.5563}{arXiv:1211.5563} [quant-ph]].
\label{Paper:FriisLeeTruongSabinSolanoJohanssonFuentes2013}}\\
\vspace*{-5.5mm}
\item{N.~Friis, A.~R.~Lee, and J.~Louko,\ \\[0.5mm]
\emph{Scalar, spinor and photon fields under relativistic cavity motion},\
(Ref.~\cite{FriisLeeLouko2013})\\[0.5mm]
\href{http://dx.doi.org/10.1103/PhysRevD.88.064028}{Phys.\ Rev.\ D\ \textbf{88}, 
064028 (2013)}\ [\href{http://arxiv.org/abs/1307.1631}{arXiv:1307.1631} [quant-ph]].
\label{Paper:FriisLeeLouko2013}}\\
\vspace*{-5.5mm}
\item{D.~E.~Bruschi, N.~Friis, I.~Fuentes, and S.~Weinfurtner,\ \\[0.5mm]
\emph{On the robustness of entanglement in analogue gravity systems},\ (Ref.~\cite{BruschiFriisFuentesWeinfurtner2013})\\[0.5mm]
\href{http://dx.doi.org/10.1088/1367-2630/15/11/113016}{New\ J.\ Phys.\ \textbf{15}, 113016 (2013)}\
[\href{http://arxiv.org/abs/arXiv:1305.3867}{arXiv:1305.3867} [quant-ph]].
\label{Paper:BruschiFriisFuentesWeinfurtner2013}}\\
\end{enumerate}

\hspace*{-7mm}\begin{Large}Theses\end{Large}\\

\begin{enumerate}[{(i)\ }]\addtocounter{enumi}{11}
\vspace*{-5.5mm}
\item{N.~Friis,\ \\[0.5mm]
\emph{Relativistic Effects in Quantum Entanglement},\ (Ref.~\cite{Friis:DiplomaThesis2010})\\[0.5mm]
\href{http://othes.univie.ac.at/8370/}{Diploma thesis,\ University of Vienna, 2010}.
\label{Thesis:Friis2010}}\\
\end{enumerate}

\section*{Aims of Relativistic Quantum Information}\label{sec:aims of RQI}
\addcontentsline{toc}{section}{Aims of Relativistic Quantum Information}

The topic of this thesis is part of the research area called \emph{Relativistic Quantum Information} (RQI). This young and thriving field aims to investigate questions 
that lie in the overlap of quantum information theory, quantum field theory, quantum optics and special as well as general relativity. The motivations and particular questions considered for such studies are numerous, but a central theme originating from the first papers dedicated to RQI (see, e.g., Refs.~\cite{Czachor1997,PeresScudoTerno2002,GingrichAdami2002,PeresTerno2004,Fuentes-SchullerMann2005}) is the observer dependence of entanglement.\\

In a series of works~\cite{GingrichAdami2002,AlsingMilburn2002,TerashimaUeda2003a,TerashimaUeda2003b,AhnLeeHwang2003,AhnLeeMoonHwang2003,JordanShajiSudarshan2007,
FriisBertlmannHuberHiesmayr2010,Friis:DiplomaThesis2010,HuberFriisGabrielSpenglerHiesmayr2011,CastroRuizNahmadAchar2011,PalgeDunningham2012,
CastroRuizNahmadAchar2012,DebarbaVianna2012,Palge:PhDThesis2013} this relativity of entanglement was first investigated for inertial observers. The preliminary conclusion drawn from these papers is the following: The entanglement between internal degrees of freedom, such as spin and momentum, in systems with a fixed number of relativistic particles depends on the chosen inertial frame. However, it remains unclear, whether this mathematical observation can be tested in any experiment, since spin measurements are not independent of the particle momenta~\cite{SaldanhaVedral2012a,SaldanhaVedral2012b}. In particular, investigations concerning the choice of relativistic spin operator are still a source of scientific debate~\cite{PalmerTakahashiWestman2013,Cabanetal2012a,Cabanetal2012b,SaldanhaVedral2013,
BaukeAhrensKeitelGrobe2013,Cabanetal2013}.\\

In parallel to the studies of Lorentz symmetry of entanglement a second branch of RQI was developed for the investigation of the effects of non-uniform motion~\cite{Fuentes-SchullerMann2005,AlsingFuentes-SchullerMannTessier2006,AdessoFuentes-SchullerEricsson2007,
Martin-MartinezFuentes2011,FriisKoehlerMartinMartinezBertlmann2011,BruschiDraganFuentesLouko2012} and spacetime curvature~
\cite{BallFuentes-SchullerSchuller2006,FuentesMannMartin-MartinezMoradi2010,Martin-MartinezGarayLeon2010a,Martin-MartinezGarayLeon2010b,
Martin-MartinezFuentes2011,Martin-MartinezGarayLeon2012} on the resources and protocols of quantum information processing~(see Ref.~\cite{AlsingFuentes2012} for a recent review). Especially the studies of non-inertial motion established a close connection between RQI and effects of quantum field theory on curved spacetimes~(see Ref.~\cite{BirrellDavies:QFbook} for an introduction), such as the \emph{Unruh effect} and \emph{Hawking effect}~\cite{BruschiLoukoMartin-MartinezDraganFuentes2010,AspachsAdessoFuentes2010}. In this context the effects of the eternal, uniform acceleration of idealized point-like observers on the entanglement between global modes was the paradigm situation of interest. While such a highly simplified toy model served well as a basis to analyze qualitative features of the effects of non-uniform motion on entanglement, it is clear that it cannot be considered to be consistent with any practical situation, see, for instance, the discussion in Ref.~\cite{DownesFuentesRalph2011}. For instance, it was left open if and how the chosen initial states could be prepared or measured for modes with support in the entire spacetime. Additionally, it seems overly restrictive to assume that accelerations need to be kept constant eternally to produce any effects. In spite of their problems the initial toy models were helpful to understand some general features and requirements of relativistic formulations of quantum information tasks.\\

\vspace*{-5mm}
To overcome the problems of the early models several systems were proposed that allow for localized preparation and measurements while maintaining high flexibility regarding the choice of trajectories in spacetime. One of these approaches is the so called \emph{Unruh--DeWitt detector}~\cite{BirrellDavies:QFbook}. This detector model describes a localized quantum system, such as a harmonic oscillator or two-level system, coupled to the quantum field along a classical trajectory. We shall not discuss the intricacies of this approach here but refer the interested reader to the relevant literature~(see, e.g., Ref.~\cite{HuLinLouko2012,LeeFuentes2012,BruschiLeeFuentes2013}). The second major attempt to construct a theoretical description of a quantum system that can serve for the storage, processing, and transmission of quantum information in a relativistic spacetime is to confine the quantum field inside a \emph{cavity}. This simple, yet rich theoretical model has many advantages, one of which is its experimental accessibility. For instance, cavities are well controlled systems in the context of quantum optics, as recognized by the \emph{Nobel Prize in Physics 2012}~(see, e.g., Ref.~\cite{RaimondBruneHaroche2001}), and they have been extensively studied in connection with the \emph{dynamical Casimir effect}~\cite{AndreataDodonov2005,Dodonov2010,WilsonDynCasNature2012}\index{dynamical Casimir effect}. Consequently, cavities were naturally considered as objects for the rigorous relativistic study of quantum information processing in RQI~\cite{DownesFuentesRalph2011,BruschiFuentesLouko2012}. Apart from these two systems also other options 
were proposed, including wave packets~\cite{DownesRalphWalk2013} and covariant formulations of single particles in curved spacetimes~\cite{PalmerTakahashiWestman2012}. In this thesis we shall follow the second path to describe the modes of quantum fields that are confined to cavities in relativistic motion by appropriate boundary conditions. An outline of the discussion is provided below.\\

Following on from the recent successes in identifying appropriate systems for RQI the field has now entered into a new phase \textemdash the connection with experiments and applications. The advances of the research on the mathematical foundations of RQI have been accompanied by breakthroughs in cutting edge experiments, such as the observation of the dynamical Casimir effect~\cite{WilsonDynCasNature2012}, or the teleportation over distances where effects of general relativity may become non-negligible~\cite{Ma-Zeilinger-tele2012}. It is thus now possible, and, moreover, feasible for theoretical research in RQI to be tested in Earth-based laboratories~\cite{FriisLeeTruongSabinSolanoJohanssonFuentes2013}, which will help to provide insight into effects of space-based experiments, for instance quantum communication between satellites~\cite{RideoutEtal2012}.\\

\vspace{-1.5mm}
Additionally, the well-developed tools of RQI are now ready for state-of-the-art applications that open up entirely new directions of research, e.g., \emph{relativistic quantum metrology} \textemdash the study of high-precision parameter estimation using relativistic settings~\cite{AspachsAdessoFuentes2010,SabinWhiteHackermuellerFuentes2013,Hosler:PhDThesis2013,HoslerKok2013,DownesMilburnCaves2012,AhmadiBruschiSabinAdessoFuentes2013}. Another area that connects to RQI is \emph{analogue gravity}~(see Ref.~\cite{BarceloLiberatiVisser2005} for a review, and Refs.~\cite{RecatiPavloffCarusotto2009,Belgiorno-Faccio2010,
WeinfurtnerTedfordPenriceUnruhLawrence2011,JaskulaPartridgeBonneauRuaudelBoironWestbrook2012,Rubino-Faccio2012,
WeinfurtnerTedfordPenriceUnruhLawrence2013} for a selection of recent advances), which aims to simulate quantum effects in curved spacetimes in compact, laboratory-based experimental setups. The techniques of RQI are here able to provide useful criteria for the presence of entanglement to identify crucial signatures of the quantumness of the expected effects.

\section*{Outline of the Thesis}\label{sec:outline}
\addcontentsline{toc}{section}{Outline of the Thesis}

This thesis aims at presenting a thorough introduction into the description of cavities as systems for quantum information processing in RQI, and the phenomena originating from this treatment. In particular, we investigate the effects of the non-uniform motion of rigid cavities in Minkowski spacetime on the entanglement between the field modes within the cavity. Several scenarios for the creation and degradation of entanglement, including applications to practical tasks such as a quantum teleportation are discussed. Due to the interdisciplinary nature of RQI, the thesis relies on background knowledge in the fields of quantum information theory, relativity, quantum field theory and quantum optics. For this reason the thesis is partitioned into two parts.\\

\vspace{-1.5mm}
Part~\ref{partI:Elements of Relativistic Quantum Information} provides the basic concepts that are needed from each of the above-mentioned fields. Chapter~\ref{chapter:QI} introduces fundamental terminology and definitions from \emph{quantum information theory}, focusing on entanglement theory and some simple applications. 
We then direct our attention to the quantization of \emph{quantum fields in relativistic spacetimes} in Chapter~\ref{chapter:Quantum Fields in Flat and Curved Spacetimes}, where we also discuss \emph{Bogoliubov transformations}, a crucial concept for this thesis. Readers familiar with the topics covered in either of the first two chapters may skip the corresponding introductory chapters, but should be aware that most of the notation and terminology of the thesis are established there.\\

With the tools of Chapters~\ref{chapter:QI} and~\ref{chapter:Quantum Fields in Flat and Curved Spacetimes} at hand we are in a position to discuss general features of \emph{quantum information theory for quantum fields} in Chapter~\ref{chapter:Entanglement in Relativistic Quantum Fields}. First we discuss the description of bosonic fields in phase space in the so-called \emph{covariance matrix formalism}. Having established the relevant techniques we further discuss the role of entanglement generation in bosonic quantum fields and we introduce criteria for entanglement resonances that are generalizing the results of \emph{Bruschi et al.} from Ref.~\cite{BruschiDraganLeeFuentesLouko2013}. 
Finally, in the last section of Part~\ref{partI:Elements of Relativistic Quantum Information} we present the intricacies of quantum information processing in fermionic Fock spaces, based on material published in Ref.~\cite[\eqref{Paper:FriisLeeBruschi2013}]{FriisLeeBruschi2013}.\\

Part~\ref{partII:Shaking Entanglement}, titled \emph{shaking entanglement}, is entirely dedicated to the rigid cavity model that was, in the context of RQI, first developed for bosonic fields by \emph{D.~Bruschi, I.~Fuentes, and J.~Louko}~\cite{BruschiFuentesLouko2012} and subsequently generalized to bosonic Gaussian states~\cite[\eqref{Paper:FriisFuentes2013}]{FriisFuentes2013} and fermionic fields~\cite[\eqref{Paper:FriisLeeBruschiLouko2012}]{FriisLeeBruschiLouko2012} by myself and collaborators. The aim is to give a pedagogical introduction to the model of perfect, rigid cavities in motion and the corresponding effects for quantum entanglement.\\

Part~\ref{partII:Shaking Entanglement} is organized as follows. In Chapter~\ref{Chapter 4 Constructing Non Uniformly Moving Cavities} the basic model is introduced, i.e., we discuss how bosonic and fermionic quantum fields are confined within \emph{rigid cavities in non-uniform motion} and how generic trajectories can be constructed. This chapter also covers the results for smoothly changing accelerations from Ref.~\cite{BruschiLoukoFaccioFuentes2013} by my collaborators. We continue in Chapter~\ref{Chapter 5 State Transformation by Non-Uniform Motion} with the \emph{state transformation} of initial Fock states and Gaussian states.\\

Having set the stage we then proceed to Chapter~\ref{Chapter 6 Motion Generates Entanglement}, where we discuss the \emph{entanglement generation} phenomena within the non-uniformly moving cavities. The bipartite case, based on my publications~\cite[\eqref{Paper:FriisBruschiLoukoFuentes2012}]{FriisBruschiLoukoFuentes2012}, and~\cite[\eqref{Paper:FriisFuentes2013}]{FriisFuentes2013}, as well as Ref.~\cite{BruschiDraganLeeFuentesLouko2013} by my collaborators, is discussed in detail in Sections~\ref{sec:Entanglement Generation in Bosonic Fock States}-\ref{sec:Entanglement Generation in Fermionic States}. The multipartite case, based on~\cite[\eqref{Paper:FriisHuberFuentesBruschi2012}]{FriisHuberFuentesBruschi2012}, is finally presented in Section~\ref{sec:Generation of Genuine Multipartite Entanglement}.\\

Finally, in Chapter~\ref{Chapter 7 Degradation of Entanglement between Moving Cavities} we analyze \emph{entanglement degradation} effects between several cavities in motion, based on the results of Ref.~\cite{BruschiFuentesLouko2012} and my publications~\cite[\eqref{Paper:FriisLeeBruschiLouko2012}]{FriisLeeBruschiLouko2012} and~\cite[\eqref{Paper:FriisLeeTruongSabinSolanoJohanssonFuentes2013}]{FriisLeeTruongSabinSolanoJohanssonFuentes2013}. In particular, we apply the formalism to study the influence on the continuous-variable teleportation protocol between cavities in motion. A scheme to test these predictions in a laboratory-based experiment using superconducting circuits is briefly discussed, before we present the conclusions. 

    \part[Elements of Relativistic Quantum Information]
    {Elements of Relativistic Quantum Information\\[10mm]
    \Large{Background, Tools \& Methods from \emph{Quantum Information},\\
    \vspace*{-0.3cm}
    \emph{Quantum Optics} and \emph{Quantum Field Theory}}}
    \label{partI:Elements of Relativistic Quantum Information}
    \newpage\ \\
	
    \chapter{Basic Concepts in Quantum Information}\label{chapter:QI}

Naturally, any relativistic study of quantum information procedures requires a formal understanding of the quantities of interest in standard quantum information theory. We will give a brief introduction into the main concepts of quantum information theory, focusing on entanglement theory. In particular, we shall restrict our attention to the concepts needed for the purpose of this thesis and refer the reader to the literature for topics that lie beyond the scope of this review chapter. For a detailed introduction to quantum information theory consult, e.g., Ref.~\cite{NielsenChuang2000}.

\section{Pure \& Mixed Quantum States}\label{sec:pure and mixed quantum states}

The essential ingredients for quantum information processing lie in the description of quantum states, and operations (state preparation, manipulation, measurements, etc.) carried out on these states. The notion of \emph{``state"} encompasses our best knowledge of the physical system and can take on various mathematical descriptions.

\subsection{Pure States}\label{subsec:pure states}

Let us begin with the idealized notion of a \emph{pure state}\index{state!pure}, where the maximal amount of information about the physical system is available. In other words, a quantum system in a pure state is perfectly controlled and it is described by a state vector~$\psi$ in a \emph{Hilbert space}\index{Hilbert space}~$\mathcal{H}$.
\begin{defi}\label{def:hilbert space}\end{defi}
    \vspace*{-1.05cm}
    \begin{tabbing} \hspace*{3.2cm}\=\hspace*{1.5cm}\=\kill
        \> \textit{A \emph{Hilbert Space} $\mathcal{H}$ is a vector space over the field $\mathbb{C}$ (or $\mathbb{R}$) equipped}\\
	    \> \textit{with an inner product $\scpr{.}{.}\,$. In addition, a Hilbert space is required}\\
        \> \textit{to be complete with respect to the norm induced by the inner product.}
    \end{tabbing}
The inner product $\scpr{.}{.}$ satisfies $\scpr{\phi}{\psi}=\scpr{\psi}{\phi}^{*}$, where $\psi,\phi\in\mathcal{H}\,$. It is \emph{\mbox{(anti-)}linear}
\newpage \noindent in the (first) second argument, i.e.,
\begin{subequations}
\label{eq:scpr anti-linearity}
\begin{align}
    \scpr{c_{1}\phi_{1}+c_{2}\phi_{2}}{\psi}    &=\,c^{*}_{1}\scpr{\phi_{1}}{\psi}+c^{*}_{2}\scpr{\phi_{2}}{\psi}\,,
    \label{eq:scpr bra antilinearity}\\
    \scpr{\phi}{c_{1}\psi_{1}+c_{2}\psi_{2}}    &=\,c_{1}\scpr{\phi}{\psi_{1}}+c_{2}\scpr{\phi}{\psi_{2}}\,,
    \label{eq:scpr ket linearity}
\end{align}
\end{subequations}
where $c_{1,2}\in\mathbb{C}\,$, $\phi_{1,2},\psi_{1,2}\in\mathcal{H}\,$, and the asterisk denotes complex conjugation. The inner product on $\mathcal{H}$ is further \emph{positive semi-definite}, i.e., \mbox{$\scpr{\psi}{\psi}\geq0\,$,} with equality if and only if $\psi=0\,$. The reflexivity of the Hilbert space, i.e., $\mathcal{H}$ coincides with its \emph{(continuous) dual space}, allows us to employ the so-called \emph{Dirac notation}, that is, we write vectors in~$\mathcal{H}$ as~$\ket{\psi}\,$, while the elements of the (continuous) dual Hilbert space~$\mathcal{H}^{*}$ are denoted as~$\bra{\psi}=\ket{\psi}^{\dagger}\,$. We further require physical states to be normalized, such that $\scpr{\psi}{\psi}=1\,$.\\

\vspace*{-1.5mm}
We can then consider (bounded) linear operators\index{linear operator}, i.e., $A:\ \ket{\psi}\mapsto \ket{\phi}=A\ket{\psi}\,$, on such a Hilbert space, where the adjoint operator~$A^{\dagger}$ is defined by the relation $\scpr{A^{\dagger}\phi}{\psi}=\scpr{\phi}{A\psi}\,$. Operators that satisfy $A^{\dagger}|_{D(A)}=A\,$, with domains $D(A^{\dagger})\supseteq D(A)\,$, are called \emph{Hermitean}\index{Hermitean operator}. Such operators represent physical observables\index{observable}\label{page:observables}, for instance, the energy of the quantum system. The eigenvalues of these operators are real and they correspond to possible outcomes of individual measurements, in which the state is projected onto the corresponding eigenstates. The expectation values\index{expectation value!pure states}
\vspace*{-2.5mm}
\begin{align}
    \expval{A}_{\psi}   &=\,\bra{\psi}A\ket{\psi}\,,
    \label{eq:expectation value pure states}
\end{align}
which represent averaged measurement outcomes, are real as well. It can be easily seen that the projection $\ket{\psi}\!\bra{\psi}$ into the state $\ket{\psi}$ is such a Hermitean operator. Operators~$U$ that satisfy $U^{\dagger}=U^{-1}$ are called \emph{unitary}\index{unitary!operator} and leave the inner product invariant, $\scpr{U\phi}{U\psi}=\scpr{\phi}{\psi}\,$. Operations such as rotations and the dynamics of closed systems are encoded in unitaries.\\

An interesting feature of quantum theory is the fact that a quantum system in a pure state can be in a coherent superposition of states that correspond to different possible measurement outcomes for a given observable. For instance, the system may be in a superposition of different energy eigenstates. This \emph{superposition principle}\index{superposition principle}\label{page:superposition principle} has far-reaching conceptual consequences, for instance for interference effects or for the notion of entanglement, see Section~\ref{sec:pure state entanglement}.

\subsection{Mixed States}\label{subsec:mixed states}
\index{state!mixed|(}

In practice the knowledge about the quantum state produced in a given preparation scheme is not perfect \textemdash\ one typically does not know for sure which pure state a quantum systems is in. Instead of a single pure state one needs to consider an ensemble of pure states, weighted with their relative probabilities. The appropriate descriptions for these \emph{mixed states} are density operators\index{density operator}~$\rho$ that we can write as convex sums of projectors on pure states, i.e.,
\vspace*{-2.5mm}
\begin{align}
    \rho    &=\,\sum\limits_{i}\pii{i}\,\ket{\psii{i}}\!\bra{\psii{i}}\,,
    \label{eq:density operator}
\end{align}
where $\sum_{i}\pii{i}=1$, and the real weights $\pii{i}$ satisfy $0\leq\pii{i}\leq1$. The operators of Eq.~\eqref{eq:density operator}, also called density matrices, are \emph{Hermitean} operators on the Hilbert space~$\mathcal{H}$ of pure states. They are \emph{normalized}, i.e., $\tr(\rho)=\sum_{i}\bra{\!\psii{i}\!}\rho\ket{\!\psii{i}\!}=1\,$, where $\{\ket{\!\psii{i}\!}\}$ is a complete orthonormal basis (CONB) of $\mathcal{H}$, satisfying $\sum_{i}\ket{\!\psii{i}\!}\!\bra{\!\psii{i}\!}=\mathds{1}\,$,\ and $\scpr{\!\psii{i}\!}{\!\psii{j}\!}=\delta_{ij}\,$. Furthermore, density operators are \emph{positive semi-definite}\index{Positive!semi-definite operator}, $\rho\geq0$, which means that their eigenvalues are non-negative. The decomposition of~$\rho$ in Eq.~\eqref{eq:density operator} into a pure state ensemble is not unique, but can always be chosen such that the $\ket{\!\psii{i}\!}$ form a CONB.\\

The expectation value\index{expectation value!mixex states} of Eq.~\eqref{eq:expectation value pure states} can readily be generalized to mixed states by considering a weighted average of the expectation values of a complete, orthonormal ensemble of pure states, $\sum_{i}\pii{i}\bra{\psii{i}}A\ket{\psii{i}}$. By inserting the identity in terms of the same CONB and writing $\rho$ in the decomposition of Eq.~\eqref{eq:density operator} one naturally arrives at
\begin{align}
    \expval{A}_{\rho}  &=\,\tr(A\rho)\,.
    \label{eq:expectation value mixed states}
\end{align}
The trace operation can further be used to define an inner product\index{Hilbert-Schmidt!inner product} $(\,.\,,\,.\,)_{\raisebox{-1.7pt}{\scriptsize{HS}}}\,$, via
\begin{align}
    (\,\rho\,,\,\sigma\,)_{\raisebox{-1.7pt}{\scriptsize{HS}}} &=\,\tr(\rho^{\dagger}\sigma)\,,
    \label{eq:hilbert schmidt inner product}
\end{align}
which promotes the space of the density operators to a Hilbert space, the so-called \emph{Hilbert-Schmidt space}\index{Hilbert-Schmidt!space}, which we are also going to denote as $\mathcal{H}$ in a slight abuse of notation. Every pure state is trivially represented in this space through its projector, for which $\rho^{2}=\rho\,$, whereas this is not the case for any mixed state that cannot be represented by a single state vector. This fact can be used to quantify the \emph{mixedness}\index{mixedness} of \textemdash\ the lack of knowledge about \textemdash\ a given density operator via the \emph{linear entropy}~$S_{L}(\rho)$\index{entropy!linear}
\begin{defi}\label{def:linear entropy}
            \hspace*{0.7cm} The \emph{linear entropy} $S_{L}(\rho)$ of a density matrix $\rho$ is defined as\\
    \vspace*{-0.4cm}
    \begin{center}$S_{L}(\rho)\,:=\,1\,-\,\tr(\rho^{2})$\,.\end{center}
    \index{entropy!linear}
\end{defi}
The linear entropy is bounded, i.e., $0\,\leq\,S_{L}(\rho)\leq\,1-\tfrac{1}{d}\,$, where $d=\dim(\mathcal{H})$, and it can be normalized by a factor $\frac{d}{d-1}$ if desired. It vanishes only for pure states, while it is strictly greater than zero for mixed states. The upper bound $(1-\tfrac{1}{d})$ is attained for the maximally mixed state $\rho_{\mathrm{mix}}=\tfrac{1}{d}\mathds{1}_{d}$\,. Another conventional measure for the mixedness is the \emph{von~Neumann entropy}\index{Von Neumann!entropy}\index{entropy!von Neumann}~$S_{\mathrm{vN}}\,$, to which $S_{L}$ is a linear approximation.
\begin{defi}\label{def:von Neumann entropy}
            \hspace*{0.7cm} The \emph{von~Neumann entropy} $S_{\mathrm{vN}}(\rho)$ of a density matrix $\rho$ is given by\\
    \vspace*{-0.4cm}
    \begin{center}$S_{\mathrm{vN}}(\rho)\,=\,-\,\tr(\rho\,\log(\rho))$\,.\end{center}
\end{defi}
The basis of the logarithm in Definition~\ref{def:von Neumann entropy} is often chosen to be $2$, but here we are going to use the natural logarithm and denote it as~\emph{``$\ln$"} in the following. We can write $S_{\mathrm{vN}}$ in terms of the eigenvalues $\lambdai{i}$ of the density operator $\rho$\,, i.e.,
\begin{align}
    S_{\mathrm{vN}}(\rho)   &=\,-\,\sum_{i}\,\lambdai{i}\,\ln(\lambdai{i})\,,
    \label{eq:von Neumann entropy}
\end{align}
from which it can be seen that $S_{\mathrm{vN}}$ is the straightforward generalization of the classical \emph{Shannon entropy}\index{Shannon entropy}\index{entropy!Shannon}. As before, the von~Neumann entropy is strictly zero if, and only if, the state is pure, while the largest value $\,\ln d\,$ is obtained for the maximally mixed state. Both $S_{L}$ and $S_{\mathrm{vN}}$ are invariant under unitary transformations on $\mathcal{H}$ \begin{align}
    S_{L,\mathrm{vN}}(U\rho\,U^{\dagger})   &=\,S_{L,\mathrm{vN}}(\rho)\,.
    \label{eq:entropies invariant under unitaries}
\end{align}
The von~Neumann entropy will be of further interest to us in the context of entanglement detection in Section~\ref{sec:mixed state entanglement}. For now we shall turn our attention to a useful parametrization of density matrices, the \emph{Bloch decomposition}\index{Bloch!decomposition}. For a single \emph{qubit}, a two-dimensional quantum system with Hilbert space~$\mathbb{C}^{2}$, a general mixed state may be written as
\begin{align}
    \rho    &=\,\frac{1}{2}(\idN{2}\,+\,a^{i}\sigma_{i})\,,
    \label{eq:bloch decomposition}
\end{align}
where $a^{i}=\tr(\rho\sigma_{i})\in\mathbb{R}$ with $a^{i}a_{i}\leq1\ (i=1,2,3)$, and we are using the \emph{Einstein summation convention}\index{Einstein summation convention} for indices that are repeated once as superscript and once as subscript. The $\sigma_{i}$ are the usual, traceless, Hermitean \emph{Pauli matrices}\index{Pauli matrices}
\begin{align}
    \sigma_{1}\,=\,\begin{pmatrix}\,0 & 1\, \\ \,1 & 0\,\end{pmatrix}\,,\ \
    \sigma_{2}\,=\,\begin{pmatrix}\,0 & -i\, \\ \,i & \,0\,\end{pmatrix}\,,\ \
    \sigma_{3}\,=\,\begin{pmatrix}\,1 & \,0\, \\ \,0 & -1\,\end{pmatrix}\,.
    \label{eq:Pauli matrices}
\end{align}
The $a^{i}$ can be interpreted as the components of a vector $\mathbf{a}=(a^{i})\in\mathbb{R}^{3}$, the \emph{Bloch vector}\index{Bloch!vector}, whose length indicates the mixedness of the state. For $|\mathbf{a}|=1$ the state is pure and lies on the surface of the so-called \emph{Bloch sphere}, while all $|\mathbf{a}|<1$ describe mixed states within the sphere. The state parametrization of Eq.~\eqref{eq:bloch decomposition} is very descriptive for spin-$\tfrac{1}{2}$ systems, where the direction of the vector $\mathbf{a}=(a^{i})$ represents the spin orientation of the state $\rho$\,. The Bloch decomposition can be extended to describe single quantum systems of (finite) dimension~$d$, called \emph{qudits}, see Ref.~\cite{BertlmannKrammer2008b}, but we shall instead turn our attention to a generalization for composite systems that is usually referred to as the \emph{generalized Bloch decomposition}\index{Bloch!decomposition; generalized} (also called \emph{Fano decomposition}~\cite{Fano1983}). Any two-qubit density operator on a Hilbert space $\mathbb{C}^{2}\otimes\mathbb{C}^{2}$ can be written as
\begin{align}
    \rho    &=  \,\frac{1}{4}(\idN{4}\,+\,a^{i}\,\sigma_{i}\otimes\idN{2}\,+\,b^{i}\,\idN{2}\otimes\sigma_{i}\,+
                \,t^{ij}\,\sigma_{i}\otimes\sigma_{j})\,.
    \label{eq:two qubit bloch decomposition}
\end{align}

In this decomposition $\mathbf{a}=(a^{i})$ and $\mathbf{b}=(b^{i})$ are the Bloch vectors of the first and second qubit, respectively, while the $t^{ij}=\tr(\rho\,\sigma_{i}\otimes\sigma_{j})$ are the components of the \emph{correlation matrix}~$t[\rho]$, which encodes correlations between the two qubits. We will encounter this object again in Section~\ref{sec:bell inequalities and nonlocality} where it plays a role for the violation of \emph{Bell inequalities}.


\index{state!mixed|)}
\section{Entanglement of Pure States}\label{sec:pure state entanglement}

Let us now turn to a more general description of composite quantum systems and their correlations. In particular, we are going to study a property called \emph{entanglement}, a fundamental resource for quantum information tasks which is a simple consequence of applying the \emph{superposition principle} (see p.~\pageref{page:superposition principle}) of quantum mechanics to composite systems. A pedagogical review of this topic can be found in Ref.~\cite{Bruss2002}, while a more extensive review is given in Ref.~\cite{HorodeckiRPMK2007}. In this section the case of bipartite pure states is discussed, before we continue with bipartite mixed states in Section~\ref{sec:mixed state entanglement} and multipartite systems in Section~\ref{sec:multipartite entanglement}. The chapter will be concluded with a brief look at applications of entanglement, such as \emph{Bell inequalities} and \emph{quantum teleportation} in Section~\ref{sec:applications of entanglement}.\\

\vspace*{-4mm}
Let us consider two Hilbert spaces, $\mathcal{H}\subtiny{0}{0}{A}$ and $\mathcal{H}\subtiny{0}{0}{B}\,$, with dimensions $\dim(\mathcal{H}\subtiny{0}{0}{A})=d\subtiny{0}{0}{A}$ and $\dim(\mathcal{H}\subtiny{0}{0}{B})=d\subtiny{0}{0}{B}\,$, and bases $\{\ket{\psii{i}}\subtiny{-1}{0}{A}\}$ and $\{\ket{\psii{i}}\subtiny{-1}{0}{B}\}\,$, respectively. Any \emph{bipartite pure state}\index{state!bipartite, pure} of the composite Hilbert space $\mathcal{H}\subtiny{0}{0}{AB}=\mathcal{H}\subtiny{0}{0}{A}\otimes\mathcal{H}\subtiny{0}{0}{B}\,$, with $d\subtiny{0}{0}{AB}=\dim(\mathcal{H}\subtiny{0}{0}{AB})=d\subtiny{0}{0}{A} d\subtiny{0}{0}{B}\,$, can be written in terms of these bases as
\vspace*{-2mm}
\begin{align}
    \ket{\psi}\subtiny{-1}{0}{AB} &=\,\sum\limits_{i,j=1}^{d\subtiny{0}{0}{A},d\subtiny{0}{0}{B}}\,c_{\,ij}\, \ket{\psii{i}}\subtiny{-1}{0}{A}\otimes\ket{\psii{j}}\subtiny{-1}{0}{B}\,,
    \label{eq:bipartite pure state}
\end{align}
such that the coefficients $c_{\,ij}\in\mathbb{C}$ satisfy $\sum_{i,j}|c_{\,ij}|^{2}=1\,$. However, for pure, bipartite states there exists a more economical choice of basis\textemdash the \emph{Schmidt basis}\textemdash than the~$d\subtiny{0}{0}{AB}$ tensor products of the basis vectors of the individual Hilbert spaces. Let us formulate this in the \emph{Schmidt decomposition theorem}, originally formulated in Ref.~\cite{Schmidt1907}.
\index{Schmidt decomposition}
\begin{theorem}\label{thm:schmidt decomposition theorem}\end{theorem}
    \vspace*{-1.30cm}
    \begin{tabbing} \hspace*{3.2cm}\=\hspace*{1.5cm}\=\kill
            \> \textit{For every pure bipartite state $\ket{\psi}\subtiny{-1}{0}{AB}$ there exist orthonormal bases}\\[1mm]
            \> \textit{$\{\ket{\chii{i}}\subtiny{-1}{0}{A}\in\mathcal{H}\subtiny{0}{0}{A}\}$ and $\{\ket{\chii{i}}\subtiny{-1}{0}{B}\in\mathcal{H}\subtiny{0}{0}{B}\}$, the \emph{Schmidt-bases}, such that}
    \end{tabbing}
    \vspace*{-6.5mm}
    \begin{center}
        $\ket{\psi}\subtiny{-1}{0}{AB}\,=\,\sum\limits_{i=1}^{d_{\mathrm{min}}}\,\sqrt{\pii{i}}\,
        \ket{\chii{i}}\subtiny{-1}{0}{A}\otimes\ket{\chii{i}}\subtiny{-1}{0}{B}$\,,
    \end{center}
    \vspace*{-7mm}
    \begin{tabbing} \hspace*{3.2cm}\=\hspace*{2cm}\=\kill
            \> \textit{where $1\leq d_{\mathrm{min}}\leq\min(d\subtiny{0}{0}{A},d\subtiny{0}{0}{B})$, and the real Schmidt numbers $\pii{i}\geq0$}\\[1mm]
            \> \textit{satisfy \,$\sum_{i}\pii{i}=1$\,.}
    \end{tabbing}

A proof of this well-known theorem can be found, for instance, in Ref.~\cite[p.~30]{Friis:DiplomaThesis2010}. For any state there is an optimal decomposition in terms of a minimal number~$d_{\mathrm{opt}}$\textemdash the \emph{Schmidt rank}\textemdash of linearly independent vectors $\ket{\chii{i}}\subtiny{-1}{0}{A}\otimes\ket{\chii{i}}\subtiny{-1}{0}{B}$\,. States of Schmidt rank~1 are called \emph{separable}\index{separability!pure states}, while those with $d_{\mathrm{opt}}=\min(d\subtiny{0}{0}{A},d\subtiny{0}{0}{B})$ are called \emph{maximally entangled}\index{entanglement!maximal}.

\index{entanglement|(}
\begin{defi}\label{def:pure bipartite separable state}\end{defi}
	\vspace*{-1.35cm}
    \begin{tabbing} \hspace*{3.2cm}\=\hspace*{1.5cm}\=\kill
			\> \textit{A bipartite pure state $\ket{\psi}\subtiny{-1}{0}{AB}\in\mathcal{H}\subtiny{0}{0}{AB}$ is called \emph{separable}
                with respect to }\\[1mm]
            \> \textit{the bipartition of $\mathcal{H}\subtiny{0}{0}{AB}$ into $\mathcal{H}\subtiny{0}{0}{A}\otimes\mathcal{H}\subtiny{0}{0}{B}$ if it can be written as}\\[1mm]
            \> \textit{$\ket{\psi}\subtiny{-1}{0}{AB}\,=\,\ket{\phi}\subtiny{-1}{0}{A}\otimes\ket{\chi}\subtiny{-1}{0}{B}$\,,\ \
                for some $\ket{\phi}\subtiny{-1}{0}{A}\in\mathcal{H}\subtiny{0}{0}{A}$ and $\ket{\chi}\subtiny{-1}{0}{B}\in\mathcal{H}\subtiny{0}{0}{B}\,$.}
	\end{tabbing}

Having established the notion of separability, the definition of entanglement follows from Definition~\ref{def:pure bipartite separable state} by negation.
\begin{defi}\label{def:entangled state}
    \hspace*{0.7cm} A state is called \emph{entangled}, if it is not separable.
\end{defi}
For entangled states, not all information about the total state can be encoded in the states of the subsystems. Consequently, the reduced state density matrices, $\rhoA{A}$ and $\rhoA{B}$ of the subsystems~$A$ and~$B$, respectively, are mixed. The reduced states are obtained from the bipartite state~$\rhoA{AB}$ by partial tracing, i.e.,
\begin{align}
    \rhoA{A}    &=\,\tr\subtiny{0}{0}{B}(\rhoA{AB})\,=\,\sum\limits_{i}
        \bigl(\idNtiny{A}\otimes\bra{\psii{i}}\subtiny{0}{0}{B}\bigr)\,
        \rhoA{AB}\,
        \bigl(\idNtiny{A}\otimes\ket{\psii{i}}\subtiny{-1}{0}{B}\bigr)\,,
    \label{eq:pure state reductions}
\end{align}
where $\{\ket{\psii{i}}\subtiny{-1}{0}{B}\}$ is a CONB of $\mathcal{H}\subtiny{0}{0}{B}$ and similarly $\rhoA{A}=\tr\subtiny{0}{0}{B}(\rhoA{AB})\,$. For compactness of notation we are going to drop the tensor product symbol, i.e., $\ket{\psi}\subtiny{-1}{0}{A}\otimes\ket{\phi}\subtiny{-1}{0}{B}=\ket{\psi}\ket{\phi}$ from now on and identify the corresponding subspaces by the ordering of the vectors. It is further convenient to indicate the subspaces operators are acting upon solely by their subscripts and drop any identity operators in a tensor product such that $\idNtiny{A}\otimes\OpNtiny{B}=\OpNtiny{B}\,$.\\

For density operators~$\rhoA{AB}$ corresponding to pure states~$\ket{\psi}\subtiny{-1}{0}{AB}$ the mixedness of the reductions can be entirely attributed to the entanglement \textemdash the quantum correlations \textemdash between the subsystems. Moreover, from Theorem~\ref{thm:schmidt decomposition theorem} it can be immediately seen that~$\rhoA{A}$ and~$\rhoA{B}$ have the same rank and the same non-zero eigenvalues, given by the Schmidt rank and the Schmidt numbers~$\pii{i}\,$, respectively. Consequently, any function of these eigenvalues alone, in particular the (von~Neumann) entropy, takes on the same value for either reduced state. This allows us to unambiguously quantify the entanglement of any bipartite pure state by the so-called \emph{entropy of entanglement}\index{entanglement!entropy of}\index{entropy!of entanglement}.

\begin{defi}\label{def:entropy of entanglement}\end{defi}
	\vspace*{-1.35cm}
    \begin{tabbing} \hspace*{3.2cm}\=\hspace*{1.5cm}\=\kill
            \> \textit{The \emph{entropy of entanglement} $\mathcal{E}$ of a bipartite pure state~$\ket{\psi}\subtiny{-1}{0}{AB}$ is}\\[2mm]
            \> \textit{defined as the von~Neumann entropy~$S_{\mathrm{vN}}$ of its reductions~$\rhoA{A}$ and~$\rhoA{B}\,$,}
    \end{tabbing}
    \vspace*{-5.5mm}
    \begin{center}
        $\mathcal{E}(\ket{\psi}\subtiny{-1}{0}{AB})\,=\,S_{\mathrm{vN}}(\rhoA{A})\,=\,S_{\mathrm{vN}}(\rhoA{B})$\,.
    \end{center}

Paradigmatic examples for entangled pure states are the two-qubit \emph{Bell states}\index{Bell!states}\index{state!Bell}
\begin{subequations}
\label{eq:Bell states}
\begin{align}
    \ket{\phi^{\pm}}    &=\,\tfrac{1}{\sqrt{2}}\Bigl(\ket{0}\ket{0}\pm\ket{1}\ket{1}\Bigr)\,,
    \label{eq:Bell states phi plus minus}\\
    \ket{\psi^{\pm}}    &=\,\tfrac{1}{\sqrt{2}}\Bigl(\ket{0}\ket{1}\pm\ket{1}\ket{0}\Bigr)\,,
    \label{eq:Bell states psi plus minus}
\end{align}
\end{subequations}
where $\ket{0}$  and $\ket{1}$ form a basis in~$\mathbb{C}^{2}$. The Bell states, on the other hand, form a complete basis of $\mathbb{C}^{2}\otimes\mathbb{C}^{2}$. They are further examples of \emph{maximally entangled} states, i.e., pure states for which the reductions have maximal rank, or, in other words, for which the reduced states are maximally mixed.

\section{Entanglement of Mixed States}\label{sec:mixed state entanglement}

For mixed states the notion of separability\index{separability!mixed states} is somewhat more involved, since it needs to leave the possibility of incoherent mixtures of uncorrelated product states $\rhoA{A}\otimes\rhoA{B}$\,.
\begin{defi}\label{def:mixed bipartite separability}\end{defi}
	\vspace*{-1.35cm}
    \begin{tabbing} \hspace*{3.2cm}\=\hspace*{1.5cm}\=\kill
			\> \textit{A bipartite mixed state $\rhoA{AB}\in\mathcal{H}\subtiny{0}{0}{AB}$ is called \emph{separable} with respect}\\[1mm]
            \> \textit{to the bipartition of $\mathcal{H}\subtiny{0}{0}{AB}$ into $\mathcal{H}\subtiny{0}{0}{A}\otimes\mathcal{H}\subtiny{0}{0}{B}$ if it can be written as}
    \end{tabbing}
    \vspace*{-5mm}
    \begin{center}$\rhoA{AB}\,=\,\sum\limits_{i}\pii{i}\,\rhoA{A}^{\,i}\otimes\rhoA{B}^{\,i}$\,,\end{center}
    \vspace*{-6.5mm}
    \begin{tabbing} \hspace*{3.2cm}\=\hspace*{2cm}\=\kill
            \> \textit{for some ensembles $\{\rhoA{A}^{\,i}\in\mathcal{H}\subtiny{0}{0}{A}\}$ and $\{\rhoA{B}^{\,i}\in\mathcal{H}\subtiny{0}{0}{B}\}\,$.}\\[1mm]
            \> \textit{As before, a state is called \emph{entangled}, if it is not separable.}
	\end{tabbing}

By this definition all separable states can be created using \emph{local operations and classical communication} (LOCC)\index{local operations \& classical communication|see{LOCC}}\index{LOCC}, i.e., any operations restricted to either of the subsystems and classical communication between the corresponding observers, usually referred to as \emph{Alice} and \emph{Bob}. However, in general it is not straightforward to determine whether a given state admits a decomposition of the form of Definition~\ref{def:mixed bipartite separability}, but we shall discuss some useful \emph{separability criteria} and \emph{measures of entanglement} in Sections~\ref{sec:detection of entanglement} and~\ref{sec:entanglement measures}, respectively.

\subsection{Detection of Entanglement}\label{sec:detection of entanglement}

\index{entanglement!detection|(}

In contrast to the pure state case, it is not unambiguously possible to attribute the mixedness of the subsystems to the overall entanglement. Neither is it conclusive to compute the entropy of entanglement for the pure states in a particular decomposition of~$\rhoA{AB}$, since there is no preferred pure state decomposition for any mixed state. Nonetheless, a sufficient (but not necessary) criterium for the presence of bipartite entanglement can be formulated in the following \emph{entropy inequalities}\index{entropy!inequalities}. A state~$\rhoA{AB}$ is entangled if the entropy of any of the reductions, $\rhoA{A}$ or~$\rhoA{B}$, is larger than the entropy of~$\rhoA{AB}\,$, i.e., $\rhoA{AB}$ is entangled if
\vspace*{-3mm}
\begin{subequations}
\label{eq:entropy inequalities}
\begin{align}
    &   S(\rhoA{AB})\,-\,S(\rhoA{A})\,<\,0\,,
    \label{eq:entropy inequality A}\\
\mbox{or}\hspace*{0.5cm}   &   S(\rhoA{AB})\,-\,S(\rhoA{B})\,<\,0\,,
    \label{eq:entropy inequality B}
\end{align}
\end{subequations}
where we have omitted the label for the chosen entropy. A proof for the von~Neumann entropy~$S_{\mathrm{vN}}$ as well as selected other entropy measures can be found in Ref.~\cite{Terhal2002}.\\

Let us now consider a more geometric method for the detection of entanglement \textemdash \emph{entanglement witnesses}\index{entanglement!witness} as in, e.g., Refs.~\cite{Bruss2002,BertlmannNarnhoferThirring2002}. From Definition~\ref{def:mixed bipartite separability} it can be easily seen that the separable states form a closed, convex subset~$\mathcal{S}\subset\mathcal{H}$ of the Hilbert space of states, while all entangled states form the complement. Since any entangled state is represented by a single point in~$\mathcal{H}$, which is trivially a compact, convex subset of~$\mathcal{H}$, the \emph{Hahn-Banach theorem} of functional analysis allows to separate this point from~$\mathcal{S}$ by a hyperplane, see Ref.~\cite[p.~75]{ReedSimon1972}. Such a hyperplane can be interpreted as a linear functional on~$\mathcal{H}$, realized by a Hermitean operator. Let us phrase this in the following \emph{entanglement witness theorem}\index{entanglement!witness theorem}, which was introduced in Ref.~\cite{HorodeckiMPR1996}.

\begin{theorem}\label{thm:entanglement witness theorem}\end{theorem}
    \vspace*{-1.35cm}
    \begin{tabbing} \hspace*{3.2cm}\=\hspace*{1.5cm}\=\kill
            \> \textit{For any entangled state~$\rho$ there exists an \emph{entanglement witness}, i.e., a}\\[2mm]
            \> \textit{Hermitean operator~$\OpNtiny{\mathrm{W}}$, such that
                $\,(\,\OpNtiny{\mathrm{W}}\,,\,\rho\,)_{\raisebox{-1.7pt}{\scriptsize{HS}}}\,<\,0\,$, while}\\[2mm]
            \> \textit{$\,(\,\OpNtiny{\mathrm{W}}\,,\,\sigma\,)_{\raisebox{-1.7pt}{\scriptsize{HS}}}\,\geq\,0\,$
                for all separable states $\sigma\in\mathcal{S}$.}
    \end{tabbing}

Although this does not directly supply an operational criterion for the detection of entanglement many operational criteria can be considered special cases of the entanglement witness theorem, see, e.g., Ref.~\cite{BertlmannKrammer2008a}. One example is the \emph{Clauser-Horne-Shimony-Holt (CHSH) criterion}\index{CHSH!criterion}\index{Clauser-Horne-Shimony-Holt}\index{criterion!CHSH} for two qubits, that we are going to discuss in Section~\ref{sec:bell inequalities and nonlocality}. \\

In a similar approach it was suggested by A.~Peres in Ref.~\cite{Peres1996} to use the \emph{partial transposition}\index{partial transposition} $\idNtiny{A}\otimes T\subtiny{-1}{-1}{B}$ to detect entanglement. Since the transposition preserves the positivity of operators, it is easy to see from Definition~\ref{def:mixed bipartite separability} that separable states remain positive under partial transposition. In Ref.~\cite{HorodeckiMPR1996} M., R., and P.~Horodecki were then able to use Theorem~\ref{thm:entanglement witness theorem} to prove that this condition for separability is sufficient only as long as $d_{AB}=d_{A}d_{B}\leq6$. Let us phrase this in the following theorem, known as \emph{positive partial transpose} (PPT) \emph{criterion}\index{PPT criterion}\index{criterion!PPT}, or \emph{Peres-Horodecki criterion}\index{Peres-Horodecki criterion|see{PPT}}.

\begin{theorem}\label{thm:PPT theorem}\end{theorem}
    \vspace*{-1.35cm}
    \begin{tabbing} \hspace*{3.2cm}\=\hspace*{1.5cm}\=\kill
            \> \textit{A bipartite state~$\rhoA{AB}\in\mathbb{C}^{2}\otimes\mathbb{C}^{2}$ or $\mathbb{C}^{2}\otimes\mathbb{C}^{3}$ is separable if, and only if,}\\[1.5mm]
            \> \textit{the partial transposition (see p.~\pageref{page:NPT entanglement}) of $\rhoA{AB}$ is positive, i.e.,}
    \end{tabbing}
    \vspace*{-2.5mm}
    \begin{center}
        $\rhoA{AB}^{\,T\subtiny{-1}{-1}{B}}\,=\,
        (\idNtiny{A}\otimes T\subtiny{-1}{-1}{B})\rhoA{AB}\,\geq\,0$\,.
    \end{center}

The subsystems~$A$ and~$B$ in the PPT theorem can of course be exchanged. We see that for two-qubit states, or states of one qubit and one qutrit, all entangled states have a \emph{negative partial transpose} (NPT), while in general there exist entangled states with positive partial transpose, called \emph{bound entangled} states. It was shown in Ref.~\cite{HorodeckiMPR1998} that these PPT entangled states are undistillable, i.e., it is not possible to obtain any pure, maximally entangled states from $n$~copies of the given mixed state by LOCC, but we shall not be further concerned with entanglement distillation in this thesis (see, e.g., Ref.~\cite{Bruss2002} for a pedagogical review of this topic). However, we shall return to the PPT criterion for the construction of useful entanglement measures \textemdash \emph{negativity measures} \textemdash in Section~\ref{sec:entanglement measures}, and, in the context of bosonic Gaussian states in Section~\ref{sec:gaussian entanglement}.

\index{entanglement!detection|)}

\subsection{Measures of Entanglement}\label{sec:entanglement measures}

\index{entanglement!measures|(}

As we have seen, it is not trivial to establish criteria for the separability of a given mixed state. Consequently, there are also many issues in the definition of mixed state entanglement measures, and many candidates have been proposed to suit the plethora of requirements. An extensive review of the various available entanglement measures, entanglement monotones, and their connections can be found in Ref.~\cite{PlenioVirmani2007}. For the purpose of this thesis we shall restrict the discussion to two representatives of these measures, the \emph{entanglement of formation} and related \emph{concurrence}\index{concurrence}, followed by an overview of the so-called \emph{negativity measures}. Let us begin with a list of requirements which are usually imposed on (bipartite) \emph{entanglement measures}~$E(\rho)$.
\begin{defi}\label{def:entanglement measure}\end{defi}
	\vspace*{-1.39cm}
    \begin{tabbing} \hspace*{3.2cm}\=\hspace*{1.5cm}\=\kill
			\> \textit{An \emph{entanglement measure} is a map from density operators~$\rho$ to the} \\
            \> \textit{non-negative real numbers $E(\rho)\in\mathbb{R}_{0}^{+}$ that satisfies:}
    \end{tabbing}
    \vspace*{-5mm}
    \begin{enumerate}[\hspace*{3.5cm}(i)]
    \item{\textit{$E(\rho)=0\ $ for all separable states $\rho\in\mathcal{S}$.}\label{req:E is zero for separable states}}
    \vspace*{-2mm}
    \item{\textit{$E(\rho)$ is non-increasing under LOCC.}\label{req:nonincreasing under LOCC}
        \index{LOCC|see{local operations \& classical communication}}\index{local operations \& classical communication}}
    \end{enumerate}

It is sometimes customary in the literature to add further requirements for genuine entanglement measures, the most popular of which are continuity, reduction to the entropy of entanglement (recall Definition~\ref{def:entropy of entanglement}) for pure states, i.e., $E(\ket{\!\psi\!}\!\bra{\!\psi\!})=\mathcal{E}(\ket{\!\psi\!})$\,, convexity, i.e., $E(\sum_{i}\pii{i}\rhoA{i})\leq\sum_{i}\pii{i}E(\rhoA{i})\,$, and (full) additivity $E(\rhoA{AB}\otimes\rhoA{CD})=E(\rhoA{AB})+E(\rhoA{CD})\,$. In the case of such additional requirements for a genuine entanglement measure, Definition~\ref{def:entanglement measure} is said to define an \emph{entanglement monotone}. For our purposes it will suffice to introduce the above convention. It should also be noted that requirement~(\ref{req:nonincreasing under LOCC}) implies that entanglement is invariant under \emph{local unitary operations}\index{local unitary operations}\index{transformation!local unitary}~$U\subtiny{-1.2}{0}{A}\otimes U\subtiny{-1.2}{0}{B}$ since the corresponding inverse transformations are also LOCC.

\subsubsection{Convex Roof Constructions}

A mathematically intuitive way of generalizing the entropy of entanglement~$\mathcal{E}$ of Definition~\ref{def:entropy of entanglement} to mixed state ensembles is the \emph{entanglement of formation} $E\sub{0.2}{0}{\mathrm{F}}(\rho)$ introduced in Ref.~\cite{BennettDiVincenzoSmolinWootters1996}.
\begin{defi}\label{def:entanglement of formation}\end{defi}\index{entanglement!of formation}
	\vspace*{-1.25cm}
    \begin{tabbing} \hspace*{3.2cm}\=\hspace*{1.5cm}\=\kill
			\> \textit{The \emph{entanglement of formation}~$E\sub{0.2}{0}{\mathrm{F}}(\rho)$ of a bipartite state
                $\rho$ is given by}
    \end{tabbing}
    \vspace*{-5mm}
    \begin{align*}
        \hspace*{2.0cm}E\sub{0.2}{0}{\mathrm{F}}(\rho)    &:=\,\inf_{\{(\pii{i},\,\ket{\psii{i}})\}}
        \sum\limits_{i}\pii{i}\,\mathcal{E}(\ket{\!\psii{i}\!})\,,
    \end{align*}
    \vspace*{-9mm}
    \begin{tabbing} \hspace*{3.2cm}\=\hspace*{2cm}\=\kill
            \> \textit{where the infimum is taken over all pure state ensembles $(\pii{i}\,,\,\ket{\!\psii{i}\!})$}\\
            \> \textit{that realize $\rho=\sum_{i}\pii{i}\ket{\!\psii{i}\!}\!\bra{\!\psii{i}\!}\,$.}
	\end{tabbing}

The entanglement of formation constitutes an entanglement measure in the sense of Definition~\ref{def:entanglement measure} and is an example for a so-called \emph{convex roof construction}\index{convex roof construction}. Therefore it is convex by construction and trivially reduces to the von~Neumann entropy for pure states. In spite of the elegance of its formal definition, the entanglement of formation is in general not a practical measure of entanglement, since the minimization in Definition~\ref{def:entanglement of formation} cannot be carried out in a closed form for arbitrary systems. However, for some situations that exhibit high symmetry, or are specifically simple, this calculation can be performed analytically. In particular, for the simple case of two qubits, a closed expression is provided by (see Refs.~\cite{BennettDiVincenzoSmolinWootters1996,Wootters1998})
\begin{align}
    E\sub{0.2}{0}{\mathrm{F}}(\rho) &=\,h\Bigl(\,\frac{1+\sqrt{1-C^{2}(\rho)}}{2}\,\Bigr)\,,
    \label{eq:entanglement of formation for 2 qubits}
\end{align}
where $h(p)$ is the \emph{Shannon entropy}~$H(\{p,1-p\})$ of the Bernoulli distribution $\{p,1-p\}$,
\begin{align}
    h(p)    &=\,-\,p\ln(p)\,-\,(1-p)\ln(1-p)\,,
    \label{eq:Bernoulli distribution Shannon entropy}
\end{align}
and $C(\rho)$ is the so-called \emph{(Wootters) concurrence}\index{concurrence} for two-qubits
\begin{align}
    C(\rho) &=\,\max\{\,0,\sqrt{\lambda_{1}}-\sqrt{\lambda_{2}}-\sqrt{\lambda_{3}}-\sqrt{\lambda_{4}}\,\}\,.
    \label{eq:wootters concurrence two qubits}
\end{align}
The $\lambda_{\mathrm{i}}\in\mathbb{R}_{0}^{+}$ are the eigenvalues of the matrix $\rho(\sigma_{2}\otimes\sigma_{2})\rho^{*}(\sigma_{2}\otimes\sigma_{2})$ in decreasing order, $\lambda_{1}\geq\lambda_{2,3,4}$. Since the entanglement of formation is a monotonous function of the concurrence only, the latter is sometimes used instead of~$E\sub{0.2}{0}{\mathrm{F}}$ even though it derives its meaning via its relation to~$E\sub{0.2}{0}{\mathrm{F}}$. In Section~\ref{sec:gaussian entanglement} we shall encounter another simple situation, symmetric two-mode Gaussian states on infinite dimensional Hilbert spaces, for which the entanglement of formation can be computed explicitly as well~\cite{GiedkeWolfKruegerWernerCirac2003}. The quantity $C^{2}(\rho)=\tau(\rho)$, where $C(\rho)$ is the two-qubit concurrence of Eq.~\eqref{eq:wootters concurrence two qubits}, can be defined as the convex roof construction\index{convex roof construction} over the linear entropy~\cite{OsborneVerstraete2006}, i.e.,
\begin{align}
    \tau(\rhoA{AB}) &:=\,\inf_{\{(\pii{i},\,\ket{\psii{i}})\}}
        \sum\limits_{i}\pii{i}\,S_{L}\bigl(\tr\subtiny{0}{0}{B}[\,\ket{\!\psii{i}\!}\!\bra{\!\psii{i}\!}\,]\bigr)\,,
    \label{eq:tangle}
\end{align}
in complete analogy to the entanglement of formation in Definition~\ref{def:entanglement of formation}. The definition of this measure\textemdash \emph{the tangle}\index{tangle} \textemdash now naturally extends to two systems of arbitrary dimensions. Beyond two qubits the tangle is generally an upper bound to \textemdash\ not identical to \textemdash\ the square of the corresponding concurrence (see, e.g., Ref.~\cite{Osborne2005}). Nonetheless, a very useful feature of the tangle is that it captures the so-called \emph{monogamy of entanglement}\index{entanglement!monogamy|see{CKW}}\label{page:monogamy of entanglement}. For three qubits $A$, $B$ and $C$ the tangle satisfies the \emph{Coffman-Kundu-Wootters} (CKW)\index{CKW monogamy inequality} \index{Coffman-Kundu-Wootters|see{CKW}} inequality
\begin{align}
    \tau(\rhoA{AB})\,+\,\tau(\rhoA{AC}) &\leq\,\tau(\rhoA{A(BC)})\,,
    \label{eq:Coffman Kundu Wootters inequality}
\end{align}
where $\rhoA{AB}=\tr\subtiny{0}{0}{C}(\rhoA{ABC})$, $\rhoA{AC}=\tr\subtiny{0}{0}{B}(\rhoA{ABC})$ and $\tau(\rhoA{A(BC)})$ is the tangle with respect to the bipartition~$A|BC$. The inequality \eqref{eq:Coffman Kundu Wootters inequality} was proven in Ref.~\cite{CoffmanKunduWootters2000}, and subsequently extended to an arbitrary number of qubits in Ref.~\cite{OsborneVerstraete2006}. Loosely speaking, the monogamy of entanglement means that a given qubit cannot be maximally entangled with more than one qubit at a time. Any gain in entanglement between qubits~$A$ and~$C$ must be compensated by a reduction in entanglement between~$A$ and~$B$. The monogamy inequalities using the tangle do not generally hold beyond qubits (see, e.g., Ref.~\cite{Ou2007}), but possible extensions to qudits relying on other measures of entanglement have been proposed~\cite{KimDasSanders2009}, for instance squashed entanglement satisfies monogamy restraints in arbitrary dimensions~\cite{KoashiWinter2004}. Monogamy can further be restored for Gaussian states in infinite dimensional systems~\cite{HiroshimaAdessoIlluminati2007}.

Despite the useful properties of the convex roof measures, it is more desirable for our purposes here to consider easily computable measures of entanglement, one of which we shall turn to now.

\vspace*{5mm}
\subsubsection{NPT Entanglement}\label{page:NPT entanglement}

The \emph{negativity} measures, based on the PPT criterion (Theorem~\ref{thm:PPT theorem}) and introduced in Ref.~\cite{VidalWerner2002}, quantify \textemdash loosely speaking\textemdash how much a given state fails to be positive after partial transposition. To implement the partial transposition we write a general bipartite mixed state~$\rho\in\mathcal{H}\subtiny{0}{0}{AB}$ in terms of local bases~$\{\ket{i}\in\mathcal{H}\subtiny{0}{0}{A}\}$ and~$\{\ket{j}\in\mathcal{H}\subtiny{0}{0}{B}\}$ as
\begin{align}
    \rho    &=\,\sum\limits_{i,i\pr\!,j,j\pr}\,\rho_{ij,i\pr j\pr}\,\ket{i}\!\bra{i^{\,\prime}\!}\otimes\ket{j}\!\bra{j^{\,\prime}\!}\,.
    \label{eq:mixed state local bases expansion}
\end{align}
The partial transposition\index{partial transposition} of~$\rhoA{AB}$ is then obtained by exchanging the indices on the operators on one of the subspaces,
\begin{align}
    \rho^{\,T\subtiny{-1}{-1}{B}}\,=\,(\idNtiny{A}\otimes T\subtiny{-1}{-1}{B})\rho    &=\,\sum\limits_{i,i\pr\!,j,j\pr}\,\rho_{ij,i\pr j\pr}\,\ket{i}\!\bra{i^{\,\prime}\!}\otimes\ket{j^{\,\prime}\!}\!\bra{j}\,.
    \label{eq:mixed state partial transposition}
\end{align}
\newpage
\begin{defi}\label{def:negativity}\end{defi}\index{negativity}
	\vspace*{-1.35cm}
    \begin{tabbing} \hspace*{3.2cm}\=\hspace*{1.5cm}\=\kill
			\> \textit{The \emph{negativity}~$\mathcal{N}(\rho)$ of a bipartite state $\rho$ is given by}
    \end{tabbing}
    \vspace*{-5mm}
    \begin{align*}
        \mathcal{N}(\rho)   &:=\,\tfrac{1}{2}\sum\limits_{i}\bigl(|\lambda_{i}|\,-\,\lambda_{i}\bigr)\,,
    \end{align*}
    \vspace*{-9mm}
    \begin{tabbing} \hspace*{3.2cm}\=\hspace*{2cm}\=\kill
            \> \textit{where the $\lambda_{i}\in[-\tfrac{1}{2},1]\ $ are the eigenvalues of $\rho^{\,T\subtiny{-1}{0}{B}}$.}
	\end{tabbing}

\vspace*{2mm}

In other words, the negativity is the modulus of the sum of the, at most~\cite{SanperaTarrachVidal1998,Rana2013}, $(d_{A}-1)(d_{B}-1)$ negative eigenvalues of the partial transposition, where as usual, $d\subtiny{0}{0}{A}=\dim(\mathcal{H}\subtiny{0}{0}{A})$ and $d\subtiny{0}{0}{B}=\dim(\mathcal{H}\subtiny{0}{0}{B})$. Alternatively, the negativity can be defined in terms of the trace norm $||\cdot||\sub{0}{0}{1}\,$, i.e., $\,\mathcal{N}(\rho)=\tfrac{1}{2}(||\rho^{\,T\subtiny{-1}{0}{B}}||\sub{0}{0}{1}-1)\,$, where $||\rho||\sub{0}{0}{1}=\tr\sqrt{\rho^{\dagger}\rho}\,$. By Definition~\ref{def:entanglement measure} the negativity is an entanglement measure, and is further convex, but the negativity does not reduce to the entropy of entanglement for pure states and it is not additive. The latter issue can be amended by defining the so-called \emph{logarithmic negativity}~$E_{\mathcal{N}}$\index{negativity!logarithmic} as
\begin{align}
    E_{\mathcal{N}}(\rho)   &=\,\log_{2}||\rho^{\,T\subtiny{-1}{0}{B}}||\sub{0}{0}{1}\,=\,\log_{2}(2\mathcal{N}(\rho)+1)\,,
    \label{eq:logarithmic negativity}
\end{align}
which is additive, but still does not reduce to the entropy of entanglement in the pure case, and it is not convex~\cite{Plenio2005}. Clearly, because of their relation to Theorem~\ref{thm:PPT theorem} neither of the negativities is able to capture bound entanglement. Nonetheless, the negativity measures are widely used because of their computational simplicity. Finally, we can relate the entanglement of formation and the negativity measures. It was shown in Ref.~\cite{VerstraeteAudenaertDehaeneDeMoor2001} that for a two-qubit state with given concurrence $C$ [recall Eq.~\eqref{eq:wootters concurrence two qubits}] the negativity is bounded by
\begin{align}
    \tfrac{1}{2}\bigl(\sqrt{(1-C)^{2}+C^{2}}-(1-C)\bigr)\,\leq\,\mathcal{N} &\leq\,\tfrac{1}{2}C\,.
    \label{eq:concurrence negativity bounds}
\end{align}
We will use the measures described above to quantify the bipartite entanglement between modes of quantum fields. However, in such scenarios we will naturally encounter also systems of more than two modes. It is therefore of interest to take a brief look at entanglement in multipartite systems, as we will do in the next section.


\index{entanglement!measures|)}

\section{Multipartite Entanglement}\label{sec:multipartite entanglement}
\index{entanglement!mulitpartite|(}

In a multipartite system with a fixed number of subsystems $A,B,\ldots,N$ the entanglement structure is much more involved than in the bipartite case, but also much richer. In general these structures are not well understood beyond three qubits, see, e.g., Refs.~\cite{Bruss2002,GuehneToth2009,GabrielHiesmayrHuber2010}, but already in the tripartite case several inequivalent classes of multipartite entanglement are known. We shall mainly be concerned with the detection of \emph{genuine multipartite entanglement} (GME), discussed in Section~\ref{sec:detection of genuine multipartite entanglement}, but first we are going to introduce this concept.

\subsection{Genuine Multipartite Entanglement}\label{sec:genuine multipartite entanglement}

We can start as before by defining $N$-partite states in $\mathcal{H}\subtiny{0}{0}{AB\cdots N}=\mathcal{H}\subtiny{0}{0}{A}\otimes\mathcal{H}\subtiny{0}{0}{B}
\otimes\cdots \otimes\mathcal{H}\subtiny{0}{0}{N}$ that do not contain any entanglement.

\begin{defi}\label{def:pure n-separable state}\end{defi}\index{separability!n-separable}
    \vspace*{-1.35cm}
	\begin{tabbing} \hspace*{3.2cm}\=\hspace*{1.5cm}\=\kill
			\> \textit{A pure, $N$-partite state $\ket{\psi}\subtiny{-1}{0}{AB\cdots N}\in\,\mathcal{H}\subtiny{0}{0}{AB\cdots N}$
                is called \emph{fully separable},}\\[2mm]
            \> \textit{if it can be written as $\ket{\psi}\subtiny{-1}{0}{AB\cdots N}=
                \,\ket{\phi\subtiny{0}{0}{A}}\otimes\ket{\phi\subtiny{0}{0}{B}}\otimes\cdots \otimes\ket{\phi\subtiny{0}{0}{N}}\,$, for}\\[2mm]
            \> \textit{some $\ket{\phi\subtiny{0}{0}{A}}\in\mathcal{H}\subtiny{0}{0}{A}\,$, $\ket{\phi\subtiny{0}{0}{B}}\in\mathcal{H}\subtiny{0}{0}{B}\,$, $\cdots$
                $\ket{\phi\subtiny{0}{0}{N}}\in\mathcal{H}\subtiny{0}{0}{N}\,$.}\\[3mm]
            \> \textit{The state is called \emph{$n$-separable} if it can be written as a tensor product}\\[2mm]
            \> \textit{with respect to some partition of $\mathcal{H}\subtiny{0}{0}{AB\cdots N}$ into~$n\leq N$ subsystems.}
	\end{tabbing}

For $n=2$, i.e., for \emph{pure bi-separable} states,\index{separability!bi-separable} one essentially obtains the pure, bipartite case where the two subsystems contain additional structure. While Definition~\ref{def:pure n-separable state} is straightforward, it is nonetheless involved to test whether a given state is $n$-separable. In contrast to the bipartite case only very specific states admit a generalized Schmidt decomposition (see Ref.~\cite{Peres1995}). To check a general given state for $n$-separability thus requires to compute the reductions of all the subsystems and verify that they are pure. We shall therefore define \emph{genuine multipartite entanglement} for pure states in the following way:
\begin{defi}\label{def:pure multipartite entangled state}\end{defi}\index{entanglement!genuine multipartite}
	\vspace*{-1.35cm}
	\begin{tabbing} \hspace*{3.2cm}\=\hspace*{1.5cm}\=\kill
            \> \textit{A pure state $\ket{\psi}\subtiny{-1}{0}{AB\cdots N}$ is called \emph{genuinely $N$-partite entangled}}\\[2mm]
            \> \textit{if it is not bi-separable in $\mathcal{H}\subtiny{0}{0}{AB\cdots N}$.}
	\end{tabbing}

For mixed states we simply extend the notion of $n$-separability to the convex sum.
\begin{defi}\label{def:mixed multipartite separability and entanglement}\end{defi}\index{entanglement!genuine multipartite}
	\vspace*{-1.35cm}
	\begin{tabbing} \hspace*{3.2cm}\=\hspace*{1.5cm}\=\kill
            \> \textit{A mixed state $\rho\subtiny{-1}{0}{AB\cdots N}$ is called \emph{$n$-separable} if it admits at least one} \\[2mm]
            \> \textit{decomposition into a convex sum of pure $n$-separable states.}\\[3mm]
            \> \textit{Conversely, a mixed state $\rho\subtiny{-1}{0}{AB\cdots N}$ is \emph{genuinely $N$-partite entangled}} \\[2mm]
            \> \textit{if it is not bi-separable $(n=2)$ in $\mathcal{H}\subtiny{0}{0}{AB\cdots N}\,$.}
	\end{tabbing}

While these definitions seem straightforward, let us consider some of their implications. Firstly, if a given state is $n$-separable, then it is automatically also $n\pr$-separable for all $n\pr<n$. The sets~$\mathcal{S}_{n}$ of $n$-separable states are thus convexly nested in each other, i.e., $\mathcal{S}_{n}\subset\mathcal{S}_{n-1}\subset\ldots\subset\mathcal{S}_{1}$. What complicates matters is the fact that the individual states in the decomposition of a bi-separable mixed state into bi-separable pure states need not be separable with respect to the same bi-partitions.\\
\newpage
Therefore, a given mixed state can be $n$-separable, even though it is entangled with respect to specific partitions. This feature makes it rather difficult to determine whether a given state is $n$-separable. However, as we will discuss shortly, one can construct witness inequalities whose violation detects GME~\cite{GabrielHiesmayrHuber2010}.

\subsection{Detection of Genuine Multipartite Entanglement}\label{sec:detection of genuine multipartite entanglement}

For the detection of genuine multipartite entanglement a seemingly mundane property of $n$-separable pure states can be used\textemdash permutational symmetry in the exchange of corresponding subsystems of two copies of the state. Let us start with two copies, $\ket{\psi}\subtiny{-1}{0}{A_{1}A_{2}}$ and $\ket{\psi}\subtiny{-1}{0}{B_{1}B_{2}}\,$, of a bi-separable pure state $\ket{\psi}\subtiny{-1}{0}{1,2}=\ket{\phi}\subtiny{-1}{0}{1}\ket{\chi}\subtiny{-1}{0}{2}\,$, where $A_{i}$ and $B_{i}\ (i=1,2)$ label the $i$-th subsystem of two otherwise identical copies $A$ and $B$, respectively. The subsystems~$A_{2}$ and~$B_{2}$ of the first and second copy may be freely exchanged,
\begin{align}
    \pi\subtiny{-1}{0}{(A_{1}|A_{2})_{2}}\ket{\psi}\subtiny{-1}{0}{A_{1}A_{2}}\otimes\ket{\psi}\subtiny{-1}{0}{B_{1}B_{2}}   &=\,
    \ket{\psi}\subtiny{-1}{0}{A_{1}B_{2}}\otimes\ket{\psi}\subtiny{-1}{0}{B_{1}A_{2}}\,,
    \label{eq:bi-separable pure state subsystem exchange}
\end{align}
where we have defined the permutation operator $\pi\subtiny{-1}{0}{(A_{1}|A_{2})_{2}}\,$, that exchanges the second subsystem of the two copies with respect to the partition $(A_{1}|A_{2})$\,. This statement trivially extends to multipartite entanglement, i.e., a pure state that is $n$-separable with respect to a $n$-partition $\mathbb{P}(n)=(A_{1}|A_{2}|\cdots|A_{n})$ is invariant under permutations~$\pi\subtiny{-0.5}{-0.5}{\mathbb{P}(n)_{i}}\,$, i.e., exchanges of the $i$-th subsystems~$A_{i}$ and~$B_{i}$ of the two copies. Using this statement the following theorem was formulated in Ref.~\cite{GabrielHiesmayrHuber2010}.

\begin{theorem}\label{thm:GME witness theorem}\end{theorem}
    \vspace*{-1.35cm}
	\begin{tabbing} \hspace*{3.2cm}\=\hspace*{1.5cm}\=\kill
            \> \textit{Every $N$-partite, $n$-separable state $\rho$ satisfies}
    \end{tabbing}
    \vspace*{-4.5mm}
    \begin{align*}
        \hspace*{3.3cm}\Bigl(\,\bra{\Phi}\rho^{\otimes2}\,\prod_{i}\pi\subtiny{-1}{0}{\mathbb{P}(n)_{i}}\ket{\Phi}\,\Bigr)^{\raisebox{-1pt}{\tiny{$\tfrac{1}{2}$}}}  &\leq\,
         \sum\limits_{\mathbb{P}(n)}\Bigl(\,\prod\limits_{i=1}^{n}
            \bra{\Phi}
                \pi^{\dagger}\subtiny{-1}{0}{\mathbb{P}(n)_{i}}\,
                \rho^{\otimes2}\,
                \pi\subtiny{-1}{0}{\mathbb{P}(n)_{i}}
            \ket{\Phi}\,\Bigr)^{\raisebox{-1pt}{\tiny{$\tfrac{1}{2n}$}}}\,,
    \end{align*}
    \vspace*{-6.5mm}
    \begin{tabbing} \hspace*{3.2cm}\=\hspace*{2cm}\=\kill
            \> \textit{for every fully separable $(2N)$-partite pure state $\ket{\Phi}\,.$}
    \end{tabbing}
\vspace*{2mm}
\begin{proof}
Let us quickly sketch the proof for this inequality. If $\rho$ is a pure, $N$-partite, $n$-separable state, then it must be $n$-separable with respect to one of the $n$-partitions $\mathbb{P}(n)$ in the sum on the right hand side. The corresponding term of the sum over all $\mathbb{P}(n)$ then cancels with the left hand side since the fully separable state $\ket{\Phi}$ can trivially be written as a tensor product of two $N$-partite states, i.e., $\ket{\Phi}=\ket{\Phi\subtiny{-1}{0}{A}}\ket{\Phi\subtiny{-1}{0}{B}}$, and $\rho=\ket{\psi}\!\bra{\psi}$ is pure. The remaining terms on the right hand side are products of $(2n)$th roots of diagonal entries $\rho_{kk}$ of the density operator~$\rho$, and therefore strictly positive. Thus the inequality is trivially satisfied. To extend the proof to mixed states one simply notes that the left hand side is the modulus \textemdash a convex function\textemdash of a density matrix element~$\rho_{kl}$, while the $(2n)$th roots on the right hand side are concave functions of the matrix elements, i.e.,
\vspace*{-2mm}
\begin{subequations}
\label{eq:convex and concave functions of rho}
\begin{align}
    |\sum\limits_{i}\,\pii{i}\,\rho_{kl}^{i}|   &\leq\,\sum\limits_{i}\,\pii{i}\,|\rho_{kl}^{i}| \,,
    \label{eq:convex function of rho}\\
    \bigl(\sum\limits_{i}\,\pii{i}\,\rho_{kk}^{i}\bigr)^{\raisebox{-1pt}{\tiny{$\tfrac{1}{2n}$}}}
    &\geq\,\sum\limits_{i}\,\pii{i}\,\bigl(\rho_{kk}^{i}\bigr)^{\raisebox{-1pt}{\tiny{$\tfrac{1}{2n}$}}}\,,
    \label{eq:concave function of rho}
\end{align}
\end{subequations}
which concludes the proof.
\end{proof}
We will use detection inequalities of this type to study GME between modes of quantum fields in cavities in Section~\ref{sec:Generation of Genuine Multipartite Entanglement}.



%

\index{entanglement!mulitpartite|)}

\section{Applications of Entanglement}\label{sec:applications of entanglement}

\subsection{The EPR Paradox}\label{sec:EPR paradox}
\index{EPR paradox|(}
\index{Einstein-Podolsky-Rosen|see{EPR}}

We have previously introduced the mathematical notion of entanglement\textemdash establishing that the superposition principle, applied to composite systems, gives rise to this intriguing property. But one might ask what distinguishes entanglement from other correlations. Fur this purpose, let us turn to the \emph{EPR thought experiment} formulated by \emph{Albert Einstein}, \emph{Boris Podolsky}, and \emph{Nathan Rosen} (EPR) in 1935. In their seminal
paper ``\emph{Can Quantum-Mechanical Description of Physical Reality Be Considered Complete?}" (Ref.~\cite{EinsteinPodolskyRosen1935}) they used an entangled state to argue that quantum theory does not meet their criteria of \emph{locality}, \emph{reality}, and \emph{completeness}.
\begin{enumerate}[\hspace*{0.5cm}(i)]
\item{  \begin{tabbing} \hspace*{2.5cm}\=\kill
        \emph{Locality:}    \> There are no instantaneous interactions between distant\\
                            \> physical systems.
        \end{tabbing}\label{req:locality}}
\item{  \begin{tabbing} \hspace*{2.5cm}\=\kill
        \emph{Realism:}     \> If the value of a physical quantity can be predicted with\\
                            \> certainty without disturbing the system, then this \\
                            \> quantity corresponds to an \emph{element of physical reality}.
        \end{tabbing}\label{req:realism}}
\item{  \begin{tabbing} \hspace*{2.5cm}\=\kill
        \emph{Completeness:}\> A theory is \emph{complete} if every element of physical reality\\
                            \> is assigned to a corresponding element in the theory.
        \end{tabbing}\label{req:completeness}}
\end{enumerate}
In short, their argument asserts that the existence of maximally entangled states in quantum mechanics gives rise to elements of reality that are not accounted for in quantum mechanics, which, according to EPR, is therefore incomplete. In spite of the far-reaching conceptual consequences of this paradox it was mostly ignored or considered to be a purely philosophical problem. In particular the reply by \emph{Niels Bohr} in Ref.~\cite{Bohr1935} supported this point of view. It was not until 1957 that the problem pointed out by EPR was appreciated, when \emph{David Bohm} and \emph{Yakir Aharonov} published their version~\cite{BohmAharonov1957} of the paradox, which we shall present here using the language of quantum information theory.\\

Let us consider two spatially distant qubits, $A$ and $B$, in the Bell state $\ket{\psi^{-}}\subtiny{-1}{0}{AB}$ of Eq.~\eqref{eq:Bell states psi plus minus}. If both subsystems are measured in the basis $\{\ket{0},\ket{1}\}$ then the measurement results are always perfectly anti-correlated. In particular, if $``0"$ is measured in subsystem~$A$, then $A$ can predict the outcome of~$B$ to be $``1"$ with certainty, endowing this result with physical reality according to requirement (\ref{req:realism}) above. However, the same argument can be made for any other single qubit basis such that all of these results should have simultaneous physical reality. Quantum mechanics on the other hand states that measurements in different bases do not generally commute. Therefore the corresponding results cannot have independent physical reality, which leads to the apparent paradox.\\

Since it seemed imprudent to remove the requirement for locality, an obvious solution was considered to be to equip quantum mechanics with a set of \emph{hidden variables} that determine the measurement outcomes, thus completing the theory in the sense of requirement~(\ref{req:completeness}). Those ``completions" of quantum mechanics in terms of so-called \emph{hidden variable theories} (HVTs) are severely constrained by no-go theorems such as \emph{Bell's theorem}~\cite{Bell1964} or \emph{contextuality} arguments (see Refs.~\cite{KochenSpecker1967,Mermin1990,Peres1991}), some of which we are going to discuss in Section~\ref{sec:bell inequalities and nonlocality}.

\index{EPR paradox|)}

\subsection{Bell Inequalities \& Non-Locality}\label{sec:bell inequalities and nonlocality}

\index{non-locality|(}

In 1964 \emph{John Stewart Bell} elevated the discussion of the EPR paradox to a new level. He formulated an inequality that allowed to decide experimentally whether or not completeness in the sense of the EPR argument can be achieved by a local HVT. We will not consider Bell's original inequality here, but instead consider a more easily testable version \textemdash the \emph{Clauser-Horne-Shimony-Holt} (CHSH) \emph{inequality} introduced in Ref.~\cite{ClauserHorneShimonyHolt1969}. However, irrespectively of the specific type of \emph{Bell inequality}\index{Bell!inequality} that is being tested we can formulate Bell's theorem in the following way.
\begin{theorem}\label{thm:bell theorem}\end{theorem}\index{Bell!theorem}
    \vspace*{-1.25cm}
    \begin{tabbing} \hspace*{3.1cm}\=\hspace*{2cm}\=\kill
            \> \textit{All HVTs that are local and realistic in the sense of the EPR requirements}\\
            \> \textit{\eqref{req:locality} and~\eqref{req:realism}, respectively, are incompatible with the predictions of quantum}\\
            \> \textit{mechanics for the outcome of certain experiments.}
    \end{tabbing}

Let us put this theorem into a mathematical framework in terms of the CHSH inequality. Let us consider two distant parties, Alice and Bob, who are measuring dichotomic quantities~$A(\mathbf{a})$ and~$B(\mathbf{b})$, respectively, where $\mathbf{a}=(a^{i})$ and $\mathbf{b}=(b^{i})$ cor-
\newpage
\noindent
respond to the measurement settings. To implement the reality requirement~\eqref{req:realism} of the HVT one can assume that the measurement results also depend on some set~$\lambda$ of hidden parameters. The locality assumption~\eqref{req:locality} simply means that the outcomes $A(B)$ are independent of the measurement settings $\mathbf{b}(\mathbf{a})$ of the other party, i.e., $A(\mathbf{a},\lambda)=\pm1,0$ and $B(\mathbf{b},\lambda)=\pm1,0$, where~$\pm1$ represent successful measurements, while $0$ corresponds to failed detections. The expectation value $\left\langle(\mathbf{a},\,\mathbf{b})\right\rangle_{\lambda}$ for the joint measurements of~$A$ and~$B$ for a hidden variable $\lambda$ with (normalized) distribution $\rho(\lambda)$ is then given by
\begin{align}
    \left\langle(\mathbf{a},\,\mathbf{b})\right\rangle_{\lambda}  &=\,\int\!d\lambda\,\rho(\lambda)\,A(\mathbf{a},\,\lambda)\,B(\mathbf{b},\,\lambda)\,.
    \label{eq:bell expvalue}
\end{align}
We can then formulate Theorem~\ref{thm:bell theorem} in terms of the CHSH inequality~\cite{ClauserHorneShimonyHolt1969}\index{CHSH!inequality}.
\begin{theorem}\label{thm:bell CHSH theorem}\end{theorem}\index{Bell!theorem}
    \vspace*{-1.35cm}
    \begin{tabbing} \hspace*{3.1cm}\=\hspace*{2cm}\=\kill
            \> \textit{The expectation values $\left\langle(\mathbf{a},\,\mathbf{b})\right\rangle_{\lambda}$ of a local, realistic theory satisfy}
    \end{tabbing}
    \vspace*{-4mm}
    \begin{center}
        \hspace*{2.0cm}$
        \left\langle\right.\!(\mathbf{a},\,\mathbf{b})\!\left.\right\rangle_{\lambda}\,-\,
        \left\langle\right.\!(\mathbf{a},\,\tilde{\mathbf{b}})\!\left.\right\rangle_{\lambda}\,+\,
        \left\langle\right.\!(\tilde{\mathbf{a}},\,\mathbf{b})\!\left.\right\rangle_{\lambda}\,+\,
        \left\langle\right.\!(\tilde{\mathbf{a}},\,\tilde{\mathbf{b}})\!\left.\right\rangle_{\lambda}\leq2$\,.
    \end{center}

For a pedagogical proof see, e.g., Ref.~\cite[pp.~47-48]{Friis:DiplomaThesis2010}. For the purpose of checking this theorem in a quantum mechanical computation the CHSH inequality can be reformulated in terms of the following two-qubit observable on $\mathbb{C}^{2}\otimes\mathbb{C}^{2}$,
\begin{align}
\mathcal{O}\subtiny{0}{0}{\mathrm{CHSH}}    &=\,a^{m}\sigma_{m}\,\otimes\,(b^{n}\,+\,\tilde{b}^{n})\sigma_{n}\,
+\,\tilde{a}^{m}\sigma_{m}\,\otimes\,(b^{n}\,-\,\tilde{b}^{n})\sigma_{n}\,,
\label{eq:bell observable}\index{CHSH!operator}
\end{align}
where summation over repeated indices is implied, and the $\sigma_{n}$ are the Pauli matrices~(\ref{eq:Pauli matrices})\index{Pauli matrices}.
The quantum mechanical expectation value of the observable in~(\ref{eq:bell observable}) can reach $\expval{\mathcal{O}\subtiny{0}{0}{\mathrm{CHSH}}}_{\mathrm{QM}}\leq2\sqrt{2}>2$, both in theory and in experiment (see, e.g., Ref.~\cite{FreedmanClauser1972}), in clear violation of the CHSH inequality. The possible violation of a Bell inequality by a given state, often referred to as \emph{non-locality}, has a profound connection with the separability of that state. In fact, one can easily construct an entanglement witness in full analogy to Theorem~\ref{thm:entanglement witness theorem}.
\begin{theorem}\label{thm:chsh criterion}\end{theorem}
    \vspace*{-1.35cm}
    \begin{tabbing} \hspace*{3.2cm}\=\hspace*{2cm}\=\kill
            \> \textit{A two-qubit state $\rho$ can violate the CHSH inequality}\\[2mm]
            \> \> $(\,\rho\,,\,2\mathds{1}\,-\,\mathcal{O}\subtiny{0}{0}{\mathrm{CHSH}}\,)_{\raisebox{-1.7pt}{\scriptsize{HS}}}\,\geq\,0$ , \\[2mm]
            \> \textit{where $\mathcal{O}\subtiny{0}{0}{\mathrm{CHSH}}$ is given by \eqref{eq:bell observable}, only if it is entangled.}
    \end{tabbing}
In fact, it was shown in Ref.~\cite{Gisin1991} that the CHSH inequality can be violated for every entangled pure state. This is in general no longer the case for mixed states.    The issue that Theorem~\ref{thm:chsh criterion} has in common with the entanglement witnesses discussed earlier, is to determine a suitable operator for a given state $\rho$\,. However, for the CHSH inequality this problem can be circumvented by the following theorem \textemdash the \emph{CHSH criterion}\index{CHSH!criterion} \textemdash proven in Ref.~\cite{HorodeckiRPM1995}.
\newpage
\begin{theorem}\label{thm:chsh criterion 2}\end{theorem}
    \vspace*{-1.25cm}
    \begin{tabbing} \hspace*{3.2cm}\=\hspace*{3cm}\=\kill
            \> \textit{The maximally possible value of the CHSH expectation value $\expval{\mathcal{O}\subtiny{0}{0}{\mathrm{CHSH}}}_{\rho}$}\\[2mm]
            \> \textit{for a given two-qubit state $\rho$ is given by
                $\ \expval{\mathcal{O}^{\mathrm{max}}\subtiny{0}{0}{\mathrm{CHSH}}}_{\rho}\,=\,2\sqrt{\mu_{1}+\mu_{1}}\,$,}\\[2mm]
            \> \textit{where $\mu_{1}$ and $\mu_{2}$ are the two largest eigenvalues of $\ M_{\rho}=t[\rho]^{T}t[\rho]\,$, and}\\[2mm]
            \> \textit{$t[\rho]$ is the correlation matrix $t[\rho]_{ij}=\tr(\rho\,\sigma_{i}\otimes\sigma_{j})\,$ of
                Eq.~\eqref{eq:two qubit bloch decomposition}.}\index{correlation matrix}
    \end{tabbing}
The criterion does not require us to determine the measurement directions $\mathbf{a},\mathbf{b},\tilde{\mathbf{a}},$ and $\tilde{\mathbf{b}}$, but instead provides us with a simple way to use Theorem~\ref{thm:chsh criterion} for the detection of entanglement. In Section~\ref{sec:quantum teleportation} we will study the protocol known as \emph{quantum teleportation} and following Ref.~\cite{HorodeckiRMP1996} we shall discover the connection this protocol has to the violation of the CHSH inequality.

\index{non-locality|)}

\subsection{Quantum Teleportation}\label{sec:quantum teleportation}

To conclude this chapter we are now going to discuss a paradigmatic protocol of quantum information processing \textemdash \emph{quantum teleportation}\index{teleportation} \textemdash introduced in Ref.~\cite{BennettBrassardCrepeauJoszaPeresWootters1993} and generalized to mixed resource states in Ref.~\cite{Popescu1994}.

\subsubsection{The Teleportation Protocol}

The setup is the following: two observers, \emph{Alice} and \emph{Bob}, share a two-qubit state $\rhoA{AB}\,$. Alice additionally has access to an unknown pure, single-qubit state $\ket{\phi}\subtiny{-1}{0}{X}$ in a subsystem that we label by $X$. Alice wishes to send the state of the third qubit to Bob using only classical communication and local operations. To this end she performs a \emph{Bell measurement}\index{Bell!measurement}, i.e., she projects the subsystems~$X$ and~$A$ into the Bell basis $\{\ket{\phi^{\pm}},\ket{\psi^{\pm}}\}$ of Eq.~\eqref{eq:Bell states}. With probability $\pii{k}$ the total state is then transformed to the state
\begin{align}
    \frac{1}{\pii{k}}   \bigl[(P\sub{-1}{-1}{k})\subtiny{-1}{0}{XA}\otimes\mathds{1}\subtiny{-1}{0}{B}\bigr]\,
                        \bigl[(P\sub{-1}{0}{\phi})\subtiny{-1}{0}{X}\otimes\rhoA{AB}\bigr]\,
                        \bigl[(P\sub{-1}{-1}{k})\subtiny{-1}{0}{XA}\otimes\mathds{1}\subtiny{-1}{0}{B}\bigr]\,,
    \label{eq:teleportation state after Bell measurement}
\end{align}
where~$k=1,2,3,4,\,$ labels the four Bell states, $P\sub{-1}{0}{\chi}$ denotes a projector on the state $\ket{\chi}\,$, $P\sub{-1}{0}{\chi}=\ket{\!\chi\!}\!\!\bra{\!\chi\!}\,$, and the probability $\pii{k}$ for the outcome~$k$ is given by
\begin{align}
    \pii{k} &=\,\tr\Bigl(\bigl[(P\sub{-1}{-1}{k})\subtiny{-1}{0}{XA}\otimes\mathds{1}\subtiny{-1}{0}{B}\bigr]\,
                        \bigl[(P\sub{-1}{0}{\phi})\subtiny{-1}{0}{X}\otimes\rhoA{AB}\bigr]\Bigr)\,.
    \label{eq:probability for kth outcome of Bell measurement}
\end{align}
Finally, Alice communicates the outcome~$k$ to Bob by sending two bits of classical information. Bob can then perform a local unitary operation~$U\subtiny{0}{0}{k}$ on subsystem~$B$ to obtain the state
\begin{align}
    \rho\subtiny{0}{0}{k}    &=\,\frac{1}{\pii{k}}\tr\subtiny{0}{0}{XA}\Bigl(
                        \bigl[(P\sub{-1}{-1}{k})\subtiny{-1}{0}{XA}\otimes(U\subtiny{0}{0}{k})\subtiny{0}{0}{B}\bigr]\,
                        \bigl[(P\sub{-1}{0}{\phi})\subtiny{-1}{0}{X}\otimes\rhoA{AB}\bigr]\,
                        \bigl[(P\sub{-1}{-1}{k})\subtiny{-1}{0}{XA}\otimes(U^{\dagger}\subtiny{0}{0}{k})\subtiny{0}{0}{B}\bigr]\Bigr)\,.
    \label{eq:teleportation final state Bob}
\end{align}
For a pure, maximally entangled resource state, e.g., for~$\rhoA{AB}=\ket{\!\psi^{\pm}\!}\!\bra{\!\psi^{\pm}\!}\,$, the transmission of the quantum state~$\ket{\phi}$ becomes perfect, i.e., $\rho\subtiny{0}{0}{k}=\ket{\!\phi\!}\!\bra{\!\phi\!}\ \forall\,k\,$.

\subsubsection{Optimized Teleportation \& Teleportation Fidelity}

For general mixed states~$\rhoA{AB}$ the teleportation is afflicted with errors and the scheme can therefore profit from variations. For instance, one can define a generic teleportation protocol $\mathbb{T}[\rhoA{X\hspace*{-0.5pt}AB}]$ that allows for (trace preserving) local operations and classical communication~\cite{HorodeckiMPR1999,VerstraeteVerschelde2003} between Alice's and Bob's subsystems, $(XA)$ and~$B$, respectively, i.e., transformations that cannot increase the shared entanglement between~$A$ and~$B\,$. For such a protocol the final state of subsystem~$B$ is given by
\begin{align}
    \rho\subtiny{0}{0}{B,\mathbb{T}}   &=\,\tr\subtiny{0}{0}{XA}\Bigl(\mathbb{T}\bigl[(P\sub{-1}{0}{\phi})\subtiny{-1}{0}{X}\otimes\rhoA{AB}\bigr] \Bigr)\,.
    \label{eq:generic teleportation final state Bob}
\end{align}
A figure of merit for this protocol is the \emph{teleportation fidelity}\index{teleportation!fidelity}~$\mathcal{F}(\mathbb{T},\rhoA{AB})$, defined as~\cite{HorodeckiMPR1999}
\begin{align}
    \mathcal{F}\subtiny{0}{0}{\mathbb{T}}(\rhoA{AB})   &=\,\int d\phi\,\bra{\phi}\rho\subtiny{0}{0}{B,\mathbb{T}}\ket{\phi}\,,
    \label{eq:generic teleportation fidelity}
\end{align}
where the integral is carried out over a uniform distribution of input states~$\ket{\phi}$\,. This teleportation fidelity can be understood as the overlap between the final state~$\rho\subtiny{0}{0}{B,\mathbb{T}}$ and the target state~$\ket{\phi}$\,, averaged over all inputs. For a separable state the teleportation fidelity cannot exceed $\tfrac{2}{3}$, in other words $\mathcal{F}\subtiny{0}{0}{\mathbb{T}}(\rho_{\mathrm{sep}})\leq\tfrac{2}{3}\,$ (see, e.g., Ref.~\cite{VerstraeteVerschelde2003}). Consequently, any state for which the fidelity can exceed the value of $\tfrac{2}{3}$ is called \emph{useful} for teleportation.
Interestingly, the fidelity of the standard teleportation protocol can be related to the violation of the CHSH inequality~\cite{HorodeckiRMP1996} \textemdash any state that violates the CHSH inequality is useful for teleportation. The corresponding fidelity, maximized over Bob's local rotations~$U\subtiny{0}{0}{k}$, is given by
\begin{align}
    \mathcal{F}_{\mathrm{max}}(\rhoA{AB})  &=\,\frac{1}{2}\bigl(1\,+\,\frac{1}{3}\sum\limits_{i}\sqrt{\mu_{i}}\bigr)\,,
    \label{eq:teleportation fidelity qubits}
\end{align}
where the $\mu_{i}$ are the eigenvalues of $\ M_{\rhoA{AB}}=t[\rhoA{AB}]^{T}t[\rhoA{AB}]\,$ from Theorem~\ref{thm:chsh criterion 2}. As the $\mu_{i}\leq1$ one finds $\sum_{i}\sqrt{\mu_{i}}\geq\sqrt{\mu_{1}+\mu_{2}}\,$, where $\mu_{3}\leq\mu_{1,2}\,$. It then immediately follows that
\begin{align}
    \mathcal{F}_{\mathrm{max}}(\rhoA{AB})  &\geq\,\frac{1}{2}\bigl(1\,+\,\frac{1}{12}
    \expval{\mathcal{O}^{\,\mathrm{max}}\subtiny{0}{0}{\mathrm{CHSH}}}_{\rhoA{AB}}\bigr)\,,
    \label{eq:teleportation fidelity vs CHSH qubits}
\end{align}
which is larger than $\tfrac{2}{3}$ only if $\expval{\mathcal{O}^{\,\mathrm{max}}\subtiny{0}{0}{\mathrm{CHSH}}}_{\rhoA{AB}}>2\,$. All mixed two-qubit states that violate the CHSH inequality are useful for teleportation. However, it should be noted that states that do not violate a Bell inequality can still be useful for some teleportation schemes (see Ref.~\cite{Popescu1994}).\\

We will consider the effects on the possible teleportation fidelity between modes of both fermionic and bosonic quantum fields confined to different cavities in non-uniform motion in Chapter~\ref{Chapter 7 Degradation of Entanglement between Moving Cavities}. Now, let us turn to the description of these quantum fields.

    \newpage\ \\

    \chapter{Quantum Fields in Flat and Curved Spacetimes}\label{chapter:Quantum Fields in Flat and Curved Spacetimes}
\vspace*{-3mm}
Let us now turn to the second pillar of RQI: \emph{relativity}, in particular, relativistic quantum field theory. We shall here consider the quantization of relativistic fields and study quantum information procedures in the corresponding Fock spaces. We note here in passing that it is possible to consider only a sector of the Fock space with a fixed number of relativistic particles in a covariant way. Such situations have been studied extensively in the context of RQI (see, e.g., Refs.~\cite{Czachor1997,PeresScudoTerno2002,GingrichAdami2002,AlsingMilburn2002,TerashimaUeda2003a,
TerashimaUeda2003b,AhnLeeHwang2003,AhnLeeMoonHwang2003,JordanShajiSudarshan2007,FriisBertlmannHuberHiesmayr2010,
HuberFriisGabrielSpenglerHiesmayr2011,PalgeDunningham2012,SaldanhaVedral2012a,
SaldanhaVedral2012b,PalmerTakahashiWestman2012,Cabanetal2012a,Cabanetal2012b,SaldanhaVedral2013,
BaukeAhrensKeitelGrobe2013}, or see Ref.~\cite{Friis:DiplomaThesis2010} for an introduction to the topic), but the discussion of these results lies beyond the scope of this thesis. Here we are interested in studying a genuine relativistic multi-particle theory that allows for particle creation phenomena.\\

\vspace*{-4mm}
Naturally, the question arises: \emph{Why is it necessary to consider quantum fields in RQI?} First and foremost, one may answer that field quantization is needed to endow the solutions of relativistic field equations with an appropriate interpretation where the usual procedure of interpreting wave functions fails. In this context quantum field theory provides a natural extension of quantum mechanics. We will further elaborate on this problem in Section~\ref{sec:the Klein Gordon field}. Another reason for the necessity to consider quantum fields in the context of RQI lies in the observer dependence of particle content~\cite{BirrellDavies:QFbook} and entanglement~\cite{AlsingFuentes2012}. Any model with a fixed particle number cannot hope to capture the intriguing effects attributed to non-uniform motion and spacetime structure, such as the \emph{Unruh effect}, the \emph{Hawking effect}, the \emph{dynamical Casimir effect}\index{dynamical Casimir effect}, or the effects of non-uniformly moving cavities that we are going to discuss in Part~\ref{partII:Shaking Entanglement}. We will now establish the basic framework of quantization in (non-interacting) quantum field theory that is needed in the following chapters. For a thorough introduction to the numerous additional aspects of quantum field theory we direct the interested reader to standard textbooks, e.g., Refs.~\cite{PeskinSchroeder1995,BirrellDavies:QFbook}.\\

Our aim in this chapter is the quantization of Lorentz invariant\index{Lorentz!invariance} field equations. We shall be concerned with two representatives of the irreducible, unitary representations of the \emph{Poincar\'{e}} group\index{Poincar\'{e} group}\index{group!Poincar\'{e}}\ \textemdash\ relativistic fields: In Section~\ref{sec:the Klein Gordon field} we will discuss the \emph{real scalar field} and in Section~\ref{sec:the dirac field} the \emph{Dirac field} will be introduced. First we are going to establish some conventions for the description of relativistic spacetimes in Section~\ref{sec:relativistic spacetimes}.

%
%
%


\section{Relativistic Spacetimes}\label{sec:relativistic spacetimes}

%



We are interested in constructing our quantum fields in a \emph{spacetime}, i.e., a smooth, connected, differentiable manifold $M$, which can be locally covered with coordinates $\{x^{\,\mu}\,|\,\mu=0,1,2,\ldots,n-1\}$ in open subsets of $\mathbb{R}^{\raisebox{1.0pt}{\scriptsize$n$}}\,$. For a more thorough introduction to curved spacetimes and general relativity see, for example, Ref.~\cite{Wald:GR}. Practically, we are here interested in the $(3+1)$-dimensional case and we shall work in units where $c=\hbar=1$ from now on. In addition the spacetime is equipped with a \emph{Lorentzian metric tensor}\index{Lorentz!metric}\index{metric tensor} $g$:
\begin{defi}\label{def:metric}\end{defi}
    \vspace*{-1.28cm}
    \begin{tabbing} \hspace*{3.0cm}\=\hspace*{1.5cm}\=\kill
        \>\textit{A \emph{Lorentzian metric} is a non-degenerate, symmetric, bilinear form $g(\,.\,,\,.\,)$}\\
	    \>\textit{with signature $(-,+,+,+)$ that maps two elements of each tangent space}\\
        \>\textit{ of $M$ to a real number. The line element is given by}
    \end{tabbing}
    \vspace*{-4mm}
    \begin{align}
        ds^{2}  &=\,g_{\mu\nu}\,dx^{\hspace*{1.0pt}\mu}dx^{\hspace*{1.0pt}\nu}\,.
        \label{eq:line element general metric}
    \end{align}
Summation is implied for repeated indices. Alternatively and equivalently, the convention $(+,-,-,-)$ may be chosen for the signature of the metric, which would result in sign changes in several of the following definitions. We shall keep the convention stated in Definition~\ref{def:metric} throughout this document.\\

\vspace*{-4mm}
A special case of the general (curved) metric $g_{\mu\nu}$ is the flat space \emph{Minkowski metric}\index{Minkowski metric}\label{page:Minkowski metric} $\eta_{\mu\nu}=\diag\{-1,1,1,1\}$ with the line element $ds^{2}=\eta_{\mu\nu}\,dx^{\hspace*{1.0pt}\mu}dx^{\hspace*{1.0pt}\nu}=-dt^{2}+dx^{2}+dy^{2}+dz^{2}\,$. The metric is not positive-definite, i.e., it is not Riemannian, but we can classify vectors and distances into three categories. Following our sign convention for the metric we define:
\begin{defi}\label{def:time-like,space-like,null}\end{defi}
    \vspace*{-1.25cm}
    \begin{tabbing} \hspace*{3.2cm}\=\hspace*{1.5cm}\=\hspace*{2.5cm}\=\kill
        \>\textit{A vector $v=v^{\,\mu}$, with $g(v,v)=g_{\mu\nu}v^{\,\mu}v^{\,\nu}=v^{\,\mu}v_{\mu}$ is called}\\[1mm]
        \> \> time-like\index{time-like}\> \textit{if\ \ $v^{\,\mu}v_{\mu}<0$\,,}\\
        \> \> null\> \textit{if\ \ $v^{\,\mu}v_{\mu}=0$\,,}\\
        \> \> space-like\index{space-like}\> \textit{if\ \ $v^{\,\mu}v_{\mu}>0$\,.}\\[1mm]
        \>\textit{Likewise, a curve $x^{\,\mu}(\lambda)$ is called time-like/null/space-like if its tangent}\\
        \>\textit{vector $v^{\,\mu}=dx^{\,\mu}(\lambda)/d\lambda$ is time-like/null/space-like at every point}.
    \end{tabbing}
All massive particles and, consequently, all observers follow time-like curves \textemdash\ worldlines\index{worldline}, while light is confined to null rays. Along any time-like curve~$\mathcal{C}_{t}$ the proper time~$\tau$\index{proper!time}, i.e., the time that elapses on a clock moving along the curve, is given by
\begin{align}
    \tau    &=\,\int_{\mathcal{C}_{t}}\sqrt{-ds^{2}}\,=\,\int_{\mathcal{C}_{t}}\sqrt{-g_{\mu\nu}\,dx^{\hspace*{1.0pt}\mu}dx^{\hspace*{1.0pt}\nu}}\,,
    \label{eq:proper time}
\end{align}
where the line element $ds^{2}$ is given by Eq.~(\ref{eq:line element general metric}). Similarly, along any space-like path $\mathcal{C}_{x}$ the proper length~$l$\index{proper!length} can be defined as
\begin{align}
    l   &=\,\int_{\mathcal{C}_{x}}\sqrt{ds^{2}}\,=\,\int_{\mathcal{C}_{x}}\sqrt{g_{\mu\nu}\,dx^{\hspace*{1.0pt}\mu}dx^{\hspace*{1.0pt}\nu}}\,.
    \label{eq:proper length}
\end{align}
For practical purposes the symmetries of a given metric~$g$ are of great interest. Such symmetries are represented by \emph{isometries}\index{isometry}~$G$, i.e., diffeomorphisms (a differentiable bijection with a differentiable inverse) that leave the metric invariant. Any one-parameter group of isometries such that
\begin{align}
    G(a)g   &=\,e^{\hspace*{1pt}a\hspace*{1pt}\xi}g\,=\,g\,,
    \label{eq:isometry}
\end{align}
is generated by a \emph{Killing vector}~$\xi$\index{Killing vector} that satisfies the \emph{Killing equation}
\begin{align}
    \nabla_{\hspace*{-1.5pt}\mu}\,\xi_{\hspace*{1pt}\nu}\,+\,\nabla_{\hspace*{-1.5pt}\nu}\,\xi_{\hspace*{1pt}\mu}  &=\,0\,.
    \label{eq:Killing equation}
\end{align}
Here $\nabla_{\hspace*{-1pt}\mu}$ is the covariant derivative with respect to the metric~$g\,$
(see Ref.~\cite{Wald:GR} for details). Alternatively, Killing vector fields may be defined via the \emph{Lie derivative}\index{Lie!derivative} of the metric, i.e., $\mathfrak{L}_{\xi}g=0$\,, see Ref.~\cite{Wald:GR}. For the analysis of quantum fields Killing vectors play an essential role in distinguishing positive and negative frequency solutions of the field equations, as we shall see in Section~\ref{sec:quantizing the Klein Gordon field}.

\section{The Klein-Gordon Field}\label{sec:the Klein Gordon field}

\subsection{The Classical Klein-Gordon Field}\label{sec:classical Klein Gordon field}

As a first representative of a quantum field let us consider the \emph{real scalar field}~$\phi(x)$. We start from the classical \emph{Lagrangian}\index{Lagrangian!real, scalar field} (density) for the free field~$\phi$ on a general (curved) background described by the metric~$g$, i.e.,
\begin{align}
    \mathcal{L} &=\,-\,\tfrac{1}{2}\sqrt{-\det g}\bigl(g^{\hspace*{1pt}\mu\nu}(\partial_{\mu}\,\phi)\,(\partial_{\nu}\,\phi)\,+\,\mathfrak{m}^{2}\phi^{2}\bigr)\,.
    \label{eq:real free scalar Lagrangian}
\end{align}
This Lagrangian describes a non-interacting field that is not coupled to the gravitational field (see, for instance, Ref.~\cite[p.~43]{BirrellDavies:QFbook}). The constant~$\mathfrak{m}$ will later be interpreted as the mass of the particles in the quantized theory. Having included the factor of $\sqrt{-\det g}$ in the Lagrangian we can straightforwardly write the action~$S$ as
\begin{align}
    S   &=\,\int\!d^{\hspace*{1pt}4}\hspace*{-1pt}x\,\mathcal{L}(\phi,\partial_{\mu}\phi)\,.
    \label{eq:action}
\end{align}
Varying the action~(\ref{eq:action}) with respect to the field, and demanding the action to be stationary, i.e., $\delta S=0$, one finds the usual \emph{Euler-Lagrange equations}\index{Euler-Lagrange equation}
\begin{align}
\frac{\partial\mathcal{L}}{\partial\phi}\,-\,
\partial_{\mu}\left(\frac{\partial\mathcal{L}}{\partial(\partial_{\mu}\phi)}\right)  &=\,0\,.
\label{eq:Euler-Lagrange equations}
\end{align}
Note that Eq.~(\ref{eq:Euler-Lagrange equations}) is covariant because $(\partial\mathcal{L}/\partial(\partial_{\mu}\phi))$ transforms as a vector density. For the Lagrangian~\eqref{eq:real free scalar Lagrangian} this \emph{action principle}\index{action principle} yields the \emph{curved spacetime Klein-Gordon equation}\index{Klein-Gordon!equation, curved spacetime}
\begin{align}
    \partial_{\mu}\bigl(\sqrt{-\det g}\,g^{\hspace*{1pt}\mu\nu}\,\partial_{\nu}\bigr)\phi\,
    -\,\sqrt{-\det g}\,\mathfrak{m}^{2}\phi  &=\,0\,.
    \label{eq:Klein Gordon equation general}
\end{align}
If the spacetime is equipped with a (translational) symmetry represented by a Killing vector~$\xi^{\mu}$, then the \emph{Noether current}\index{Noether current}~$J^{\hspace*{1pt}\mu}$, given by
\begin{align}
    J^{\mu} &=\,\xi^{\nu}\,T^{\hspace*{1pt}\mu}_{\hspace*{8pt}\nu}\,=\,\frac{1}{\sqrt{-\det g}}\,\xi^{\nu}
    \bigl(\frac{\partial\mathcal{L}}{\partial(\partial_{\mu}\phi)}\partial_{\nu}\phi\,-\,
    \delta^{\hspace*{1pt}\mu}_{\hspace*{8pt}\nu}\mathcal{L}\bigr)\,,
    \label{eq:Noether current curved spacetime}
\end{align}
where $T^{\hspace*{1pt}\mu}_{\hspace*{8pt}\nu}$ is the \emph{stress-energy-momentum tensor}\index{stress-energy-momentum tensor}, is conserved, i.e., $\nabla_{\mu}J^{\hspace*{1pt}\mu}=0\,$.

%
%
%



\subsection{Quantizing the Klein-Gordon Field}\label{sec:quantizing the Klein Gordon field}

Let us now turn to the solutions $\phii{n}$ \textemdash\ the \emph{mode functions} \textemdash\ of the Klein-Gordon equation~(\ref{eq:Klein Gordon equation general}). Following Ref.~\cite{BirrellDavies:QFbook} we define the (pseudo) inner product\index{Klein-Gordon!inner product}
\begin{align}
    (\,\phii{1}\,,\,\phii{2}\,)_{\raisebox{-1.7pt}{\scriptsize{KG}}}    &=\,
    -i\int_{\Sigma}d\Sigma^{\hspace*{1pt}\mu}\,
    \bigl(\phii{1}\partial_{\mu}\phii{2}^{*}\,-\,\phii{2}^{*}\partial_{\mu}\phii{1}\bigr)\,,
    \label{eq:KG inner product}
\end{align}
where $\Sigma$ is a spacelike Cauchy surface (assuming global hyperbolicity of the spacetime) \textemdash\ a surface that is intersected by every inextendible, causal (time-like or null) curve exactly once \textemdash\  and $d\Sigma^{\hspace{1pt}\mu}=g^{\hspace*{1pt}\mu\nu}d\Sigma_{\hspace{1pt}\nu}\,$, with $g_{\mu\lambda}\,g^{\hspace*{1pt}\lambda\nu}=\delta_{\mu}^{\hspace*{4pt}\nu}\,$. The volume form\index{volume form} $d\Sigma_{\hspace{1pt}\mu}$ for the three-surface~$\Sigma\,$ is given by~\cite[p.~10]{Takagi1986}
\begin{align}
    d\Sigma_{\hspace{1pt}\mu}   &=\, \frac{1}{3!}\sqrt{-\det g}\, \,\epsilon_{\mu\mu_{1}\mu_{2}\mu_{3}}\,dx^{\mu_{1}}\wedge dx^{\mu_{2}}\wedge dx^{\mu_{3}}\,,
    \label{eq:volume form}
\end{align}
where $\epsilon_{\mu\mu_{1}\mu_{2}\mu_{3}}$, with $\epsilon_{0123}=1$, is the totally antisymmetric Levi-Civita symbol.
Note that the inner product is independent of the chosen hypersurface~$\Sigma\,$, see Ref.~\cite{BirrellDavies:QFbook}. One can choose a complete set of orthonormal solutions with respect to the pseudo inner product~(\ref{eq:KG inner product}) such that $(\,\phii{m}\,,\,\phii{n}\,)_{\raisebox{-1.7pt}{\scriptsize{KG}}}=
-(\,\phii{m}^{*}\,,\,\phii{n}^{*}\,)_{\raisebox{-1.7pt}{\scriptsize{KG}}}=\delta_{mn}\,$, and $(\,\phii{m}\,,\,\phii{n}^{*}\,)_{\raisebox{-1.7pt}{\scriptsize{KG}}}=0\,$. The positive and negative frequency solutions (see below) to the Klein-Gordon equation then form Hilbert spaces, respectively. We should note here that we have assumed a discrete set of solutions for the sake of simplicity, but the treatment can easily be reformulated for a continuous spectrum.\\

In a general spacetime there is no preferred splitting into the solutions $\{\phii{n}\}$ and $\{\phii{n}^{*}\}$. Such a distinction can be uniquely made with respect to a time-like Killing vector field~$\xi^{\mu}\,$ and the corresponding conservation of energy [see Eq.~(\ref{eq:Noether current curved spacetime})]. The solutions can then be naturally split into positive frequency solutions $\phii{n}$ and negative frequency solutions $\phii{n}^{*}$ according to the signs of their eigenvalues,
\begin{subequations}
\label{eq:pos and neg freq solutions}
\begin{align}
    i\,\xi\,\phii{n}   &=\,+\,\omega_{n}\,\phii{n}\,,
    \label{eq:pos freq solutions}\\
    i\,\xi\,\phii{n}^{*}   &=\,-\,\omega_{n}\,\phii{n}^{*}\,,
    \label{eq:neg freq solutions}
\end{align}
\end{subequations}
where $\omega_{n}>0\,$. Because the Klein-Gordon product of Eq.~(\ref{eq:KG inner product}) is not positive-definite, linear combinations of \emph{mode functions} $\phii{n}$ and $\phii{n}^{*}$ cannot be interpreted as single particle wave functions. It thus becomes necessary to promote the \emph{Klein Gordon field}\index{Klein-Gordon!field} to an operator of the form
\begin{align}
    \phi    &=\,\sum\limits_{n}\bigl(\phii{n}\,\an{n}\,+\,\phii{n}^{*}\,\adn{n}\bigr)\,,
    \label{eq:Klein Gordon field}
\end{align}
where the \emph{annihilation operators}~$\an{n}$ and the \emph{creation operators}~$\adn{n}$ satisfy the \emph{commutation relations}
\begin{subequations}
\label{eq:canonical commutation relations}
\begin{align}
\comm{\an{m}}{\adn{n}}  &=\,\delta_{mn}\,,
\label{eq:canonical commutation relations delta}\\
\comm{\an{m}}{\an{n}}   &=\,\comm{\adn{m}}{\adn{n}}=0\,.
\label{eq:canonical commutation relations vanishing}
\end{align}
\end{subequations}
In Chapter~\ref{Chapter 4 Constructing Non Uniformly Moving Cavities} we are going to consider explicit solutions to the Klein-Gordon equation in both Minkowski and Rindler coordinates, both subject to appropriate cavity boundary conditions.

\subsection{The Bosonic Fock Space}\label{sec:bosonic Fock space}

We now turn to the physical interpretation of the annihilation and creation operators. As their names suggest, the operators $\an{n}$ annihilate a particle in the state $\phii{n}\,$, while the operators $\adn{n}$ create such a particle. The corresponding \emph{Fock space} is a Hilbert space that is constructed from a vacuum state~$\ket{0}$\index{vacuum!bosons} that does not contain any particles. Mathematically, the vacuum state is defined via the relation
\begin{align}
\an{n}\,\ket{0} &=\,0\ \ \forall\,n\,.
\end{align}
Any single boson~$(\operatorname{1-b})$ state~$\ket{\!\phi^{\operatorname{1-b}}\!}$ can be simply obtained by acting on the vacuum with a linear combination of creation operators, i.e.,
\begin{align}
    \ket{\!\phi^{\operatorname{1-b}}\!}    &=\,\sum\limits_{i}\,\theta_{i}\,\ket{\!\phii{i}\!}\,,
    \label{eq:general single boson state}
\end{align}
where $\theta_{i}\in\mathbb{C}\,$, $\ket{\!\phii{i}\!}=\adn{i}\ket{0}\,$ and $\sum_{i}\,|\theta_{i}|^{2}=1\,$. The states~$\ket{\!\phii{n}\!}$ form a complete basis of the single boson Hilbert space~$\mathcal{H}_{\operatorname{1-b}}\,$. The vacuum state, on the other hand, is an element of the Hilbert space $\mathcal{H}_{\operatorname{0-b}}=\mathds{C}$. When a second particle is added we need to keep in mind that the particles are indistinguishable from each other, such that we have to symmetrize the tensor product with respect to the exchange of the two particles
\begin{align}
    \ket{\!\phii{m},\phii{n}\!} &=\,\adn{m}\adn{n}\,\ket{0}\,=\,\ket{\!\phii{m}\!}\circ\ket{\!\phii{n}\!}\,:=\,
    \tfrac{1}{\sqrt{2}}\Bigl(\ket{\!\phii{m}\!}\otimes\ket{\!\phii{n}\!}\,+
    \,\ket{\!\phii{n}\!}\otimes\ket{\!\phii{m}\!}\Bigr)\,.
    \label{eq:two boson state definition}
\end{align}
The two-boson states are thus elements of the \emph{symmetrized} tensor product space of two single-boson Hilbert spaces, i.e.,
\begin{align}
    \mathcal{H}_{\operatorname{2-b}}   &=\,S\Bigl(\mathcal{H}_{\operatorname{1-b}}\otimes\mathcal{H}_{\operatorname{1-b}}\Bigr)\,.
    \label{eq:two boson space}
\end{align}
Similarly, states with higher particle content need to be symmetrized as well, and we can write the 
\emph{bosonic Fock space}\index{Fock!space, bosonic}~$\mathbb{F}$ as the direct sum over all boson numbers of the symmetrized Hilbert spaces, i.e.,
\begin{align}
    \label{eq:bosonic Fock space}
    \mathbb{F}(\mathcal{H}_{\operatorname{1-b}})    &=\,
    \bigoplus\limits_{m=0}^{\infty}S\Bigl(\mathcal{H}_{\operatorname{1-b}}^{\otimes\hspace*{0.5pt}m}\Bigr)
    =\,
    \mathcal{H}_{\operatorname{0-b}}\oplus\mathcal{H}_{\operatorname{1-b}}\oplus S\Bigl(\mathcal{H}_{\operatorname{1-b}}\otimes\mathcal{H}_{\operatorname{1-b}}\Bigr)\oplus\ldots\,,
\end{align}
where $
    \mathcal{H}^{\otimes\hspace*{0.5pt} m} 
$ denotes the $m$-fold tensor product and we write $\mathcal{H}_{\operatorname{0-b}}$ as $\mathcal{H}_{\operatorname{1-b}}^{\otimes\hspace*{0.5pt} 0}$. A general state in the space~$\mathbb{F}$ can be written as
\begin{align}
    \ket{\Phi^{\mathbb{F}}}    &=\,
    \theta_{\mbox{\tiny{$0$}}}\ket{0}\,+\,\sum\limits_{i\neq0}\theta_{i}\ket{\!\phii{i}\!}
    +
    \sum\limits_{j,k}\theta_{jk}\ket{\!\phii{j},\phii{k}\!}\,+\,\ldots\ ,
    \label{eq:bosonic Fock space general state}
\end{align}
where $\theta_{\mbox{\tiny{$0$}}},\theta_{i},\theta_{jk},\ldots\in\mathbb{C}\,$, and the vectors ``$\ket{\cdot}$" in all sectors of fixed particle content are understood as being extended to the total Fock space~$\mathbb{F}$ via the direct sum with zero vectors for all other sectors, e.g., $\ket{\!\phii{i}\!}=0_{\operatorname{0-b}}\oplus\ket{\!\phii{i}\!}\oplus0_{\operatorname{2-b}}\oplus\!\ldots\ $. For ease of notation we are going to make some adjustments to the way we denote these states. We shall use the \emph{occupation number}\index{Fock!states}\index{state!Fock} \emph{notation} where non-zero numbers of particles in each mode are indicated by integers with the corresponding mode labels as subscripts, for instance, $\ket{\!\phii{m},\phii{n}\!}=\ket{\!1_{m}\!}\ket{\!1_{n}\!}$, such that
\begin{align}
    \bra{\!1_{m}\!}\scpr{1_{n}}{1_{i}}\ket{\!1_{j}\!}    &=\,
    \delta_{ni}\delta_{mj}\,+\,\delta_{nj}\delta_{mi}\,,
    \label{eq:boson state normalization}
\end{align}
and we have dropped the symbol~$``\circ"$ for the symmetrized tensor product. For several excitations in the same mode~$i$ our conventions imply
\begin{subequations}
\label{eq:bosonic creation annihilation factors}
\begin{align}
    \an{i}\ket{\!n_{\hspace*{1pt}i}\!}   &=\,\sqrt{n_{\hspace*{1pt}i}}\ket{\!n_{\hspace*{1pt}i}-1\!}\,,
    \label{eq:bosonic annihilation factors}\\
    \adn{i}\ket{\!n_{\hspace*{1pt}i}\!}  &=\,\sqrt{n_{\hspace*{1pt}i}+1}\ket{\!n_{\hspace*{1pt}i}+1\!}\,.
    \label{eq:bosonic creation factors}
\end{align}
\end{subequations}
We note here in passing that we have chosen the split notation $\ket{\!1_{m}\!}\ket{\!1_{n}\!}$, rather than $\ket{\!1_{m},1_{n}\!}$ because this proves to be a more useful notation for computations with fermions in Part~\ref{partII:Shaking Entanglement}. 
Thus, we can rewrite Eq.~(\ref{eq:bosonic Fock space general state}) as
\begin{align}
    \ket{\Phi^{\mathbb{F}}}    &=\,
    \theta_{\mbox{\tiny{$0$}}}\ket{0}\,+\,\sum\limits_{i\neq0}\theta_{i}\ket{\!1_{i}\!}
    \,+\,
    \sum\limits_{j,k}\theta_{jk}\ket{\!1_{j}\!}\ket{\!1_{k}\!}\,+\,\ldots\ .
    \label{eq:bosonic Fock space general state new notation}
\end{align}

\subsection{Bosonic Bogoliubov Transformations}\label{sec:Bogoliubov transformations}

In Sections~\ref{sec:quantizing the Klein Gordon field} and~\ref{sec:bosonic Fock space} we have quantized the scalar field for a particular set of mode functions $\phii{n}\,$ and we have further classified them into positive and negative frequency solutions using a time-like Killing vector field. However, typically such choices are not unique, in other words, a different basis $\{\phiitilde{n},\phiitilde{n}^{*}\}$ can be chosen. Furthermore, if a different time-like Killing vector is chosen to separate particles (positive frequency) and antiparticles (negative frequency) then also the particle content of a given state will be affected. The transformations that connect the two choices of solutions are called \emph{Bogoliubov transformations}\index{Bogoliubov transformation!bosons}\index{transformation!Bogoliubov}.
\begin{defi}\label{def:Bogoliubov transformation bosons}\end{defi}
    \vspace*{-1.15cm}
    \begin{tabbing} \hspace*{3.2cm}\=\hspace*{1.5cm}\=\kill
        \> \textit{A \emph{bosonic Bogoliubov transformation} is an isomorphism (a bijective}\\
	    \> \textit{map) between two representations of the commutation relation algebra
            \index{algebra!commutation relation} of}\\
        \> \textit{Eq.~(\ref{eq:canonical commutation relations}). The transformation is unitary with respect
            to the (pseudo)}\\
        \> \textit{inner product $(\,.\,,\,.\,)_{\raisebox{-1.7pt}{\scriptsize{KG}}}$ of
            Eq.~(\ref{eq:KG inner product}) and the inner product of the bosonic}\\
        \> \textit{Fock space.}
    \end{tabbing}
For two given sets of mode functions $\{\phii{n},\phii{n}^{*}\}$ and $\{\phiitilde{n},\phiitilde{n}^{*}\}$ with mode operators $\{\an{n},\adn{n}\}$ and $\{\antilde{n},\adntilde{n}\}$, respectively, we can write the Bogoliubov transformation as a linear transformation of the mode functions and mode operators,
\begin{subequations}
\label{eq:bosonic Bogo transformation}
    \begin{align}
        \phiitilde{m}   &=\,\sum\limits_{n}\bigl(\alphamn{mn}\,\phii{n}\,+\,\betamn{mn}\,\phii{n}^{*}\bigr)\,,
        \label{eq:bosonic Bogo transformation mode functions}\\
        \antilde{m}   &=\,\sum\limits_{n}\bigl(\alpha^{*}_{mn}\,\an{n}\,-\,\beta^{\,*}_{mn}\,\adn{n}\bigr)\,,
        \label{eq:bosonic Bogo transformation operators}
    \end{align}
\end{subequations}
respectively, where the complex numbers
\begin{subequations}
\label{eq:bosonic bogo coefficients definition}
\begin{align}
    \alpha_{mn} &=\,(\,\tilde{\phi}_{m}\,,\,\phi_{n}\,)_{\raisebox{-1.7pt}{\scriptsize{KG}}}\,,
    \label{eq:bosonic bogo coefficients alpha def}\\[1mm]
    \beta_{mn}  &=\,-\,(\,\tilde{\phi}_{m}\,,\,\phi^{*}_{n}\,)_{\raisebox{-1.7pt}{\scriptsize{KG}}}\,,
    \label{eq:bosonic bogo coefficients beta def}
\end{align}
\end{subequations}
are called the \emph{Bogoliubov coefficients}\index{Bogoliubov coefficients bosons!general}. If the $\beta$-type coefficients are absent, the remaining $\alpha$ coefficients do not change the particle content of a given state, but simply shift excitations between different modes. The coefficients $\alpha_{mn}$ can hence be understood as a form of generalized rotation in the space of positive or negative frequency solutions, respectively. The coefficients $\beta_{mn}\,$, on the other hand, change the particle content, which can be easily seen from Eq.~(\ref{eq:bosonic Bogo transformation operators}) since $\beta_{mn}$ relates annihilation and creation operators. The unitarity of the transformation demands that the Bogoliubov coefficients satisfy\index{Bogoliubov identities bosons!general}
\begin{subequations}
\label{eq:bosonic Bogo unitarity}
    \begin{align}
        \sum\limits_{n}\bigl(\alpha_{nm}^{*}\,\alphamn{nl}\,-\,\betamn{nm}\,\beta_{nl}^{\,*}\,\bigr)  &=\,\delta_{ml}\,,
        \label{eq:bosonic Bogo unitarity 1}\\
        \sum\limits_{n}\bigl(\alphamn{nm}\,\beta_{nl}^{\,*}\,-\,\beta_{nm}^{\,*}\,\alphamn{nl}\,\bigr)   &=\,0\,.
        \label{eq:bosonic Bogo unitarity zero}
    \end{align}
\end{subequations}
\newpage
The linear transformations of the mode operators can alternatively be written as unitary operations on the states in the Fock space. These transformations are realized by exponentials of Hermitean operators that are quadratic in the mode operators. In other words, all linear, unitary transformations on the Fock space can be represented by Bogoliubov transformations. Throughout this thesis we will make extensive use of Bogoliubov transformations to describe physical transformations of states in the Fock space.

\section{The Dirac Field}\label{sec:the dirac field}

Now, let us turn to the description of fermionic fields, in particular, the \emph{Dirac field}.

\subsection{Quantizing the Dirac Field}\label{sec:quantizing the Dirac field}

Mirroring the approach when introducing the scalar field we start from a Lagrangian\index{Lagrangian!Dirac field} density for the Dirac field in a curved spacetime, given by
\begin{align}
    \mathcal{L} &=\,\sqrt{-\det g}\,
    \Bigl(\tfrac{i}{2}\bigl[\bar{\psi}\gamma^{\hspace*{1pt}\mu}\nabla_{\hspace*{-1.5pt}\mu}\psi\,-
        \,(\nabla_{\hspace*{-1.5pt}\mu}\bar{\psi})\gamma^{\hspace*{1pt}\mu}\psi\bigr]\,-\,\mathfrak{m}\bar{\psi}\psi\Bigr)\,.
    \label{eq:Dirac Lagrangian curved space}
\end{align}
where $\nabla_{\hspace*{-1.5pt}\mu}$ is the appropriate covariant derivative (see, e.g., Ref.~\cite{Weldon2001,AlsingStephensonKilian2009} for details), $\mathfrak{m}$ is the mass of the field excitations, $\gamma^{\hspace*{1pt}\mu}$ are the curved space \emph{Dirac~$\gamma$ matrices}\index{Dirac!gamma matrices} satisfying the anticommutation relation
\begin{align}
    \{\gamma^{\hspace*{1pt}\mu},\gamma^{\hspace*{1pt}\nu}\}   &=\,-\,2\hspace*{1pt}g^{\hspace*{1pt}\mu\nu}\,,
    \label{eq:curved space Dirac gammas}
\end{align}
and $\bar{\psi}$ denotes the {Dirac conjugate}~$\bar{\psi}=\psi^{\dagger}\gamma^{0}\,$. For a more detailed construction of this Lagrangian see Ref.~\cite[p.~85]{BirrellDavies:QFbook}. 
\\

As in Section~\ref{sec:classical Klein Gordon field} we invoke an action principle to obtain the Euler-Lagrange equations\index{Euler-Lagrange equation} [see Eqs.~(\ref{eq:action}) and (\ref{eq:Euler-Lagrange equations})] which here yields the \emph{curved space Dirac equation}\index{Dirac!equation, curved space}
\begin{align}
    \bigl(i\gamma^{\hspace*{1pt}\mu}\nabla_{\hspace*{-1.5pt}\mu}\,-\,\mathfrak{m}\bigr)\,\psi &=\,0\,.
    \label{eq:curved space Dirac equation}
\end{align}
One is then interested in a complete set of solutions that are orthonormal with respect to the inner product
\begin{align}
    (\,\psi_{1}\,,\,\psi_{2}\,)_{\raisebox{-1.7pt}{\scriptsize{D}}}    &=\,
    \int_{\Sigma}d\Sigma^{\hspace*{1pt}\mu}\,
    \bar{\psi}_{1}\gamma_{\mu}\psi_{2}\,,
    \label{eq:Dirac inner product}
\end{align}
with the conventions for $d\Sigma^{\hspace*{1pt}\mu}$ as used in Eq.~(\ref{eq:KG inner product}). In the presence of a time-like Killing vector field the solutions can be meaningfully classified into positive and negative frequency solutions. However, since we will allow the excitations of the Dirac field to carry electric charge we will choose different symbols for the mode operators that annihilate or create particles or antiparticles. In addition we use non-negative (negative) numbers to label the (anti)particle solutions, such that the quantized field can be written as
\begin{align}
\psi    &=\,\sum\limits_{n\geq0}\,b_{n}\psi_{n}\,+\,\sum\limits_{n<0}\,c_{n}^{\dagger}\psi_{n}\,,
\label{eq:Dirac field}
\end{align}
where the operators $b_{n}$ annihilate a Dirac fermion\index{Dirac!fermions} in the state $\psi_{n}\,$, while the $c_{n}$ annihilate an antifermion. As before we have simplified the discussion to a discrete spectrum for further convenience. The mode operators satisfy the \emph{anticommutation relations}
\begin{subequations}
\label{eq:anticomm relations}
    \begin{align}
        \anticomm{b_{m}}{b_{n}^{\dagger}} &=\,\anticomm{c_{m}}{c_{n}^{\dagger}}\,=\,\delta_{mn}\,,
        \label{eq:anticomm non-vanishing}\\[1.5mm]
        \anticomm{b_{m}}{b_{n}} &=\,\anticomm{c_{m}}{c_{n}}\,=\,
        \anticomm{b_{m}}{c_{n}}\,=\,\anticomm{b_{m}}{c_{n}^{\dagger}}\,=\,0\,,
        \label{eq:anticomm vanishing}
    \end{align}
\end{subequations}
where $\anticomm{.}{.}$ denotes the anticommutator. We will study some explicit examples for solutions to the Dirac equation in Chapter~\ref{Chapter 4 Constructing Non Uniformly Moving Cavities} when we consider Dirac fields contained in cavities. For now we are more interested in the construction of the fermionic Fock space.

\subsection{The Fermionic Fock Space}\label{sec:fermionic Fock space}

\index{Fock!space, fermionic|(}


In analogy to our approach in the bosonic case, let us now construct the fermionic Fock space. As previously, we start from a vacuum state\index{vacuum!fermions} that is annihilated by all annihilation operators, $b_{m}\ket{0}=c_{n}\ket{0}=0\ \, \forall\,m\geq0,n<0$\,. The creation operators $b_{m}^{\dagger}$ and $c_{n}^{\dagger}$ acting upon the vacuum state $\ket{0}$ will populate the vacuum with single excitations of particles and antiparticles, respectively, i.e.,
\begin{subequations}
\label{eq:single fermion excitations}
    \begin{align}
    \ket{\!\psi_{m}\!}   &=\,b_{m}^{\dagger}\,\ket{0}\,,
    \label{eq:single fermion excitation}\\
    \ket{\!\psi_{n}\!}   &=\,c_{n}^{\dagger}\,\ket{0}\,,
    \label{eq:single antifermion excitation}
    \end{align}
\end{subequations}
where we assume that the mode labels~$m\geq0$ and~$n<0$ distinguish the particle and antiparticle mode solutions. As can be quickly seen from this property and Eq.~(\ref{eq:anticomm non-vanishing}), the states $\ket{\!\psi_{i}\!}$ are orthonormal and they 
further form a complete basis of the single-fermion Hilbert space $\mathcal{H}_{\operatorname{1-f}}$, whereas $\ket{0}\in\mathcal{H}_{\operatorname{0-f}}\neq\mathcal{H}_{\operatorname{1-f}}$. We write a general state in the fermionic single-excitation space $\mathcal{H}_{\operatorname{1-f}}$ as
\begin{align}
    \ket{\!\psi^{\operatorname{1-f}}\!}    &=\,\sum\limits_{i}\,\mu_{i}\,\ket{\!\psi_{i}\!}\,,
    \label{eq:general single fermion state}
\end{align}
with $\mu_{i}\in\mathbb{C}$ and $\sum_{i}|\mu_{i}|^{2}=1$ such that $\scpr{\psi^{\operatorname{1-f}}}{\psi^{\operatorname{1-f}}}=1\,$. The form of the states of Eq.~(\ref{eq:general single fermion state}) may be further restricted by superselection rules. For instance, conservation of charge would exclude any superpositions of states of different charge. We shall not explicitly include such restrictions, but any superselection rule can be applied to the Fock space construction we present here. Let us now turn to states of multiple fermions. A second fermion can be added to the state~(\ref{eq:single fermion excitations}) by the action of another creation operator $b_{m}^{\dagger}$ or $c_{n}^{\dagger}\,$, e.g.,
\begin{align}
    b_{m}^{\dagger}b_{n}^{\dagger}\,\ket{0} &\propto\,\ket{\!\psi_{m},\psi_{n}\!}\,.
    \label{eq:two fermion excitation}
\end{align}
Clearly, the anticommutation relations (\ref{eq:anticomm relations}) require the two-fermion state to be
antisymmetric with respect to the exchange of the mode labels $m$ and $n$. We therefore define
\begin{align}
    \ket{\!\psi_{m},\psi_{n}\!} &=\,b_{m}^{\dagger}b_{n}^{\dagger}\,\ket{0}\,=\,
    \ket{\!\psi_{m}\!}\wedge\ket{\!\psi_{n}\!}\,=\,
    \tfrac{1}{\sqrt{2}}\Bigl(\ket{\!\psi_{m}\!}\otimes\ket{\!\psi_{n}\!}\,-
    \,\ket{\!\psi_{n}\!}\otimes\ket{\!\psi_{m}\!}\Bigr)\,.
    \label{eq:two fermion state definition}
\end{align}
The two-fermion states are thus elements of the antisymmetrized tensor product space of two single-fermion Hilbert spaces, i.e.,
\begin{align}
    \mathcal{H}_{\operatorname{2-f}}   &=\,\bar{S}\Bigl(\mathcal{H}_{\operatorname{1-f}}\otimes\mathcal{H}_{\operatorname{1-f}}\Bigr)\,,
    \label{eq:two fermion space}
\end{align}
and a general state within this space can be written as
\begin{align}
    \ket{\!\psi^{\operatorname{2-f}}\!}    &=\,\sum\limits_{i,j}\,\mu_{ij}\,\ket{\!\psi_{i},\psi_{j}\!}\,,
    \label{eq:general two fermion state}
\end{align}
where the coefficients $\mu_{ij}\in\mathbb{C}$ form an antisymmetric matrix. States with more than two fermions can then be
constructed by antisymmetrizing over the corresponding number of single-fermion states. Finally, the \emph{fermionic Fock space} $\bar{\mathbb{F}}$ is simply given as the direct sum over all fermion numbers of the antisymmetrized Hilbert spaces, i.e.,
\begin{align}
    \bar{\mathbb{F}}(\mathcal{H}_{\operatorname{1-f}})    &=\,
    \bigoplus\limits_{m=0}^{\infty}\bar{S}\Bigl(\mathcal{H}_{\operatorname{1-f}}^{\otimes m}\Bigr)\,=\,
    \mathcal{H}_{\operatorname{0-f}}\oplus\mathcal{H}_{\operatorname{1-f}}\oplus
    \bar{S}\Bigl(\mathcal{H}_{\operatorname{1-f}}\otimes\mathcal{H}_{\operatorname{1-f}}\Bigr)\oplus\ldots\,.
    \label{eq:fermionic Fock space}
\end{align}
where, as before, $
    \mathcal{H}^{\otimes \hspace*{0.5pt}m} 
$ denotes the $m$-fold tensor product and we write $\mathcal{H}_{\operatorname{0-f}}$ as $\mathcal{H}_{\operatorname{1-f}}^{\otimes 0}$. A~general state in the space $\bar{\mathbb{F}}$ can be written as
\begin{align}
    \ket{\Psi^{\bar{\mathbb{F}}}}    &=\,
    \mu_{\mbox{\tiny{$0$}}}\ket{0}\,+\,\sum\limits_{i\neq0}\mu_{i}\ket{\!\psi_{i}\!}\,+\,
    \sum\limits_{j,k}\mu_{jk}\ket{\!\psi_{j}\!}\wedge\ket{\!\psi_{k}\!}\,+\,\ldots\,,
    \label{eq:fermionic Fock space general state}
\end{align}
where $\mu_{\mbox{\tiny{$0$}}}, \mu_{i}, \mu_{jk},\ldots\in\mathbb{C}\,$, and the vectors ``$\ket{\cdot}$" in all sectors of fixed particle content are understood as being extended to the total Fock space~$\bar{\mathbb{F}}$ via the direct sum with zero vectors for all other sectors, e.g., $\ket{\!\psi_{i}\!}=0_{\operatorname{0-f}}\oplus\ket{\!\psi_{i}\!}\oplus0_{\operatorname{2-f}}\oplus\!\ldots\ $. Let us now simplify the notation. To distinguish more clearly from the bosonic case we will from now on denote states in the fermionic Fock space by double-lined Dirac notation, i.e., $\fket{.}$ instead of $\ket{.}$, where the antisymmetric ``wedge'' product is implied when two vectors are multiplied, i.e., $\fket{.}\fket{.}=\fket{.}\wedge\fket{.}$. Furthermore, let us again use the common \emph{occupation number notation} and write~$1_{n}$ instead of $\psi_{n}$ to denote an excitation in the mode~$n$. 
With this convention in mind we can rewrite Eq.~(\ref{eq:fermionic Fock space general state}) as
\begin{align}
    \fket{\Psi}    &=\,
    \mu_{\mbox{\tiny{$0$}}}\fket{0}\,+\,\sum\limits_{i\neq0}\mu_{i}\fket{\!1_{i}\!}\,+\,
    \sum\limits_{j,k}\mu_{jk}\fket{\!1_{j}\!}\fket{\!1_{k}\!}\,+\,\ldots\ .
    \label{eq:fermionic Fock space general state rewritten}
\end{align}
For the adjoint space we use the convention [compare to Eq.~(\ref{eq:two fermion state definition})]
\begin{align}
    \fbra{\!1_{n}\!}\fbra{\!1_{m}\!}    &:=\,
    \fbra{0}b_{n}b_{m}\,=\,\Bigl(b_{m}^{\dagger}b_{n}^{\dagger}\,\fket{0}\Bigr)^{\dagger}\,=\,-\,
    \tfrac{1}{\sqrt{2}}\Bigl(\bra{\!\psi_{n}\!}\otimes\bra{\!\psi_{m}\!}\,-
    \,\bra{\!\psi_{m}\!}\otimes\bra{\!\psi_{n}\!}\Bigr)\,,
    \label{eq:adjoint space convention}
\end{align}
\vspace*{-1mm}
which allows us to write
\begin{align}
    \fbra{\!1_{m}\!}\fscpr{1_{n}}{1_{i}}\fket{\!1_{j}\!}    &=\,
    \delta_{ni}\delta_{mj}\,-\,\delta_{nj}\delta_{mi}\,,
    \label{eq:fermion state normalization}
\end{align}
which is convenient for computations in the fermionic Fock space. It should be noted that, in standard quantum information notation, e.g., as used in Chapter~\ref{chapter:QI}, the position of a ``ket'' corresponds to a particular ordering of the subspaces with respect to the tensor product structure of the total space. Here, however, there is no tensor product structure corresponding to different modes according to which the vectors $\fket{.}$ can be naturally ordered. We shall return to this issue in Section~\ref{sec:Entanglement in Fermionic Quantum Fields} in Chapter~\ref{chapter:Entanglement in Relativistic Quantum Fields}.

\index{Fock!space, fermionic|)}

\subsection{Bogoliubov transformations of the Dirac Field}\label{sec:Bogoliubov transformations of the Dirac Field}

Similarly as for the bosonic case (see Definition~\ref{def:Bogoliubov transformation bosons}) one can define a change of basis in the set of mode solutions to the Dirac equation as a fermionic Bogoliubov transformation\index{Bogoliubov transformation!fermions}\index{transformation!Bogoliubov}.
\begin{defi}\label{def:Bogoliubov transformation fermions}\end{defi}
    \vspace*{-1.00cm}
    \begin{tabbing} \hspace*{3.0cm}\=\hspace*{1.5cm}\=\kill
        \> \textit{A \emph{fermionic Bogoliubov transformation} is an isomorphism (a bijective}\\
	    \> \textit{map) between two representations of the anticommutation relation
            algebra\index{algebra!anticommutation relation}}\\
        \> \textit{of Eq.~(\ref{eq:anticomm relations}). The transformation is unitary with respect to the inner product}\\
        \> \textit{$(\,.\,,\,.\,)_{\raisebox{-1.7pt}{\scriptsize{D}}}$ of Eq.~(\ref{eq:Dirac inner product})
            and the inner product of the fermionic Fock space.}
    \end{tabbing}
Given two sets of mode solutions to the Dirac equation~(\ref{eq:Dirac field}), $\{\psi_{m}\}$ and $\{\tilde{\psi}_{n}\}$ we can write the \emph{fermionic Bogoliubov transformation} as
\begin{align}
\tilde{\psi}_{m}    &=\,\sum\limits_{n}\,A_{mn}\psi_{n}\,.
\label{eq:Dirac field bogo transformation}
\end{align}
Although our notation allows us to use only a single symbol for all fermionic Bogoliubov coefficients\index{Bogoliubov coefficients fermions!general}, it is obvious that coefficients $A_{mn}=(\,\psi_{n}\,,\,\tilde{\psi}_{m}\,)_{\raisebox{-1.7pt}{\scriptsize{D}}}$ with $m,n\geq0$ or $m,n<0$ are $\alpha$-type coefficients, while those with subscripts with mixed signs represent $\beta$-type coefficients responsible for particle creation phenomena. The unitarity of the transformation is expressed as the unitarity of the matrix $A=(A_{mn})$, i.e.,\index{Bogoliubov identities fermions!general}
\begin{align}
    A^{\dagger}\,A  &=\,\mathds{1}\,.
    \label{eq:fermionic bogo identity}
\end{align}

%
%
%
%
%
%
%
%

%

    \newpage\ \\
    \chapter{Entanglement in Relativistic Quantum Fields}\label{chapter:Entanglement in Relativistic Quantum Fields}

With the definitions and methodology of Chapters~\ref{chapter:QI} and~\ref{chapter:Quantum Fields in Flat and Curved Spacetimes} at our disposal we now turn to some elementary concerns in relativistic quantum information (RQI)\ \textemdash\ entanglement in relativistic quantum fields. The appropriate relativistic treatment of quantum correlations requires to work with quantum fields and the corresponding Fock spaces. As in the previous chapter we separate the discussion of bosonic and fermionic quantum fields into Sections~\ref{sec:entanglement in bosonic quantum fields} and~\ref{sec:Entanglement in Fermionic Quantum Fields}, respectively. Starting with bosons we review the tools available (see, e.g., Ref.~\cite{Adesso:PhDThesis2007} for a detailed introduction) for the important class of \emph{Gaussian states} in Sections~\ref{sec:gaussian states}--\ref{sec:gaussian entanglement}. We include a short discussion of teleportation with Gaussian states and the construction of \emph{entanglement resonances} as introduced in Ref.~\cite{BruschiDraganLeeFuentesLouko2013} and Ref.~\cite[(\ref{Paper:BruschiFriisFuentesWeinfurtner2013})]{BruschiFriisFuentesWeinfurtner2013}
. Finally, quantum information techniques for \emph{fermionic Fock spaces} originally presented in Ref.~\cite[(\ref{Paper:FriisLeeBruschi2013})]{FriisLeeBruschi2013} are examined in Section~\ref{sec:Entanglement in Fermionic Quantum Fields}.

\section{Entanglement in Bosonic Quantum Fields}\label{sec:entanglement in bosonic quantum fields}

The first question encountered when studying entanglement in bosonic Fock spaces is the choice of bipartitions. In other words, one has to address the question of selecting appropriate subsystems. Since the particle content is not fixed it becomes necessary to consider instead the entanglement between different modes of the quantum field. The Fock space is not naturally equipped with a tensor product structure with respect to different modes. Nonetheless, the symmetrization in Eq.~(\ref{eq:bosonic Fock space}) allows a unique one-to-one mapping between states in $\mathbb{F}(\mathcal{H}_{\operatorname{1-b}})$ and a tensor product space $\mathcal{H}_{1}\otimes\mathcal{H}_{2}\otimes\mathcal{H}_{3}\otimes\ldots$, where the subscripts~$(1,2,3,\ldots)$ label the modes of the quantum field. Note that such an argument cannot be made in a straightforward way for fermions, as discussed in Section~\ref{sec:Entanglement in Fermionic Quantum Fields}.

In addition, working in a bosonic Fock space naturally raises the question how the infinite dimensions are handled. Practically there are two simple ways to circumvent this complication. The first option is to justify a truncation to finite dimensions by considering density operators in $\mathbb{F}(\mathcal{H}_{\operatorname{1-b}})$ that have finite rank. We shall consider such an approach in Section~\ref{sec:Bosonic State Transformation}. The other option is to switch from the Fock space to the \emph{phase space}\index{phase space}, which we shall explore in the following sections.

\subsection{Continuous Variables: Gaussian States}\label{sec:gaussian states}

\index{state!Gaussian|(}\index{Gaussian!state|(}

Instead of the Hilbert space description we have used in Chapter~\ref{chapter:QI} quantum systems can be represented by a \emph{characteristic function}\index{characteristic function} in \emph{phase space}\index{phase space}, e.g., the \emph{Wigner function}\index{Wigner function}~$W(q,p)$ for $n$-modes, given by~(see, e.g., \cite[p.~173]{GarrisonChiao:QuantumOptics})
\begin{align}
    W(q,p)  &=\,\frac{1}{\pi^{n}}\int_{\mathbb{R}^{\raisebox{1pt}{\tiny{$n$}}}}d^{\hspace*{1pt}n}
    \hspace*{-1pt}\tilde{q}\,\bra{q-\tilde{q}}\rho\ket{q+\tilde{q}}\,
    e^{i\,\tilde{q}^{\hspace*{0.5pt}T}\hspace*{-1pt}p}\,,
    \label{eq:Wigner function}
\end{align}
where $q,p\in\mathbb{R}^{\raisebox{1pt}{\tiny{$n$}}}$ and $\ket{q}$ are the eigenstates of the quadrature operator $\hat{q}_{i}:=\frac{1}{\sqrt{2}}(a_{i}+a^{\dagger}_{i})$, i.e., $\hat{q}_{i}\ket{q}=q_{i}\ket{q}$ with $q=(q_{1},q_{2},\ldots,q_{i},\ldots,q_{n})^{T}$. The Wigner function is a \emph{quasi probability distribution} since it can take on negative values. Note that the quadratures~$q$ and~$p$ are phase space variables, but do not necessarily correspond to positions and momenta in spacetime. For the class of \emph{Gaussian states} the defining feature is that the Wigner function, or other characteristic functions of choice (see Ref.~\cite[p.~30]{Adesso:PhDThesis2007}), are multivariate Gaussian distributions. Such distributions are completely determined by the vector of first moments $\left\langle\mathbb{X}\right\rangle_{\!\rho}\,$, where
\begin{align}
    \mathbb{X}:=\bigl(  \hat{q}_{1},\hat{p}_{1},
                        \hat{q}_{2},\hat{p}_{2},\ldots,
                        \hat{q}_{n},\hat{p}_{n}\bigr)^{\mathrm{T}}\,,
    \label{eq:first moments}
\end{align}
and the real, symmetric \emph{covariance matrix}\index{covariance!matrix}~$\Gamma$ with components
\begin{align}
    \Gamma_{ij}:=\bigl<\mathbb{X}_{i}\mathbb{X}_{j}+\mathbb{X}_{j}\mathbb{X}_{i}\bigr>_{\!\rho}
    -2\bigl<\mathbb{X}_{i}\bigr>_{\!\rho}\bigl<\mathbb{X}_{j}\bigr>_{\!\rho}\,.
    \label{eq:covariance matrix}
\end{align}
Here $\left\langle\,.\,\right\rangle_{\!\rho}$ is the expectation value in the state $\rho$, see Eq.~(\ref{eq:expectation value mixed states}), and the operators $a_{n}$ and $a_{n}^{\dagger}$ have been combined into the \emph{quadrature operators}\index{quadratures}
\begin{subequations}
\label{eq:quadratures}
\begin{align}
    \hat{q}_{n} &:=\,\frac{1}{\sqrt{2}}(a_{n}+a^{\dagger}_{n})\,,
    \label{eq:generalized position}\\
    \hat{p}_{n} &:=\,\frac{-i}{\sqrt{2}}(a_{n}-a^{\dagger}_{n})\,.
    \label{eq:generalized momentum}
\end{align}
\end{subequations}
From Eq.~(\ref{eq:canonical commutation relations}) it immediately follows that the Hermitean quadrature operators\index{quadratures} satisfy the canonical commutation relations\index{canonical!commutation relation}
\begin{subequations}
\label{eq:canonical comm relations quadratures}
\begin{align}
    \comm{\hat{q}_{k}}{\hat{p}_{l}} &=\,i\,\delta_{kl}\,,
    \label{eq:canonical comm relations quadratures nonvanishing}\\
    \comm{\hat{q}_{k}}{\hat{q}_{l}} &=\,\comm{\hat{p}_{k}}{\hat{p}_{l}} \,=\,0\,.
    \label{eq:canonical comm relations quadratures zero}
\end{align}
\end{subequations}

Let us now consider some examples for Gaussian states. The simplest and most fundamental representative is the vacuum state~$\ket{0}$\index{vacuum!bosons}, for which the first moments vanish and, in our conventions, the covariance matrix is proportional to the identity matrix, $\Gamma_{\hspace*{-1pt}\mathrm{vac}}=\mathds{1}\,$. The family of \emph{coherent states}\index{state!coherent}\index{coherent state}~$\ket{\alpha}$ is obtained from the vacuum by displacements in phase space. Physically, coherent states can be used, for instance, for the description of the electromagnetic field of a laser beam. Note that the notation using the symbol $\alpha$ is customary in the literature (see, e.g., Ref.~\cite[p.~150]{GarrisonChiao:QuantumOptics}) and is not to be confused with the notation for the matrix of Bogoliubov coefficients $\alpha_{mn}$ from Eq.~(\ref{eq:bosonic bogo coefficients alpha def}). For any mode~$k$ the displacement operator\index{displacement operator}
\begin{align}
    D_{k}(\alpha)    &=\,e^{\,\alpha\hspace*{1pt}a_{k}^{\dagger}\,+\,\alpha^{*}a_{\raisebox{-2pt}{\tiny{$k$}}}}
    \label{eq:displacement operator}
\end{align}
with $\alpha\in\mathbb{C}$, takes the vacuum state to a coherent state, $\hat{D}_{k}(\alpha)\ket{0}=\ket{\alpha_{k}}\,$. The displacement shifts the corresponding first moments to $\left\langle\,\hat{q}_{k}\,\right\rangle=\sqrt{2}\operatorname{Re}(\alpha)$ and $\left\langle\,\hat{p}_{k}\,\right\rangle=\sqrt{2}\operatorname{Im}(\alpha)$, respectively, while it leaves the covariance matrix unchanged, i.e., $\Gamma_{\hspace*{-1pt}\alpha}=\mathds{1}\,$. A \emph{single-mode squeezed state}\index{state!single-mode squeezed}\index{squeezed state!single-mode}, on the other hand, is obtained by a quadratic combination of creation and annihilation operators acting on the vacuum, i.e., for mode~$k$ we have
\begin{align}
    U_{\raisebox{0pt}{\tiny{$\mathrm{S}$}}}(s_{k})   &=\,e^{\raisebox{1.5pt}{\scriptsize{$\,
    \tfrac{s_{k}}{2}(\hspace*{1pt}a^{\dagger\hspace*{0.5pt}2}_{k}\,-\,a_{k}^{\raisebox{1.5pt}{\tiny{$2$}}})$}}}
    \label{eq:single mode squeezing Fock space}
\end{align}
where $s_{k}\in\mathbb{R}$ is the single-mode squeezing parameter\index{squeezing parameter!single-mode} for mode~$k\,$. While coherent states can be thought of as quasi-classical, i.e., they approximate the classical description of light as closely as possible (see, e.g., Ref.~\cite[p.~148]{GarrisonChiao:QuantumOptics}), squeezed states are considered to be truly non-classical. For our discussion it is sufficient to consider a real squeezing parameter, since squeezing along other quadratures can be achieved by applying additional local rotations. The covariance matrix for a single-mode squeezed state $\ket{s_{k}}=U_{\raisebox{0pt}{\tiny{$\mathrm{S}$}}}(s_{k})\ket{0}$ is given by
\begin{align}
    \Gamma_{\raisebox{0pt}{\tiny{$\mathrm{S}$}}}(s_{k})    &=\,
    \begin{pmatrix}e^{2s_{k}}   &   0   \\  0   &   e^{-2s_{k}}\end{pmatrix}\,.
    \label{eq:single mode squeezed covariance matrix}
\end{align}
As can be easily seen from the form of~$\Gamma_{\raisebox{0pt}{\tiny{$\mathrm{S}$}}}(s_{k})\,$, a positive squeezing parameter indicates squeezing in the $p$-quadrature, while the orthogonal $q$-quadrature is broadening, such that the product of the covariances remains constant, satisfying the Heisenberg bound. All the Gaussian states we have mentioned so far are pure, i.e., $\det(\Gamma)=1\,$. As a last example in this section let us consider the family of mixed \emph{thermal states}\index{state!thermal}\index{thermal state}\label{page:thermal state}. The covariance matrix for a thermal state of mode~$k$ with frequency~$\omega_{k}$ at temperature~$T$ is given by~$\Gamma_{\hspace*{-1pt}\mathrm{th}}=\coth\bigl(\tfrac{\hbar\hspace*{0.8pt}\omega_{k}}
{2\hspace*{0.5pt}\raisebox{-0.3pt}{\scriptsize{$k_{\mathrm{B}}T$}}}\bigr)\mathds{1}$, where we have explicitly inserted Planck's constant~$\hbar$ and Boltzmann's constant~$k_{\mathrm{B}}$ for clarity. For vanishing first moments, the average particle number $\bar{n}_{k}$ of mode~$k$ is given by
\begin{align}
    \bar{n}_{k}=\bigl<a_{k}^{\dagger}a_{k}\bigr>=\tfrac{1}{4}\bigl[\tr(\Gamma)-2\bigr]\,,
    \label{eq:average particle number mode k}
\end{align}
such that the average particle number of the thermal state is distributed according to Bose-Einstein statistics, i.e., $\bar{n}_{k}=\bigl(\exp[\tfrac{\hbar\hspace*{0.8pt}\raisebox{0.4pt}{\scriptsize{$\omega_{k}$}}}{k_{\mathrm{B}}T}]-1\bigr)^{-1}$.

\subsection{Symplectic Operations}\label{sec:symplectic operations}

To formalize our treatment of the phase space the canonical commutation relations~(\ref{eq:canonical comm relations quadratures}) can be conveniently combined to define the \emph{symplectic form}~$\Omega$ via the relation
\begin{align}
    \comm{\mathbb{X}_{k}}{\mathbb{X}_{l}}   &=\,i\,\Omega_{kl}\,,
    \label{eq:symplectic form definition}
\end{align}
such that the symplectic form\index{symplectic!form} for~$n$ modes has the matrix representation
\begin{align}
    \Omega  &=\,\bigoplus\limits_{i=1}^{n}\Omega_{i}\ \ \ \ \mbox{with}\ \ \ \
    \Omega_{i}\,=\,\begin{pmatrix}\,0 & 1\,\, \\ -1 & 0\,\,\end{pmatrix}\,.
    \label{eq:symplectic form}
\end{align}
The symplectic form can be used to express a \emph{bona fide} condition
, i.e., $\Gamma+i\Omega\geq0$, which is satisfied by any covariance matrix~$\Gamma$ representing a physical state. Linear transformations~$S$ that leave~$\Omega$ invariant, i.e.,
\begin{align}
    S\,\Omega\, S^{T}   &=\,\Omega\,,
    \label{eq:def of symplectic transformations}
\end{align}
are called \emph{symplectic transformations}\index{symplectic!transformation}\index{transformation!symplectic} and they correspond to unitaries on the Fock space that are generated by Hamiltonians that are quadratic in the quadrature operators, see, e.g., Eq.~(\ref{eq:single mode squeezing Fock space}). Such transformations, along with displacements~(\ref{eq:displacement operator}) and partial tracing over any number of modes preserve the Gaussian character of states. The symplectic transformations on $n$~modes form the \emph{real, symplectic group}\index{symplectic!group}\index{group!symplectic} $\operatorname{Sp}(2n,\mathbb{R})$\,. It is precisely this group of operations that can be realized by Bogoliubov transformations of the kind of Eq.~(\ref{eq:bosonic Bogo transformation operators}), and displacements can be incorporated by adding a constant offset~$\underline{\alpha}$ to each mode
\begin{align}
        \antilde{m}   &=\,\sum\limits_{n}\bigl(\alpha^{*}_{mn}\,\an{n}\,-\,\beta^{\,*}_{mn}\,\adn{n}\bigr)
        \,+\,\underline{\alpha}_{m}\,,
        \label{eq:bosonic Bogo transformation with displacement}
\end{align}
where the parameters~$\underline{\alpha}_{m}$ correspond to the displacements [see Eq.~(\ref{eq:displacement operator})] of each mode. Symplectic transformations can be written explicitly in terms of the Bogoliubov coefficients\index{Bogoliubov transformation!bosons} in a straightforward fashion~\cite[(\ref{Paper:FriisFuentes2013})]{FriisFuentes2013},
\begin{align}
    \begin{scriptsize}
    S\,=\,\begin{pmatrix}
        \mathcal{M}_{11}  &   \mathcal{M}_{12}  &   \mathcal{M}_{13}   &   \ldots  \\
        \mathcal{M}_{21}  &   \mathcal{M}_{22}  &   \mathcal{M}_{23}   &   \ldots  \\
        \mathcal{M}_{31}  &   \mathcal{M}_{32}  &   \mathcal{M}_{33}   &   \ldots  \\
        \vdots            &   \vdots            &   \vdots             &   \ddots  \\
    \end{pmatrix}
    \end{scriptsize}\,,
    \label{eq:Gaussian n-mode Bogo transformation}
\end{align}
i.e., we decompose the transformation matrix into the $2\times2$ sub-blocks $\mathcal{M}_{mn}$ given by
\begin{align}
    \mathcal{M}_{mn}\,=\,
        \begin{pmatrix}
            \operatorname{Re}(\alpha_{mn}\,-\,\beta_{mn})     &
            \operatorname{Im}(\alpha_{mn}\,+\,\beta_{mn}) \\[1.5mm]
            -\operatorname{Im}(\alpha_{mn}\,-\,\beta_{mn})    &
            \operatorname{Re}(\alpha_{mn}\,+\,\beta_{mn}) \\
        \end{pmatrix}\,.
    \label{eq:M Bogo matrix}
\end{align}
The transformed covariance matrix $\tilde{\Gamma}$ is then simply obtained as
\begin{align}
    \tilde{\Gamma}\,=\,S\,\Gamma\,S^{T}\,,
    \label{eq:general bogo transformed covariance matrix}
\end{align}
and partial tracing over any modes is achieved by simply removing the corresponding rows and columns from the covariance matrix. It is convenient to characterize different types of symplectic transformations, following Ref.~\cite{WolfEisertPlenio2003} we distinguish:
\begin{enumerate}[\hspace*{0.5cm}(i)]
    \item{\emph{Passive symplectic transformations}\index{symplectic!transformation, passive}
    \index{passive (optical) transformation}~$S_{\raisebox{0pt}{\tiny{$\mathrm{P}$}}}$ are represented by orthogonal, symplectic matrices $S_{\raisebox{0pt}{\tiny{$\mathrm{P}$}}}^{\hspace*{1pt}T}
    S_{\raisebox{-1pt}{\tiny{$\mathrm{P}$}}}=\mathds{1}\,$ and they form a subgroup of $\operatorname{Sp}(2n,\mathbb{R})$. Practically, passive transformations can be realized, for instance, by passive/linear optical elements, such as (ideal) \emph{beam splitters} or \emph{phase space rotations}.\label{eq:passive sympl trafo}}
    \item{\emph{Active symplectic transformations}\index{symplectic!transformation, active}
    \index{active (optical) transformation}~$S_{\raisebox{0pt}{\tiny{$\mathrm{A}$}}}$ are represented by symmetric, symplectic matrices $S_{\raisebox{0pt}{\tiny{$\mathrm{A}$}}}^{\hspace*{1pt}T}=
    S_{\raisebox{-1pt}{\tiny{$\mathrm{A}$}}}\,$. Active transformations, such as single- and two-mode squeezing, can be realized by active/non-linear optical elements and they change the energy and average particle number, as opposed to passive transformations.\label{eq:active sympl trafo}}
\end{enumerate}
\label{page:passive and active transformations}
Every symplectic transformation can be decomposed into passive and active transformations, in particular we may decompose any symplectic matrix as~$S=S_{\raisebox{0pt}{\tiny{$\mathrm{P}$}}}
S_{\raisebox{0pt}{\tiny{$\mathrm{A}$}}}$, see Ref.~\cite{ArvindDuttaMukundaSimon1995}. We have already encountered the single-mode squeezing operation~(\ref{eq:single mode squeezing Fock space}) as an example for an active symplectic transformation and it gains additional significance via the \emph{Boch-Messiah reduction}~\cite{Braunstein2005}\index{Boch-Messiah reduction}.

\begin{theorem}\label{thm:Bloch-Messiah reduction}\end{theorem}
    \vspace*{-1.28cm}
    \begin{tabbing} \hspace*{3.0cm}\=\hspace*{1.5cm}\=\kill
            \> \textit{Every $n$-mode symplectic transformation~$S$ can be written in the} Bloch-\\[2mm]
            \> -Messiah decomposition\\[2mm]
            \> \> \textit{$S=S_{\raisebox{0pt}{\tiny{$\mathrm{P}$}}}\bigl(
            S_{\raisebox{0pt}{\tiny{$\mathrm{S}$}}}(s_{1})\oplus
            S_{\raisebox{0pt}{\tiny{$\mathrm{S}$}}}(s_{2})\oplus\cdots\oplus
            S_{\raisebox{0pt}{\tiny{$\mathrm{S}$}}}(s_{n})\bigr)
            S^{\hspace*{1pt}\prime}_{\raisebox{0pt}{\tiny{$\mathrm{P}$}}}$\,,}\\[2mm]
            \> \textit{where $S_{\raisebox{0pt}{\tiny{$\mathrm{S}$}}}(s_{i})$ is the symplectic representation of the single-mode squeezing}\\[2mm]
            \> \textit{of Eq.~(\ref{eq:single mode squeezing Fock space}) in mode~$i\,$, while $S_{\raisebox{0pt}{\tiny{$\mathrm{P}$}}}$ and $S^{\hspace*{1pt}\prime}_{\raisebox{0pt}{\tiny{$\mathrm{P}$}}}$ are passive $n$-mode operations.}
    \end{tabbing}
A proof of Theorem~\ref{thm:Bloch-Messiah reduction} can be found in Ref.~\cite{Braunstein2005}. Finally, let us consider the diagonalization in phase space. Every~$n$ mode covariance matrix can be brought to the so-called \emph{Williamson normal form}\index{Williamson normal form}~$\Gamma_{\mathrm{W}}$, given by
\begin{align}
    \Gamma_{\mathrm{W}}&=\,\bigoplus\limits_{i=1}^{n}
    \begin{pmatrix}\,\nu_{i} & 0\,\, \\ 0 & \nu_{i}\,\,\end{pmatrix}\,,
    \label{eq:Williamson normal form}
\end{align}
by a symplectic transformation, $\Gamma_{\mathrm{W}}=S\hspace*{1pt}\Gamma S^{T}\,$, see Ref.~\cite{Williamson1936}. The \emph{symplectic eigenvalues}\index{symplectic!eigenvalues}~$\nu_{i}\geq1$, which are invariant under global symplectic transformations, form the symplectic spectrum\index{symplectic!spectrum} of the covariance matrix and they can be computed as the eigenvalues of $|i\hspace*{0.5pt}\Omega\hspace*{0.5pt}\Gamma|$\,. In addition to the symplectic spectrum the determinant of the covariance matrix, $\det(\Gamma)$, is a global symplectic invariant. This can be easily seen from Eq.~(\ref{eq:def of symplectic transformations}), which implies that $\det(S)=1\,$ for all symplectic operations~$S\,$. We now have all the ingredients for the discussion of entanglement in phase space.

\index{state!Gaussian|)}\index{Gaussian!state|)}

\subsection{Two-Mode Squeezed States}\label{sec:two-mode squeezed states}

\index{state!Gaussian|(}\index{Gaussian!state|(}

The powerful tools of Gaussian states can be used to study the entanglement between different bosonic modes. As we have mentioned, a complete description of Gaussian states is provided by the first and second moments. However, the first moments can be arbitrarily adjusted using the displacements of Eq.~(\ref{eq:displacement operator}) \textemdash\ operations that act \emph{locally} on the phase spaces of particular modes. Therefore, all necessary information about the entanglement between Gaussian states is encoded solely in the \emph{covariance matrix}~$\Gamma$. For the intents and purposes of this thesis it is sufficient to consider two-mode Gaussian states only, but the analysis can be extended to more modes if so desired~\cite{AdessoSerafiniIlluminati2006}. The paradigm for an entangled Gaussian state is the \emph{two-mode squeezed state}\index{state!two-mode squeezed}\index{squeezed state!two-mode}, which can be produced, for example, in parametric down conversion using non-linear optical crystals, see, e.g., Ref.~\cite[pp.~391]{GarrisonChiao:QuantumOptics}. For two modes~$k$ and~$k\pr$ we can create such a state by acting on the vacuum with the operator
\begin{align}
    U_{\raisebox{0pt}{\tiny{$\mathrm{TMS}$}}}(r)   &=\,e^{\raisebox{1.5pt}{\scriptsize{$\,
    r(\hspace*{1pt}a^{\dagger}_{k}a^{\dagger}_{k\pr}\,-\,a_{k}a_{k\pr})$}}}\,,
    \label{eq:two mode squeezing Fock space}
\end{align}
where $r\in\mathbb{R}$ is called two-mode squeezing parameter\index{squeezing parameter!two-mode}. Similar as in the case of single-mode squeezing it is possible to redefine the operator in Eq.~(\ref{eq:two mode squeezing Fock space}) using a complex squeezing parameter. However, this is equivalent to applying local rotations \textemdash\ squeezing along different directions. For a real squeezing parameter the two-mode covariance matrix~$\Gamma_{\raisebox{0pt}{\tiny{$\mathrm{TMS}$}}}(r)$ representing the state~$U_{\raisebox{0pt}{\tiny{$\mathrm{TMS}$}}}(r)\ket{0}$ is given by
\begin{align}
    \Gamma_{\raisebox{0pt}{\tiny{$\mathrm{TMS}$}}}(r)   &=\,
    \begin{pmatrix}
    \cosh(2r)   &   0   &   \sinh(2r)   &   0   \\
    0   &   \ \cosh(2r)   &   0   &   -\sinh(2r)   \\
    \sinh(2r)   &   0   &   \cosh(2r)   &   0   \\
    0   &   -\sinh(2r)  &   0   &   \ \cosh(2r)
    \end{pmatrix}\,.
    \label{eq:two mode squeezed covariance matrix}
\end{align}
The corresponding symplectic transformation~$S_{\raisebox{0pt}{\tiny{$\mathrm{TMS}$}}}(r)\,$, such that $S_{\raisebox{0pt}{\tiny{$\mathrm{TMS}$}}}(r)\Gamma_{\hspace*{-1pt}\mathrm{vac}}\hspace*{1pt}
S^{\hspace*{1pt}T}_{\raisebox{0pt}{\tiny{$\mathrm{TMS}$}}}(r)=\Gamma_{\raisebox{0pt}{\tiny{$\mathrm{TMS}$}}}(r)\,$, is given by $S_{\raisebox{0pt}{\tiny{$\mathrm{TMS}$}}}(r)=
\Gamma_{\raisebox{0pt}{\tiny{$\mathrm{TMS}$}}}(\tfrac{r}{2})\,$.
Alternatively, one may initially prepare the two modes in an antisymmetrically (i.e., $s_{k}=-s_{k\pr}=s$) single-mode squeezed state $\Gamma_{\raisebox{0pt}{\tiny{$\mathrm{S}$}}}(s)\oplus\Gamma_{\raisebox{0pt}{\tiny{$\mathrm{S}$}}}(-s)$ by applying local squeezing operations and combine the two modes on an ideal, balanced beam splitter\label{page:beam splitter}\index{beam splitter} to obtain a two-mode squeezed state $\Gamma_{\raisebox{0pt}{\tiny{$\mathrm{TMS}$}}}(r=s)\,$ \textemdash\ a straightforward application of the Bloch-Messiah reduction\index{Boch-Messiah reduction} (Theorem~\ref{thm:Bloch-Messiah reduction}). The ideal beam splitter for the modes~$k$ and~$k\pr$ is realized by a passive symplectic transformation represented~[see Eqs.~(\ref{eq:Gaussian n-mode Bogo transformation}) and~(\ref{eq:M Bogo matrix})] by the Bogoliubov coefficients $\alpha_{kk}=-\alpha_{k\pr k\pr}=\cos\Theta$ and $\alpha_{kk\pr}=\alpha_{k\pr k}=\sin\Theta\,$, while all other coefficients are zero, and the beam splitter is called balanced for $\Theta=\tfrac{\pi}{4}\,$. This construction is in fact even more profound, every pure, two-mode Gaussian state is locally equivalent to a two-mode squeezed state. In other words, for pure, two-mode Gaussian states the entanglement is fully characterized by the squeezing parameter~$r$\index{squeezing parameter!two-mode} and every such state can be brought to the form of Eq.~(\ref{eq:two mode squeezed covariance matrix}) by local rotations and single-mode squeezings that do not change the entanglement.

\subsubsection{The Standard Form of Two-mode Gaussian States}

For general, mixed states of two modes~$k$ and~$k\pr$, given by the covariance matrix
\begin{align}
    \Gamma   &=\,
    \begin{pmatrix}
    \Gammak{k}   &   C   \\
    C^{T}   &   \Gammak{k\pr}
    \end{pmatrix}\,,
    \label{eq:two mode covariance matrix}
\end{align}
where $\Gammak{k}\,$, $\Gammak{k\pr}$ and $C$ are real $2\times2$ matrices, it is customary to introduce the standard form~$\Gamma_{\raisebox{0pt}{\tiny{$\mathrm{St}$}}}\,$, given by
\vspace*{-3mm}
\begin{align}
    \Gamma_{\raisebox{0pt}{\tiny{$\mathrm{St}$}}}   &=\,
    \begin{pmatrix}
    \GammaStk{k}   &   C_{\raisebox{0pt}{\tiny{$\mathrm{St}$}}}   \\
    C_{\raisebox{0pt}{\tiny{$\mathrm{St}$}}}   &   \GammaStk{k\pr}
    \end{pmatrix}\,,
    \label{eq:two mode standard form}
\end{align}
where $\GammaStk{k}=\diag\{\gamma_{k},\gamma_{k}\}\,$, $\GammaStk{k\pr}=\diag\{\gamma_{k\pr},\gamma_{k\pr}\}\,$, and $C_{\raisebox{0pt}{\tiny{$\mathrm{St}$}}}=\diag\{\gamma_{+},\gamma_{-}\}\,$, with $\gamma_{k},\gamma_{k\pr}, \gamma_{\pm}\in\mathbb{R}\,$. Every covariance matrix can be converted to its standard form by local symplectic operations\index{symplectic!local operations}\index{transformation!local symplectic}~$S_{k}\oplus S_{k\pr}\,$~\cite{Simon2000}, and the standard form is unique if an ordering is specified for $\gamma_{\pm}$, e.g., $\gamma_{+}\geq|\gamma_{-}|$. The determinants of the $2\times2$ sub-blocks, i.e., $\det(\GammaStk{k})=\det(\Gammak{k})$, $\det(\GammaStk{k\pr})=\det(\Gammak{k\pr})$, and $\det(C_{\raisebox{0pt}{\tiny{$\mathrm{St}$}}})=\det(C)$, are \emph{local symplectic invariants}\index{symplectic!local invariants}. Two-mode Gaussian states for which $\GammaStk{k}=
\GammaStk{k\pr}$ are called \emph{symmetric}\index{state!symmetric two-mode Gaussian}.

\subsection{Entanglement of Gaussian States}\label{sec:gaussian entanglement}

We now turn to the quantification of entanglement of Gaussian states. As mentioned, the entanglement between two modes in a pure, Gaussian state is completely characterized by the two-mode squeezing parameter~$r$. However, we wish to find a quantification that also relates to our previous treatment of non-Gaussian states in Chapter~\ref{chapter:QI}. Fortunately, we can directly connect to the tools introduced in Section~\ref{sec:entanglement measures}. As shown in Ref.~\cite{Simon2000} the Peres-Horodecki criterion\index{PPT criterion} (Theorem~\ref{thm:PPT theorem}) provides a necessary and sufficient condition for entanglement of two-mode Gaussian states.\\

The partial transposition\index{partial transposition} is implemented on the phase space by a mirror operation \textemdash\ a sign flip \textemdash\ of the $p$-quadrature of one of the modes. The ``partially transposed" covariance matrix $\oversmile{\Gamma}$ is then simply $\oversmile{\Gamma}=\oversmile{T}_{\hspace*{-1pt}k\pr}\Gamma\hspace*{0.8pt}\oversmile{T}_{\hspace*{-1pt}k\pr}\,$, where $\oversmile{T}_{\hspace*{-1pt} k\pr}=\mathds{1}\oplus\diag\{1,-1\}\,$. In complete analogy to the usual partial transposition, the symplectic eigenvalues [see Eq.~(\ref{eq:Williamson normal form})] of~$\oversmile{\Gamma}$ do not necessarily correspond to a physical state anymore. The smallest eigenvalue $\num{-}\geq0$ of $|i\hspace{0.3pt}\Omega\hspace*{0.3pt}\oversmile{\Gamma}|\,$, where $\,\spectr(i\hspace{0.3pt}\Omega\hspace*{0.3pt}\oversmile{\Gamma})=\{\pm\num{-},\pm\num{+}\}\,$ with $\num{+}\geq\num{-}\,$, can be smaller than~1, see Ref.~\cite{Simon2000}.
\begin{theorem}\label{thm:PPT Gaussian}\end{theorem}
    \vspace*{-1.35cm}
    \begin{tabbing} \hspace*{3.0cm}\=\hspace*{4.0cm}\=\kill
            \> \textit{A two-mode Gaussian state represented by the covariance matrix~$\Gamma$ is}\\[2mm]
            \> \textit{entangled if, and only if the smallest eigenvalue of $|i\hspace*{0.3pt}\Omega\oversmile{\Gamma}|$ is smaller than~1\,,}\\[2mm]
            \> \> \textit{\ $0\,\leq\,\num{-}\,<1\,$}.
    \end{tabbing}
The smallest symplectic eigenvalue~$\num{-}$ (we will omit the suffix ``of the partial transpose" from now on and rely on the distinction made by the ``$\,\oversmile{\,}\,$" symbol) can then be used to construct the usual negativity measures.

\subsubsection{Entanglement Measures for Gaussian States}

Both the logarithmic negativity~$E_{\mathcal{N}}$ [see Eq.~(\ref{eq:logarithmic negativity})] and the negativity~$\mathcal{N}$ (see Definition~\ref{def:negativity}) are monotonously decreasing functions of~$\num{-}$~\cite{AdessoSerafiniIlluminati2004} such that we can write the simple expressions
\begin{subequations}
\label{eq:Gaussian negativities}
\begin{align}
    E_{\mathcal{N}}(\Gamma)   &=\,\operatorname{max}\{0,-\,\log_{2}(\num{-})\}\,,
    \label{eq:logarithmic negativity Gaussian}\\[2mm]
    \mathcal{N}(\Gamma) &=\,\operatorname{max}\{0,(1\,-\,\num{-})/2\num{-}\}\,.
    \label{eq:negativity Gaussian}
\end{align}
\end{subequations}
For symmetric two-mode states, i.e., for which the local symplectic invariants $\GammaStk{k}$ and $\GammaStk{k\pr}$ are the same, it is even possible to compute the entanglement of formation~$E\sub{0.2}{0}{\mathrm{F}}$ (see Definition~\ref{def:entanglement of formation}). The involved minimization procedure reveals that the corresponding state decomposition is realized within the set of two-mode Gaussian states~\cite{GiedkeWolfKruegerWernerCirac2003} and the entanglement of formation can be expressed as
\begin{align}
    E\sub{0.2}{0}{\mathrm{F}}  &=\,
    \left\{\,  \begin{matrix}
    \mathfrak{h}(\num{-})  &   \ \ \mbox{if}\ \ 0\,\leq\,\num{-}\,<\,1\\[0.5mm]
    0   &   \ \ \mbox{if}\ \ \num{-}\geq\,1
    \end{matrix}\right.\,,
    \label{eq:EoF for symmetric Gaussian states}
\end{align}
where the entropic function~$\mathfrak{h}(\num{-})$ is defined as
\begin{align}
\mathfrak{h}(\num{-})    &:=\,
\frac{(1+\num{-})^{2}}{4\,\num{-}}
\ln\frac{(1+\num{-})^{2}}{4\,\num{-}}\,-\,
\frac{(1-\num{-})^{2}}{4\,\num{-}}
\ln\frac{(1-\num{-})^{2}}{4\,\num{-}}\,.
    \label{eq:entropy measure}
\end{align}
Operationally it is also quite straightforward to check if a given covariance matrix is a symmetric two-mode Gaussian state \textemdash\ a necessary and sufficient condition is~$\det(\Gammak{k})=\det(\Gammak{k\pr})$\,. Moreover, for symmetric states the smallest symplectic eigenvalue~$\num{-}$ provides a unique characterization of the entanglement, i.e., all (known) entanglement measures are monotonously decreasing functions of~$\num{-}$ and they provide the same ordering of entangled states~\cite{AdessoIlluminati2005b}. Unfortunately, this is no longer true for non-symmetric two-mode Gaussian states \textemdash\ the answer to the question \emph{``Is one state more entangled than another?"} generally depends on the chosen measure of entanglement~\cite{AdessoIlluminati2005b}.\\

As an example, let us consider again the two-mode squeezed state of Eq.~(\ref{eq:two mode squeezed covariance matrix}). The smallest symplectic eigenvalue for this state is directly related to the two-mode squeezing parameter~$r$ via the relation $\num{-}(r)=e^{-2|r|}\,$. It can be easily seen that maximal entanglement can only be achieved in the limit $r\rightarrow\infty\,$. Practically, squeezing parameters of approximately $\tfrac{1}{2}\ln(2)$ can be reached in the current experiments with optical squeezing in the microwave regime, see, e.g., Refs.~\cite{EichlerEtal2011,FlurinRochMalletDevoretHuard2012,MenzelEtal2012}.

\subsubsection{Teleportation with Gaussian States}\label{page:Gaussian teleportation}

\index{teleportation}
Let us illustrate the role of continuous variable entanglement with the help of the teleportation protocol. In Section~\ref{sec:quantum teleportation} we have discussed this fascinating application of entangled resource states for two qubits. A continuous variable teleportation protocol for the teleportation of coherent states~(\ref{eq:displacement operator}) may be introduced in complete analogy to the qubit scenario, see Refs.~\cite{Vaidmann1994,BraunsteinKimble1998}. In this version of the teleportation scheme the observers share an entangled two-mode Gaussian state with vanishing first moments. Alice wishes to teleport an unknown coherent state, i.e., its first moments, to Bob. To this end she mixes the unknown state with her mode of the resource state on a balanced beam splitter (see p.~\pageref{page:beam splitter}) and performs \emph{homodyne detection}\index{homodyne detection} \textemdash\ projective measurements in quadrature eigenstates (see Ref.~\cite[pp.~49]{FerraroOlivaresParis:GaussianReview2005} for details) \textemdash\  on the two outputs. As usual she sends the measurement results to Bob via a classical channel. Bob, in turn, can then perform the necessary displacements to retrieve, approximately, the unknown input state.\\

A crucial difference to the qubit teleportation lies in the imperfection of the shared entanglement. Perfect correlations between two modes of a continuous variable state \textemdash\  EPR correlations \textemdash\ would require an infinite amount of squeezing. The \emph{teleportation fidelity}\index{teleportation!fidelity, Gaussian states}~$\mathcal{F}(\Gamma)$ for the continuous variable teleportation protocol with an entangled resource state represented by the covariance matrix~$\Gamma$ is given by~\cite{MariVitali2008}
\begin{align}
    \mathcal{F} &=\, \frac{2}{\sqrt{4\,+\,2\,\tr(N)+\det(N)}}\,,
    \label{eq:Gaussian teleportation fidelity}
\end{align}
where the $2\times2$ matrix $N=Z\Gammak{k}Z+ZC+C^{T}Z+\Gammak{k\pr}$ is given in terms of the sub-blocks of the two-mode covariance matrix from Eq.~(\ref{eq:two mode covariance matrix}) and $Z=\diag\{1,-1\}\,$. The fidelity is strictly smaller than~1 for finite squeezing. We note in passing a typographical error in Eq.~(1) of Ref.~\cite[(\ref{Paper:FriisLeeTruongSabinSolanoJohanssonFuentes2013})]{FriisLeeTruongSabinSolanoJohanssonFuentes2013} as compared to Eq.~(\ref{eq:Gaussian teleportation fidelity}) above. As in the qubit case the teleportation fidelity may be optimized over local operations that do not increase the shared entanglement~\cite{AdessoIlluminati2005a}. The fidelity~$\mathcal{F}_{\mathrm{opt}}\,$, optimized over all local Gaussian operations, can be bounded by functions that depend only on the smallest symplectic eigenvalue of the partial transpose~\cite{MariVitali2008}, i.e.,
\begin{align}
    \frac{1\,+\,\num{-}}{1\,+\,3\num{-}}    &\leq\,\mathcal{F}_{\mathrm{opt}}\,\leq\,\frac{1}{1\,+\,\num{-}}\,.
    \label{eq:bounds on Gaussian fidelity}
\end{align}
The upper bound becomes tight for symmetric two-mode Gaussian states, see Ref.~\cite{AdessoIlluminati2005a}. We shall make use of these tools to study the effects of non-uniform motion on the continuous variable teleportation protocol in Chapter~\ref{Chapter 7 Degradation of Entanglement between Moving Cavities}.

\subsubsection{Entanglement Resonances}\label{sec:entanglement resonances}

The last sections have demonstrated that the entanglement of Gaussian states can be easily described, quantified and used for tasks such as quantum teleportation. In addition it is useful to understand how entanglement can be enhanced by successive symplectic transformations, for instance in applications in analogue gravity systems~\cite[(\ref{Paper:BruschiFriisFuentesWeinfurtner2013})]{BruschiFriisFuentesWeinfurtner2013}, or non-uniform cavity motion, see Refs.~\cite{BruschiDraganLeeFuentesLouko2013} and~\cite[(\ref{Paper:FriisHuberFuentesBruschi2012})]{FriisHuberFuentesBruschi2012}.\\

As we have established in Section~\ref{sec:symplectic operations} any symplectic transformation can be decomposed into a passive, orthogonal transformation $S_{\raisebox{0pt}{\tiny{$\mathrm{P}$}}}$ and an active, symmetric transformation $S_{\raisebox{0pt}{\tiny{$\mathrm{A}$}}}=S_{\raisebox{0pt}{\tiny{$\mathrm{A}$}}}^{\,T}\,$, i.e., $S=S_{\raisebox{0pt}{\tiny{$\mathrm{P}$}}}S_{\raisebox{0pt}{\tiny{$\mathrm{A}$}}}$ (see Ref.~\cite{ArvindDuttaMukundaSimon1995}). For two modes the passive transformations include rotations and beam splitters, while the active transformations can involve single-mode and two-mode squeezing. Let us now consider a symplectic transformation~$S$ for two modes that leaves the quadratures of the individual modes on equal footing, i.e., without any overall single-mode squeezing. The active part of the transformation thus consists only of pure two-mode squeezing, $S_{\raisebox{0pt}{\tiny{$\mathrm{A}$}}}=S_{\raisebox{0pt}{\tiny{$\mathrm{TMS}$}}}(r)\,$.\\

In addition, we assume that the initial covariance matrix of the two modes is proportional to the identity, $\Gamma=\mathds{1}\,$. This is the case if the initial state is the vacuum state\index{vacuum!bosons} or, given that the two modes have the same frequency, a thermal state at temperature~$T$ (see p.~\pageref{page:thermal state}). For simplicity let us pick the vacuum state. If the physical transformation that is embodied by the symplectic matrix $S=S_{\raisebox{0pt}{\tiny{$\mathrm{P}$}}}S_{\raisebox{0pt}{\tiny{$\mathrm{TMS}$}}}(r)$ can be repeated, then the entanglement will grow with the number of repetitions if the \emph{resonance condition}\index{resonance condition} (see~\cite{BruschiDraganLeeFuentesLouko2013} and \cite[(\ref{Paper:BruschiFriisFuentesWeinfurtner2013})]{BruschiFriisFuentesWeinfurtner2013})
\begin{align}
    \comm{S}{S^{\,T}}   &=\,0\,.
    \label{eq:resonance condition}
\end{align}
is satisfied. This condition has a very intuitive interpretation. The condition is fulfilled if the state $\Gamma_{\raisebox{0pt}{\tiny{$\mathrm{TMS}$}}}$ that is created by the two-mode squeezing, $\Gamma_{\raisebox{0pt}{\tiny{$\mathrm{TMS}$}}}=
S_{\raisebox{0pt}{\tiny{$\mathrm{TMS}$}}}
S_{\raisebox{0pt}{\tiny{$\mathrm{TMS}$}}}^{\,T}$
is invariant under the passive transformation $S_{\raisebox{0pt}{\tiny{$\mathrm{P}$}}}$, i.e.,
\begin{align}
    S_{\raisebox{0pt}{\tiny{$\mathrm{P}$}}}\,
    \Gamma_{\raisebox{0pt}{\tiny{$\mathrm{TMS}$}}}\,
    S_{\raisebox{0pt}{\tiny{$\mathrm{P}$}}}^{\,T}
    &=\,\Gamma_{\raisebox{0pt}{\tiny{$\mathrm{TMS}$}}}\,.
    \label{eq:resonance condition new}
\end{align}
At this stage it is essential to note that the two-mode squeezing operations form a one parameter subgroup of the symplectic group $\operatorname{Sp}(2n,\mathbb{R})$,
\begin{align}
    S_{\raisebox{0pt}{\tiny{$\mathrm{TMS}$}}}(r_{1})\,S_{\raisebox{0pt}{\tiny{$\mathrm{TMS}$}}}(r_{2})    &=\,
    S_{\raisebox{0pt}{\tiny{$\mathrm{TMS}$}}}(r_{1}+r_{2})\,.
    \label{eq:two mode squeezing subgroup}
\end{align}
It is then straightforward to see that the resonance condition of Eq.~(\ref{eq:resonance condition}) indicates when the repeated symplectic transformation realizes consecutive squeezing along the same direction and, consequently, accumulates entanglement. Such procedures have been suggested for various physical systems, including entanglement generation in BECs for analogue gravity experiments~\cite[(\ref{Paper:BruschiFriisFuentesWeinfurtner2013})]{BruschiFriisFuentesWeinfurtner2013} and for modes of quantum fields in non-uniformly moving cavities, see Refs.~\cite{BruschiDraganLeeFuentesLouko2013} and~\cite[(\ref{Paper:FriisHuberFuentesBruschi2012})]{FriisHuberFuentesBruschi2012}. We will turn our attention back on entanglement resonances in Section~\ref{sec:Resonances of Entanglement Generation}.\\

\index{state!Gaussian|)}\index{Gaussian!state|)}


For now, let us return to fermionic systems and analyze the description of quantum information tasks for anticommuting field operators.

\section{Entanglement in Fermionic Quantum Fields}\label{sec:Entanglement in Fermionic Quantum Fields}

Fermionic systems have been analyzed as agents for quantum information processing in a multitude of
studies, ranging from discussions of fermionic modes of relativistic quantum fields~\cite{Shi2004,AlsingFuentes-SchullerMannTessier2006,FuentesMannMartin-MartinezMoradi2010,
BruschiLoukoMartin-MartinezDraganFuentes2010,Martin-MartinezFuentes2011,FriisKoehlerMartinMartinezBertlmann2011,
SmithMann2012,FriisLeeBruschiLouko2012,FriisBruschiLoukoFuentes2012,FriisHuberFuentesBruschi2012,
MonteroMartin-Martinez2012b}, and fermionic lattices~\cite{Zanardi2002},
to discussions of the entanglement between fixed numbers of indistinguishable particles~\cite{SchliemannLossMacDonald2000,SchliemannCiracKusLewensteinLoss2001,PaskauskasYou2001,
LiZengLiuLong2001,EckertSchliemannBrussLewenstein2002,Shi2003,WisemanVaccaro2003,WisemanBartlettVaccaro2004,GhirardiMarinatto2004,
CabanPodlaskiRembielinskiSmolinskiWalczak2005,IeminiVianna2012}. In the latter case, only pure states of fixed particle numbers are considered and a selection of entanglement measures are available, see, e.g., Ref.~\cite{WisemanVaccaro2003}. However, these restrictions seem to be much more limiting than required. From the point of view of quantum information theory it is natural to ask for an extension to incoherent mixtures of quantum states, see Section~\ref{subsec:mixed states}. Furthermore, from the perspective of a relativistic description particle numbers are not usually conserved, i.e., the particle content of a given pure state is observer dependent (see, e.g., the discussion in Section~\ref{sec:Bogoliubov transformations} or~\ref{sec:Bogoliubov transformations of the Dirac Field}). The description of fermionic entanglement should therefore include coherent and incoherent mixtures of different particle numbers. Any required superselection rules, e.g., for (electric) charge~\cite{StrocchiWightman1974} or parity, can then be considered as special cases of such a framework.\\

In the light of this fact it is therefore reasonable to consider the entanglement between fermionic modes, in a similar way as is conventionally done for bosonic modes, see, for instance, the treatment in Section~\ref{sec:entanglement in bosonic quantum fields}. In this section we give an account of the material published in Ref.~\cite[(\ref{Paper:FriisLeeBruschi2013})]{FriisLeeBruschi2013}, albeit with slightly altered notations to better fit the framework of this thesis. We show that the entanglement of a system of fermionic modes can be defined unambiguously by enforcing a physically reasonable definition of its subsystems. This procedure is completely independent of any superselection rules.\\

A central question that appears in practical situations is: \emph{Can fermionic modes be considered as qubits?} The short answer to this question is ``No." Due to the \emph{Pauli exclusion principle}\index{Pauli exclusion principle}, fermionic modes are naturally restricted to two degrees of freedom, i.e., each mode can be unoccupied or contain a single excitation. This has provided many researchers with an \emph{ad hoc} justification for the comparison with qubits \textemdash\ two-level systems used in quantum information, which has incited debates among scientists, see, e.g., the exchange in Refs.~\cite{MonteroMartin-Martinez2011b,BradlerJauregui2012,MonteroMartin-Martinez2012a,Bradler2011}. In limited situations certain techniques from the study of qubits can indeed be applied to fermionic systems. However, while mappings between fermionic systems and qubits are possible in principle, e.g., via the Jordan-Wigner transformation~\cite{BanulsCiracWolf2007}, the problem lies in the consistent mapping between the subsystems. In the following we shall give a more precise answer to the question above, along with a detailed description of the problem as published in Ref.~\cite[(\ref{Paper:FriisLeeBruschi2013})]{FriisLeeBruschi2013}.\\

Any superselection rules further restrict the possible operations that can be performed on single-mode subsystems, and it was argued that this should lead to a modified definition of the entanglement between modes~\cite{WisemanVaccaro2003}. At least for fixed particle content this problem can be circumvented~\cite{HeaneyVedral2009}. Moreover, even if quantum correlations are not directly accessible, a transfer of the entanglement to systems that are not encumbered by such restrictions should be possible. In other words, entanglement may be swapped from the fermionic modes to systems that are not subject to superselection rules, thus justifying the use of unmodified measures for mode entanglement.\\

The main aim of Section~\ref{sec:Entanglement in Fermionic Quantum Fields} is establishing a clear framework for the implementation of fermionic field modes as vessels for quantum information tasks. To this end we present an analysis of the problem at hand, i.e., how the modes in a fermionic Fock space can be utilized as subsystems for quantum information processing. We present a framework that is based on simple physical requirements in which this can be achieved.
We further discuss the issues and restrictions in mapping fermionic modes to qubits and we show how previous work and proposed solutions, e.g., invoking superselection rules~\cite{BradlerJauregui2012}, fit into this framework.\\

Section~\ref{sec:Entanglement in Fermionic Quantum Fields} is structured as follows: In Section~\ref{sec:fermionic density operators} we start with a brief discussion of the implementation of density operators in the fermionic Fock space introduced in Section~\ref{sec:fermionic Fock space}. We then go on to formulate the \emph{``fermionic ambiguity''} that has been pointed out in Ref.~\cite{MonteroMartin-Martinez2011b} in Section~\ref{sec:the fermionic ambiguity}. Subsequently, we reinterpret this as an ambiguity in the definition of mode subsystems, which can be resolved by physical consistency conditions, in Section~\ref{sec:partial trace ambiguity}. Finally, we discuss the implications for the quantification of entanglement between two fermionic modes in Section~\ref{sec:entanglement of fermionic modes}, before we investigate situations beyond two modes in Section~\ref{sec:fermionic entanglement beyond 2 modes}.

\subsection{Density Operators in the Fermionic Fock Space}\label{sec:fermionic density operators}

In complete analogy to the usual case of mixed states~(\ref{eq:density operator}) on tensor product spaces we can now construct incoherent mixtures of pure states in a fermionic Fock space. For simplicity, we now restrict our analysis to a finite dimensional $n$\emph{-mode fermionic Fock space}~$\bar{\mathbb{F}}_{n}$ [see Eq.~(\ref{eq:fermionic Fock space})]
\begin{align}
    \bar{\mathbb{F}}_{n}(\mathcal{H}_{\operatorname{1-f}})    &=\,
    \bigoplus\limits_{m=0}^{n}\bar{S}\Bigl(\mathcal{H}_{\operatorname{1-f}}^{\otimes m}\Bigr)\,.
    \label{eq:n mode fermionic Fock space}
\end{align}
Let us first consider the projector on the state $\fket{\Psi}$ from Eq.~(\ref{eq:fermionic Fock space general state rewritten}), i.e.,
\begin{align}
    \fket{\Psi}\!\fbra{\Psi}  &=\,
    |\mu_{\mbox{\tiny{$0$}}}|^{2}\fket{0}\!\fbra{0}\,+\,
    \sum\limits_{i,i\pr}\mu_{i}\,\mu_{i\pr}^{*}\,\fket{\!1_{i}\!}\!\fbra{\!1_{i\pr}\!}
    \,+\,\sum\limits_{i}\Bigl(\mu_{i}\,\mu_{\mbox{\tiny{$0$}}}^{*}\,\fket{\!1_{i}\!}\!\fbra{0}\,+\,\mathrm{H.~c.}\Bigr)
    \nonumber\\[1.0mm]
    &\ +\,\sum\limits_{j,j\pr,k,k\pr}\mu_{jk}\,\mu_{j\pr k\pr}^{*}\,\fket{\!1_{j}\!}\fket{\!1_{k}\!}\!\fbra{\!1_{j\pr}\!}\fbra{\!1_{k\pr}\!}\,+\,\ldots\, ,
    \label{eq:projector on general fermion state}
\end{align}
where ``$\mathrm{H.~c.}$" denotes the Hermitean conjugate, $(\mathcal{O}+\mathrm{H.~c.})=(\mathcal{O}+\mathcal{O}^{\dagger})$. We can check that such an object satisfies the criteria for a density operator:
\begin{enumerate}[\hspace*{0.5cm}(i)]
    \item{It can be immediately noticed that~(\ref{eq:projector on general fermion state}) provides a \emph{Hermitean} operator.
        \label{item:hermitean fermion density operator}}
    \item{The \emph{normalization}, i.e., $\tr\bigl(\fket{\Psi}\!\fbra{\Psi}\bigr)=1\,$, is guaranteed by the normalization of $\fket{\Psi}\,$. In other words,
        the trace of~(\ref{eq:projector on general fermion state}) is well defined and independent of the chosen (complete, orthonormal) basis in $\bar{\mathbb{F}}_{n}\,$.\label{item:normalization fermion density operator}}
    \item{\emph{Positivity}: Finally, the eigenvalues of $\fket{\Psi}\!\fbra{\Psi}$ are well defined, i.e.,~(\ref{eq:projector on general fermion state}) can be represented as a diagonal matrix with diagonal entries $\{1,0,0,\ldots\}$, which clearly is a positive semidefinite spectrum.\label{item:positivity fermion density operator}}
\end{enumerate}
Incoherent mixtures of such pure states can then simply be formed using convex sums, i.e.,
\begin{align}
    \varrho &=\,\sum\limits_{n}\pii{n}\fket{\Psi_{n}}\!\fbra{\Psi_{n}}\,,
    \label{eq:fermionic mixed states}
\end{align}
where $\sum_{n}\pii{n}=1\,$, to construct the elements of the \emph{Hilbert-Schmidt space} $\mathcal{H}_{S}(\bar{\mathbb{F}}_{n})$ over the fermionic Fock space. Properties~$(\textrm{\ref{item:hermitean fermion density operator}})$ and~$(\textrm{\ref{item:normalization fermion density operator}})$ can trivially be seen to be satisfied for such \emph{mixed} states. The positivity of~(\ref{eq:fermionic mixed states}) \textemdash\ condition~(\ref{item:positivity fermion density operator}), however,
requires some additional comments. The operator~$\varrho$ can be diagonalized by a unitary transformation $U$ on $\bar{\mathbb{F}}_{n}$, which in turn can be
constructed from exponentiation of Hermitean or anti-Hermitean operators formed from algebra elements~$b_{n}\,$, $b_{m}^{\dagger}\,$, $c_{n}\,$ and $c_{m}^{\dagger}\,$. Operationally this procedure is rather elaborate. A simpler approach is the diagonalization of a matrix representation of $\varrho$\,. As we shall see in Section~\ref{sec:the fermionic ambiguity}, the matrix representation of $\varrho$ is not unique, but all possible representations $\pi_{i}(\varrho)$ are unitarily equivalent, such that their
eigenvalues all coincide with those of $\varrho\,$, i.e.,
\begin{align}
    \spectr\bigl(\pi_{i}(\varrho)\bigr) &=\,\spectr\bigl(\varrho\bigr)\ \ \forall\,i\,.
    \label{eq:spectrum of fermionic density matrix representation}
\end{align}

\subsection{The Fermionic Ambiguity}\label{sec:the fermionic ambiguity}

Let us now turn to the apparent ambiguity in such fermionic systems when quantum information tasks are considered. It was pointed out
in Ref.~\cite{MonteroMartin-Martinez2011b} that the anticommutation relations~(\ref{eq:anticomm relations}) do not suggest a natural
choice for the basis vectors of the fermionic Fock space for the multi-particle sector, i.e., for two fermions in the modes $m$ and $n$,
either
\begin{align}
    \fket{\!1_{m}\!}\fket{\!1_{n}\!}\ \ \ \mbox{or}\ \ \ \fket{\!1_{n}\!}\fket{\!1_{m}\!}\,=\,-\,\fket{\!1_{m}\!}\fket{\!1_{n}\!}
    \label{eq:fermionic ambiguity basis}
\end{align}
could be used to represent the physical state. This becomes of importance when we try to map the states in a fermionic $n$-mode Fock space to vectors in an $n$-fold tensor product space, i.e.,
\begin{subequations}
\label{eq:fermionic qubit mapping}
\begin{align}
    \pi_{i}:\ \bar{\mathbb{F}}_{n}\ &\longrightarrow\ \mathcal{H}_{1}\otimes\ldots\otimes\mathcal{H}_{n}\,
    \label{eq:fermionic qubit mapping spaces}\\[1.5mm]
    \fket{\psi}\ &\stackrel{\pi_{i}}{\longmapsto}\ \ket{\psi_{(i)}}
    \label{eq:fermionic qubit mapping pure states}\\[1.5mm]
    \varrho\ &\stackrel{\pi_{i}}{\longmapsto}\ \pi_{i}(\varrho)
    \label{eq:fermionic qubit mapping mixed states}
\end{align}
\end{subequations}
where the spaces $\mathcal{H}_{i}=\mathbb{C}^{2}$~$(i=1,\ldots,n)$ are identical, single-qubit Hilbert spaces. The mappings $\pi_{i}$ are unitary, i.e., $\fscpr{\phi}{\psi}=\scpr{\phi_{(i)}}{\psi_{(i)}}$ and $\tr(\varrho\sigma)=\tr(\pi_{i}(\varrho)\pi_{i}(\sigma))\,$. This implies that the maps~$\pi_{i}$ for different~$i$ are unitarily equivalent. In particular, the different matrix representations $\pi_{i}(\varrho)$ are related by multiplication of selected rows and columns of the matrix by $(-1)\,$. In the language of quantum information theory the states~$\psi_{(i)}$ are related by \emph{global unitary} transformations\index{transformation!global unitary}. It thus becomes apparent that the entanglement of $\pi_{i}(\varrho)$ with respect to a bipartition
\begin{align}
\mathcal{H}_{\mu_{1}}\otimes\ldots\otimes\mathcal{H}_{\mu_{m}}|\mathcal{H}_{\mu_{m+1}}\otimes\ldots\otimes\mathcal{H}_{\mu_{n}}
    \label{eq:bipartition}
\end{align}
will generally depend on the chosen mapping. Clearly, this is an unfavorable situation, but the inequivalence of entanglement measures for different such mappings has been noted before (see, e.g., Refs.~\cite{BoteroReznik2004,CabanPodlaskiRembielinskiSmolinskiWalczak2005,BradlerJauregui2012}), while other investigations~\cite{FriisLeeBruschiLouko2012,FriisBruschiLoukoFuentes2012,FriisHuberFuentesBruschi2012} did not suffer from any problems due to this ambiguity. Recently, the authors of Ref.~\cite{BradlerJauregui2012} suggested that the ambiguity can be resolved by restrictions imposed by charge superselection rules, while Refs.~\cite{MonteroMartin-Martinez2011b,MonteroMartin-Martinez2012b} suggested a solution by enforcing a particular operator ordering. We will discuss both of these approaches in Section~\ref{sec:partial trace ambiguity}, where we present simple and physically intuitive criteria for quantum information processing on a fermionic Fock space.
Most importantly, we will show in Sections~\ref{sec:partial trace ambiguity} and~\ref{sec:fermionic entanglement beyond 2 modes} that mappings of the type of (\ref{eq:fermionic qubit mapping}) can only be considered to be consistent for special cases, e.g., when the analysis is limited to two fermionic modes obeying charge superselection.

\subsection{The Partial Trace Ambiguity}\label{sec:partial trace ambiguity}

While the sign ambiguity in the sense of the different mappings $\pi_{i}$ is the superficial cause of the issue, we want to
discuss now a separate, and in some sense more fundamental problem: partial traces over ``mode subspaces." We are interested
in the entanglement between modes of a fermionic quantum field. However, in the structure of the Fock space, there is no
tensor product decomposition into Hilbert spaces for particular modes [see, e.g., Eq.~(\ref{eq:two fermion state definition})].
Only a tensor product structure with respect to individual fermions is available, but since the particles are indistinguishable,
the entanglement between two particles in this sense has to be defined very carefully~\cite{WisemanVaccaro2003}. This issue is not
unique for fermions and is sometimes referred to as \emph{``fluffy bunny''} entanglement (see Ref.~\cite{WisemanBartlettVaccaro2004}).\\

For the decomposition into different modes we only have a wedge product structure available. In Ref.~\cite{BradlerJauregui2012}
the authors suggest that entanglement should be considered with respect to this special case of the ``braided tensor product."
As far as the construction of the density operators with respect to such a structure is concerned, we agree with this view (see
Section~\ref{sec:fermionic density operators}), and no ambiguities arise regarding the description of the total $n$-mode system.
However, the crucial problem lies in the definition of the partial tracing over a subset of the $n$~modes. This is best
illustrated for a simple example: Consider a system of two fermionic modes labelled~$\kappa$ and~$\kappa\pr$, where we assume without loss of generality that both are positive frequency modes. A general, mixed state of these two modes can be written as
\begin{align}
    \varrho_{\kappa\kappa\pr}   &=\,
    c_{1}\,\fket{0}\!\fbra{0}\,+\,
    c_{2}\,\fket{\!1_{\kappa\pr}\!}\!\fbra{\!1_{\kappa\pr}\!}
    \,+\,
    c_{3}\,\fket{\!1_{\kappa}\!}\!\fbra{\!1_{\kappa}\!}\,+\,
    c_{4}\,\fket{\!1_{\kappa}\!}\fket{\!1_{\kappa\pr}\!}\!\fbra{\!1_{\kappa\pr}\!}\fbra{\!1_{\kappa}\!}
    \nonumber\\[1.5mm]
    &\ +\,
    \Bigl(
    d_{1}\,\fket{0}\!\fbra{\!1_{\kappa\pr}\!}\,+\,
    d_{2}\,\fket{0}\!\fbra{\!1_{\kappa}\!}\,+\,
    d_{3}\,\fket{0}\!\fbra{\!1_{\kappa\pr}\!}\fbra{\!1_{\kappa}\!}\,+\,
    d_{4}\,\fket{\!1_{\kappa\pr}\!}\!\fbra{\!1_{\kappa}\!}
    \nonumber\\[1.5mm]
    &\ +\,
    d_{5}\,\fket{\!1_{\kappa\pr}\!}\!\fbra{\!1_{\kappa\pr}\!}\fbra{\!1_{\kappa}\!}\,+\,
    d_{6}\,\fket{\!1_{\kappa}\!}\!\fbra{\!1_{\kappa\pr}\!}\fbra{\!1_{\kappa}\!}\,+\,\mathrm{H.c.}\,\Bigr)\,,
    \label{eq:general two-mode fermion state}
\end{align}
where appropriate restrictions on the coefficients $c_{i}\in\mathbb{R}$ and $d_{j}\in\mathbb{C}$ apply to ensure the positivity
and normalization of $\varrho_{\kappa\kappa\pr}\,$. Here we have, for now, disregarded superselection rules. Let us now determine
the corresponding reduced density operators (on the Fock space) for the individual modes $\kappa$ and $\kappa\pr$. Usually one would
select a basis of the subsystem that is being traced over, e.g., for tracing over mode $\kappa\pr$ one could choose
$\{\fket{0},\fket{\!1_{\kappa\pr}\!}\}$. This clearly cannot work since basis vectors with different numbers of excitations are orthogonal. We thus have to define the partial trace in a different way. This is equally true for bosonic fields as well. However, in contrast to the fermionic case, no ambiguities arise in such a redefinition for bosonic fields. For the diagonal elements of the reduced fermionic states the redefinition of the partial trace is straightforward as well. These elements are obtained from
\begin{subequations}
\label{eq:partial trace diagonal elements}
\begin{align}
    \tr_{m}\bigl(\fket{0}\!\fbra{0}\bigr) &:=\,\fket{0}\!\fbra{0}\,,
    \label{eq:partial trace diagonal element vacuum}\\[1.0mm]
    \tr_{m}\bigl(\fket{\!1_{n}\!}\!\fbra{\!1_{n}\!}\bigr) &:=\,(1-\delta_{mn})\fket{\!1_{n}\!}\!\fbra{\!1_{n}\!}\,+\,\delta_{mn}\fket{0}\!\fbra{0}\,,
    \label{eq:partial trace diagonal element single particle}\\[1.0mm]
    \tr_{m}\bigl(\fket{\!1_{m}\!}\fket{\!1_{n}\!}\!\fbra{\!1_{n}\!}\fbra{\!1_{m}\!}\bigr) &:=\,\fket{\!1_{n}\!}\!\fbra{\!1_{n}\!}\
    (m\neq n)\,,
    \label{eq:partial trace diagonal element two particles}
\end{align}
\end{subequations}
where $n,m=\kappa,\kappa\pr\,$. While the diagonal elements are unproblematic and do not suffer from any ambiguities, we have to be more careful with the off-diagonal elements. Three of these will not contribute, i.e.,
\begin{align}
    \tr_{m}\bigl(\fket{\!1_{m}\!}\!\fbra{\!1_{n}\!}\bigr) &=\,
    \tr_{m}\bigl(\fket{0}\!\fbra{\!1_{m}\!}\fbra{\!1_{n}\!}\bigr)\,=\,
    \tr_{m}\bigl(\fket{\!1_{n}\!}\!\fbra{\!1_{m}\!}\fbra{\!1_{n}\!}\bigr)\,=\,0\,,
    \label{eq:partial trace offdiagonal elements vanishing}
\end{align}
and two more are unproblematic as well, i.e.,
\begin{align}
    \tr_{m}\bigl(\fket{0}\!\fbra{\!1_{n}\!}\bigr) &:=\,(1-\delta_{mn})\fket{0}\!\fbra{\!1_{n}\!}\,.
    \label{eq:partial trace offdiagonal elements trivial}
\end{align}
The last element,
\begin{align}
    \tr_{m}\bigl(\fket{\!1_{m}\!}\!\fbra{\!1_{m}\!}\fbra{\!1_{n}\!}\bigr)   &=\,
    -\,\tr_{m}\bigl(\fket{\!1_{m}\!}\!\fbra{\!1_{n}\!}\fbra{\!1_{m}\!}\bigr)
    \,=\,\pm\,\fket{0}\!\fbra{\!1_{n}\!}\,,
    \label{eq:partial trace offdiagonal ambiguous}
\end{align}
however, presents an ambiguity. If a mapping $\pi_{i}$ to a two-qubit Hilbert space is performed, the choice of map will
determine the corresponding sign in the partial trace over either of the qubits. The differences in entanglement related
to the fact that $\pi_{i}(\varrho)$ and $\pi_{j}(\varrho)$ are related by a global unitary are thus explained by the
relative sign between the contributions of Eq.~(\ref{eq:partial trace offdiagonal elements trivial}) and
Eq.~(\ref{eq:partial trace offdiagonal ambiguous}) to the same element of the reduced density matrix.\\

However, simple \emph{physical requirements} restrict the choice in this relative sign. Any reduced state formalism has
to satisfy the simple criterion that the reduced density operator contains all the information about the subsystem that
can be obtained from the global state when measurements are performed only on the respective subsystem alone.
Let us put this statement in more mathematical terms. For any bipartition $A|B$ of a Hilbert space~$\mathcal{H}$ (with respect to any
braided tensor product structure on~$\mathcal{H}$) and any state $\rho\in\mathcal{H}$ the partial trace operation $\tr\subtiny{0}{0}{B}$ must satisfy
\begin{align}
    \left\langle\,\mathcal{O}_{n}(A)\,\right\rangle_{\rho}  &=\,\left\langle\,\mathcal{O}_{n}(A)\,\right\rangle_{\tr_{B}(\rho)}\,,
    \label{eq:consistency condition}
\end{align}
where $\left\langle\mathcal{O}\right\rangle_{\rho}$ denotes the expectation value of the operator $\mathcal{O}$ in the state $\rho$
and $\{\mathcal{O}_{n}(A)\}$ is the set of all (Hermitean) operators that act on the subspace $A$ only. For the operator
$\varrho_{\kappa\kappa\pr}$ from Eq.~(\ref{eq:general two-mode fermion state}) the condition~(\ref{eq:consistency condition}) can be written as
\begin{align}
    \tr\bigl(\mathcal{O}_{n}(\kappa)\varrho_{\kappa\kappa\pr}\bigr)  &=\,
    \tr\bigl(\mathcal{O}_{n}(\kappa)\varrho_{\kappa}\bigr)\,,
    \label{eq:consistency condition 2 modes}
\end{align}
where $\varrho_{\kappa}=\tr_{\kappa\pr}(\varrho_{\kappa\kappa\pr})\,$. This consistency condition uniquely determines the relative signs
between different contributions to the same elements of $\varrho_{\kappa}\,$. Let us consider the (Hermitean) operators
$(b_{\kappa}+b_{\kappa}^{\dagger})$ and $i(b_{\kappa}-b_{\kappa}^{\dagger})\,$. Their expectation values for the global state
$\varrho_{\kappa\kappa\pr}$ are given by
\begin{subequations}
\label{eq:consistencty condition global exp values 2 modes kappa}
    \begin{align}
    \tr\bigl(\,(b_{\kappa}+b_{\kappa}^{\dagger})\varrho_{\kappa\kappa\pr}\,\bigr)   &=\,
    2\,\mathrm{Re}(\,d_{2}+d_{5}\,)\,,
    \label{eq:consistencty condition global exp values 2 modes kappa x}\\[1.0mm]
    \tr\bigl(\,i(b_{\kappa}-b_{\kappa}^{\dagger})\varrho_{\kappa\kappa\pr}\,\bigr)   &=\,
    2\,\mathrm{Im}(\,d_{2}+d_{5}\,)\,.
    \label{eq:consistencty condition global exp values 2 modes kappa p}
\end{align}
\end{subequations}
For the mode $\kappa\pr$, on the other hand, we compute
\begin{subequations}
\label{eq:consistencty condition global exp values 2 modes kappa prime}
    \begin{align}
    \tr\bigl(\,(b_{\kappa\pr}+b_{\kappa\pr}^{\dagger})\varrho_{\kappa\kappa\pr}\,\bigr)   &=\,
    2\,\mathrm{Re}(\,d_{1}-d_{6}\,)\,,
    \label{eq:consistencty condition global exp values 2 modes kappa pr x}\\[1.0mm]
    \tr\bigl(\,i(b_{\kappa\pr}-b_{\kappa\pr}^{\dagger})\varrho_{\kappa\kappa\pr}\,\bigr)   &=\,
    2\,\mathrm{Im}(\,d_{1}-d_{6}\,)\,.
    \label{eq:consistencty condition global exp values 2 modes kappa pr p}
    \end{align}
\end{subequations}
Equations~(\ref{eq:consistencty condition global exp values 2 modes kappa}) and
(\ref{eq:consistencty condition global exp values 2 modes kappa prime}) determine the sign in
Eq.~(\ref{eq:partial trace offdiagonal ambiguous}) and we find the reduced states
\begin{subequations}
\label{eq:reduced states for kappa and kappa prime}
    \begin{align}
        \varrho_{\kappa}\,=\,\tr_{\kappa\pr}\bigl(\varrho_{\kappa\kappa\pr}\bigr) &=\,
        (c_{1}+c_{2})\,\fket{0}\!\fbra{0}\,+\,
        (c_{3}+c_{4})\,\fket{\!1_{\kappa}\!}\!\fbra{\!1_{\kappa}\!}
        \label{eq:reduced state for kappa}\\[1.0mm]
        &\ +\,
        \Bigl((d_{2}+d_{5})\,\fket{0}\!\fbra{\!1_{\kappa}\!}\,+\,\mathrm{H.c.}\Bigr)\,,
        \nonumber\\[1.0mm]
        \varrho_{\kappa\pr}\,=\,\tr_{\kappa}\bigl(\varrho_{\kappa\kappa\pr}\bigr) &=\,
        (c_{1}+c_{3})\,\fket{0}\!\fbra{0}\,+\,
        (c_{2}+c_{4})\,\fket{\!1_{\kappa\pr}\!}\!\fbra{\!1_{\kappa\pr}\!}
        \label{eq:reduced state for kappa prime}\\[1.0mm]
        &\ +\,\Bigl((d_{1}-d_{6})\,\fket{0}\!\fbra{\!1_{\kappa\pr}\!}\,+\,\mathrm{H.c.}\Bigr)\,,
        \nonumber
    \end{align}
\end{subequations}
for the modes~$\kappa$ and~$\kappa\pr$, respectively. Notice that this formally corresponds to tracing \emph{``inside out,"}
that is, first (anti)commuting operators towards the projector on the vacuum state before removing them, such that
\begin{align}
    \tr_{m}\bigl(b_{m}^{\dagger}\fket{0}\!\fbra{0}b_{m}b_{n}\bigr)   &=\,
    \fket{0}\!\fbra{\!1_{n}\!}\,.
    \label{eq:partial trace offdiagonal ambiguous inside out}
\end{align}
We have now arrived at a point where we can make a general statement about the consistency conditions. Let us formulate
this in the following theorem.

\begin{theorem}\label{theorem:consistency condition}\end{theorem}
    \vspace*{-1.22cm}
    \begin{tabbing} \hspace*{3.2cm}\=\hspace*{1.5cm}\=\kill
            \> \textit{Given a density operator $\varrho_{1,\ldots,n}\in\mathcal{H}_{S}(\bar{\mathbb{F}}_{n})$ for $n$~fermionic modes (labelled}\\[1mm]
            \> \textit{$1,\ldots,n$) the consistency conditions~\rm{(\ref{eq:consistency condition})} completely determine the }\\[1mm]
            \> \textit{reduced states on $\mathcal{H}_{S}(\bar{\mathbb{F}}_{m})$ for any~$m$ with $1<m<n$.}
    \end{tabbing}

\begin{proof}
This can be seen in the following way: for any matrix element
\begin{align}
\varphi\,b^{\dagger}_{\mu_{1}}\ldots b^{\dagger}_{\mu_{i}}\fket{0}\!\fbra{0}b_{\nu_{1}}\ldots b_{\nu_{j}}
\label{eq:general matrix element n-1 mode state}
\end{align}
of an $(n-1)$-mode reduced state $\varrho_{1,\ldots,(n-1)}=\tr_{n}(\varrho_{1,\ldots,n})$, where $\varphi\in\mathbb{C}$ and the sets
\begin{subequations}
\label{eq:n-1 mode subsets}
    \begin{align}
        \mu\,:=\,\{\mu_{1},\ldots,\mu_{i}\}  &\subseteq\,\{1,2,\ldots,(n-1)\}
        \label{eq:n-1 mode subset mu}\\[1.5mm]
    \mbox{and}\ \  \nu\,:=\,\{\nu_{1},\ldots,\nu_{j}\}   &\subseteq\,\{1,2,\ldots,(n-1)\}
        \label{eq:n-1 mode subset nu}
    \end{align}
\end{subequations}
label subsets of the mode operators for the $(n-1)$ modes, can have contributions from at most two matrix elements of $\varrho_{1,\ldots,n}\,$, i.e.,
\begin{subequations}
\label{eq:contributions to n-1 modes element from n mode state}
    \begin{align}
        &\ \tr_{n}\bigl(\varphi_{1}\,b^{\dagger}_{\mu_{1}}\ldots b^{\dagger}_{\mu_{i}}
            \fket{0}\!\fbra{0}b_{\nu_{1}}\ldots b_{\nu_{j}}\bigr)\,,
        \label{eq:contribution no particle to n-1 modes element from n mode state}\\[1.5mm]
    \mbox{and}\ \
        &\ \tr_{n}\bigl(\varphi_{2}\,b^{\dagger}_{\mu_{1}}\ldots b^{\dagger}_{\mu_{i}}b^{\dagger}_{n}
            \fket{0}\!\fbra{0}b_{n}b_{\nu_{1}}\ldots b_{\nu_{j}}\bigr)\,.
        \label{eq:contribution one particle to n-1 modes element from n mode state}
    \end{align}
\end{subequations}
The composition of $\varphi$ into $\varphi_{1}\in\mathbb{C}$ and $\varphi_{2}\in\mathbb{C}$, i.e., $\varphi=\varphi_{1}\pm\varphi_{2}\,$, is determined by the
consistency conditions of Eq.~(\ref{eq:consistency condition}). For every matrix element~(\ref{eq:general matrix element n-1 mode state}) with corresponding partial trace contributions from (\ref{eq:contributions to n-1 modes element from n mode state}) there exists a pair of Hermitean operators
\begin{subequations}
\label{eq:general element consistency condition operators}
\begin{align}
\mathcal{O}_{x}(\lambda,\tau)   &=\,b_{\lambda_{1}}\ldots b_{\lambda_{k}}b^{\dagger}_{\tau_{1}}\ldots b^{\dagger}_{\tau_{l}}
    \,+\,b_{\tau_{l}}\ldots b_{\tau_{1}}b^{\dagger}_{\lambda_{k}}\ldots b^{\dagger}_{\lambda_{1}}\,,
    \label{eq:general element consistency condition operator x}\\[1.5mm]
\mathcal{O}_{p}(\lambda,\tau)   &=\,b_{\lambda_{1}}\ldots b_{\lambda_{k}}b^{\dagger}_{\tau_{1}}\ldots b^{\dagger}_{\tau_{l}}
    \,-i\,b_{\tau_{l}}\ldots b_{\tau_{1}}b^{\dagger}_{\lambda_{k}}\ldots b^{\dagger}_{\lambda_{1}}\,,
    \label{eq:general element consistency condition operator p}
\end{align}
\end{subequations}
with $\lambda:=\{\lambda_{1},\ldots,\lambda_{k}\}=\mu/\nu$ and $\tau:=\{\tau_{1},\ldots,\tau_{l}\}=\nu/\mu\,$, that
uniquely determine the relative sign of~$\varphi_{1}$ and~$\varphi_{2}\,$.
These operators are unique up to an overall multiplication with scalars.
The tracing procedure can be repeated when any other of the $(n-1)$ remaining modes
are traced over. Since the order of the partial traces is of no importance for the final reduced state, all reduced
density operators are completely determined.
\end{proof}

Consequently, the reduced density matrices
in the fermionic Fock space can be considered as proper density operators, i.e., they are Hermitean, normalized, and
their eigenvalues are well defined and non-negative. Moreover, since the eigenvalues are free of ambiguities, all
functions of these eigenvalues, in particular, all entropy measures for density operators, are well defined. Also,
the operator ordering that was suggested in Ref.~\cite{MonteroMartin-Martinez2011b} is consistent with our consistency
condition. Let us stress here that this analysis does not depend on any superselection rules that might be imposed in addition.
We will see how these enter the problem when mappings to qubits are attempted in Section~\ref{sec:entanglement of fermionic modes}.

\subsection{Entanglement of Fermionic Modes}\label{sec:entanglement of fermionic modes}

We are now in a position to reconsider a measure of entanglement between fermionic modes. We can define the entanglement of formation
$\bar{E}\sub{0.2}{0}{\mathrm{F}}$ for fermionic systems with respect to a chosen bipartition $A|B$ as
\begin{align}
    \bar{E}\sub{0.2}{0}{\mathrm{F}}(\varrho)    &=\,
    \inf_{\{(\pii{n},\fket{\!\Psi_{n}\!})\}}\,\sum\limits_{n}\,\pii{n}\,\mathcal{E}(\fket{\!\Psi_{n}\!})\,,
    \label{eq:fermionic entanglement of formation}
\end{align}
in complete analogy to the previous Definition~\ref{def:entanglement of formation}. Here the minimum is taken over all pure
state ensembles $\{(\pii{n},\fket{\!\Psi_{n}\!})\}$ that realize $\varrho$ according to
Eq.~(\ref{eq:fermionic mixed states}) and $\mathcal{E}(\fket{\!\Psi\!})$ denotes the entropy of entanglement (Definition~\ref{def:entropy of entanglement}) of the
pure state $\fket{\!\Psi\!}$. Since the entropy of entanglement is a function of the eigenvalues of the reduced states
$\tr\subtiny{0}{0}{B}\bigl(\fket{\!\Psi\!}\!\fbra{\!\Psi\!}\bigr)$ or $\tr\subtiny{0}{0}{A}\bigl(\fket{\!\Psi\!}\!\fbra{\!\Psi\!}\bigr)$ alone, we
can conclude that this is a well-defined quantity. As pointed out in Ref.~\cite{CabanPodlaskiRembielinskiSmolinskiWalczak2005},
the minimization in Eq.~(\ref{eq:fermionic entanglement of formation}) can be restricted to pure state decompositions that respect
superselection rules. Since this restriction limits the set of states over which the minimization is carried out, the quantity without this
restriction will be a lower bound to the ``physical'' entanglement of formation. For two fermionic modes the minimization over all states
that respect superselection rules can indeed be carried out (see Ref.~\cite{CabanPodlaskiRembielinskiSmolinskiWalczak2005}). However, in
general this step will be problematic.\\

Let us now turn to some computable entanglement measures, in particular, let us investigate if and how the \emph{negativity}~$\mathcal{N}$\index{negativity} (see Definition~\ref{def:negativity} or Ref.~\cite{VidalWerner2002}) and the \emph{concurrence}~$C$\index{concurrence} (see Eq.~(\ref{eq:wootters concurrence two qubits}) or Ref.~\cite{BennettDiVincenzoSmolinWootters1996}) can be computed to quantify fermionic mode entanglement. Both of these measures are operationally based on the tensor product structure of qubits,
since the partial transposition\index{partial transposition} is a map that is well defined only for basis vectors on a tensor product space. To employ this measure, let us therefore try to find a mapping $\pi_{i}$ to such a tensor product structure that is consistent with the conditions of Eq.~(\ref{eq:consistency condition}). Starting with the two-mode state $\varrho_{\kappa\kappa\pr}$ of Eq.~(\ref{eq:general two-mode fermion state}), we are
looking for a map $\pi$ that takes $\{\fket{0},\fket{\!1_{\kappa}\!},\fket{\!1_{\kappa\pr}\!},\fket{\!1_{\kappa}\!}\fket{\!1_{\kappa\pr}\!}\}$ to
$\{\ket{00},\ket{01\!},\ket{\!10},\ket{\!11\!}\}$, where $\ket{mn}=\ket{m}\otimes\ket{n}\in\mathcal{H}_{\kappa}\otimes\mathcal{H}_{\kappa\pr}$, such that
\begin{subequations}
\label{eq:two fermion mode consistent mapping}
    \begin{align}
    \varrho &\ \longmapsto\    \pi(\varrho)\,,
    \label{eq:two fermion mode consistent mapping total state}\\[1.0mm]
    \varrho_{\kappa} &\ \longmapsto\    \pi(\varrho_{\kappa})\,=\,\tr_{\kappa\pr}\bigl(\pi(\varrho)\bigr)\,,
    \label{eq:two fermion mode consistent mapping total state mode kappa}\\[1.0mm]
    \varrho_{\kappa\pr} &\ \longmapsto\    \pi(\varrho_{\kappa\pr})\,=\,\tr_{\kappa}\bigl(\pi(\varrho)\bigr)\,.
    \label{eq:two fermion mode consistent mapping total state mode kappa prime}
    \end{align}
\end{subequations}
The condition for a consistent mapping can be represented in the following diagram:
\begin{eqnarray}
\varrho_{\kappa\kappa\pr}\                        &\  \stackrel{\pi}{\longmapsto}\    &\ \pi(\varrho_{\kappa\kappa\pr})
\nonumber\\[2.0mm]
\tr_{\kappa\pr}\ \downarrow\ \ \ &                                   &\ \ \ \downarrow\ \ \tr_{\kappa\pr}
\label{eq:graphical representation of consitent map}\\[2.0mm]
\varrho_{\kappa}          \  \    &\  \stackrel{\pi}{\longmapsto}\    &\ \pi(\varrho_{\kappa})
\nonumber
\end{eqnarray}
In other words, a mapping $\pi: \varrho\mapsto\pi(\varrho)$ from the space $\mathcal{H}_{S}(\bar{\mathbb{F}}_{2})$ to $\mathcal{H}_{\kappa}\otimes\mathcal{H}_{\kappa\pr}$ is considered to be consistent if it commutes with the partial trace operation. It is quite simple to check that these requirements generally cannot be met, i.e., writing $\varrho_{\kappa\kappa\pr}$ of Eq.~(\ref{eq:general two-mode fermion state}) as a matrix with respect to the basis $\{\fket{0},\fket{\!1_{\kappa}\!},\fket{\!1_{\kappa\pr}\!},\fket{\!1_{\kappa}\!}\fket{\!1_{\kappa\pr}\!}\}$ we get
\begin{align}
\varrho &=\,
    \begin{pmatrix}
        c_{1}      &   d_{1}       &   d_{2}       &   d_{3}   \\
        d_{1}^{*}   &   c_{2}      &   d_{4}       &   d_{5}   \\
        d_{2}^{*}   &   d_{4}^{*}   &   c_{3}      &   d_{6}   \\
        d_{3}^{*}   &   d_{5}^{*}   &   d_{6}^{*}   &   c_{4}
    \end{pmatrix}\,.
\label{eq:general two mode state matrix representation}
\end{align}
A mapping of the desired type should be obtained by multiplying any number of rows and the corresponding columns by $(-1)$ and considering the resulting
matrix as the representation $\pi(\varrho)$ on $\mathcal{H}_{\kappa}\otimes\mathcal{H}_{\kappa\pr}$. The desired result should have a relative
sign switch between $d_{1}$ and $d_{6}$, while the signs in front of $d_{2}$ and $d_{5}$ should be the same. This clearly is not
possible unless some of the coefficients vanish identically, e.g., by imposing superselection rules. For example, conservation of charge would require the coefficients $d_{1},d_{2},d_{5}$, $d_{6}$, and, depending on the charge of the modes $\kappa$ and $\kappa\pr$, either $d_{3}$ or $d_{4}$ to vanish identically. In this way only incoherent mixtures of pure states with different charge are allowed, but no coherent superpositions.\\

We thus find that \emph{two fermionic modes can only be consistently represented as two qubits when charge superselection is respected}. In that case only one off-diagonal element can be non-zero and the sign of this element is insubstantial, i.e., it does not influence the reduced states or the value of any entanglement measure. In particular, the results for entanglement generation and degradation between two fermionic modes presented in Refs.~\cite[(\ref{Paper:FriisKoehlerMartinMartinezBertlmann2011}-\ref{Paper:FriisBruschiLoukoFuentes2012})]
{FriisLeeBruschiLouko2012,FriisBruschiLoukoFuentes2012,
FriisKoehlerMartinMartinezBertlmann2011} respect both charge superselection and the consistency conditions of Eq.~(\ref{eq:consistency condition}).\\

Let us return to the choice of entanglement measure for the permitted mappings to two qubits. We now restrict the entanglement of formation $\bar{E}\sub{0.2}{0}{\mathrm{F}}$ as defined in Eq.~(\ref{eq:fermionic entanglement of formation}) to states that obey charge superselection, as suggested in Ref.~\cite{CabanPodlaskiRembielinskiSmolinskiWalczak2005}. As discussed earlier, this means the usual entanglement of formation $E\sub{0.2}{0}{\mathrm{F}}$ of Definition~\ref{def:entanglement of formation} provides a lower bound to $\bar{E}\sub{0.2}{0}{\mathrm{F}}$, i.e.,
\begin{align}
E\sub{0.2}{0}{\mathrm{F}}  &\leq\,\bar{E}\sub{0.2}{0}{\mathrm{F}}\,.
\label{eq:lower bound to fermionic EoF}
\end{align}
For two qubits $E\sub{0.2}{0}{\mathrm{F}}=E\sub{0.2}{0}{\mathrm{F}}(C)$ is a monotonically increasing function of the concurrence~$C$\index{concurrence}. We propose an analogous functional dependence of $\bar{E}\sub{0.2}{0}{\mathrm{F}}=\bar{E}\sub{0.2}{0}{\mathrm{F}}(\bar{C})$ on a parameter~$\bar{C}$, that we call \emph{``fermionic concurrence."}\index{fermionic concurrence} Evidently, the function $\bar{C}(\varrho)$ is an entanglement monotone that is bounded from below by the usual concurrence~$C$. As can be seen from Eq.~(\ref{eq:concurrence negativity bounds}) (see also Ref.~\cite{VerstraeteAudenaertDehaeneDeMoor2001}), the negativity~$\mathcal{N}$ further provides a lower bound to the concurrence, i.e., in our convention of Definition~\ref{def:negativity}, $2\mathcal{N}\leq C$. Consequently, the negativity provides a lower bound to~$\bar{C}$, i.e.,
\begin{align}
2\mathcal{N}    &\leq\,C\,\leq\,\bar{C}\,.
\label{eq:negativity and ferm concurrence bounds}
\end{align}
For two modes it is thus at least possible to compute lower bounds to entanglement measures explicitly. It was suggested in Ref.~\cite{WisemanVaccaro2003}
that conventional entanglement measures overestimate the quantum correlations that can physically be extracted from fermionic systems. The operations that can be performed on each single-mode subsystem are limited by (charge) superselection as well. However, we conjecture that the inaccessible entanglement between the fermionic modes can always be swapped to two (uncharged) bosonic modes for which the local bases can be chosen arbitrarily.

\subsection{Fermionic Entanglement Beyond Two Modes}\label{sec:fermionic entanglement beyond 2 modes}

Finally, let us consider the entanglement between more than two fermionic modes. In principle, any measure
of entanglement that is based on entropies of the subsystems is well defined on the fermionic Fock space, as we
have discussed. However, we would like to employ computable measures. Let us therefore start by attempting a
consistent mapping from three fermionic modes to three qubits, in analogy to the two-mode case in
Section~\ref{sec:entanglement of fermionic modes}. For simplicity we assume that the modes $\kappa$, $\kappa\pr$, and
$\kappa\prpr$ all have equal charge such that the most general mixed state of these modes can be written as
\begin{align}
    \varrho_{\kappa\kappa\pr\hspace*{-1pt}\kappa\prpr}   &=\,
    \mu_{1}\,\fket{0}\!\fbra{0}
    \,+\,
    \mu_{2}\,\fket{\!1_{\kappa\prpr}\!}\!\fbra{\!1_{\kappa\prpr}\!}
    \,+\,
    \mu_{3}\,\fket{\!1_{\kappa\pr}\!}\!\fbra{\!1_{\kappa\pr}\!}
    \,+\,
    \mu_{4}\,\fket{\!1_{\kappa\pr}\!}\fket{\!1_{\kappa\prpr}\!}\!\fbra{\!1_{\kappa\prpr}\!}\fbra{\!1_{\kappa\pr}\!}
    \nonumber\\[1.0mm]
    &\ +\,
    \mu_{5}\,\fket{\!1_{\kappa}\!}\!\fbra{\!1_{\kappa}\!}
    \,+\,
    \mu_{6}\,\fket{\!1_{\kappa}\!}\fket{\!1_{\kappa\prpr}\!}\!\fbra{\!1_{\kappa\prpr}\!}\fbra{\!1_{\kappa}\!}
    \,+\,
    \mu_{7}\,\fket{\!1_{\kappa}\!}\fket{\!1_{\kappa\pr}\!}\!\fbra{\!1_{\kappa\pr}\!}\fbra{\!1_{\kappa}\!}
    \nonumber\\[1.5mm]
    &\ +\,
    \mu_{8}\,\fket{\!1_{\kappa}\!}\fket{\!1_{\kappa\pr}\!}\fket{\!1_{\kappa\prpr}\!}\!
        \fbra{\!1_{\kappa\prpr}\!}\fbra{\!1_{\kappa\pr}\!}\fbra{\!1_{\kappa}\!}
    \,+\,
    \Bigl(\,
    \nu_{1}\,\fket{\!1_{\kappa\prpr}\!}\!\fbra{\!1_{\kappa\pr}\!}
    \,+\,
    \nu_{2}\,\fket{\!1_{\kappa\prpr}\!}\!\fbra{\!1_{\kappa}\!}
    \nonumber\\[1.5mm]
    &\ +\,
    \nu_{3}\,\fket{\!1_{\kappa\pr}\!}\!\fbra{\!1_{\kappa}\!}
    \,+\,
    \nu_{4}\,\fket{\!1_{\kappa\pr}\!}\fket{\!1_{\kappa\prpr}\!}\!\fbra{\!1_{\kappa\prpr}\!}\fbra{\!1_{\kappa}\!}
    \,+\,
    \nu_{5}\,\fket{\!1_{\kappa\pr}\!}\fket{\!1_{\kappa\prpr}\!}\!\fbra{\!1_{\kappa\pr}\!}\fbra{\!1_{\kappa}\!}
    \nonumber\\[1.5mm]
    &\ +\,
    \nu_{6}\,\fket{\!1_{\kappa}\!}\fket{\!1_{\kappa\prpr}\!}\!\fbra{\!1_{\kappa\pr}\!}\fbra{\!1_{\kappa}\!}\,+\,\mathrm{H.c.}\,\Bigr)\,.
    \label{eq:general three-mode fermion state supseselection}
\end{align}
The relevant consistency conditions to construct the three different reduced two-mode density matrices
$\varrho_{\kappa\kappa\pr}\,$, $\varrho_{\kappa\kappa\prpr}$ and $\varrho_{\kappa\pr\kappa\prpr}$ are given by
\begin{subequations}
\label{eq:three mode consistency conditions}
    \begin{align}
        \tr\bigl(\,(b_{\kappa}^{\dagger}b_{\kappa\pr}\,+\,b_{\kappa\pr}^{\dagger}b_{\kappa})\,
        \varrho_{\kappa\kappa\pr\kappa\prpr}\bigr)  &=\,2\,\mathrm{Re}(\nu_{3}\,+\,\nu_{4})\,,
        \label{eq:three mode consistency conditions kappa kappa pr}\\[1.5mm]
        \tr\bigl(\,(b_{\kappa}^{\dagger}b_{\kappa\prpr}\,+\,b_{\kappa\prpr}^{\dagger}b_{\kappa})\,
        \varrho_{\kappa\kappa\pr\kappa\prpr}\bigr)  &=\,2\,\mathrm{Re}(\nu_{2}\,-\,\nu_{5})\,,
        \label{eq:three mode consistency conditions kappa kappa prpr}\\[1.5mm]
        \tr\bigl(\,(b_{\kappa\pr}^{\dagger}b_{\kappa\prpr}\,+\,b_{\kappa\prpr}^{\dagger}b_{\kappa\pr})\,
        \varrho_{\kappa\kappa\pr\kappa\prpr}\bigr)  &=\,2\,\mathrm{Re}(\nu_{1}\,+\,\nu_{6})\,.
        \label{eq:three mode consistency conditions kappa pr kappa prpr}
    \end{align}
\end{subequations}

Again, the correct partial traces are obtained by tracing ``inside out" [see Eq.~(\ref{eq:partial trace offdiagonal ambiguous inside out})].
This is not a coincidence. The prescription for the partial trace to anticommute operators towards the projector of the vacuum state before
eliminating them takes into account the number of anticommutations occurring in computations of the expectation values of
Eq.~(\ref{eq:consistency condition}). A matrix representation of the three-mode state $\varrho_{\kappa\kappa\pr\kappa\prpr}$ is given by
\begin{align}
\varrho_{\kappa\kappa\pr\kappa\prpr}    &=\,
    \begin{pmatrix}
        \mu_{1}     &   0           &   0           &   0           &   0       &   0           &   0       &   0       \\
        0           &   \mu_{2}     &   \nu_{1}     &   0           &   \nu_{2} &   0           &   0       &   0       \\
        0           &   \nu_{1}^{*} &   \mu_{3}     &   0           &   \nu_{3} &   0           &   0       &   0       \\
        0           &   0           &   0           &   \mu_{4}     &   0       &   \nu_{4}     &   \nu_{5} &   0       \\
        0           &   \nu_{2}^{*} &   \nu_{3}^{*} &   0           &   \mu_{5} &   0           &   0       &   0       \\
        0           &   0           &   0           &   \nu_{4}^{*} &   0       &   \mu_{6}     &   \nu_{6} &   0       \\
        0           &   0           &   0           &   \nu_{5}^{*} &   0       &   \nu_{6}^{*} &   \mu_{7} &   0       \\
        0           &   0           &   0           &   0           &   0       &   0           &   0       &   \mu_{8}
    \end{pmatrix}\,.
\label{eq:general three-mode fermion state supseselection matrix rep}
\end{align}

Similar as before, one can try to interpret Eq.~(\ref{eq:general three-mode fermion state supseselection matrix rep}) as a matrix
representation of a three-qubit state and exchange the signs of the basis vectors in the three qubit state such that the consistency
conditions of Eq.~(\ref{eq:three mode consistency conditions}) are met, i.e., opposite signs in front of $\nu_{2}$ and $\nu_{6}$, while the signs
in front of the pairs $\nu_{3},\nu_{4}$ and $\nu_{1},\nu_{6}$ are each the same. This is not possible, even though superselection
rules are respected. This suggests that the superselection rules only coincidentally aid the fermionic qubit mapping for two modes.
They simply force all the problematic coefficients to disappear. However, for more than two modes we find here that a mapping to a
tensor product space cannot be performed consistently in general. Therefore, computing a measure like the negativity to determine the
entanglement between more than two modes appears to be meaningless. Due to the lack of practical alternatives, the minimization over
all states consistent with charge superselection to find $\bar{E}\sub{0.2}{0}{\mathrm{F}}$ of Eq.~(\ref{eq:fermionic entanglement of formation}) should
be considered since the restriction of the set of permissable states could make this computation feasible.\\


Let us briefly summarize the key aspects of Section~\ref{sec:Entanglement in Fermionic Quantum Fields}. We have discussed the implementation of fermionic modes as fundamental objects for quantum information tasks. The foundation of this
task is the rigorous construction of the notion of mode subsystems in a fermionic Fock space. We have demonstrated that this can
be achieved despite the absence of a simple tensor product structure. Our simple consistency conditions give a clear picture of
this process, which can be easily executed operationally by performing partial traces ``inside out." Thus we show that fermionic
mode entanglement, quantified by the (fermionic) entanglement of formation or any other function of the eigenvalues of the reduced
states, is indeed a well-defined concept, free of any ambiguities and independent of any superselection rules.\\

However, problems arise when mappings from the fermionic Fock space to qubit spaces are attempted. We have explicitly demonstrated in
two examples, for two and three modes, that such mappings cannot generally succeed. Only in the limited case where only two modes are
considered and the quantum states obey charge superselection can one meaningfully speak of an equivalence between the two fermionic
modes and two qubits. In this case the application of tools such as the negativity or concurrence is justified. We have argued that
these measures will at least provide a lower bound to genuine measures of fermionic mode entanglement.\\

Nonetheless, open questions remain. In particular, it is not clear if any practically computable measures exist for situations beyond two qubits. In
Ref.~\cite[(\ref{Paper:FriisHuberFuentesBruschi2012})]{FriisHuberFuentesBruschi2012}, which we will discuss in Section~\ref{sec:Generation of Genuine Multipartite Entanglement}, we have employed the witnesses for genuine multipartite entanglement presented in Theorem~\ref{thm:GME witness theorem}~(see also Refs.~\cite{GabrielHiesmayrHuber2010,HuberMintertGabrielHiesmayr2010}) for fermionic modes. These witnesses are completely compatible with
the framework we have presented here, but they can only provide lower bounds to entropic entanglement measures.\\

Finally, we have conjectured that the entanglement in fermionic modes is accessible even in spite of superselection rules that restrict
the possible operations performed on single modes by means of entanglement swapping. The investigation of this question, while beyond the scope
of this thesis, will certainly be of future interest.

\index{entanglement|)} 
    \newpage\ \\
%
%

    \part[Shaking Entanglement]
    {Shaking Entanglement\\[5mm]
    \Large{Quantum Correlations in \emph{Non-Uniformly} Moving Cavities}\label{partII:Shaking Entanglement}}
    \newpage\ \\
    \chapter{Constructing Non-Uniformly Moving Cavities}
\label{Chapter 4 Constructing Non Uniformly Moving Cavities}

It is the aim of this chapter to establish the mathematical model for relativistically rigid cavities that has been introduced by \emph{David Bruschi}, \emph{Ivette Fuentes}, and \emph{Jorma Louko} in Ref.~\cite{BruschiFuentesLouko2012} in the context of quantum information procedures, but shares features with earlier work, see e.g., Refs.~\cite{DodonovKlimovManko1990,Dodonov2001,Dodonov2010,DalvitMazzitelli1999,AlvesGranhenPires2010}. The framework has later been extensively expanded, including extensions to $(1+1)$ dimensional cavities for massless fermionic fields~\cite[(\ref{Paper:FriisLeeBruschiLouko2012})]{FriisLeeBruschiLouko2012} and smoothly varying accelerations for cavities containing bosonic fields~\cite{BruschiLoukoFaccioFuentes2013}. Recently, the cavity model has been further extended in Ref.~\cite[(\ref{Paper:FriisLeeLouko2013})]{FriisLeeLouko2013} to allow for fully $(3+1)$ dimensional quantum fields, including massive and massless scalar fields, massive and massless Dirac fields, as well as the electromagnetic field.\\

The initial motivation for a relativistic cavity model originates in RQI. Relativistic quantum fields are affected by the kinematics of spacetime, changes in boundary conditions, and the presence of horizons. Therefore, well known phenomena such as the Hawking-effect, or the Unruh-effect, associated to black holes and accelerated motion respectively (see, e.g., Ref.~\cite{BirrellDavies:QFbook}), are expected to influence relativistic quantum information processing~\cite{AlsingFuentes2012}. However, for a meaningful, operational description of RQI it is essential to enforce some notion of localization. In other words, the ``local" (in the sense of the tensor product) observer needs unrestricted control over his quantum system, which, in turn, requires the system to be spatially localizable with respect to the observer. This is certainly not the case for global modes of a quantum field in the whole Minkowski spacetime. Such solutions can be regarded as a means to handle a scattering theory, but for the purpose of RQI other approaches have to be considered.\\

The ideas for localization in RQI are numerous, e.g., by considering wave-packets~\cite{DownesRalphWalk2013,DraganDoukasMartinMartinez2013}, or Unruh-DeWitt type detectors~\cite{LeeFuentes2012,BruschiLeeFuentes2013,BrownMartinMartinezMenicucciMann2013}. The confinement of a quantum field in a cavity of finite length is the method that we shall discuss here. This framework will not remove issues inherent to quantum field theory, for instance, we are not proposing our method as a solution to conceptual problems of relativistic quantum measurement theories~\cite{Sorkin1993}. However, it seems reasonable to assume that measurements in a laboratory involve length scales of, say, centimeters, rather than lightyears, which may well also practically remove the conceptual issues raised in~\cite{Sorkin1993}. With these restrictions in mind, the relativistic cavity model provides a conceptually satisfying theoretical apparatus to study the fundamental connection between non-uniform motion, particle creation and quantum correlations.\\

Recently, the relativistic cavity model, using a scalar field as representative for electromagnetic radiation~\cite[(\ref{Paper:FriisLeeLouko2013})]{FriisLeeLouko2013}, has generated interest also as a possible system for experimental tests employing \emph{superconducting circuits}. The conceptual similarity to the \emph{dynamical Casimir effect}\index{dynamical Casimir effect} (see Section~\ref{sec:smoothly varying accelerations}), which has recently been verified for such materials~\cite{WilsonDynCasNature2012,LaehteenmaekiParaoanuHasselHakonen2013}, in principle allows for analogous tests of more general effects of non-uniform motion. Such an experimental setup was proposed in Ref.~\cite[(\ref{Paper:FriisLeeTruongSabinSolanoJohanssonFuentes2013})]{FriisLeeTruongSabinSolanoJohanssonFuentes2013} and we shall discuss the setup in Chapter~\ref{Chapter 7 Degradation of Entanglement between Moving Cavities}. The fermionic cavity model, on the other hand, is motivated by the prospect of simulating effects of non-uniform motion in analogue fermionic solid state systems, see, e.g., Refs.~\cite{BoadaCeliLatorreLewenstein2011,ZhangWangZhu2012,Iorio2012}.\\

This chapter is structured as follows: in Section~\ref{sec:relativistically rigid cavity} we discuss the geometric aspects of rigid cavities in Minkowski spacetime. In Sections~\ref{sec:Scalar Fields in Rigid Cavities} and~\ref{sec:Dirac Fields in Rigid Cavities} we then go on to study respectively the quantized scalar field and Dirac field in inertial and uniformly accelerated cavities. We match the segments of uniform motion to construct rigid cavities that are moving non-uniformly. In Section~\ref{sec:Grafting Generic Cavity Trajectories} different trajectories \textemdash\ travel scenarios \textemdash\ are constructed, including smoothly varying accelerations.\\

\vspace*{-3mm}
Note that we are using unit where $\hbar=c=1$ throughout.

\section{The Relativistically Rigid Cavity}\label{sec:relativistically rigid cavity}

Before we start to consider quantum fields in cavities, let us ask about the cavity itself. In particular we have to inquire ``\emph{How can we describe a rigid cavity in relativity?}" We attempt to answer this question by explaining the notion of rigidity we have chosen for our model. Ultimately, every model needs to be compared with empirical data, but we will argue here that the construction introduced in Ref.~\cite{BruschiFuentesLouko2012} is a conceptually satisfying approach.

\subsubsection{Inertial Rigid Cavity}

\begin{figure}[h]
\centering
\includegraphics[width=0.75\textwidth]{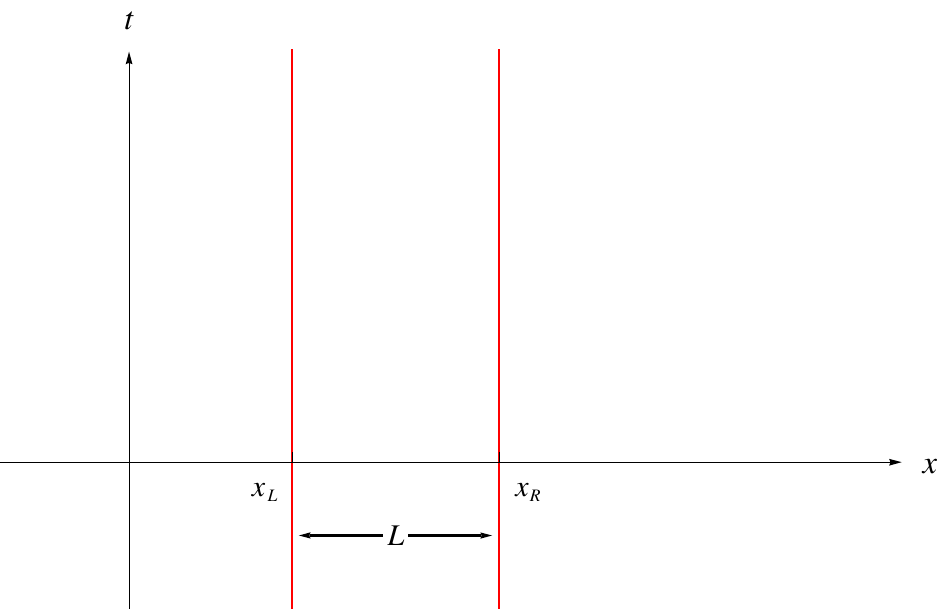}
\caption{
\textbf{Inertial rigid cavity:} From the point of view of a co-moving observer with coordinates $(t,x)$, the inertial, rigid cavity has boundaries at $x=\xLR{L}$ and $x=\xLR{R}\,$, such that the cavity has the proper length $L=\xLR{R}-\xLR{L}\,$.
\label{fig:inertial rigid cavity}}
\end{figure}

The starting point is an ideally lossless, inertial cavity of fixed length~$L$ in a $(1+1)$ dimensional Minkowski spacetime. We pick a co-moving inertial frame with coordinates $(t,x)$ such that for all times~$t$ the boundaries of the cavity are located at~$\xLR{L}$ and $\xLR{R}\,$, respectively, where $\xLR{R}-\xLR{L}=L>0\,$, see Fig.~\ref{fig:inertial rigid cavity}.

From the point of view of inertial observers that are moving with a constant velocity~$v$ with respect to this cavity its length $L_{v}$ is seen to be Lorentz contracted, i.e., $L_{v}=L\sqrt{1-v^{2}}$, where~$L$ is the proper length [see Eq.~(\ref{eq:proper length})] as measured by the observer co-moving with the cavity. In technical terms, $x(t)=\xLR{L}$ and $x(t)=\xLR{R}$ are integral curves of the global time-like Killing vector~$\partial_{t}\,$,\index{Killing vector} see Definition~\ref{def:time-like,space-like,null} and Eq.~\ref{eq:Killing equation}. In other words, the cavity walls are ``dragged along" by the Killing vector $\partial_{t}$\,. In principle the coordinates could have been picked such that $\xLR{L}=0$ and $\xLR{R}=L$ but the choice of $\xLR{L}>0$ will be more convenient for the accelerating cavity.

\subsubsection{Accelerated Motion \textemdash\ Rindler Coordinates}

\begin{figure}[hb]
\centering
\includegraphics[width=0.85\textwidth]{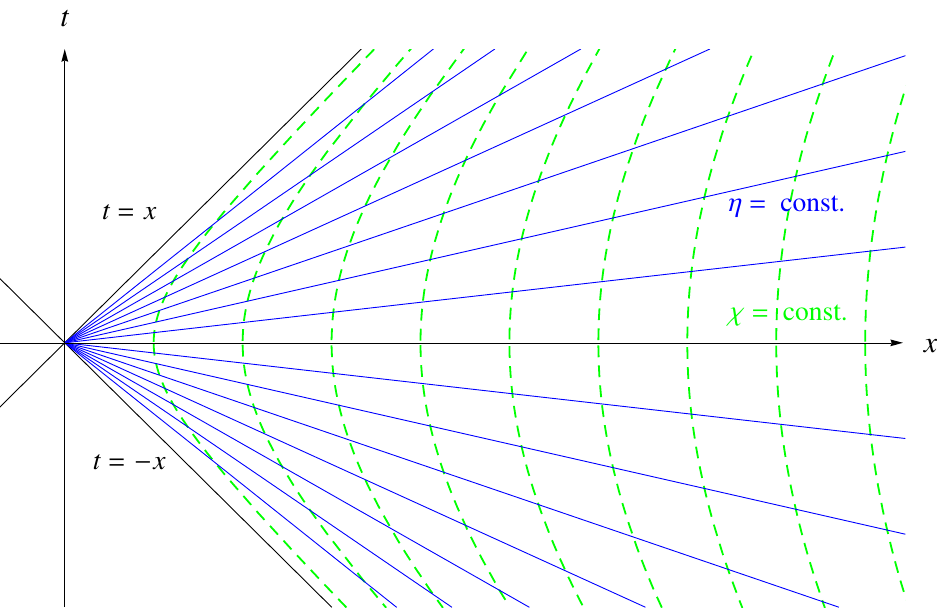}
\caption{
\textbf{Rindler coordinates:} Lines of constant~$\chi$ (dashed green confocal hyperbolae) and constant~$\eta$ (solid blue radial lines) are shown for selected values of the coordinates in the right Rindler wedge~$|t|<x$. The Rindler horizon is indicated by the solid black lines $t=x$ and $t=-x\,$. The hyperbolae~$\chi=\operatorname{const.}$ describe a family of (point-like) observers that are eternally uniformly accelerated.
\label{fig:right Rindler wedge}}
\end{figure}

In order to accelerate the cavity walls we consider appropriate coordinates \textemdash\ \emph{Rindler coordinates}\index{Rindler coordinates}~$(\eta,\chi)$, see Ref.~\cite{Takagi1986}. For the quadrant $|t|<x\,$ (without loss of generality we accelerate towards increasing~$x$) we choose the hyperbolic Rindler coordinates
\begin{subequations}
\label{eq:right wedge rindler coordinates}
\begin{align}
t   &=\,\chi\sinh(\eta)\,,\label{eq:right rindler t}\\
x   &=\,\chi\cosh(\eta)\,,\label{eq:right rindler x}
\end{align}
\end{subequations}
where $0<\chi<\infty$ and $-\infty<\eta<\infty\,$, see Fig.~\ref{fig:right Rindler wedge}. Let us see why these coordinates are suitable for accelerated motion. From Eqs.~(\ref{eq:right wedge rindler coordinates}) one can easily see that the lines of constant~$\chi$ are time-like (see Definition~\ref{def:time-like,space-like,null}) and can therefore be used to describe an (ideally point-like) observer. From Eq.~(\ref{eq:proper time}) it can be straightforwardly verified that the proper time~$\tau$\index{proper!time} along a worldline $\chi=\operatorname{const.}$ is given by $\tau=\chi\eta\,$. The coordinate time~$\eta$ is thus proportional to the proper time for fixed~$\chi$. Parameterizing the worldline by~$\tau$ and taking the second derivative with respect to the proper time one arrives at
\begin{align}
a^{\,\mu}(\tau) &=\,\frac{d^{2}}{d\tau^{2}}x^{\,\mu}(\tau)\,=\,
    \frac{1}{\chi}\begin{pmatrix}   \sinh(\tau/\chi)   \\   \cosh(\tau/\chi)\end{pmatrix}\,.
    \label{eq:four acceleration}
\end{align}
The magnitude~$\mathbf{a}$ of this vector (with respect to the $(1+1)$ dimensional Minkowski metric~$\diag\{-1,+1\}$), i.e., where $\mathbf{a}^{2}=a^{\,\mu}a_{\,\mu}$, is called the \emph{proper acceleration}\label{page:proper acceleration}\index{proper!acceleration} and it is here given by $\mathbf{a}=1/\chi\,$. Physically the proper acceleration is the Newtonian acceleration along the worldline as measured in the instantaneous rest frame. In conclusion, lines of constant~$\chi$ correspond to worldlines of (ideally point-like) observers with fixed proper acceleration~$1/\chi$ towards increasing values of~$x$ and proper time~$\chi\eta$. If so chosen, leftward acceleration can be described by a second set of Rindler coordinates for the quadrant $|t|<-x$ with the replacement $x\rightarrow-x$ in Eqs.~(\ref{eq:right wedge rindler coordinates}), see p.~\pageref{page:leftward acceleration}.

\subsubsection{Accelerated Rigid Cavity}

To use Rindler coordinates in the construction of an accelerating cavity it is still necessary to give some thought to the notion of \emph{rigidity}. Let us consider one cavity wall that is following a worldline of constant $\chi=\chiLR{L}\,$. At every instant of the coordinate time~$\eta$ the plane of simultaneity from the perspective of an observer identified with the cavity wall is the line of constant~$\eta\,$. Note that the proper distance\index{proper!length} [Eq.~(\ref{eq:proper length})] between two different hyperbolae $\chi=\chiLR{L}=\operatorname{const.}$ and $\chi=\chiLR{R}=\operatorname{const.}$ along lines of fixed~$\eta$ is constant as well. Thus, a cavity of length $L=\chiLR{R}-\chiLR{L}$ with walls that are uniformly accelerating with \emph{different} proper accelerations $1/\chiLR{L}$ and $1/\chiLR{R}\,$, respectively, can be considered to be rigid. The argument can be extended to any part of the cavity between the two walls, e.g., an observer placed in the centre of the cavity whose proper acceleration is given by $2/(\chiLR{R}+\chiLR{L})$ experiences the walls at fixed proper distance $L/2$ throughout the journey. In analogy to the inertial case the uniformly accelerated boundaries can now be considered to be ``dragged along" by the Killing vector~$\partial_{\eta}=x\partial_{t}+t\partial_{x}\,$.\index{Killing vector}

\begin{figure}[h]
\centering
\includegraphics[width=0.75\textwidth]{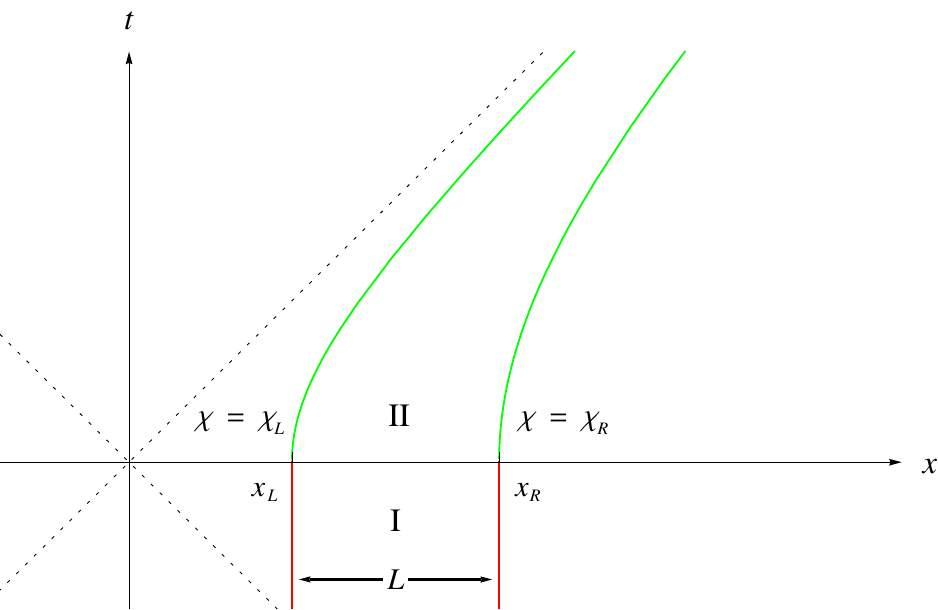}
\caption{
\textbf{Relativistically rigid cavity:} The rigid inertial (region~I) and uniformly accelerated (region~II) cavities can be combined by pasting together the boundaries $x=\xLR{L}$ and $x=\xLR{R}\,$, with $\chi=\chiLR{L}$ and $\chi=\chiLR{R}$, respectively, at $\eta=t=0\,$. The proper length with respect to a co-moving observer is $L=\xLR{R}-\xLR{L}\,$. The dashed lines indicate the light cone at the origin.
\label{fig:rigid cavity}}
\end{figure}

At last, the inertial cavity depicted in Fig~\ref{fig:inertial rigid cavity} can be uniformly accelerated by pasting the inertial and uniformly accelerated trajectories of the cavity walls along slices of fixed~$\eta\,$, such that $\chiLR{L}=\xLR{L}$ and $\chiLR{R}=\xLR{R}$. The tangent vectors of the trajectories are orthogonal to the corresponding line~$\eta=\operatorname{const.}$ in the sense of the Minkowski metric (see p.~\pageref{page:Minkowski metric}), see Fig.~\ref{fig:rigid cavity}. We shall extend this framework to generic trajectories in Section~\ref{sec:Grafting Generic Cavity Trajectories}. 


\section{Scalar Fields in Rigid Cavities}
\label{sec:Scalar Fields in Rigid Cavities}

For the quantization of the fields we adopt the same strategy as for the geometric construction. We first consider the quantization procedure for the scalar field individually for an inertial and a uniformly accelerated cavity before we match the two. The whole programme is then repeated for the Dirac field in Section~\ref{sec:Dirac Fields in Rigid Cavities}.

\subsection{Cavity in Uniform Motion \textemdash\ Scalar Field}\label{sec:uniform motion scalar field}

\vspace*{2mm}
\subsubsection{Scalar Field in Inertial Cavity}

Let us consider a real, scalar field~$\phi$\index{Klein-Gordon!field} in a $(1+1)$ dimensional Minkowski spacetime with metric $ds^{2}=\eta_{\hspace*{0.5pt}\mu\nu}\hspace*{0.5pt}dx^{\mu}dx^{\nu}=-dt^{2}+dx^{2}\,$. The field satisfies the \emph{Klein-Gordon equation}\index{Klein-Gordon!equation, Minkowski coord.} [see Eq.~(\ref{eq:Klein Gordon equation general})]
\begin{align}
    \left(-\square\,+\,\mathfrak{m}^{2}\right) \phi(t,x) &=\,0\,,
    \label{eq:Klein Gordon Minkowski}
\end{align}
where $\square=\eta^{\,\mu\nu}\partial_{\mu}\partial_{\nu}$ is the scalar Laplacian\index{Laplacian} and~$\mathfrak{m}\geq0$ is the mass to be associated with the excitations of the quantum field. To confine the mode solutions $\phii{n}$ to the inertial cavity discussed in Section~\ref{sec:relativistically rigid cavity} we impose the \emph{Dirichlet boundary conditions}\index{Dirichlet boundary conditions}\index{boundary conditions!Dirichlet}
\begin{align}
    \phii{n}(t,\xLR{L})   &=\,\phii{n}(t,\xLR{R})\,=\,0\,.
    \label{eq:Dirichlet boundary cond}
\end{align}
Alternatively, other boundary conditions, for instance Neumann boundary conditions may be chosen. This is of relevance when a Maxwell field is considered in a cavity, for which the two polarization degrees of freedom behave like Dirichlet and Neumann scalar fields, respectively~\cite[(\ref{Paper:FriisLeeLouko2013})]{FriisLeeLouko2013}. The qualitative features of the scalar field cavity model under these two types of conditions are the same, and the Dirichlet conditions seem to be the intuitively most natural restrictions to model perfectly reflecting cavity walls for the scalar field. Therefore, we shall be content to focus our discussion on the Dirichlet boundary condition of Eq.~(\ref{eq:Dirichlet boundary cond}) above. For the situation here the Klein-Gordon (pseudo) inner product\index{Klein-Gordon!inner product} of Eq.~(\ref{eq:KG inner product}) reads
\begin{align}
    (\,\phii{m}\,,\,\phii{n}\,)_{\raisebox{-1.7pt}{\scriptsize{KG}}}    &=\,
    -i\int\limits_{\xLR{L}}^{\xLR{R}}dx\,
    \bigl(\phii{m}\hspace*{0.5pt}\partial_{t}\hspace*{0.5pt}\phii{n}^{*}\,-\,
        \phii{n}^{*}\hspace*{0.5pt}\partial_{t}\hspace*{0.5pt}\phii{m}\bigr)\,.
    \label{eq:KG inner product Minkowski}
\end{align}
A standard basis of orthonormal [w.r.t. the inner product~(\ref{eq:KG inner product Minkowski})] solutions to Eq.~(\ref{eq:Klein Gordon Minkowski}), subject to the boundary conditions of Eq.~(\ref{eq:Dirichlet boundary cond}), can be found to be
\begin{align}
    \phii{n} (t,x) &=\,\frac{1}{\sqrt{\omega_{n}L}}\sin \! \left(\frac{n\pi (x-\xLR{L})}{L}\right)
    e^{-i\hspace*{0.5pt}\omega_{n}t}\,.
    \label{Minkowski solutions Dirichlet}
\end{align}
The field modes $\phii{n}$ are labelled by the discrete index $n=1,2,3,\ldots,$\, and are of positive frequency \begin{align}
    \omega_{n} &= \frac{1}{L}\sqrt{M^{2} + (\pi n)^{2}}\,,
    \label{eq:inertial modes frequencies}
\end{align}
where we have introduced the dimensionless parameter $M:=\mathfrak{m}\hspace*{0.5pt}L\,$, with respect to the Minkowski time-translation Killing vector $\partial_{t}\,$.\index{Killing vector} The phase in Eq.~(\ref{Minkowski solutions Dirichlet})
has been chosen such that $\partial_{x}\phii{n}>0$ at $(t,x)=(0,\xLR{L})\,$. In the inertial region the field can be decomposed as
\begin{align}
    \phi    &=\,\sum\limits_{n}\bigl(\phii{n}\,\an{n}\,+\,\phii{n}^{*}\,\adn{n}\bigr)\,,
    \label{eq:Klein Gordon field inertial}
\end{align}
where the field operators~$\an{n}$ and~$\adn{n}$ satisfy the commutation relations of Eqs.~(\ref{eq:canonical commutation relations}).

\subsubsection{Scalar Field in Uniformly Accelerated Cavity}

To quantize the field in the accelerated region~II of Fig.~\ref{fig:rigid cavity} we again employ the Rindler coordinates\index{Rindler coordinates} $(\eta,\chi)$ of Eq.~(\ref{eq:right wedge rindler coordinates}) for which the line element is
\begin{align}
    ds^{2}  &=\,-\chi^{2}\hspace*{0.5pt}d\eta^{2}\,+\,d\chi^{2}\,.
    \label{eq:Rindler metric}
\end{align}
Since now $\sqrt{-\det g}=\chi$ and $g^{\hspace*{1pt}\mu\nu}=\diag\{-1/\chi^{2},1\}$ we can rewrite the Klein-Gordon equation~(\ref{eq:Klein Gordon equation general}) in Rindler coordinates\index{Klein-Gordon!equation, Rindler coordinates},
\begin{align}
    \bigl(-\partial_{\eta}^{2}\,+\,(\chi\partial_{\chi})^{2}\,-\,\mathfrak{m}^{2}\chi^{2}\bigr)\phi(\eta,\chi)&=\,0\,.
    \label{eq:KG equation Rindler coord}
\end{align}
Before applying the boundary conditions it is useful to make the ansatz $\phi(\eta,\chi)=f(\chi)e^{-i\hspace*{0.5pt}\Omega\eta}$ for the solutions such that Eq.~(\ref{eq:KG equation Rindler coord}) can be cast into the form
\begin{align}
    \bigl(\chi^{2}\partial_{\chi}^{2}\,+\,\chi\partial_{\chi}\,-\,[\mathfrak{m}^{2}\chi^{2}\,+\,(i\hspace*{0.5pt}\Omega)^{2}]\bigr)f(\chi)&=\,0\,.
    \label{eq:KG equation Rindler coord mod Bessel}
\end{align}
With a simple coordinate re-scaling $\chi\rightarrow \mathfrak{m}\chi\,$, where $\partial_{\chi}\rightarrow \mathfrak{m}^{-1}\partial_{\chi}\,$ and we assume $\mathfrak{m}>0$, it becomes apparent that Eq.~(\ref{eq:KG equation Rindler coord mod Bessel}) is the \emph{modified Bessel equation}\index{Bessel!equation, modified}~\cite{nist-dig-library}. At this stage we enforce the Dirichlet boundary conditions\index{boundary conditions!Dirichlet}
\begin{align}
    \phiitilde{n}(\eta,\chiLR{L})    &=\,\phiitilde{n}(\eta,\chiLR{R})\,=\,0\,,
    \label{eq:Dirichlet boundary cond Rindler}
\end{align}
for the region~II field modes $\phiitilde{n}$ in complete analogy to Eq.~(\ref{eq:Dirichlet boundary cond}). The inner product of Eq.~(\ref{eq:KG inner product}) for the metric~(\ref{eq:Rindler metric}) now reads~\cite{Takagi1986}
\begin{align}
    (\,\phiitilde{m}\,,\,\phiitilde{n}\,)_{\raisebox{-1.7pt}{\scriptsize{KG}}}    &=\,
    -i\int\limits_{\chiLR{L}}^{\chiLR{R}}d\chi\,\chi^{-1}\,
    \bigl(\phiitilde{m}\hspace*{0.5pt}\partial_{\eta}\hspace*{0.5pt}\phiitilde{n}^{*}\,-\,
        \phiitilde{n}^{*}\hspace*{0.5pt}\partial_{\eta}\hspace*{0.5pt}\phiitilde{m}\bigr)\,.
    \label{eq:KG inner product Rindler}
\end{align}
A basis of mode functions that are orthogonal in the inner product~(\ref{eq:KG inner product Rindler}) and that are of positive frequency with respect to the time-like boost Killing vector $\partial_{\eta}$\index{Killing vector} are given by
\begin{align}
    \phiitilde{n}(\eta,\chi)    &=\,N_{n}\,e^{-i\hspace*{0.5pt}\Omega_{n} \eta}\,\bigl[\,
    I_{-i\hspace*{0.5pt}\Omega_{n}}(\mathfrak{m}\chiLR{L})\,I_{i\hspace*{0.5pt}\Omega_{n}}(\mathfrak{m}\chi)\,-\,
    I_{i\hspace*{0.5pt}\Omega_{n}}(\mathfrak{m}\chiLR{L})\,I_{-i\hspace*{0.5pt}\Omega_{n}}(\mathfrak{m}\chi)\,\bigr]\,,
    \label{eq:Rindler mode solutions}
\end{align}
where $N_{n}$ is a normalization constant, the modes are labelled $n=1,2,3,\ldots$\,, and the quantities $I_{\pm i\hspace*{0.5pt}\Omega_{n}}(\mathfrak{m}\chi)$ are the \emph{modified Bessel functions of the first kind}\index{Bessel!functions, modified, 1$^{st}$ kind}~\cite{nist-dig-library}. Finally, the Rindler frequencies $\Omega_{n}\,$, which are ordered by ascending mode label, i.e., $\Omega_{m}>\Omega_{n}$ for $m>n\,$, are determined by the second boundary condition $\phiitilde{n}(\eta,\chiLR{R})=0\,$. The particular form of $\Omega_{n}$ and the choice of $N_{n}$ are best discussed in the context of the transition between the inertial region~I and the accelerated region~II, which we shall do in the following Section~\ref{sec:matching scalar field}. The quantum field in region~II is now naturally decomposed into the modes $\phiitilde{n}$ as
\vspace*{-2mm}
\begin{align}
    \phi    &=\,\sum\limits_{n}\bigl(\phiitilde{n}\,\antilde{n}\,+\,\phiitilde{n}^{*}\,\adntilde{n}\bigr)\,,
    \label{eq:Klein Gordon field accelerated}
\end{align}
where the field operators~$\antilde{n}$ and~$\adntilde{n}$ satisfy
$\comm{\antilde{m}}{\adntilde{n}}=\delta_{mn}$ and $\comm{\antilde{m}}{\antilde{n}}=0$\,.

\subsection{Matching: Inertial to Rindler \textemdash\ Scalar Field}\label{sec:matching scalar field}

\begin{figure}[ht]
\centering
\includegraphics[width=0.75\textwidth]{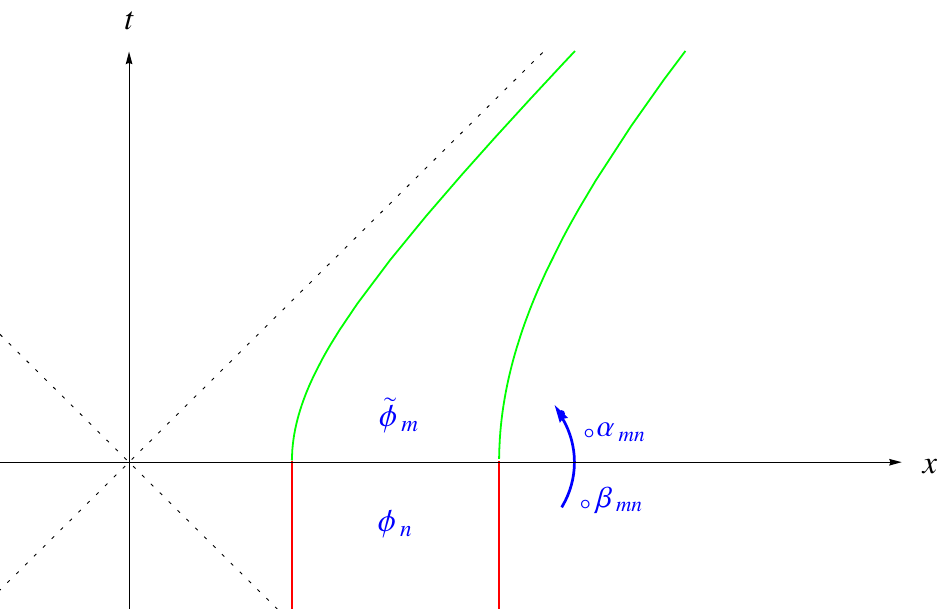}
\caption{
\textbf{Minkowski to Rindler Bogoliubov transformation \textemdash\ scalar field:} The cavity modes~$\phii{n}$ in the inertial region and the modes~$\phiitilde{m}$ in the uniformly accelerated region are related by a Bogoliubov transformation with coefficients $\MRalpha{mn}$ and $\MRbeta{mn}\,$, see Eq.~(\ref{eq:bosonic Bogos Mink to Rindler}).
\label{fig:Mink to Rindler Bogos}}
\end{figure}

Consider now the transition between the inertial region~I and the uniformly accelerated region~II. At $t=0$ the cavity walls suddenly accelerate, such that their velocity changes smoothly but their proper accelerations have finite jumps, $0\rightarrow1/\chiLR{L}$ and $0\rightarrow1/\chiLR{R}\,$, respectively. We model the instantaneous change in the mode structure by a linear transformation \textemdash\ a \emph{Bogoliubov transformation}\index{Bogoliubov transformation!bosons}, see Definition~\ref{def:Bogoliubov transformation bosons}, which is of the form
\begin{align}
        \phiitilde{m}   &=\,\sum\limits_{n}\bigl(\MRalpha{mn}\,\phii{n}\,+\,\MRbeta{mn}\,\phii{n}^{*}\bigr)\,,
        \label{eq:bosonic Bogos Mink to Rindler}
\end{align}
as illustrated in Fig.~\ref{fig:Mink to Rindler Bogos}. From the field decompositions~(\ref{eq:Klein Gordon field inertial}) and~(\ref{eq:Klein Gordon field accelerated}) the Minkowski to Rindler Bogoliubov coefficients~$\MRalpha{mn}$ and $\MRbeta{mn}$\index{Bogoliubov coefficients bosons!Minkowski to Rindler} can be written as
\begin{subequations}
\label{eq:bosonic Mink to Rindler Bogo coefficients}
\begin{align}
    \MRalpha{mn} &=\,(\,\tilde{\phi}_{m}\,,\,\phi_{n}\,)_{\raisebox{-1.7pt}{\scriptsize{KG}}}\,,
    \label{eq:bosonic Mink to Rindler Bogo coefficients alpha}\\[1mm]
    \MRbeta{mn}  &=\,-\,(\,\tilde{\phi}_{m}\,,\,\phi^{*}_{n}\,)_{\raisebox{-1.7pt}{\scriptsize{KG}}}\,,
    \label{eq:bosonic Mink to Rindler Bogo coefficients beta}
\end{align}
\end{subequations}

where one may either evaluate the inner product~(\ref{eq:KG inner product Minkowski}) at $t=0$ or, equivalently,~(\ref{eq:KG inner product Rindler}) at $\eta=0$\,. Unfortunately, even though it is straightforward to write the abstract definitions of Eq.~(\ref{eq:bosonic Mink to Rindler Bogo coefficients}), the corresponding integrals do not yield expressions in terms of known elementary functions. However, it is convenient to perform a suitable power expansion of the integrand. To this end it is useful to parameterize the cavity geometry by the quantities~$L$ and~$h$, where the dimensionless parameter
\begin{align}
h   &=\,\mathbf{a}_{\mathrm{c}}\,L
\end{align}
is the product of the proper acceleration~$\mathbf{a}_{\mathrm{c}}=2/(\chiLR{R}+\chiLR{L})$\index{proper!acceleration} (see p.~\pageref{page:proper acceleration}) at the centre of the cavity and its width $L=(\chiLR{R}-\chiLR{L})$. The cavity boundaries, expressed through~$h$ and $L$ read
\begin{subequations}
\label{eq:boundaries through h and L}
\begin{align}
    \chiLR{L}   &=\,\left(\frac{1}{h}\,-\,\frac{1}{2}\right)L\,,
    \label{eq:left boundary through L and h}\\
    \chiLR{R}   &=\,\left(\frac{1}{h}\,+\,\frac{1}{2}\right)L\,,
    \label{eq:right boundary through L and h}
\end{align}
\end{subequations}
where $0<h<2$ such that the acceleration at both ends remains finite. We shall work perturbatively in~$h\,$ from now on, i.e., we are interested in finding Taylor-Maclaurin expansions\index{Taylor-Maclaurin expansion} around $h=0\,$ for all quantities of interest. First, noting that the coordinate time~$\eta$ is dimensionless we find that the proper time observed at the centre of the cavity is given by $L\eta/h\,$. The angular frequencies $\tilde{\omega}_{n}$ with respect to this proper time are then obtained from the dimensionless Rindler frequencies~$\Omega_{n}$ as
\begin{align}
\tilde{\omega}_{n}  &=\,\frac{h}{L}\Omega_{n}\,=\,\frac{n\,\pi\,h}{2\,L\,\artanh(h/2)}\,=\,\omega_{n}\,(1\,+\,O(h^{2}))\,,
\label{eq:cavity centre proper freq}
\end{align}
where $O(x)$ denotes a quantity for which $O(x)/x$ is finite for $x\rightarrow0\,$. To leading order in the expansion the Minkowski and Rindler mode functions, $\phii{n}$ and $\phiitilde{n}\,$, must be equal up to a phase factor, which we set to unity in the normalization constant $N_{n}$ of Eq.~(\ref{eq:Rindler mode solutions}) such that $\partial_{\chi}\phiitilde{n}>0$ at $(\eta,\chi)=(0,\chiLR{L})\,$. Since both the order and the arguments of the modified Bessel functions~$I_{\pm i\hspace*{0.5pt}\Omega_{n}}(m\chiLR{L})$ in Eq.~(\ref{eq:Rindler mode solutions}) diverge at $h=0$ we have to use the corresponding uniform asymptotic expansions~\cite{Dunster1990} to obtain the perturbative expansion of the inner products in~(\ref{eq:bosonic Mink to Rindler Bogo coefficients}). With some computational effort expansions of the Bogoliubov coefficients are obtained as
\begin{subequations}
\label{eq:Mink to Rindler alphas and betas small h expansion}
\begin{align}
    \MRalpha{mn}  &=  \MRalphahmn{0}{mn}\,+\,\MRalphahmn{1}{mn}\,h\,+\,\MRalphahmn{2}{mn}\,h^{2}\,+\,O(h^{3})\,,
    \label{eq:Mink to Rindler alphas small h expansion}\\
    \MRbeta{mn}   &=  \MRbetahmn{1}{mn}\,h\,+\,\MRbetahmn{2}{mn}\,h^{2}\,+\,O(h^{3})\,,
    \label{eq:Mink to Rindler betas small h expansion}
\end{align}
\end{subequations}
where the superscripts $^{(n)}$ indicate the coefficients of $h^{n}\,$. The leading order is $\MRalphahmn{0}{mn}=\delta_{mn}\,$, while $\MRalphahmn{1}{nn}=\MRbetahmn{1}{nn}=0\,$. For $m\neq n$ we find the linear terms
\begin{subequations}
\label{eq:bosonic Bogo coeffs linear}
\begin{align}
    \MRalphahmn{1}{mn} &=\,-\,\frac{\pi^{2} m n \bigl(1 - {(-1)}^{m+n}\bigr)}
        {L^{4}\,\sqrt{\omega_{m}\omega_{n}}\,\left(\omega_{m}\,-\,\omega_{n}\right)^{3}}\,,
        \label{eq:bosonic Bogo coeffs linear alpha}\\[2mm]
    \MRbetahmn{1}{mn} &=\,\frac{\pi^{2} m n \bigl(1 - {(-1)}^{m+n}\bigr)}
        {L^{4}\,\sqrt{\omega_{m}\omega_{n}}\,\left(\omega_{m}\,+\,\omega_{n}\right)^{3}}\,,
        \label{eq:bosonic Bogo coeffs linear beta}
\end{align}
\end{subequations}
with $\omega_{n}$ given by Eq.~(\ref{eq:inertial modes frequencies}). Note that the linear coefficients vanish for mode pairs $(m,n)$ with equal parity, i.e., $\MRalphahmn{1}{mn}=\MRbetahmn{1}{mn}=0$ if $(m+n)$ is even. The second order coefficients can be obtained with the same procedure, but we will not need their explicit form in the following. However, we shall note that it has been verified that the Bogoliubov coefficients up to and including second order are satisfying the Bogoliubov identities~(\ref{eq:bosonic Bogo unitarity}) when terms proportional to~$h^{2}$ are kept. In addition, the second order coefficients $\MRalphahmn{2}{mn}$ and $\MRbetahmn{2}{mn}$ are proportional to $(1 + {(-1)}^{m+n})$ and, consequently, vanish for index pairs $(m,n)$ with opposite parity, i.e., if $(m+n)$ is odd.
\vspace*{-2.8mm}
\begin{figure}[hb!]
\centering
\includegraphics[width=0.73\textwidth]{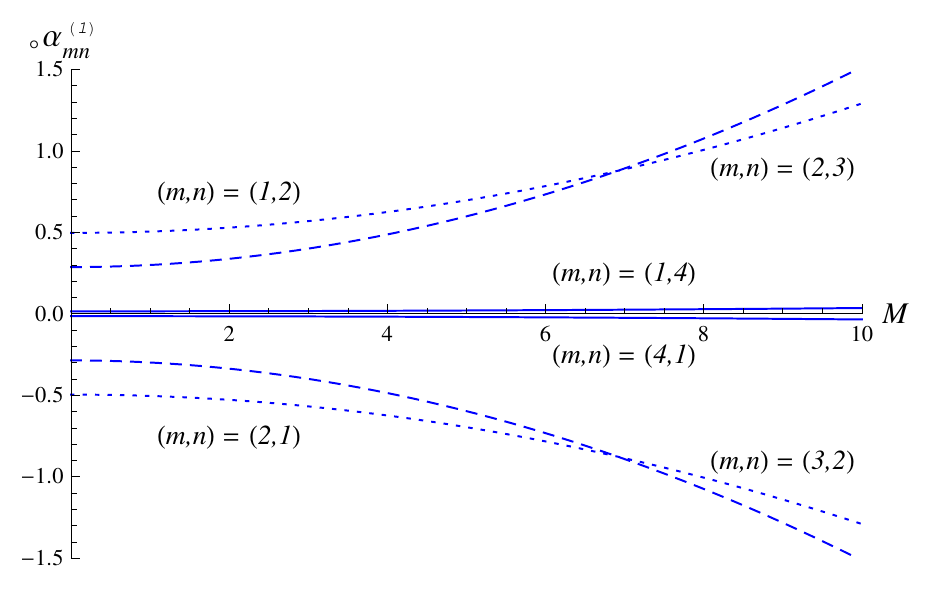}
\caption{
\textbf{Minkowski to Rindler Bogoliubov coefficients \textemdash\ scalar field $\alpha$'s:} The behaviour of the leading order ``mode mixing" Bogoliubov coefficients $\MRalphahmn{1}{mn}$ from~(\ref{eq:bosonic Bogo coeffs linear alpha}) is shown for increasing mass of the $(1+1)$-dimensional real scalar field with Dirichlet boundary conditions. A selection of the coefficients $\MRalphahmn{1}{mn}$ is plotted against the dimensionless combination $M := \mathfrak{m}L$. The coefficients are proportional to $M^{2}$ as $M\rightarrow\infty\,$, while the intersections with the vertical axis give the massless limit, $\mathfrak{m}\rightarrow0\,$, of Eq.~(\ref{eq:bosonic Bogo coeffs linear alpha}).\hspace*{-6mm}
\label{fig:KG alphas}}
\end{figure}

\newpage
\begin{figure}[ht!]
\centering
\includegraphics[width=0.75\textwidth]{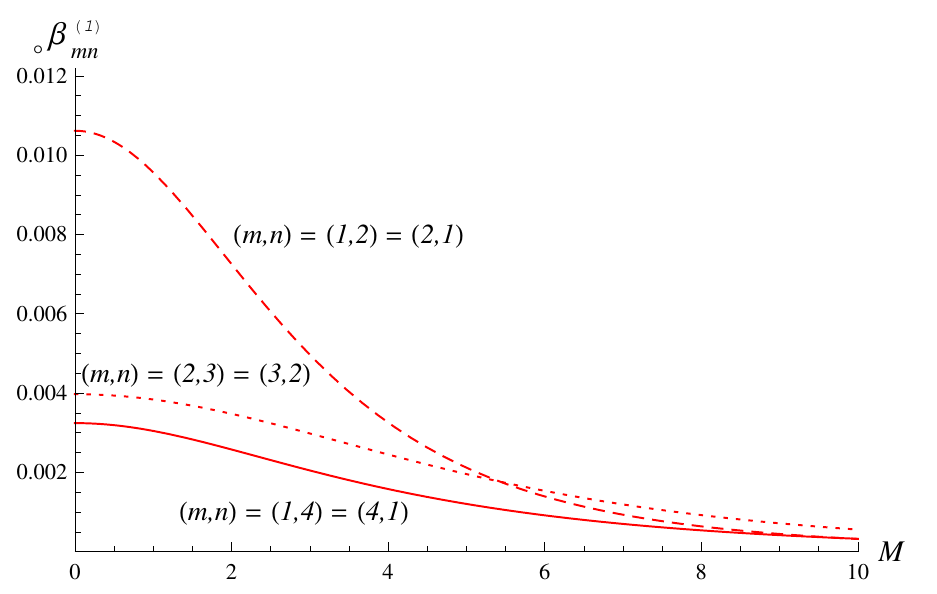}
\caption{
\textbf{Minkowski to Rindler Bogoliubov coefficients \textemdash\ scalar field $\beta$'s:} The behaviour of the leading order ``particle creation" Bogoliubov coefficients $\MRbetahmn{1}{mn}$ from~(\ref{eq:bosonic Bogo coeffs linear beta}) is shown for increasing mass of the $(1+1)$-dimensional real scalar field with Dirichlet boundary conditions. A selection of the coefficients $\MRbetahmn{1}{mn}$ is plotted against the dimensionless combination $M := \mathfrak{m}L$. The coefficients are proportional to $M^{-4}$ as $M\rightarrow\infty\,$, while the intersections with the vertical axis give the massless limit, $\mathfrak{m}\rightarrow0\,$, of Eq.~(\ref{eq:bosonic Bogo coeffs linear beta}).
\label{fig:KG betas}}
\end{figure}
We trust the perturbative expansion for $h\ll1$ when the indices of the coefficients are bounded from above by any constant. For non-zero mass we additionally require that $M\hspace*{0.3pt}h\ll1$ but within this regime we allow for $M\gg1$ such that $M^{2}h\lesssim1\,$. In that case the dominant contributions to the coefficients $\MRalphahmn{1}{mn}$ behave as~$M^{2}$, while the coefficients $\MRbetahmn{1}{mn}$ are suppressed as~$M^{-4}\,$. Further considerations regarding the perturbative regime will be presented as demanded by the applications, for instance in Chapter~\ref{Chapter 5 State Transformation by Non-Uniform Motion}. Finally, even though it was assumed that $\mathfrak{m}>0$ to obtain~(\ref{eq:Rindler mode solutions}) it can be verified that the limit $\mathfrak{m}\rightarrow0$ in~(\ref{eq:Mink to Rindler alphas and betas small h expansion}) coincides with the results obtained if the mass~$\mathfrak{m}$ is set to zero from the start, see, e.g., Ref.~\cite{BruschiFuentesLouko2012}.

\subsubsection{Leftward Acceleration}
\label{page:leftward acceleration}
\begin{figure}[ht]
\centering
\includegraphics[width=0.75\textwidth]{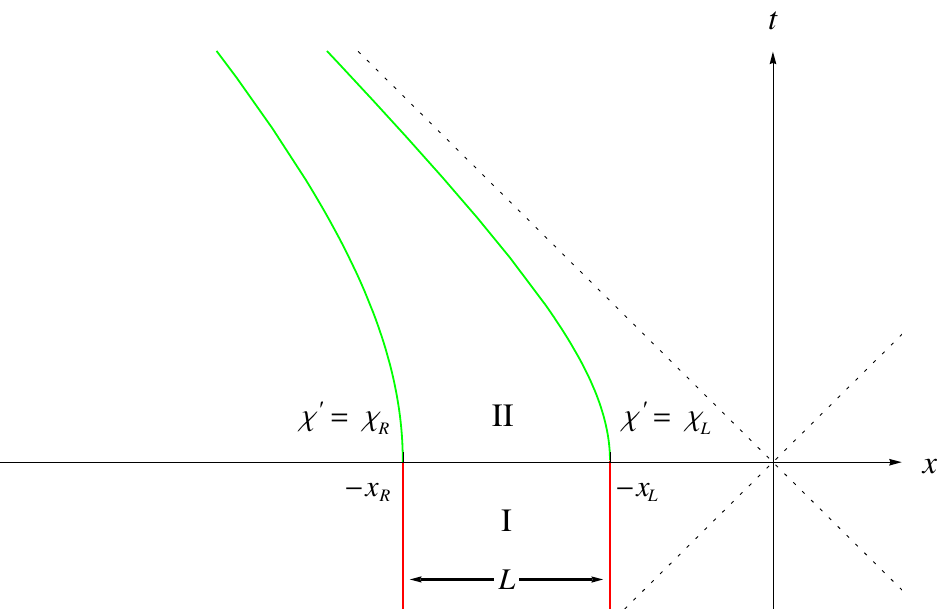}
\caption{
\textbf{Acceleration to the left:} The cavity is mirrored to the left Rindler wedge $|t|<-x\,$. The spatial reflection leaves the Bogoliubov coefficients unchanged but the inverted signs of the odd modes have to be taken into account when matching the phases of the modes.
\label{fig:leftward acceleration}}
\end{figure}

As explained in Section~\ref{sec:relativistically rigid cavity} we have so far considered acceleration towards increasing values of~$x\,$. For accelerations towards decreasing~$x$ we may repeat the whole procedure laid out in the previous sections in a similar way. Instead of~(\ref{eq:right wedge rindler coordinates}) we may introduce Rindler coordinates $(\eta\pr,\chi\pr)$ for the quadrant $|t|<-x\,$ via
\begin{subequations}
\label{eq:left wedge rindler coordinates}
\begin{align}
t   &=\,\chi\pr\sinh(\eta\pr)\,,
    \label{eq:left rindler t}\\
x   &=\,-\,\chi\pr\cosh(\eta\pr)\,.
    \label{eq:left rindler x}
\end{align}
\end{subequations}
The metric now reads $ds^{2}=-\chi^{\prime\,2}\hspace*{0.5pt}d\eta^{\prime\,2}\,+\,d\chi^{\prime\,2}$, as before in~(\ref{eq:Rindler metric}) but with the primed Rindler coordinates. For the inertial cavity the positions of the left and right cavity boundaries are now at $-\xLR{R}$ and $-\xLR{L}\,$, respectively, i.e., the cavity geometry has been mirrored with respect to $x=0\,$. For the leftward accelerated region the left and right wall now follow segments of the hyperbolae $\chi\pr=\chiLR{R}$ and $\chi\pr=\chiLR{L}$, respectively, see Fig.~\ref{fig:leftward acceleration}.
The Bogoliubov coefficients, i.e., the inner products~(\ref{eq:bosonic Mink to Rindler Bogo coefficients}) of the mirrored cavity modes~$\phii{n}\pr$ and $\phiitilde{m}\pr$, are left unchanged by such a reflection. However, to match our previous phase convention we require $\partial_{x}\phii{n}|_{t=0}>0$ and $\partial_{\eta\pr}\phiitilde{m}\pr|_{\eta=0}>0$ at the left boundaries. For the even modes $m,n=1,3,5,\ldots,$ this is automatically satisfied, but we have to compensate for the sign flip of the odd modes $m,n=2,4,\ldots,$ acquired due to the reflection. Therefore we include factors of $(-1)^{n+1}$ and $(-1)^{m+1}$ for the Minkowski and (left wedge) Rindler modes, respectively. In conclusion we find that the coefficients for leftward acceleration are obtained from the rightward acceleration coefficients $\MRalpha{mn}$ and $\MRbeta{mn}$ in~(\ref{eq:bosonic Mink to Rindler Bogo coefficients}) by inclusion of a factor $(-1)^{m+n}\,$. Practically this may be implemented by assuming the expansions~(\ref{eq:Mink to Rindler alphas and betas small h expansion}), including $h^{2}$ contributions, to hold for both cases with positive (negative)~$h$ indicating acceleration towards increasing (decreasing) values of~$x$.\\

\vspace*{-5mm}
As a last comment before we turn to the Dirac field in Section~\ref{sec:Dirac Fields in Rigid Cavities} we note that the case of linear acceleration in $(1+1)$ dimensions immediately generalizes to higher dimensions. The momenta in the spatial directions transverse to the acceleration simply contribute to the mass in the $(1+1)$ dimensional analysis, see Eq.~(\ref{eq:extra dimensions mass}).

\section{Dirac Fields in Rigid Cavities}
\label{sec:Dirac Fields in Rigid Cavities}

It is the aim of this section to consider a \emph{Dirac spinor} field that is confined to an accelerating cavity. The motivation for this approach is two-fold. First, we wish to gain insight into the influence of particle statistics on the mechanisms of the transformation to identify features of the effects of non-uniform motion that are independent of the chosen quantum field. Second, possible applications in solid state systems~\cite{BoadaCeliLatorreLewenstein2011,Iorio2012,ZhangWangZhu2012} may offer the possibility of experimental verification of particle creation effects in fermionic systems. We proceed in a similar way as in Section~\ref{sec:uniform motion scalar field}, presenting the results obtained in Refs.~\cite[(\ref{Paper:FriisLeeBruschiLouko2012})]{FriisLeeBruschiLouko2012} and \cite[(\ref{Paper:FriisLeeLouko2013})]{FriisLeeLouko2013}.

\subsection{Cavity in Uniform Motion \textemdash\ Dirac Field}\label{sec:uniform motion dirac field}

\subsubsection{Dirac Field in Inertial Cavity}

\begin{figure}[hb!]
\centering
\includegraphics[width=0.93\textwidth]{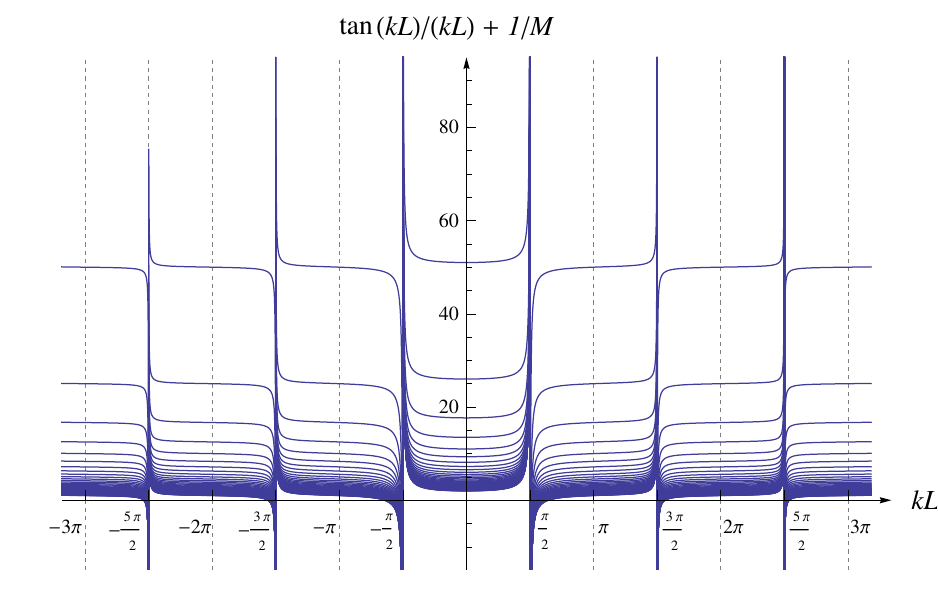}
\caption{
\textbf{Transcendental equation:} For fixed parameters~$\mathfrak{m}$ and~$L$ the allowed frequencies~$\omega_{k_{n}}$ from Eq.(\ref{eq:Dirac Mink frequencies}) for the Dirac field modes in the cavity are determined by the positive and negative numbers~$k_{n}$ that satisfy the transcendental equation~(\ref{eq:Mink modes transcendental equation}). The roots of the function $\tan(kL)/(kL)+1/M$ determine the possible values of~$(k_{n}L)\,$. Curves are shown here for discrete steps, $M\in\{\tfrac{2l}{100}|l=1,2,\ldots,50\}$ ($l$ increasing from top to bottom), of the dimensionless combination $M=\mathfrak{m}L\,$.
\label{fig:transcendal eq Mink spinor modes}}
\end{figure}

Let us consider a Dirac field\index{Dirac!field}~$\psi$ in $(1+1)$ dimensional Minkowski spacetime with metric $ds^{2}=\eta_{\hspace*{0.5pt}\mu\nu}\hspace*{0.5pt}dx^{\mu}dx^{\nu}=-dt^{2}+dx^{2}\,$. The Dirac equation\index{Dirac!equation, Minkowski coord.} [see Eq.~(\ref{eq:curved space Dirac equation})] now reads
\begin{align}
    \bigl(i\,\gamma^{\hspace*{1pt}\mu}\,\partial_{\mu}\,-\,\mathfrak{m}\bigr)
    \psi &=\,0\,,
    \label{eq:Minkowsi Dirac equation}
\end{align}
where the Dirac gamma matrices satisfy $\{\gamma^{\hspace*{1pt}\mu},\gamma^{\hspace*{1pt}\nu}\} =-2\hspace*{1pt}\eta^{\hspace*{1pt}\mu\nu}\,$, and $\mathfrak{m}$ again denotes the mass of the field quanta. As for the scalar field, additional spatial dimensions can be added, but the formalism reduces to the $(1+1)$ dimensional case by Fourier decomposition, such that the momenta transverse to the chosen direction supply strictly positive contributions to the mass~$\mathfrak{m}>0\,$, see Eq.~(\ref{eq:extra dimensions mass}). Working in the $(1+1)$ dimensional case we can work only with $2\times2$ representations of $\gamma^{\hspace*{1pt}0}$ and $\gamma^{\hspace*{1pt}1}\,$, for instance
\begin{align}
    \gamma^{\hspace*{1pt}0}\,=\,\begin{pmatrix} \ 1   &   \ 0   \\  \ 0   &   -1  \end{pmatrix}\,,
    \hspace*{0.5cm}
    \gamma^{\hspace*{1pt}1}\,=\,\begin{pmatrix} \ 0   &   1\    \\  -1   &   0\   \end{pmatrix}\,.
    \label{eq:1+1 dim gammas Dirac basis}
\end{align}
The matrices further satisfy $(\gamma^{\hspace*{1pt}0})^{2}=-(\gamma^{\hspace*{1pt}1})^{2}=\mathds{1}$ and it is convenient to multiply~(\ref{eq:Minkowsi Dirac equation}) by $\gamma^{\hspace*{1pt}0}$ to rewrite the Dirac equation as
\begin{align}
    i\,\partial_{t}\,\psi  &=\,H_{\mathrm{D}}\,\psi\,=\,
    \bigl(-i\,\gamma^{\hspace*{1pt}0}\gamma^{\hspace*{1pt}1}\,\partial_{x}\,+\,
    \mathfrak{m}\,\gamma^{\hspace*{1pt}0}\bigr)
    \psi\,.
    \label{eq:Minkowsi Dirac equation Schroedinger style}
\end{align}
We then introduce a basis $\{u_{\pm}\}$ consisting of eigenspinors of $\gamma^{\hspace*{1pt}0}\gamma^{\hspace*{1pt}1}$ such that
\begin{subequations}
\label{eq:spinor basis}
\begin{align}
    \gamma^{\hspace*{1pt}0}\gamma^{\hspace*{1pt}1}u_{\pm}    &=\,\pm\,u_{\pm}\,,
    \label{eq:spinor basis alpha 1}\\
    \gamma^{\hspace*{1pt}0}u_{\pm}    &=\,u_{\mp}\,.
    \label{eq:spinor basis beta}
\end{align}
\end{subequations}
The basis is orthonormal in the sense that $u_{\pm}^{\dagger}u_{\raisebox{-1pt}{\tiny{$\pm$}}}=1\,$, and $u_{\pm}^{\dagger}u_{\raisebox{-1pt}{\tiny{$\pm$}}}=0\,$.
For the representation of Eq.~(\ref{eq:1+1 dim gammas Dirac basis}) the basis takes the specific form
\begin{align}
    u_{\pm} &=\,\frac{1}{\sqrt{2}}\begin{pmatrix}   \ 1\,   \\  \pm1\,    \end{pmatrix}\,.
    \label{eq:spinor basis representation}
\end{align}
As for the scalar field we separate the variables to find the linearly
independent solutions of Eq.~(\ref{eq:Minkowsi Dirac equation Schroedinger style}), which can be expressed as
\begin{subequations}
\label{eq:Mink solutions without boundaries}
\begin{align}
\psi_{+,k}  &=\, \bigl(\,\cos[\xi(k)]\,u_{+}\,+\,\sin[\xi(k)]\,u_{-}\,\bigr)
    e^{ \hspace*{0.3pt}i\hspace*{0.3pt}k\hspace*{0.3pt}x\hspace*{0.3pt}-
        \hspace*{0.3pt}i\hspace*{0.3pt}\omega_{k}\hspace*{0.3pt}t}\,,
    \label{eq:Mink solutions psi plus k}\\
\psi_{-,k}  &=\, \bigl(\,\sin[\xi(k)]\,u_{+}\,+\,\cos[\xi(k)]\,u_{-}\,\bigr)
    e^{ -\hspace*{0.3pt}i\hspace*{0.3pt}k\hspace*{0.3pt}x\hspace*{0.3pt}-
        \hspace*{0.3pt}i\hspace*{0.3pt}\omega_{k}\hspace*{0.3pt}t}\,,
    \label{eq:Mink solutions psi minus k}
\end{align}
\end{subequations}
where $k$ is a non-zero real number, $\xi(k):=\tfrac{1}{2}\arctan(\tfrac{\mathfrak{m}}{k})\,$, and
the eigenvalues $\omega_{k}$ of the Dirac Hamiltonian~$H_{\mathrm{D}}$ of Eq.~(\ref{eq:Minkowsi Dirac equation Schroedinger style}) are given by
\begin{align}
    \omega_{k}\,=\,\operatorname{sgn}(k)\,\sqrt{\mathfrak{m}^{2}\,+\,k^{2}}\,.
    \label{eq:Dirac Mink frequencies}
\end{align}
The functions $\psi_{+,k}$ and $\psi_{-,k}$ represent right-movers and left-movers respectively. We are now in a position to introduce the cavity for the Dirac field with boundaries at $x=\xLR{L}$ and $x=\xLR{R}$ as discussed in Section~\ref{sec:relativistically rigid cavity}. A natural way to restrict the fermions to this region is to require the (spatial) probability current to vanish at the boundaries, i.e.,
\begin{align}
    \bar{\psi}_{1}\,\gamma^{\hspace*{1pt}1}\,\psi_{2}\bigr|_{x\hspace*{2pt}=\hspace*{2pt}\xLR{L}}
    \,=\,
    \bar{\psi}_{1}\,\gamma^{\hspace*{1pt}1}\,\psi_{2}\bigr|_{x\hspace*{2pt}=\hspace*{2pt}\xLR{R}}
    &=\,0\,,
    \label{eq:vanishing Dirac current boundaries}
\end{align}
where $\bar{\psi}=\psi^{\dagger}\gamma^{\hspace*{1pt}0}$ as in Section~\ref{sec:the dirac field}. Following the procedure laid out in Ref.~\cite{BonneauFarautValent2001} to obtain the deficiency indices for the Dirac Hamiltonian~$H_{\mathrm{D}}$ on the finite interval $[\xLR{L},\xLR{R}]$ we find that the self-adjoint extensions of~$H_{\mathrm{D}}$ are determined by two independent phases. Physically these represent the phase shifts at the reflections on the cavity walls. Imposing the boundary conditions of Eq.~(\ref{eq:vanishing Dirac current boundaries}) individually at each wall gives the solutions
\begin{subequations}
\label{eq:solutions individual boundaries}
\begin{align}
    \xLR{L}:\ \ \ \psi    &=\,\bigl[
    e^{-i\tfrac{\pi}{4}}\cos(\xi(k)-\xiLR{L})\,-\,
    e^{i\tfrac{\pi}{4}}\sin(\xi(k)+\xiLR{L})\bigr]\,
    e^{-i\hspace*{0.5pt}k\hspace*{0.5pt}\xLR{L}}\,\psi_{+,k}
    \label{eq:solutions boundary xL}\\
    &\ +\,\bigl[e^{i\tfrac{\pi}{4}}\cos(\xi(k)-\xiLR{L})\,-\,
    e^{-i\tfrac{\pi}{4}}\sin(\xi(k)+\xiLR{L})\bigr]\,
    e^{i\hspace*{0.5pt}k\hspace*{0.5pt}\xLR{L}}\ \psi_{-,k}\,,
    \nonumber\\[2mm]
    \xLR{R}:\ \ \ \psi    &=\,\bigl[
    e^{-i\tfrac{\pi}{4}}\cos(\xi(k)-\xiLR{R})\,-\,
    e^{i\tfrac{\pi}{4}}\sin(\xi(k)+\xiLR{R})\bigr]\,
    e^{-i\hspace*{0.5pt}k\hspace*{0.5pt}\xLR{R}}\,\psi_{+,k}
    \label{eq:solutions boundary xR}\\
    &\ +\,\bigl[e^{i\tfrac{\pi}{4}}\cos(\xi(k)-\xiLR{R})\,-\,
    e^{-i\tfrac{\pi}{4}}\sin(\xi(k)+\xiLR{R})\bigr]\,
    e^{i\hspace*{0.5pt}k\hspace*{0.5pt}\xLR{R}}\ \psi_{-,k}\,.
    \nonumber
\end{align}
\end{subequations}
The real parameters $\xiLR{L}$ and $\xiLR{R}\,$, parameterizing the $U(1)$ phases mentioned above, specify the boundary conditions at the left and right cavity wall, respectively. To single out physically significant choices of these parameters we turn to the \emph{MIT bag boundary conditions}~\cite{ChodosJaffeJohnsonThornWeisskopf1974,ElizaldeBordagKirsten1997}\index{boundary conditions! MIT bag}, named after the affiliation of the authors of~\cite{ChodosJaffeJohnsonThornWeisskopf1974} \textemdash\ the Massachusetts Institute of Technology. These boundary conditions, originally developed for the description of composite hadrons, emerge when the field inside the ``bag" is matched to a field with a different mass outside the boundaries and the latter mass is subsequently taken to infinity. In this sense the MIT bag boundary conditions are the analogue of the Dirichlet boundary conditions in (non-relativistic) quantum mechanics, which arise in the limit when the height of the walls of a potential well are taken to infinity, see Ref.~\cite{BelchevWalton2010}. Using our notation here the MIT bag boundary conditions read
\begin{align}
    \bigl(1\,-\,i\,\gamma^{\hspace*{1pt}1}\bigr)\,\psi\bigr|_{x\hspace*{2pt}=\hspace*{2pt}\xLR{L}}
    \,=\,
    \bigl(1\,+\,i\,\gamma^{\hspace*{1pt}1}\bigr)\,\psi\bigr|_{x\hspace*{2pt}=\hspace*{2pt}\xLR{R}}
    &=\,0\,.
    \label{eq:MIT bag boundary conditions}
\end{align}
Applying the conditions~(\ref{eq:MIT bag boundary conditions}) to Eqs.~(\ref{eq:solutions boundary xL}) and~(\ref{eq:solutions boundary xR}), respectively, we find that the MIT bag boundary conditions correspond to the choices~$\xiLR{L}=0$ and $\xiLR{R}=\frac{\pi}{2}\,$. When the Dirac field is confined to the cavity by application of the boundary conditions at both walls the corresponding (normalized) mode function spinors are found to be
\begin{align}
    \psii{k_{n}}    &=\,\sqrt{\frac{\omega_{k_{n}}^{2}}{2L \bigl(\omega_{k_{n}}^{2}\,+\,[\mathfrak{m}/L]\bigr)}}\,
    \left(e^{-i\hspace*{0.5pt}\xi(k_{n})}\,e^{-i\hspace*{0.5pt}k_{n}\hspace*{0.5pt}\xLR{L}}\,\psi_{+,k_{n}}\,+\,
    i\,e^{i\hspace*{0.5pt}\xi(k_{n})}\,e^{i\hspace*{0.5pt}k_{n}\hspace*{0.5pt}\xLR{L}}\,\psi_{-,k_{n}}\right)\,,
    \label{eq:Mink spinor cavity modes}
\end{align}
where $\psi_{\pm,k_{n}}$ and $\xi(k_{n})$ are as in~(\ref{eq:Mink solutions without boundaries}), the frequencies are given by Eq.~(\ref{eq:Dirac Mink frequencies}), and the~$k_{n}\in\mathbb{R}$, labelled by consecutive integers~$n$, take on the discrete values that satisfy the \emph{transcendental equation}\index{transcendental equation}
\vspace*{-2mm}
\begin{align}
    \frac{\tan(k_{n} L)}{k_{n} L} &=\,-\,\frac{1}{\mathfrak{m} L}\,.
    \label{eq:Mink modes transcendental equation}
\end{align}
The positive and negative frequencies appear symmetrically in the spectrum, see Fig.~\ref{fig:transcendal eq Mink spinor modes}.
\vspace*{-3mm}
\newpage
The phase in Eq.(\ref{eq:Mink spinor cavity modes}) has been chosen such that the spinors at~$(t,x)=(0,\xLR{L})$ are positive multiples of $(u_{+}+iu_{-})\,$. The Dirac field within the cavity can now be decomposed as
\begin{align}
    \psi    &=\,\sum\limits_{m\geq0}\,\psii{k_{m}}\,b_{m}\,+\,\sum\limits_{n<0}\,\psii{k_{n}}\,c_{n}^{\dagger}\,,
    \label{eq:Dirac field Mink decomposition}
\end{align}
where we have chosen the convention that $m\geq0$ and $n<0$ label positive and negative frequency solutions, respectively. The solutions appear symmetrically in the spectrum even though they are not symmetrically labelled, i.e., the lowest energy solutions are labelled by $m=0$ and $n=-1\,$. The operators $b_{m}^{\dagger}$ and $c_{n}^{\dagger}$ create particles and antiparticles, respectively, in the modes $m\geq0$ and $n<0\,$, respectively, and they satisfy the anticommutation relations of~(\ref{eq:anticomm relations}). Finally, we consider the massless limit, $\mathfrak{m}\rightarrow0\,$, for which the possible values of~$k_{n}$ coincide with the frequencies of Eq.~(\ref{eq:Dirac Mink frequencies}), such that
\begin{align}
    \omega_{k_{n}}   &=\,k_{n}\,=\,\frac{\pi}{L}\bigl(n\,+\,\tfrac{1}{2}\bigr)\,,
    \label{eq:massless Mink spinor frequencies}
\end{align}
which corresponds to the case $(s,\theta)=(\tfrac{1}{2},\tfrac{\pi}{2})$ discussed in Ref.~\cite[(\ref{Paper:FriisLeeBruschiLouko2012})]{FriisLeeBruschiLouko2012}. Note that there is no zero mode. With the notation of~(\ref{eq:massless Mink spinor frequencies}) the cavity spinors~(\ref{eq:Mink spinor cavity modes}) for the massless Dirac field take the form
\begin{align}
    \psii{k_{n}}    &=\,\frac{1}{\sqrt{2L}}\,\left(\,
    e^{\hspace*{0.3pt}i\hspace*{0.5pt}\omega_{k_{n}}(x\,-\,\xLR{L})}\,u_{+}\,+\,
    i\,e^{\hspace*{0.3pt}-i\hspace*{0.5pt}\omega_{k_{n}}(x\,-\,\xLR{L})}\,u_{-}\,\right)\,
    e^{\hspace*{0.3pt}-i\hspace*{0.5pt}\omega_{k_{n}}\hspace*{0.3pt}t}\,.
    \label{eq:Mink spinor cavity modes massless}
\end{align}

\subsubsection{Dirac Field in Uniformly Accelerated Cavity}

For the Dirac field in the accelerated region we have to determine the form of the Dirac equation~(\ref{eq:curved space Dirac equation}) in Rindler coordinates\index{Rindler coordinates}\,. To this end we express the covariant derivative $\nabla_{\mu}\psi=(\partial_{\mu}-\Gamma_{\mu})\psi$ (see, e.g., Ref.~\cite{AlsingStephensonKilian2009} and mind our sign convention for the metric) in terms of the \emph{spin connection coefficients}\index{spin connection} $\Gamma_{\mu}\,$ (Note that we have chosen the symbol $\Gamma_{\mu}$ here to adhere to usual conventions for this object even though we have used $\Gamma$ to denote the covariance matrix in Section~\ref{sec:entanglement in bosonic quantum fields} and the two concepts are unrelated.). The spin connection coefficients for the Rindler coordinates~(\ref{eq:right wedge rindler coordinates}) can be obtained from a straightforward procedure~\cite{McMahonAlsingEmbid2006} which yields $\Gamma_{\chi}=0$ and
\begin{align}
    \Gamma_{\eta}   &=\,-\frac{1}{2}\gamma^{\hspace*{1pt}0}\gamma^{\hspace*{1pt}1}\,,
    \label{eq:spin connection Rindler}
\end{align}
where the $\gamma^{\hspace*{1pt}\mu}$ are the Minkowski space gamma matrices from~(\ref{eq:1+1 dim gammas Dirac basis}). With this the Dirac equation for the right Rindler\index{Dirac!equation, Rindler coordinates} wedge becomes
\begin{align}
    i\,\partial_{\eta}\,\tilde{\psi}  &=\,
    \bigl(-i\,\gamma^{\hspace*{1pt}0}\gamma^{\hspace*{1pt}1}\,[\chi\partial_{\chi}\,+\,\tfrac{1}{2}]\,+\,
    \mathfrak{m}\,\gamma^{\hspace*{1pt}0}\chi\bigr)
    \tilde{\psi}\,.
    \label{eq:Rindler Dirac equation Schroedinger style}
\end{align}
For a formal derivation of~(\ref{eq:Rindler Dirac equation Schroedinger style}) involving the explicit construction of the dyads see, e.g., Refs.~\cite{AlsingStephensonKilian2009,Langlois2004,Langlois:PhdThesis2005}.
We proceed, as in the inertial case, by finding the linearly independent solutions to this equation, given by
\begin{subequations}
\label{eq:Rindler solutions without boundaries}
\begin{align}
\tilde{\psi}_{+,\Omega}  &=\, \bigl(\,
    I_{i\Omega-\raisebox{0pt}{\scriptsize{$\frac{1}{2}$}}}(\mathfrak{m}\chi)\,u_{+}\,+\,
    I_{i\Omega+\raisebox{0pt}{\scriptsize{$\frac{1}{2}$}}}(\mathfrak{m}\chi)\,u_{-}\,\bigr)
    e^{\hspace*{0.3pt}-i\hspace*{0.3pt}\Omega\hspace*{0.3pt}\eta}\,,
    \label{eq:Rindler solutions psi plus Omega}\\[1mm]
\tilde{\psi}_{-,\Omega}  &=\, \bigl(\,
    I_{-i\Omega+\raisebox{0pt}{\scriptsize{$\frac{1}{2}$}}}(\mathfrak{m}\chi)\,u_{+}\,+\,
    I_{-i\Omega-\raisebox{0pt}{\scriptsize{$\frac{1}{2}$}}}(\mathfrak{m}\chi)\,u_{-}\,\bigr)
    e^{\hspace*{0.3pt}-i\hspace*{0.3pt}\Omega\hspace*{0.3pt}\eta}\,,
    \label{eq:Rindler solutions psi minus Omega}
\end{align}
\end{subequations}
where $\Omega$ are the real dimensionless Rindler frequencies, while the $I_{\pm i\Omega\pm\raisebox{0pt}{\scriptsize{$\frac{1}{2}$}}}(\mathfrak{m}\chi)$ and $I_{\pm i\Omega\mp\raisebox{0pt}{\scriptsize{$\frac{1}{2}$}}}(\mathfrak{m}\chi)$ are the \emph{modified Bessel functions of the first kind}\index{Bessel!functions, modified, 1$^{st}$ kind}~\cite{nist-dig-library}. Let us briefly illustrate how to arrive at this form of the solutions by concentrating on the solutions~(\ref{eq:Rindler solutions psi plus Omega}). We insert the ansatz
\begin{align}
\tilde{\psi}_{+,\Omega}  &=\, \bigl(\,
    \tilde{I}_{+}(\chi)\,u_{+}\,+\,i\,\tilde{I}_{-}(\chi)\,u_{-}\,\bigr)
    e^{\hspace*{0.3pt}-i\hspace*{0.3pt}\Omega\hspace*{0.3pt}\eta}\,,
    \label{eq:Rindler solutions psi plus Omega ansatz}
\end{align}
where $ \tilde{I}_{\pm}(\chi)$ are yet unknown functions, into Eq.~(\ref{eq:Rindler Dirac equation Schroedinger style}). Consecutively, the orthonormality of the spinors $u_{\pm}$ from~(\ref{eq:spinor basis}) is used, i.e., $u^{\dagger}_{+}$ or $u^{\dagger}_{-}$ are applied from the left to arrive at either of the equations
\begin{subequations}
\label{eq:Rindler Dirac Bessel derivation}
\begin{align}
    \bigl(i\Omega\,-\,\tfrac{1}{2}\,-\,\chi\partial_{\chi}\,\bigr)\,\tilde{I}_{+}(\chi)\,+\,
    \mathfrak{m}\chi\,\tilde{I}_{-}(\chi)  &=\,0\,,
    \label{eq:Rindler Dirac Bessel derivation 1}\\
    \bigl(i\Omega\,+\,\tfrac{1}{2}\,+\,\chi\partial_{\chi}\,\bigr)\,\tilde{I}_{-}(\chi)\,-\,
    \mathfrak{m}\chi\,\tilde{I}_{+}(\chi)  &=\,0\,.
    \label{eq:Rindler Dirac Bessel derivation 2}
\end{align}
\end{subequations}
Assuming $(\mathfrak{m}\chi)\neq0\,$, the function $\tilde{I}_{-}$ can be expressed from~(\ref{eq:Rindler Dirac Bessel derivation 1}) and inserted into~(\ref{eq:Rindler Dirac Bessel derivation 2}), or vice versa for $\tilde{I}_{+}$, to obtain the \emph{modified Bessel equation}\index{Bessel!equation, modified}~\cite{nist-dig-library}
\begin{align}
    \left(\chi^{2}\partial_{\chi}^{2}\,+\,\chi\partial_{\chi}\,-\,
    [\mathfrak{m}^{2}\chi^{2}\,+\,(i\hspace*{0.5pt}\Omega\,\mp\,\tfrac{1}{2})^{2}]\right)\,
    \tilde{I}_{\pm}(\chi)&=\,0\,,
    \label{eq:Dirac equation Rindler coord mod Bessel}
\end{align}
revealing $\tilde{I}_{\pm}(\chi)=I_{i\Omega\mp\raisebox{0pt}{\scriptsize{$\frac{1}{2}$}}}(\mathfrak{m}\chi)\,$, as claimed in~(\ref{eq:Rindler solutions psi plus Omega}). As before, we apply the vanishing current boundary conditions~(\ref{eq:vanishing Dirac current boundaries}) at the cavity walls $\chi=\chiLR{L}$ and $\chiLR{R}$ individually, i.e.,
\begin{subequations}
\label{eq:Rindler solutions individual boundaries}
\begin{align}
\chiLR{L}:   &\label{eq:Rindler solutions boundary chiL}\\
    &\tilde{\psi}\,=\,\bigl[
    (1+e^{i\tildexiLR{L}}\tanh[\mathfrak{m}\chiLR{L}])\,
    I_{-i\Omega-\raisebox{0pt}{\scriptsize{$\frac{1}{2}$}}}(\mathfrak{m}\chiLR{L})
    \,-\,
    (e^{i\tildexiLR{L}}+\tanh[\mathfrak{m}\chiLR{L}])\,
    I_{-i\Omega+\raisebox{0pt}{\scriptsize{$\frac{1}{2}$}}}(\mathfrak{m}\chiLR{L})
    \bigr]\,\tilde{\psi}_{+,\Omega}
    \nonumber \\
    &\ \ \ \ \ \ +\,\bigl[
    (e^{i\tildexiLR{L}}+\tanh[\mathfrak{m}\chiLR{L}])\,
    I_{i\Omega-\raisebox{0pt}{\scriptsize{$\frac{1}{2}$}}}(\mathfrak{m}\chiLR{L})
    \,-\,
    (1+e^{i\tildexiLR{L}}\tanh[\mathfrak{m}\chiLR{L}])\,
    I_{i\Omega+\raisebox{0pt}{\scriptsize{$\frac{1}{2}$}}}(\mathfrak{m}\chiLR{L})
    \bigr]\,\tilde{\psi}_{-,\Omega}\,,
    \nonumber\\[2mm]
\chiLR{R}:    &\label{eq:Rindler solutions boundary chiR} \\
    &\tilde{\psi}\,=\,\bigl[
    (1+e^{i\tildexiLR{R}}\tanh[\mathfrak{m}\chiLR{R}])\,
    I_{-i\Omega-\raisebox{0pt}{\scriptsize{$\frac{1}{2}$}}}(\mathfrak{m}\chiLR{R})
    \,-\,
    (e^{i\tildexiLR{R}}+\tanh[\mathfrak{m}\chiLR{R}])\,
    I_{-i\Omega+\raisebox{0pt}{\scriptsize{$\frac{1}{2}$}}}(\mathfrak{m}\chiLR{R})
    \bigr]\,\tilde{\psi}_{+,\Omega}
    \nonumber\\
    &\ \ \ \ \ \ +\,\bigl[
    (e^{i\tildexiLR{R}}+\tanh[\mathfrak{m}\chiLR{R}])\,
    I_{i\Omega-\raisebox{0pt}{\scriptsize{$\frac{1}{2}$}}}(\mathfrak{m}\chiLR{R})
    \,-\,
    (1+e^{i\tildexiLR{R}}\tanh[\mathfrak{m}\chiLR{R}])\,
    I_{i\Omega+\raisebox{0pt}{\scriptsize{$\frac{1}{2}$}}}(\mathfrak{m}\chiLR{R})
    \bigr]\,\tilde{\psi}_{-,\Omega}\,,
    \nonumber
\end{align}
\end{subequations}
where the real parameters $\tildexiLR{L}$ and $\tildexiLR{R}$ specify the boundary conditions at $\chiLR{L}$ and $\chiLR{R}\,$, respectively. Once again we specialize to the \emph{MIT bag boundary conditions}~\cite{ChodosJaffeJohnsonThornWeisskopf1974,ElizaldeBordagKirsten1997}\index{boundary conditions! MIT bag}, here of the form
\begin{align}
    \bigl(1\,-\,i\,\gamma^{\hspace*{1pt}1}\bigr)\,\tilde{\psi}\bigr|_{\chi\hspace*{2pt}=\hspace*{2pt}\chiLR{L}}
    \,=\,
    \bigl(1\,+\,i\,\gamma^{\hspace*{1pt}1}\bigr)\,\tilde{\psi}\bigr|_{\chi\hspace*{2pt}=\hspace*{2pt}\chiLR{R}}
    &=\,0\,,
    \label{eq:MIT bag boundary conditions Rindler}
\end{align}
which singles out $\tildexiLR{L}=0$ and $\tildexiLR{R}=\pi\,$. Enforcing both boundary conditions we arrive at the cavity spinor solutions
\begin{align}
    \tilde{\psi}_{\raisebox{-1.5pt}{\scriptsize{$\Omega_{n}$}}}    &=\,N_{\Omega_{n}}\,\left(\bigl[
    I_{-i\Omega_{n}-\raisebox{0pt}{\scriptsize{$\frac{1}{2}$}}}(\mathfrak{m}\chiLR{L})\,-\,
    I_{-i\Omega_{n}+\raisebox{0pt}{\scriptsize{$\frac{1}{2}$}}}(\mathfrak{m}\chiLR{L})\bigr]\,
    \tilde{\psi}_{+,\Omega_{k(n)}}\right.
    \label{eq:Rindler spinor modes}\\
    &\ \ +\,\left.\bigl[
    I_{i\Omega_{n}-\raisebox{0pt}{\scriptsize{$\frac{1}{2}$}}}(\mathfrak{m}\chiLR{L})\,-\,
    I_{i\Omega_{n}+\raisebox{0pt}{\scriptsize{$\frac{1}{2}$}}}(\mathfrak{m}\chiLR{L})\bigr]\,
    \tilde{\psi}_{-,\Omega_{n}}\right)\,,
    \nonumber
\end{align}
where the discrete frequencies $\Omega_{n}\,$, satisfying the transcendent equation
\begin{align}
    \hspace*{-2mm}\operatorname{Re}\!\left(
    \bigl[
        I_{-i\Omega_{n}-\raisebox{0pt}{\scriptsize{$\frac{1}{2}$}}}(\mathfrak{m}\chiLR{L})\,-\,
        I_{-i\Omega_{n}+\raisebox{0pt}{\scriptsize{$\frac{1}{2}$}}}(\mathfrak{m}\chiLR{L})
    \bigr]\bigl[
        I_{-i\Omega_{n}-\raisebox{0pt}{\scriptsize{$\frac{1}{2}$}}}(\mathfrak{m}\chiLR{R})\,+\,
        I_{-i\Omega_{n}+\raisebox{0pt}{\scriptsize{$\frac{1}{2}$}}}(\mathfrak{m}\chiLR{R})
    \bigr]\right)  \,=\,0\,,\label{eq:Rindler transcendent equation}
\end{align}
are labelled by consecutive integers~$n\,$. The normalization constant $N_{\Omega_{n}}$ appearing in Eq.~(\ref{eq:Rindler spinor modes}) is determined from the inner product\index{Dirac!inner product}
\begin{align}
    (\,\psi_{1}\,,\,\psi_{2}\,)_{\raisebox{-1.7pt}{\scriptsize{D}}}    &=\,
    \int\limits_{\chiLR{L}}^{\chiLR{R}}d\chi\,
    \psi_{1}^{\dagger}\psi_{2}\,,
    \label{eq:Dirac inner product Rindler}
\end{align}
which follows from~(\ref{eq:Dirac inner product}) by noting that $\gamma^{\hspace*{1pt}\eta}=(1/\chi)\gamma^{\hspace*{1pt}0}\,$. The forms of $\Omega_{n}$ and $N_{\Omega_{n}}$ become more apparent when we match the accelerated cavity to the inertial one. We further note that the exchange $\Omega\rightarrow-\Omega$ takes the order of the modified Bessel functions to their complex conjugates. Consequently, Eq.(\ref{eq:Rindler transcendent equation}) is invariant under this mapping and the spectrum is again symmetric with respect to positive and negative frequency modes. We select the mode labelling such that integers $n\geq0$ $(n<0)$ indicate solutions of positive (negative) frequency with respect to the time-like Kiling vector $\partial_{\eta}\,$, such that we can decompose the field as
\begin{align}
    \psi    &=\,\sum\limits_{n\geq0}\,\tilde{\psi}_{\raisebox{-1.5pt}{\scriptsize{$\Omega_{n}$}}}\,\tilde{b}_{n}\,+\,
    \sum\limits_{n<0}\,\tilde{\psi}_{\raisebox{-1.5pt}{\scriptsize{$\Omega_{n}$}}}\,\tilde{c}_{n}^{\dagger}\,.
    \label{eq:Dirac field Rindler decomposition}
\end{align}
The operators $\tilde{b}_{n},\tilde{c}_{n}$ and their Hermitean conjugates satisfy the anticommutation relations from~(\ref{eq:anticomm relations}). As a last consideration here we take the limit $\mathfrak{m}\rightarrow0$ and obtain
\begin{align}
    \tilde{\psi}_{\raisebox{-1.5pt}{\scriptsize{$\Omega_{n}$}}} &=\,
        \frac{e^{-i\hspace*{0.5pt}\Omega_{n}\hspace*{0.5pt}\eta}}{\sqrt{2\chi\ln(\chiLR{R}/\chiLR{L})}}\,
        \left(\left(\frac{\chi}{\chiLR{L}}\right)^{i\Omega_{n}}\,u_{+}\,+\,i\,
        \left(\frac{\chi}{\chiLR{L}}\right)^{-i\Omega_{n}}\,u_{-}\right)\,,
    \label{eq:Rindler spinor modes massless}
\end{align}
where the massless Rindler frequencies $\Omega_{n}$ are given by
\begin{align}
    \Omega_{n}   &=\,\frac{\pi}{\ln(\chiLR{R}/\chiLR{L})}\bigl(n\,+\,\tfrac{1}{2}\bigr)\,=\,
    \frac{(n\,+\,\tfrac{1}{2})\pi}{2\,\artanh(h/2)}\,,
    \label{eq:massless Rindler spinor frequencies}
\end{align}
where $h$ is as in~(\ref{eq:boundaries through h and L}), and the normalization $N_{\Omega_{n}}$ in~(\ref{eq:Rindler spinor modes}) was chosen so that the phases of the Minkowski and Rindler modes match at $t=\eta=0\,$, i.e., at $(\eta,\chi)=(0,\chiLR{L})$ the modes~(\ref{eq:Rindler spinor modes massless}) are positive multiples of $(u_{+}+iu_{-})$. Equations~(\ref{eq:Rindler spinor modes massless}) and~(\ref{eq:massless Rindler spinor frequencies}) again reproduce the case $(s,\theta)=(\tfrac{1}{2},\tfrac{\pi}{2})$ analyzed in Ref.~\cite[(\ref{Paper:FriisLeeBruschiLouko2012})]{FriisLeeBruschiLouko2012}.

\subsection{Matching: Inertial to Rindler \textemdash\ Dirac Field}\label{sec:matching Dirac field}

\begin{figure}[ht]
\centering
\includegraphics[width=0.75\textwidth]{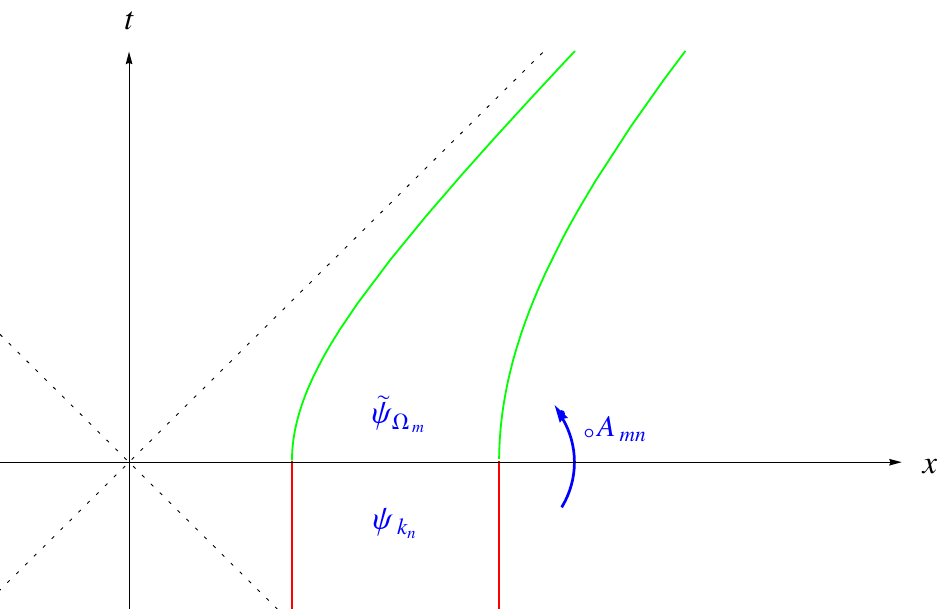}
\caption{
\textbf{Minkowski to Rindler Bogoliubov transformation \textemdash\ Dirac field:} The cavity modes~$\psii{k_{n}}$ in the inertial region and the modes~$\tilde{\psi}_{\protect\raisebox{-1.5pt}{\scriptsize{$\Omega_{m}$}}}$ in the uniformly accelerated region are related by a Bogoliubov transformation with coefficients $\MRA{mn}\,$, see Eq.~(\ref{eq:Dirac Bogo Inertial to Rindler}).
\label{fig:Mink to Rindler Bogos Dirac}}
\end{figure}

We match the inertial and accelerated cavity containing the Dirac field at the junction $t=\eta=0\,$, as laid out in Section~\ref{sec:relativistically rigid cavity}, where we assume the acceleration to be towards increasing values of~$x$. The Minkowski (region~I in Fig.~\ref{fig:rigid cavity}) spinors~(\ref{eq:Mink spinor cavity modes}) and the Rindler (region~II in Fig.~\ref{fig:rigid cavity}) spinors~(\ref{eq:Rindler spinor modes}) are related by a Bogoliubov transformation\index{Bogoliubov transformation!fermions} (see Section~\ref{sec:Bogoliubov transformations of the Dirac Field} and Fig.~\ref{fig:Mink to Rindler Bogos Dirac})
\begin{align}
    \tilde{\psi}_{\raisebox{-1.5pt}{\scriptsize{$\Omega_{m}$}}} &=\,\sum\limits_{n}\,\MRA{mn}\,\psii{k_{n}}
    \label{eq:Dirac Bogo Inertial to Rindler}
\end{align}
where the Bogoliubov coefficients are of the form
\begin{align}
    \MRA{mn}    &=\,(\,\psii{k_{n}} \,,\,\tilde{\psi}_{\raisebox{-1.5pt}{\scriptsize{$\Omega_{m}$}}})_{\raisebox{-1.7pt}{\scriptsize{D}}}\,,
     \label{eq:Dirac Bogo coefficients}
\end{align}
and the Dirac inner product is given by~(\ref{eq:Dirac inner product Rindler}). Since both sets of modes are normalized the matrix ${}_{\text{o}}A=(\MRA{mn})$ is unitary\index{unitary!transformation}\index{transformation!unitary}
\begin{align}
    \sum\limits_{j}\,\MRAstar{jk}\,\MRA{jl}   &=\,\delta_{kl}\,.
    \label{eq:Dirac bogos unitarity}
\end{align}

As in Section~\ref{sec:matching scalar field} we now turn to the small~$h$ approximation. In other words, we wish to find the Taylor-Maclaurin expansion\index{Taylor-Maclaurin expansion} of the Bogoliubov coefficients $\MRA{mn}$ as functions of the parameter $h=\mathbf{a}_{\mathrm{c}}L$ around the value $h=0\,$. In this regime a comparison of~(\ref{eq:Mink spinor cavity modes}) and~(\ref{eq:Rindler spinor modes}) reveals that the leading order of the Rindler frequencies is proportional to $h^{-1}\,$. This complicates the uniform expansion of the modified Bessel functions because their order approaches the imaginary axis as $h$ goes to zero. For details on the intricacies of the uniform expansions see Ref.~\cite{Olver1954}. We therefore choose a slightly different procedure than in Section~\ref{sec:matching scalar field}, i.e., we perform the power expansion in~$h$ directly for the differential equation to which Eq.~(\ref{eq:Rindler solutions without boundaries}) provides the solutions. This task is simplified by the introduction of a new dimensionless variable~$\lambda$, where
\begin{align}
    \chi    &=\,\frac{L}{h}(1\,+\,h\,\lambda)\,,
\end{align}
such that at $\chiLR{L}$ and $\chiLR{R}$ we have $\vLR{L}=-\tfrac{1}{2}$ and $\vLR{R}=\tfrac{1}{2}$, respectively. This procedure reveals
\begin{align}
    \Omega_{n}  &=\,\frac{L}{h}\,\omega_{k_{n}}\,(1\,+\,O(h^{2}))\,,
    \label{eq:Rindler freq power expansion}
\end{align}
where the $k_{n}$ are determined by Eq.~(\ref{eq:Mink modes transcendental equation}). Note that we have used the same symbol~$\Omega_{n}$ for the Rindler frequencies of both the scalar~[see~(\ref{eq:cavity centre proper freq})] and Dirac field [see~(\ref{eq:Rindler freq power expansion})], but they do not generally match and neither do the corresponding Minkowski frequencies $\omega_{n}$ [see~(\ref{eq:inertial modes frequencies})] and $\omega_{k_{n}}$ [see~(\ref{eq:Dirac Mink frequencies})]. Finally, a lengthy but straightforward computation provides the Bogoliubov coefficients for the Dirac cavity field between the inertial region~I and the accelerated region~II (see Fig.~\ref{fig:rigid cavity})\index{Bogoliubov coefficients fermions!Minkowski to Rindler}
\begin{align}
    \MRA{mn}    &=\,\MRAhmn{0}{mn}\,+\,\MRAhmn{1}{mn}\,h\,+\,\MRAhmn{2}{mn}\,h^{2}\,+\,O(h^{3})\,,
    \label{eq:Dirac inertial to Rindler bogos expansion}
\end{align}
where the superscript $^{(n)}$ in brackets indicates the coefficients of~$h^{n}$, and $\MRAhmn{0}{mn}=\delta_{mn}\,$. The non-vanishing coefficients linear in~$h$ are
\begin{align}
    \MRAhmn{1}{mn}    &=\,
    \frac{2\bigl({(-1)}^{m+n}\,-\,1\bigr)
    \,|k_{m} k_{n}|\, C_{k_{m}}^{\,2} C_{k_{n}}^{\,2} (C_{k_{m}}+C_{k_{n}})
    (C_{k_{m}}C_{k_{n}} + \mathfrak{m}^{2})}
    {\sqrt{L^{2} \omega_{k_{m}}^{2} + \mathfrak{m} L}\,
    \sqrt{L^{2} \omega_{k_{n}}^{2} +\mathfrak{m} L}\,
    (C_{k_{m}}-C_{k_{n}})^{3}\,(C_{k_{m}}C_{k_{n}} - \mathfrak{m}^{2})^{3}}\,,
    \label{eq:Dirac Mink to Rindler bogos linear}
\end{align}
for $m\neq n\,$, and $C_{k_{n}}=k_{n}+\omega_{k_{n}}\,$, where $\omega_{k_{n}}$ is given by~(\ref{eq:Dirac Mink frequencies}). The consecutive indices $m,n\geq0$ $(<0)$ label positive (negative) frequency modes. Coefficients that relate modes of the same frequency sign correspond to $\alpha$-type coefficients for bosonic fields, while those that connect positive and negative frequency modes are $\beta$-type coefficients, responsible for particle creation. It is interesting to note that the MIT bag boundary conditions prevent particle creation in pairs of modes with equal energies, i.e., the leading order coefficients $\MRAhmn{1}{mn}$ vanish identically for modes~$m$ and~$n$ with $\omega_{k_{m}}=-\omega_{k_{n}}\,$. The linear coefficients in~(\ref{eq:Dirac Mink to Rindler bogos linear}) form an anti-Hermitean matrix, as required, and they consistently reduce to the case $s=\tfrac{1}{2}$ of Ref.~\cite[(\ref{Paper:FriisLeeBruschiLouko2012})]{FriisLeeBruschiLouko2012} in the massless limit, i.e., for $m\neq n$ and $\mathfrak{m}\rightarrow0$ we have
\begin{align}
\MRAhmn{1}{mn}    &=\,
    \frac{\bigl({(-1)}^{m+n}\,-\,1\bigr)\,(m\,+\,n\,+\,1)}
    {2\,\pi^{2}\,(m\,-\,n)^{3}}\,.
    \label{eq:Dirac Mink to Rindler bogos massless linear}
\end{align}

\begin{figure}[ht!]
\centering
(a)\includegraphics[width=0.8\textwidth]{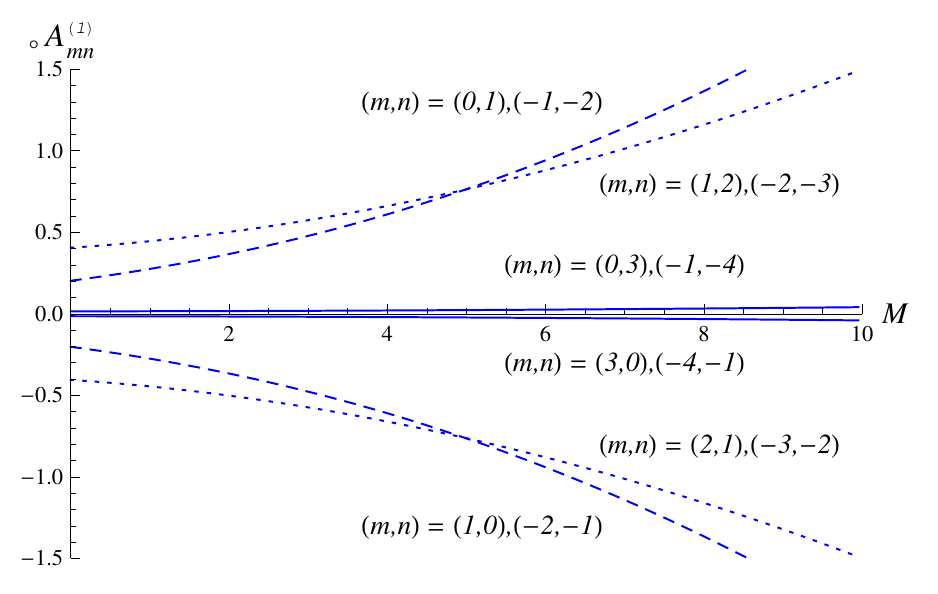}
(b)\hspace*{-3mm}\includegraphics[width=0.8\textwidth]{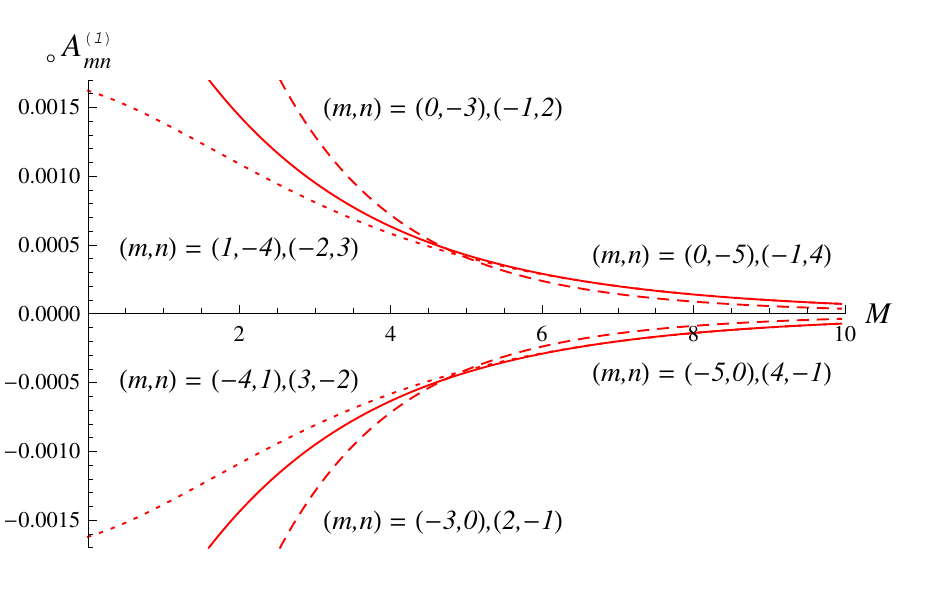}
\caption{
\textbf{Minkowski to Rindler Bogoliubov coefficients \textemdash\ Dirac field:} The behaviour of the leading order Bogoliubov coefficients $\MRAhmn{1}{mn}$ from~(\ref{eq:Dirac Mink to Rindler bogos linear}) for the $(1+1)$-dimensional Dirac field with MIT bag boundary conditions is shown for increasing mass. The coefficient $\MRAhmn{1}{mn}$ is plotted against the dimensionless combination $M := \mathfrak{m}L$. Figure \ref{fig:Dirac alphas and betas}~(a) shows a selection of Bogoliubov coefficients that relate modes with the same sign of the frequency ($\alpha$-type coefficients): These mode-mixing coefficients are proportional to $M^{2}$ as $M\rightarrow\infty\,$. Figure \ref{fig:Dirac alphas and betas}~(b) shows a selection of Bogoliubov coefficients that relate positive frequency modes with negative frequency modes ($\beta$-type coefficients). These particle creation coefficients are proportional to $M^{-6}$ as $M\rightarrow\infty\,$. For all curves the intersections with the vertical axis are given by the massless coefficients of Eq.~(\ref{eq:Dirac Mink to Rindler bogos massless linear}).
\label{fig:Dirac alphas and betas}}
\end{figure}

On the other hand, for large mass, taking the limit $M=\mathfrak{m}L\to\infty$, it can be shown from (\ref{eq:Dirac Mink to Rindler bogos linear})
that the mode-mixing $\alpha$-type coefficients behave as~$M^{2}$
(Fig.~\ref{fig:Dirac alphas and betas}~(a)), while the $\beta$-type coefficients
decrease as~$M^{-6}$ (Fig.~\ref{fig:Dirac alphas and betas}~(b)). As in the bosonic case we emphasize that the perturbative expansions can be trusted for $h\ll1$ when the indices of the coefficients are bounded from above by any constant. For non-zero mass we additionally require $M\hspace*{0.3pt}h\ll1$ but we allow for $M\gg1$ within this regime as long as $M^{2}h\lesssim1\,$.\\

The second order coefficients $\MRAhmn{2}{mn}$ are complicated and won't be needed explicitly in this work, but we note that they are proportional to $(1+(-1)^{m+n})$ as their scalar field counterparts (see Section~\ref{sec:matching scalar field}) and it has been verified~\cite[(\ref{Paper:FriisLeeLouko2013})]{FriisLeeLouko2013} that the unitarity conditions~(\ref{eq:Dirac bogos unitarity}) are satisfied when terms of order~$h^{2}$ are included. A similar argument (see Ref.~\cite[(\ref{Paper:FriisLeeLouko2013})]{FriisLeeLouko2013}) as for the bosonic field (see pp.~\ref{page:leftward acceleration}) can be applied to consider leftward acceleration. For this procedure it is required to keep in mind that a spatial reflection changes the sign of the spatial components of $\gamma^{\hspace*{1pt}\mu}$ in the Dirac equation, reversing the roles of $u_{+}$ and $u_{-}\,$. As before, matching the conventions established for the phases leads to the conclusion that leftwards acceleration can be described to second order in~$h$ by the exchange $h\rightarrow-h\,$.\\

\section{Grafting Generic Cavity Trajectories}
\label{sec:Grafting Generic Cavity Trajectories}

With Eqs.~(\ref{eq:bosonic Bogo coeffs linear}) and~(\ref{eq:Dirac Mink to Rindler bogos linear}) we have established, to leading order in~$h\,$, the Bogoliubov transformations for the real scalar field and Dirac field, respectively, for an instantaneous transition from an inertial to a uniformly accelerated cavity according to the geometry depicted in Fig.~\ref{fig:rigid cavity}. This allows us to study the effects on the states of the quantum field when the cavity is suddenly accelerated. Conceptually the role of the observer is clearly laid out. Without loss of generality we may consider the observer at the centre of the cavity, who experiences excitations with frequencies $\tilde{\omega}_{n}$ from~(\ref{eq:cavity centre proper freq}). Nonetheless, the fact that the cavity in Fig.~\ref{fig:rigid cavity} is accelerated eternally evokes the question how more general non-uniform motion of the cavity can be described. We are now going to investigate exactly this issue.

\subsection{The Basic Building Block}\label{sec:basic building block}

\begin{figure}[ht]
\centering
\includegraphics[width=0.75\textwidth]{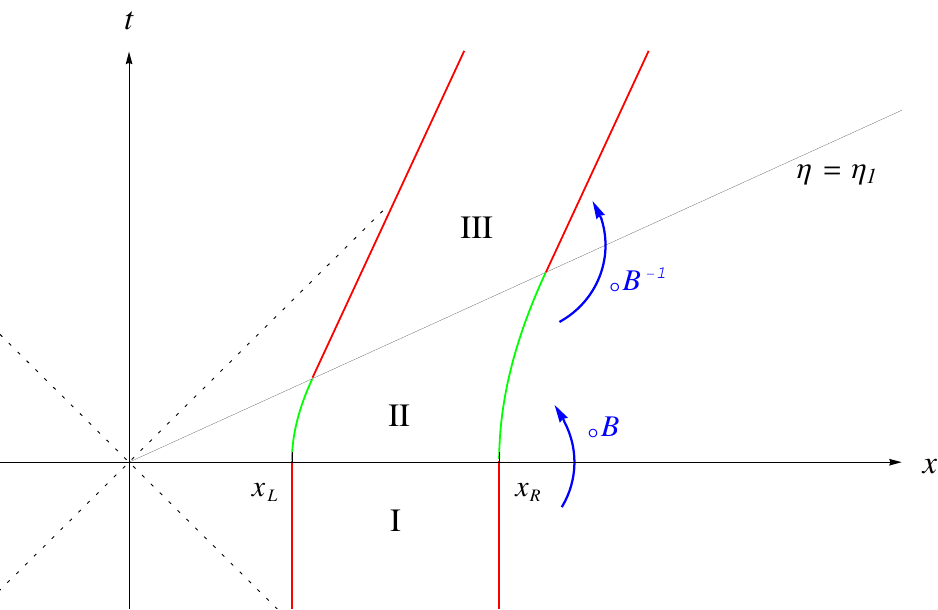}
\caption{
\textbf{Basic building block:} The rigid cavity is at rest initially (region~$\mathrm{I}$),
then undergoes a period of uniform acceleration from $t=0$ to Rindler coordinate time $\eta=\eta_{1}$
(region $\mathrm{II}$) and is thereafter again inertial (region $\mathrm{III}$). The transitions between periods of inertial motion and uniform acceleration induce Bogoliubov transformations~$_{\text{o}}B$ (I$\rightarrow$ II) and~$_{\text{o}}B^{-1}$ (II$\rightarrow$ III)\,.
\label{fig:basic building block}}
\end{figure}

The key to understanding more general cavity trajectories (see Section~\ref{sec:generalized travel scenarios}) lies in the transition from an inertial cavity, back to an inertial cavity, with a single intermediate period of uniform acceleration \textemdash\ the \emph{basic building block}. Let us first briefly return to the geometry of the rigid cavity as inspected in Section~\ref{sec:relativistically rigid cavity}. As we have noted there the inertial and uniformly accelerated cavities are connected along a slice of fixed Rindler coordinate time~$\eta\,$. We follow this recipe also when inverting the procedure. Stopping the acceleration of the cavity at $\eta=\eta_{1}=\operatorname{const.}$ we maintain rigidity, see Fig.~\ref{fig:basic building block}. The cavity walls of the inertial cavity after the acceleration (region~III in Fig.~\ref{fig:basic building block}) are again parallel and at a distance~$L$, as measured by the co-moving observer. For an observer that is at rest with respect to the initial cavity in region~I the cavity in region~III is moving at a constant speed such that the final cavity's length is Lorentz contracted. It is interesting to note that the requirement of rigidity in relativity suggests that different parts of the cavity need to accelerate at different rates and for different durations (in terms of proper time).

To construct the Bogoliubov transformation (see Sections~\ref{sec:Bogoliubov transformations} and \ref{sec:Bogoliubov transformations of the Dirac Field}) corresponding to the basic building block let us denote the abstract transformation between the inertial region~I and the accelerated region~II by~$_{\text{o}}B\,$. For now, we postpone to distinguish between bosons and fermions. For each case all the Bogoliubov coefficients can be combined into formally infinite-dimensional matrix representations of~$_{\text{o}}B\,$. From the Lorentz symmetry between regions~I and~III it becomes evident that the transformation at the junction between regions~II and~III can be chosen to be simply the inverse transformation~$_{\text{o}}B^{-1}\,$. This means that the same phase conventions are chosen for the region~III solutions at~$\eta=\eta_{1}$ as we have previously established at~$\eta=0$ for the region~I solutions and the phases acquired by the Rindler modes in region~II are accounted for separately by a diagonal matrix $\tilde{G}(\eta_{1})\,$, such that\index{Bogoliubov transformation!basic building block}
\begin{align}
   _{\text{B}}B   &=\,_{\text{o}}B^{-1}\,\tilde{G}(\eta_{1})\,_{\text{o}}B\,.
   \label{eq:basic building block general}
\end{align}
Let us now consider the representations of~(\ref{eq:basic building block general}) for bosons and fermions separately.

\subsubsection{The Bosonic Building Block}

\index{Bogoliubov coefficients bosons!basic building block|(}
For the scalar field we represent~$_{\text{o}}B$ and its inverse by the matrices~$_{\text{o}}\mathcal{A}$ and~$_{\text{o}}\mathcal{A}^{-1}$, respectively, in which we combine the matrices $_{\text{o}}\alpha=(\MRalpha{mn})$ and $_{\text{o}}\beta=(\MRbeta{mn})\,$, i.e.,
\begin{align}
    _{\text{o}}\mathcal{A}  &=\,\begin{pmatrix}
    _{\text{o}}\alpha   &   _{\text{o}}\beta    \\
    _{\text{o}}\beta^{*}    &   _{\text{o}}\alpha^{*}
    \end{pmatrix}\,,\hspace*{3mm}
    _{\text{o}}\mathcal{A}^{-1}\,=\,\begin{pmatrix}
    \ _{\text{o}}\alpha^{\dagger}   &   -_{\text{o}}\beta^{T}\,    \\
    -_{\text{o}}\beta^{\dagger}    &   \ _{\text{o}}\alpha^{T}\,
    \end{pmatrix}\,.
    \label{eq:boson Bogo matrix decomposition}
\end{align}
From the Bogoliubov identities~(\ref{eq:bosonic Bogo unitarity}) one can easily verify that the inverse of $_{\text{o}}\mathcal{A}$ is formed by the map $(\ _{\text{o}}\alpha,\ _{\text{o}}\beta)\rightarrow(\ _{\text{o}}\alpha^{\dagger},\ -_{\text{o}}\beta^{T})$ such that $_{\text{o}}\mathcal{A}^{-1} _{\text{o}}\mathcal{A}=\mathds{1}\,$. The bosonic basic building block transformation can then be written as
\begin{align}
   _{\text{B}}\mathcal{A}   &=\,{}_{\text{o}}\mathcal{A}^{-1}\,\tilde{G}(\eta_{1})\,{}_{\text{o}}\mathcal{A}\,,
   \label{eq:bosonic basic building block general}
\end{align}
where $\tilde{G}(\eta_{1})=G(\eta_{1})\oplus G^{*}(\eta_{1})\,$, and the diagonal matrix $G=\diag\{G_{n}|n=1,2,\ldots\}$ has entries $G_{n}(\eta_{1})=\exp(i\Omega_{n}\eta_{1})\,$. The transformation $_{\text{B}}\mathcal{A}$ can be decomposed into matrices $_{\text{B}}\alpha$ and $_{\text{B}}\beta\,$, i.e.,
\vspace*{-2mm}
\begin{subequations}
\label{eq:bosonic building block alphas and betas}
\begin{align}
{}_{\text{B}}\alpha     &=\,{}_{\text{o}}\alpha^{\dagger}\,G\,{}_{\text{o}}\alpha\,-\,
                                {}_{\text{o}}\beta^{T}\,G^{*}\,{}_{\text{o}}\beta^{*}\,,
                                \label{eq:basic building block alphas}\\
{}_{\text{B}}\beta      &=\,{}_{\text{o}}\alpha^{\dagger}\,G\,{}_{\text{o}}\beta\,-\,
                                {}_{\text{o}}\beta^{T}\,G^{*}\,{}_{\text{o}}\alpha^{*}\,,
                                \label{eq:basic building block betas}
\end{align}
\end{subequations}
analogously to the decomposition~(\ref{eq:boson Bogo matrix decomposition}), such that the modes~$\phii{n}$ in region~I and the Minkowski modes in region~III, denoted by $\phiihat{m}\,$, are related by
\begin{align}
    \phiihat{m} &=\,\sum\limits_{n}\bigl(\,\BBBalpha{mn}\,\phii{n}\,+\,\BBBbeta{mn}\,\phii{n}^{*}\,\bigr)\,.
    \label{eq:boson region I to region III bogo modes}
\end{align}
We now wish to obtain the small~$h$ expansion of the Bogoliubov coefficients $\BBBalpha{mn}$ and $\BBBbeta{mn}\,$. The phase factors $G_{n}(\eta_{1})$ can be conveniently written in such an expansion as
\begin{align}
    G_{n}(\eta_{1}) &=\,\exp(i\Omega_{n}\eta_{1})\,=\,\exp(i\omega_{n}\tilde{\tau}_{1})\,+\,O(h^{2})\,=\,
    \Ghn{0}{n}(\tilde{\tau}_{1})\,+\,O(h^{2})
    \label{eq:boson phase factors expanded}
\end{align}
by substituting the proper time $\tilde{\tau}=L\eta/h$ at the centre of the cavity and the Minkowski frequencies $\omega_{n}$ [see Eq.~(\ref{eq:cavity centre proper freq})]. Given Eq.~(\ref{eq:boson phase factors expanded}) as well as the expansions~(\ref{eq:Mink to Rindler alphas and betas small h expansion}) and~(\ref{eq:bosonic Bogo coeffs linear}) of the Minkowski to Rindler coefficients we obtain
\begin{subequations}
\label{eq:bosonic building block alphas and betas expansion}
\begin{align}
{}_{\text{B}}\alpha     &=\,\BBBalphah{0}\,+\,\BBBalphah{1}\,h\,+\,O(h^{2})\,
                            \label{eq:basic building block alphas expansion}\\
                        &=\,\Gh{0}\,+\,\bigl(\Gh{0}\,\MRalphah{1}\,+\,
                            \MRalphahsymb{1}{\dagger}\,\Gh{0}\bigr)\,h\,+\,O(h^{2})\,,
                        \nonumber\\[1.5mm]
{}_{\text{B}}\beta      &=\,\BBBbetah{1}\,h\,+\,O(h^{2})\,=\,
                            \bigl(\Gh{0}\,\MRbetah{1}\,-\,\MRbetahsymb{1}{T}\,
                            \Ghstar{0}\bigr)\,h\,+\,O(h^{2})\,.
                                \label{eq:basic building block betas expansion}
\end{align}
\end{subequations}
Using the antisymmetry and symmetry, respectively, of the real matrices $\MRalphah{1}$ and $\MRbetah{1}$~(\ref{eq:bosonic Bogo coeffs linear}) we can conveniently write the coefficients of $\BBBalphah{1}$ and $\BBBbetah{1}$ as
\begin{subequations}
\label{eq:bosonic building block alphas and betas linear}
\begin{align}
\BBBalphahmn{1}{mn}     &=\,\MRalphahmn{1}{mn}\,\bigl(\Ghn{0}{m}\,-\,\Ghn{0}{n}\bigr)\,,
                                \label{eq:basic building block alphas linear}\\
\BBBbetahmn{1}{mn}     &=\,\MRbetahmn{1}{mn}\,\bigl(\Ghn{0}{m}\,-\,\Ghnstar{0}{n}\bigr)\,.
                                \label{eq:basic building block betas linear}
\end{align}
\end{subequations}
\index{Bogoliubov coefficients bosons!basic building block|)}

\subsubsection{The Fermionic Building Block}

\index{Bogoliubov coefficients fermions!basic building block|(}
For the fermionic representation ${}_{\text{B}}A$ of the basic building block transformation in Fig.~\ref{fig:basic building block} we write
\begin{align}
   _{\text{B}}A   &=\,_{\text{o}}A^{-1}\,G(\eta_{1})\,_{\text{o}}A\,
                =\,_{\text{o}}A^{\dagger}\,G(\eta_{1})\,_{\text{o}}A\,,
   \label{eq:fermionic basic building block general}
\end{align}
where the phases for the spinor modes are encoded in the matrix
\begin{align}
    G   &=\,\diag\{G_{n}|\ldots,-2,-1,0,1,\ldots\}\,,
    \label{eq:fermionic G matrix}
\end{align}
and the individual phases~$G_{n}$ are as in~(\ref{eq:boson phase factors expanded}) but with the appropriate frequencies $\Omega_{n}$ and $\omega_{k_{n}}$ for the Dirac spinors. With this the spinor mode solutions of the final inertial region~III, denoted by $\psiihat{m}\,$, are obtained from the region~I solutions $\psii{n}$ by
\begin{align}
    \psiihat{m} &=\,\sum\limits_{n}\,\BBBA{mn}\,\psii{n}\,.
    \label{eq:fermion region I to region III bogo modes}
\end{align}
The power expansion of $_{\text{B}}A$ for $h\ll1$ is of the form
\begin{align}
    _{\text{B}}A   &=\,\Gh{0}\,+\,\bigl(\Gh{0}\,\MRAh{1}\,-\,\MRAh{1}\,\Gh{0}\bigr)\,h\,+\,O(h^{2})\,,
    \label{eq:fermionic BBB expansion}
\end{align}
where we have used the unitarity of the Minkowski to Rindler transformation, which implies that $\MRAh{1}$ is anti-Hermitean. In components the linear order of the fermionic Bogoliubov coefficients for the basic building block reads
\begin{align}
    \BBBAhmn{1}{mn}  &=\,\MRAhmn{1}{mn}\,\bigl(\Ghn{0}{m}\,-\,\Ghn{0}{n}\bigr)\,.
    \label{eq:fermionic building block coefficients linear}
\end{align}


As a special case, let us consider the massless quantum fields. If $\mathfrak{m}=0$ the frequencies of both the scalar and Dirac field are equally spaced. Consequently, the Bogoliubov coefficients for the basic building block are periodic in the duration of the acceleration. In Chapter~\ref{Chapter 6 Motion Generates Entanglement} we shall reconsider this periodicity for select examples of more generic travel scenarios, which are investigated in the following Section~\ref{sec:generalized travel scenarios}.
\index{Bogoliubov coefficients fermions!basic building block|)}

\subsection{Generalized Travel Scenarios}\label{sec:generalized travel scenarios}

With the basic building block transformations of Eqs.~(\ref{eq:bosonic basic building block general}) and~(\ref{eq:fermionic basic building block general}) at hand we are now in a position to construct more complicated trajectories. The key ingredient is to notice that two basic building blocks can be straightforwardly connected by an intermediate period of inertial coasting of proper time~$\tau\,$. The inertial segment is represented by a matrix $\mathcal{G}(\tau)$, composed as $\tilde{G}$ in~(\ref{eq:bosonic basic building block general}) for the scalar field or as~(\ref{eq:fermionic G matrix}) for the Dirac field, but the basic phase factors $G_{n}$ are replaced by their leading order terms~$\Ghn{0}{n}\,$. With this notation the Bogoliubov transformation~$\mathcal{B}$ for a generic travel scenario connecting two inertial regions with~$n$ intermediate periods of uniform acceleration can be written as
\begin{align}
    \mathcal{B} &=\,{}_{\text{B}}B_{n}\,\mathcal{G}(\tau_{n-1})\,{}_{\text{B}}B_{n-1}\,\ldots\,
        \mathcal{G}(\tau_{2})\,{}_{\text{B}}B_{2}\,\mathcal{G}(\tau_{1})\,{}_{\text{B}}B_{1}\,.
    \label{eq:bogo generic travel scenario}
\end{align}
The individual building block transformations~${}_{\text{B}}B_{i}$ are given by their representatives~${}_{\text{B}}\mathcal{A}_{i}={}_{\text{B}}\mathcal{A}(\eta_{\,i},h_{i})$ from~(\ref{eq:bosonic basic building block general}) and ${}_{\text{B}}A_{i}={}_{\text{B}}A(\eta_{\,i},h_{i})$ from~(\ref{eq:fermionic basic building block general}) for the scalar and Dirac field, respectively. Assuming that the accelerations~$h_{i}$ of all segments are small, i.e., $|h_{i}|\ll1$ we can perform a power expansion for all of these parameters. The leading order of~(\ref{eq:bogo generic travel scenario}) is then given by
\begin{align}
\GBh{0} &=\,\mathcal{G}(\tau_{\mathrm{tot}})
\,=\,\mathcal{G}(\sum\limits_{i=1}^{n-1}\tau_{i}+\sum\limits_{j=1}^{n}\tilde{\tau}_{j})\,,
\end{align}
where $\tau_{\mathrm{tot}}$ is the total proper time as measured at the centre of the cavity between the initial and final inertial segment. The coefficients $\GBhmn{1_{i}}{mn}$ of the linear term $\sum_{i}\GBh{1_{i}}h_{i}$ in the expansion are all proportional to the linear Minkowski to Rindler coefficients, $\MRalphahmn{1}{mn}\,$, and $\MRbetahmn{1}{mn}\,$, or $\MRAhmn{1}{mn}\,$ from (\ref{eq:bosonic Bogo coeffs linear}) and~(\ref{eq:Dirac Mink to Rindler bogos linear}), respectively, and therefore share their basic structure. The linear terms $\GBhmn{1_{i}}{mn}$ vanish identically for pairs of modes $(m,n)$ that have the same parity, i.e., for which $(m+n)$ is even, in particular for $m=n\,$. Up to and including second order terms in the expansion the direction of the acceleration in the $i$-th building block may be controlled by the sign of~$h_{i}\,$ (see pp.~\pageref{page:leftward acceleration}). Assuming for simplicity of notation that every $h_{i}$ can be written as $\epsilon_{i}\,h\,$, for a fixed~$h$ and $\epsilon_{i}\in\mathbb{R}$,  we can thus conclude that the coefficients for a generic travel scenario have a power expansion of the form\label{page:general travel scenario coefficients}\index{Bogoliubov coefficients bosons!general expansion}\index{Bogoliubov coefficients fermions!general expansion}
\begin{subequations}
\label{eq:generic travel scenario expansion}
\begin{align}
    \alphamn{mn}    &=\,\Ghn{0}{m}\delta_{mn}\,+\,\alphahmn{1}{mn}\,h\,+\,\alphahmn{2}{mn}\,h^{2}\,+\,O(h^{3})\,,
    \label{eq:generic travel scenario expansion alphas}\\[1mm]
    \betamn{mn}     &=\,\betahmn{1}{mn}\,h\,+\,\betahmn{2}{mn}\,h^{2}\,+\,O(h^{3})\,,
    \label{eq:generic travel scenario expansion betas}\\[2mm]
    \Amn{mn}        &=\,\Ghn{0}{m}\delta_{mn}\,+\,\Ahmn{1}{mn}\,h\,+\,\Ahmn{2}{mn}\,h^{2}\,+\,O(h^{3})\,,
    \label{eq:generic travel scenario expansion A}
\end{align}
\end{subequations}
where the appropriate forms of $\Ghn{0}{m}=\Ghn{0}{m}(\tau_{\mathrm{tot}})$ apply for the scalar and Dirac field, and the diagonal first order coefficients vanish, i.e., $\alphahmn{1}{nn}=\betahmn{1}{nn}=0$ and $\Ahmn{1}{nn}=0$. We \mbox{insert} the expansions~(\ref{eq:generic travel scenario expansion}) into the \emph{Bogoliubov identities} for bosonic~(\ref{eq:bosonic Bogo unitarity}) and fermionic fields~(\ref{eq:fermionic bogo identity}), respectively, to express these unitarity requirements\index{unitarity} for the linear coefficients in the perturbative expansion as
\begin{subequations}\index{Bogoliubov identities bosons!expansion order $h$}\index{Bogoliubov identities fermions!expansion order $h$}
\label{eq:generic travel scenario bogo identities linear}
\begin{align}
    \Ghnstar{0}{m}\,\alphahmn{1}{mn}\,+\,\Ghn{0}{n}\,\alphahmnstar{1}{nm}    &=\,0\,,
    \label{eq:generic travel scenario bogo identities bosons alphas 1}\\[1.5mm]
    \Ghnstar{0}{m}\,\betahmn{1}{mn}\,-\,\Ghnstar{0}{n}\,\betahmn{1}{nm} &=\,0\,,
    \label{eq:generic travel scenario bogo identities bosons betas 1}\\[1.5mm]
    \Ghnstar{0}{m}\,\Ahmn{1}{mn}\,+\,\Ghn{0}{n}\,\Ahmnstar{1}{nm}    &=\,0\,.
    \label{eq:generic travel scenario bogo identities fermions 1}
\end{align}
\end{subequations}
It is convenient to consider also the second order of the Bogoliubov identities, i.e.,\index{Bogoliubov identities bosons!expansion order $h^{2}$}\index{Bogoliubov identities fermions!expansion order $h^{2}$}
\begin{subequations}
\label{eq:generic travel scenario bogo identities quadratic}
\begin{align}
    \Ghnstar{0}{m}\,\alphahmn{2}{mn}\,+\,\Ghn{0}{n}\,\alphahmnstar{2}{nm}    &=\,-\,
    \sum\limits_{l}\,\bigl(\alphahmnstar{1}{lm}\alphahmn{1}{ln}\,-\,\betahmn{1}{lm}\betahmnstar{1}{ln}\bigr)\,,
    \label{eq:generic travel scenario bogo identities bosons alphas 2}\\[1mm]
    \Ghnstar{0}{m}\,\betahmn{2}{mn}\,-\,\Ghnstar{0}{n}\,\betahmn{2}{nm} &=\,-\,
    \sum\limits_{l}\,\bigl(\alphahmnstar{1}{lm}\betahmn{1}{ln}\,-\,\betahmn{1}{lm}\alphahmnstar{1}{ln}\bigr)\,,
    \label{eq:generic travel scenario bogo identities bosons betas 2}\\[1mm]
    \Ghnstar{0}{m}\,\Ahmn{2}{mn}\,+\,\Ghn{0}{n}\,\Ahmnstar{2}{nm}    &=\,-\,\sum\limits_{l}\,\Ahmnstar{1}{lm}\Ahmn{1}{ln}\,,
    \label{eq:generic travel scenario bogo identities fermions 2}
\end{align}
\end{subequations}
which will be helpful in the following chapters. Let us now illustrate the construction of generic trajectories for a specific example in the next section.

\subsubsection{Trip to Alpha-Centauri}

A particular example of interest is a travel scenario where the cavity undergoes two periods of uniform acceleration such that the cavity comes to rest in the same inertial frame it started from, but is possibly located at a remote location in spacetime. To illustrate this travel scenario one may think of a spaceship carrying the cavity on a \emph{one-way trip to Alpha Centauri}~\cite{BruschiFuentesLouko2012}.

\begin{figure}[ht!]
\centering
\includegraphics[width=0.75\textwidth]{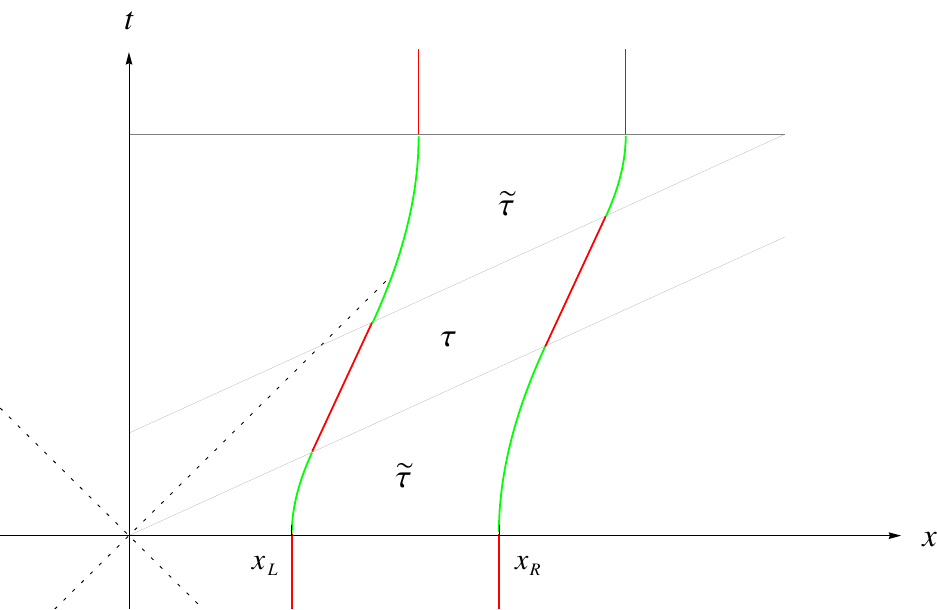}
\caption{
\textbf{Trip to Alpha Centauri:} The travel scenario contains two periods of uniform acceleration of the same duration~$\tilde{\tau}$ and proper acceleration as measured at the centre of the cavity. One of these segments is towards increasing, the other towards decreasing values of~$x\,$. The periods of uniform acceleration are separated by a segment of inertial coasting at fixed velocity for the (proper) time~$\tau\,$, allowing the cavity to reach a (possibly) remote location such as \emph{Alpha Centauri}~\cite{BruschiFuentesLouko2012}.
\label{fig:alpha centauri}}
\end{figure}

We decompose the Bogoliubov transformation into two basic building blocks of the same duration~$\tilde{\tau}$ as measured at the centre of the cavity and equal, but opposite accelerations, here represented by~$\pm h$ in a slight abuse of notation. In between we leave a period of inertial coasting for the (proper) time~$\tau$, such that the total transformation is of the form\index{Bogoliubov transformation!Alpha Centauri}
\begin{align}
{}_{\alpha\text{c}}\mathcal{B} &=\,{}_{\text{B}}B(\tilde{\tau},-h)\,\mathcal{G}(\tau)\,{}_{\text{B}}B(\tilde{\tau},h)\,.
\label{eq:trip to alpha centauri general}
\end{align}
For the scalar field we substitute ${}_{\text{B}}\mathcal{A}$ from~(\ref{eq:bosonic basic building block general}) for ${}_{\text{B}}B$ and immediately obtain the decompositions
\begin{subequations}
\label{eq:trip to alpha centauri}
\begin{align}
    {}_{\alpha\text{c}}\alpha    &=\,
        {}_{\text{B}}\alpha(\tilde{\tau},-h)\,\Gh{0}(\tau)\,{}_{\text{B}}\alpha(\tilde{\tau},h)\,+\,
        {}_{\text{B}}\beta(\tilde{\tau},-h)\,\Ghstar{0}(\tau)\,{}_{\text{B}}\beta^{*}(\tilde{\tau},h)\,,
    \label{eq:trip to alpha centauri alphas}\\[1mm]
    {}_{\alpha\text{c}}\beta     &=\,
        {}_{\text{B}}\alpha(\tilde{\tau},-h)\,\Gh{0}(\tau)\,{}_{\text{B}}\beta(\tilde{\tau},h)\,+\,
        {}_{\text{B}}\beta(\tilde{\tau},-h)\,\Ghstar{0}(\tau)\,{}_{\text{B}}\alpha^{*}(\tilde{\tau},h)\,,
    \label{eq:trip to alpha centauri betas}
\end{align}
\end{subequations}
from which we obtain the power expansion coefficients\index{Bogoliubov coefficients bosons!Alpha Centauri}
\begin{subequations}
\label{eq:trip to alpha centauri expansion}
\begin{align}
    \ACalphahmn{0}{mn}    &=\,\delta_{mn}\,\Ghn{0}{m}(2\tilde{\tau}\,+\,\tau)\,,
    \label{eq:trip to alpha centauri expansion alphas 0}\\[1mm]
    \ACalphahmn{1}{mn}    &=\,\MRalphahmn{1}{mn}\,\bigl[
    \Ghn{0}{m}(2\tilde{\tau}+\tau)\,-\,
    \Ghn{0}{m}(\tilde{\tau}+\tau)\Ghn{0}{n}(2\tilde{\tau})
    \label{eq:trip to alpha centauri expansion alphas 1}\\
    &\hspace*{1.2cm} -\,\Ghn{0}{m}(\tilde{\tau})\Ghn{0}{n}(\tilde{\tau}+\tau)\,+\,
    \Ghn{0}{n}(2\tilde{\tau}+\tau)\bigr]\,,
    \nonumber\\[1mm]
    \ACbetahmn{1}{mn}    &=\,\MRbetahmn{1}{mn}\,\bigl[
    \Ghn{0}{m}(2\tilde{\tau}+\tau)\,-\,
    \Ghn{0}{m}(\tilde{\tau}+\tau)\Ghnstar{0}{n}(2\tilde{\tau})
    \label{eq:trip to alpha centauri expansion betas 1}\\
    &\hspace*{1.2cm} -\,\Ghn{0}{m}(\tilde{\tau})\Ghnstar{0}{n}(\tilde{\tau}+\tau)\,+\,
    \Ghnstar{0}{n}(2\tilde{\tau}+\tau)\bigr]\,.
    \nonumber
\end{align}
\end{subequations}
A similar computation for the fermionic Bogoliubov transformation reveals\index{Bogoliubov coefficients fermions!Alpha Centauri}
\begin{subequations}
\label{eq:trip to alpha centauri expansion fermions}
\begin{align}
    \ACAhmn{0}{mn}    &=\,\delta_{mn}\,\Ghn{0}{m}(2\tilde{\tau}\,+\,\tau)\,,
    \label{eq:trip to alpha centauri expansion fermions 0}\\[1mm]
    \ACAhmn{1}{mn}    &=\,\MRAhmn{1}{mn}\,\bigl[
    \Ghn{0}{m}(2\tilde{\tau}+\tau)\,-\,
    \Ghn{0}{m}(\tilde{\tau}+\tau)\Ghn{0}{n}(2\tilde{\tau})
    \label{eq:trip to alpha centauri expansion fermions 1}\\
    &\hspace*{1.2cm} -\,\Ghn{0}{m}(\tilde{\tau})\Ghn{0}{n}(\tilde{\tau}+\tau)\,+\,
    \Ghn{0}{n}(2\tilde{\tau}+\tau)\bigr]\,,
    \nonumber
\end{align}
\end{subequations}
with the appropriate frequencies for the Dirac spinor modes in the phase factors. Finally, let us consider the possibility of smooth transitions between segments of inertial motion and uniform acceleration in Section~\ref{sec:smoothly varying accelerations}.

%

\subsection{Smoothly Varying Accelerations}\label{sec:smoothly varying accelerations}

As a last step in the construction of generic travel scenarios let us reconsider the assumptions of sharp transitions between inertial motion and uniform accelerations. Certainly, accelerations can be thought of as switching instantaneously if the change in acceleration occurs much faster than the characteristic timescale of the cavity. Such a timescale can be constructed from the propagation speed in the cavity and its length.\\

However, it is also of interest to study the effects of non-uniform motion for which the acceleration is allowed to vary smoothly in time, without sharp jumps. Let us construct the corresponding Bogoliubov transformation as a limit of Eq.~(\ref{eq:bogo generic travel scenario}), following Ref.~\cite{BruschiLoukoFaccioFuentes2013}. Since the acceleration can now take on arbitrary values we can remove the inertial segments and we fix only an initial time~$\tau_{0}$ and a final time~$\tau\,$, such that we have
\begin{align}
    \mathcal{B}(\tau,\tau_{0}) &=\,\underset{\tilde{\tau}\rightarrow0}{\underset{N\rightarrow\infty}{\lim}}\
    {}_{\text{B}}B(\tilde{\tau}_{N},h_{N})\ {}_{\text{B}}B(\tilde{\tau}_{N-1},h_{N-1})\ \ldots\ {}_{\text{B}}B(\tilde{\tau}_{1},h_{1})\,
    \label{eq:bogo generic travel scenario smooth}
\end{align}
where the total time is fixed
\begin{align}
\sum\limits_{n=1}^{N}\,\tilde{\tau}_{n}   &=\,\tau-\tau_{0}\,=\,\operatorname{const.}\,.
    \label{eq:total travel time smooth acceleration}
\end{align}
An infinitesimal increase in time, i.e., $\tau\rightarrow\tau+d\tau\,$, is then achieved by applying the transformation ${}_{\text{B}}B(d\tau,h)$ to $\mathcal{B}(\tau,\tau_{0})$ of~(\ref{eq:bogo generic travel scenario smooth}) from the left,
\begin{align}
    \mathcal{B}(\tau+d\tau,\tau_{0})  &=\,{}_{\text{B}}B(d\tau,h)\,\mathcal{B}(\tau,\tau_{0})\,.
    \label{infinitesimal increas in tau}
\end{align}
Neglecting terms of $O(d\tau^{2})$ the infinitesimal basic building block transformation can be written as
\begin{align}
    {}_{\text{B}}B(d\tau,h) &=\,{}_{\text{o}}B^{-1}(h)\,\tilde{G}(h,d\tau)\ {}_{\text{o}}B(h)\,=\,
    {}_{\text{o}}B^{-1}(h)\,\bigl(\mathds{1}\,+\,i\,\tilde{\Omega}(h)\,d\tau\bigr)\ {}_{\text{o}}B(h)\,,
    \label{eq:infinitesimal basic building block}
\end{align}
where we define the matrix $\tilde{\Omega}$ individually for the scalar and Dirac field,
\begin{subequations}
\label{eq:frequency matrices smooth accelerations}
\begin{align}
    \mbox{scalar:}\ \ \ \tilde{\Omega}(h)   &:=\,\diag\{\Omega_{n}|n=1,2,\ldots\}\,\oplus\,\diag\{-\Omega_{n}|n=1,2,\ldots\}\,,
    \label{eq:frequency matrix smooth acceleration bosons}\\[2mm]
    \mbox{Dirac:}\ \ \ \tilde{\Omega}(h)   &:=\,\diag\{\Omega_{n}|n=\ldots,-2,-1,0,1,\ldots\}\,.
    \label{eq:frequency matrix smooth acceleration fermions}
\end{align}
\end{subequations}
This enables us to write the derivative of $\mathcal{B}(\tau,\tau_{0})$ with respect to the proper time~$\tau$ as
\begin{align}
    \frac{\partial}{\partial\tau}\,\mathcal{B}(\tau,\tau_{0}) &=\,
    \lim_{d\tau\rightarrow0}\ \frac{\mathcal{B}(\tau+d\tau,\tau_{0})\,-\,\mathcal{B}(\tau,\tau_{0})}{d\tau}\,
    =\,i\,{}_{\text{o}}B^{-1}(h)\ \tilde{\Omega}(h)\ {}_{\text{o}}B(h)\ \mathcal{B}(\tau,\tau_{0})\,.
    \label{eq:derivative of smooth travel bogos}
\end{align}
\newpage
The simple differential equation~(\ref{eq:derivative of smooth travel bogos}) can be immediately recognized to be of the form of the \emph{Schr{\"o}dinger equation}. Since $h=h(\tau)$ is now time-dependent the solution to Eq.~(\ref{eq:derivative of smooth travel bogos}) is given in terms of a \emph{time-ordered integral} (see, e.g., Ref.~\cite[pp.~84]{PeskinSchroeder1995})\index{Bogoliubov transformation!smooth acceleration}
\begin{align}
    \mathcal{B}(\tau,\tau_{0})  &=\,T\,\exp\left(\,i\,\int\limits_{\tau_{0}}^{\tau}
    {}_{\text{o}}B^{-1}(h(\tau\pr))\ \tilde{\Omega}(h(\tau\pr))\ {}_{\text{o}}B(h(\tau\pr))\,d\tau\pr\,\right)\,.
    \label{eq:time ordered solution smooth acceleration}
\end{align}
From~(\ref{eq:time ordered solution smooth acceleration}) it can be seen that the transformation reduces to the acquisition of phases for any time interval for which $h=\operatorname{const.}\,$. In addition, no small~$h$ approximation has been performed yet and it can therefore be argued quite generally, that any effects of mode mixing and particle creation are due to the \emph{changes} in acceleration. In other words, the non-uniformity of the acceleration is responsible for any effects. In the case of the small~$h$ approximation we find the leading terms of the expansion for the scalar field from~(\ref{eq:derivative of smooth travel bogos}) and~(\ref{eq:Mink to Rindler alphas and betas small h expansion}) as
\begin{subequations}
\label{eq:smooth acceleration expansion bosons}
\begin{align}
    \Salphamn{nn}   &=\,\Ghn{0}{n}(\tau-\tau_{0})\,+\,O(h^{2})\,,
    \label{eq:smooth acceleration expansion alphas 0}
\end{align}
and for $m\neq n$ the leading order coefficients are\index{Bogoliubov coefficients bosons!smooth acceleration}
\begin{align}
    \Salphamn{mn}   &=\,i\,L\,(\omega_{m}-\omega_{n})\ \MRalphahmn{1}{mn}\,\Ghn{0}{m}(\tau-\tau_{0})\,
    \int\limits_{\tau_{0}}^{\tau}e^{i(\omega_{m}-\omega_{n})(\tau\pr-\tau_{0})}\,\mathbf{a}_{\mathrm{c}}(\tau\pr)\,d\tau\pr\,,
    \label{eq:smooth acceleration expansion alphas 1}\\[1mm]
    \Sbetamn{mn}   &=\,i\,L\,(\omega_{m}+\omega_{n})\ \MRbetahmn{1}{mn}\,\Ghn{0}{m}(\tau-\tau_{0})\,
    \int\limits_{\tau_{0}}^{\tau}e^{i(\omega_{m}+\omega_{n})(\tau\pr-\tau_{0})}\,\mathbf{a}_{\mathrm{c}}(\tau\pr)\,d\tau\pr\,,
    \label{eq:smooth acceleration expansion betas 1}
\end{align}
\end{subequations}
where $\Ghn{0}{m}(\tau)=\exp(i\omega_{m}\tau)\,$, $\omega_{m}$ is given by~(\ref{eq:inertial modes frequencies}), and we have assumed the cavity length~$L$ to be fixed, while the proper acceleration $\mathbf{a}_{\mathrm{c}}(\tau)$ at the centre of the cavity varies smoothly. The product $h(\tau)=\mathbf{a}_{\mathrm{c}}(\tau)L\ll1$ is assumed to be small throughout the journey. The linear order in the expansion of the Bogoliubov coefficients is thus given by a \emph{Fourier transform} of the time-dependent acceleration. A similar calculation involving~(\ref{eq:derivative of smooth travel bogos}) and~(\ref{eq:Dirac inertial to Rindler bogos expansion}) supplies the fermionic counterparts~\cite[(\ref{Paper:FriisLeeLouko2013})]{FriisLeeLouko2013}\index{Bogoliubov coefficients fermions!smooth acceleration}
\begin{subequations}
\label{eq:smooth acceleration expansion fermions}
\begin{align}
    \SAmn{nn}   &=\,\Ghn{0}{n}(\tau-\tau_{0})\,+\,O(h^{2})\,,
    \label{eq:smooth acceleration expansion fermions 0}\\[1mm]
    \SAmn{mn}   &=\,i\,L\,(\omega_{m}-\omega_{n})\ \MRAhmn{1}{mn}\,\Ghn{0}{m}(\tau-\tau_{0})\,
    \int\limits_{\tau_{0}}^{\tau}e^{i(\omega_{m}-\omega_{n})(\tau\pr-\tau_{0})}\,\mathbf{a}_{\mathrm{c}}(\tau\pr)\,d\tau\pr\,,
    \label{eq:smooth acceleration expansion fermions 1}\\[-3mm]
    &   \hspace*{8cm}(m\neq n)\nonumber
\end{align}
\end{subequations}
with the frequencies determined by~(\ref{eq:Dirac Mink frequencies}) and~(\ref{eq:Mink modes transcendental equation}). Since all the linear coefficients for the scalar and Dirac field are proportional to the Minkowski to Rindler coefficients we can conclude that the linear coefficients vanish for mode pairs of equal parity, i.e., if $(m+n)$ is even, regardless of the travel scenario or the smoothness of the acceleration.\\

The magnitude of the coefficients for a given travel scenario depends on the jumps in the acceleration. If~$\mathbf{a}_{\mathrm{c}}(\tau)$ changes much slower than the oscillating terms in~(\ref{eq:smooth acceleration expansion bosons}) or~(\ref{eq:smooth acceleration expansion fermions}) the effects governed by the magnitude of the coefficients are significantly reduced. One such effect is the \emph{dynamical Casimir effect}\index{dynamical Casimir effect} (see, e.g., Refs.~\cite{Moore1970,LambrechtJaekelReynaud1996,Dodonov2001,Dodonov2010}), in which one or both of the boundaries of a cavity undergo periodic motion at a resonance frequency to produce pairs of particles. The cavity model as described in this chapter accounts for this effect in its incarnation where the two walls are kept at a fixed distance throughout this oscillation, e.g., by letting the acceleration in~(\ref{eq:smooth acceleration expansion bosons}) be sinusoidal~\cite{BruschiLoukoFaccioFuentes2013,BruschiLoukoFaccio2013}.

    \newpage\ \\
    \chapter{State Transformation by Non-Uniform Motion}
\label{Chapter 5 State Transformation by Non-Uniform Motion}
\vspace*{-1mm}
The analysis of Chapter~\ref{Chapter 4 Constructing Non Uniformly Moving Cavities} has provided the Bogoliubov transformations between the mode functions and mode operators of a rigid cavity undergoing non-uniform motion. The cavity is assumed to confine massive scalar or Dirac fields in $(1+1)$ dimensions and additional spatial dimensions can be included by their strictly positive contributions to the mass. The motion of the cavity is assumed to be inertial at the start and finish of the journey, but is non-uniform in between, possibly including smooth as well as sharp transitions between different accelerations. For practical reasons a perturbative approach is adopted. The coefficients of the Bogoliubov transformation between the initial ``in-region" and the final ``out-region" are obtained as Taylor-Maclaurin expansions\index{Taylor-Maclaurin expansion} in the parameter $h=\mathbf{a}_{\mathrm{c}}L\ll1\,$, where $\mathbf{a}_{\mathrm{c}}$ is the proper acceleration\index{proper!acceleration} at the centre of the cavity, and the length $L$ of the cavity is considered to be fixed. A discussion of the numerical values for the expansion parameter can be found in Section~\ref{sec:Simulations in Superconducting Circuits}. For a generic travel scenario the expansions of the coefficients are of the form given in~(\ref{eq:generic travel scenario expansion}). The in-region mode functions and annihilation operators for bosons and fermions are denoted by $\phii{n}\,$, $\an{n}\,$, and $\psii{n}\,$, $\bn{n}\,$, $\cn{n}\,$, respectively, while the out-region quantities are denoted as $\phiihat{n}\,$, $\anhat{n}\,$, $\psiihat{n}\,$, $\bnhat{n}\,$, and $\cnhat{n}\,$.\\

\vspace*{-3mm}
The purpose of the present chapter is to implement these Bogoliubov transformations on the corresponding \emph{Fock spaces} (see Sections~\ref{sec:bosonic Fock space} and~\ref{sec:fermionic Fock space}) as well as in \emph{phase space} (see Section~\ref{sec:entanglement in bosonic quantum fields}). We separate the description of the bosonic and fermionic transformations. In Sections~\ref{sec:Bosonic State Transformation} and~\ref{sec:Transformation of Bosonic Gaussian States} we study the transformation in the bosonic Fock space and phase space, respectively, before we turn to the fermionic Fock space transformations in Section~\ref{sec:Fermionic State Transformation}.\\

\vspace*{-3mm}
This chapter combines results that were derived as part of the research conducted for related investigations by myself~\cite[(\ref{Paper:FriisLeeBruschiLouko2012})-(\ref{Paper:FriisFuentes2013}),
(\ref{Paper:FriisLeeTruongSabinSolanoJohanssonFuentes2013})]{FriisLeeBruschiLouko2012,FriisBruschiLoukoFuentes2012,
FriisHuberFuentesBruschi2012,FriisFuentes2013,FriisLeeTruongSabinSolanoJohanssonFuentes2013} and others~\cite{BruschiFuentesLouko2012}.

\section{Bosonic Fock State Transformation}\label{sec:Bosonic State Transformation}

\subsection{Bosonic Vacuum Transformation}\label{sec:Bosonic Vacuum Transformation}

To construct the Bogoliubov transformation on the bosonic Fock space\index{Fock!space, bosonic} (see Section~\ref{sec:bosonic Fock space}) the natural starting point is the vacuum, i.e., relating the in-region vacuum state $\ket{0}$ and the out-region vacuum $\kethat{0}\,$. Since the Bogoliubov transformation is linear in the mode operators it can be represented by exponentials of quadratic combinations of the operators $\an{m}$ and $\adn{n}\,$, see Section~\ref{sec:symplectic operations}. Such a transformation can further be split into passive and active transformations (see p.~\pageref{page:passive and active transformations}). The former leave the vacuum invariant, while the active transformations are generated by quadratic combinations of the form
\begin{align}
    \sum\limits_{p,q}\bigl(\Vmn{pq}\,\adnhat{p}\,\adnhat{q}\,-\,\Vmnstar{pq}\,\anhat{p}\,\anhat{q}\bigr)\,,
    \label{eq:boson vac trafo ansatz}
\end{align}
with $V_{pq}\in\mathbb{C}\,$. Since $\an{n}\kethat{0}=0\,$\index{vacuum!bosons}, and a quick application of the commutation relations~(\ref{eq:canonical commutation relations}) shows that also $\comm{\adnhat{p}\,\adnhat{q}}{\anhat{p\pr}\,\anhat{q\pr}}\kethat{0}=0\,$, the transformation between the vacua can be written as~\cite{FabbriNavarro-Salas2005}
\begin{align}
    \ket{0} &=\,N_{\text{vac}}\,\exp(W)\,\kethat{0}\,=\,
    N_{\text{vac}}\,\exp\bigl(\tfrac{1}{2}\sum\limits_{p,q}\,\Vmn{pq}\,\adnhat{p}\,\adnhat{q}\bigr)\,
    \kethat{0}\,,
    \label{eq:boson vac trafo}
\end{align}
where we have included a factor of $\tfrac{1}{2}$ for convenience. If the in-region and out-region Fock spaces are unitarily equivalent the state is normalized by a finite constant $N_{\text{vac}}\,$. We shall return to this criterion on page~\pageref{page:unitarity bosons}. The next step of our investigation is to determine the symmetric matrix~$V=(\Vmn{pq})$ in~(\ref{eq:boson vac trafo}). We exploit the property that the in-region vacuum~$\ket{0}$ is annihilated by all operators~$\an{n}\,$, where we insert the Hermitean conjugate of the inverse Bogoliubov transformation from~(\ref{eq:bosonic Bogo transformation operators}) to write
\begin{align}
    \an{n}\,\ket{0} &=\,N_{\text{vac}}\,
    \sum\limits_{m}\bigl(\alphamn{mn}\,\anhat{m}\,+\,\betamnstar{mn}\,\adnhat{m}\bigr)\,
    \exp(W)\,\kethat{0}\,=\,0\,.
    \label{eq:derivation of bosonic V}
\end{align}
The \emph{Hadamard Lemma}\index{Hadamard Lemma} of the Baker-Campbell-Hausdorff formula, i.e.,
\begin{align}
    X\,e^{Y}    &=\,e^{Y}\,\bigl(X\,+\,\comm{X}{Y}\,+\,\tfrac{1}{2!}\comm{\comm{X}{Y}}{Y}\,+\ldots\bigr)\,,
    \label{eq:Hadamard lemma}
\end{align}
is then used to commute $\an{n}$ and $\exp(W)\,$. One straightforwardly obtains the commutators
\begin{subequations}
\label{eq:derivation of bosonic V commutators}
\begin{align}
    \comm{\anhat{m}}{W} &=\,\sum\limits_{p}\,\Vmn{pm}\,\adnhat{p}\,,
    \label{eq:derivation of bosonic V commutators nonzero}\\[1mm]
    \comm{\adnhat{m}}{W} &=\,\comm{\comm{\anhat{m}}{W}}{W}\,=\,0\,.
    \label{eq:derivation of bosonic V commutators zero}
\end{align}
\end{subequations}
Combining~(\ref{eq:derivation of bosonic V})-(\ref{eq:derivation of bosonic V commutators}) we arrive at the condition
\begin{align}
    \sum\limits_{m}\,\Vmn{pm}\,\alphamn{mn}\,+\,\betamnstar{pn}   &=\,0\,.
    \label{eq:condition for V components}
\end{align}
Since the matrix $\alpha=(\alphamn{mn})$ is invertible~\cite{Wald1975} we can rephrase~(\ref{eq:condition for V components}) to directly express~$V$ as
\begin{align}
    V   &=\,-\beta^{*}\,\alpha^{-1}\,.
    \label{eq:condition for V matrix}
\end{align}
Returning to the perturbative treatment we employ the expansions~(\ref{eq:generic travel scenario expansion}) to get
\begin{subequations}
\label{eq:bosonic V expansion}
\begin{align}
    \Vhmn{0}{pq}    &=\,0\,,
    \label{eq:bosonic V expansion 0}\\[1mm]
    \Vhmn{1}{pq}   &=\,-\,\Ghnstar{0}{q}\,\betahmnstar{1}{pq}\,,
    \label{eq:bosonic V expansion 1}\\[1.5mm]
    \Vhmn{2}{pq}   &=\,\Ghnstar{0}{p}\,\Ghnstar{0}{q}\,\sum\limits_{m}\,
    \betahmnstar{1}{mp}\,\alphahmn{1}{mq}\,-\,\Ghnstar{0}{q}\,\betahmnstar{2}{pq}\,,
    \label{eq:bosonic V expansion 2}
\end{align}
\end{subequations}
where we have used the identity~(\ref{eq:generic travel scenario bogo identities bosons betas 1}) to rewrite the second order terms~(\ref{eq:bosonic V expansion 2}).

\subsubsection{Unitarity of the Transformation \textemdash\ Bosons}\label{page:unitarity bosons}
\index{unitarity!bosons|(}
A subtlety in the transformation is the question whether the in-region and out-region Fock spaces are unitarily equivalent. In other words, it is not guaranteed that Eq.~(\ref{eq:boson vac trafo}) is well-defined. The condition for the unitarity of the transformation is that the matrix~$V$ from~(\ref{eq:condition for V matrix}) is \emph{Hilbert-Schmidt}\index{Hilbert-Schmidt!condition}~(see, e.g., Ref.~\cite{Shale1962,ShaleStinespring1965,HoneggerRieckers1996,Labonte1974}), i.e., the norm induced by the inner product~(\ref{eq:hilbert schmidt inner product}) is finite. Indeed, assuming that $\alpha$ is bounded it is enough~\cite{Labonte1974,Wald1975,FabbriNavarro-Salas2005} to require this for the matrix~$\beta=(\betamn{mn})$
such that the normalization constant in~(\ref{eq:boson vac trafo}) is finite. We examine this condition perturbatively to leading order in~$h\,$, where only the coefficients~$\betahmn{1}{mn}$ contribute, see~(\ref{eq:bosonic V expansion 1}). The Hilbert-Schmidt condition to leading order is then
\begin{align}
    \sum\limits_{m,n}\,|\betahmn{1}{mn}|^{2}  &<\,\infty\,.
    \label{eq:unitarity condition bosons leading order}
\end{align}
In spite of the plethora of available travel scenarios there are essentially two cases of interest, the sharp transition from Minkowski to Rindler solutions with coefficients~$\MRbetahmn{1}{mn}\,$ from~(\ref{eq:bosonic Bogo coeffs linear beta}), and the smoothly changing accelerations with coefficients~$\Sbetahmn{1}{mn}$ from~(\ref{eq:smooth acceleration expansion betas 1}). Starting with the sharp transitions, we first consider the case $M=0\,$, for which we get
\begin{align}
    \sum\limits_{m,n}\,|\MRbetahmn{1}{mn}(M=0)|^{2} &=\,
    \frac{8}{\pi^{4}}\,\sum\limits_{\substack{m\,\text{even}\\ n\,\text{odd}}}\,\frac{m\,n}{(m+n)^{6}}\,=\,\frac{1}{48\pi^{4}}\bigl(28\,\zeta(3)\,-\,31\,\zeta(5)\bigr)\,,
    \label{eq:unitarity conditions bosons zero mass}
\end{align}
where $\zeta(z)$ is the \emph{Riemann zeta function}\index{Riemann!zeta function}, which is finite for $\operatorname{Re}(z)>1$ and real for $z\in\mathbb{R}\,$. For $M>0$ elementary estimates for the left hand side of~(\ref{eq:unitarity condition bosons leading order}) are obtained by treating the sum as a \emph{Riemann sum}\index{Riemann!sum}. Substituting~$x=m\pi/M$ and~$y=n\pi/M$ the Riemann sum can be written as~\cite[(\ref{Paper:FriisLeeLouko2013})]{FriisLeeLouko2013}
\begin{align}
    M^{2}\,\sum\limits_{m,n}\,|\MRbetahmn{1}{mn}(M)|^{2} &\rightarrow\,
    \frac{2}{\pi^{2}}\,\int_{\substack{x>0\\ y>0}}\,\frac{x^{2}\,y^{2}\,dx\,dy}
    {\sqrt{1+x^{2}}\,\sqrt{1+y^{2}}\,\bigl(\sqrt{1+x^{2}}+\sqrt{1+y^{2}}\bigr)^{6}}\,=\,\frac{1}{90\pi^{2}}\,,
    \label{eq:unitarity conditions bosons nonzero mass}
\end{align}
suggesting the Hilbert-Schmidt condition is satisfied for the $(1+1)$ dimensional scalar cavity field for all~$M\,$. To extend this line of argument to additional spatial dimensions we note that the transverse momenta $k_{\perp}$ enter into the dimensionless parameter~$M$ via the substitution
\begin{align}
    M\,\rightarrow\,\sqrt{\mathfrak{m}^{2}\,+\,k_{\perp}^{2}}\,L\,.
    \label{eq:extra dimensions mass}
\end{align}
For the Hilbert-Schmidt condition it is necessary to sum over the transverse momenta of the additional~$n$ dimensions and the estimate of~(\ref{eq:unitarity conditions bosons nonzero mass}) involves integrals of the form
\begin{align}
    \int_{x_{i}>0}dx_{1}\ldots dx_{n}\,\frac{1}{1\,+\,\sum_{i=1}^{n}x_{i}^{2}}\,,
    \label{eq:extra dimensions Riemann sum estimations}
\end{align}
which diverge for $n\geq2\,$. These contributions suggest that the Hilbert-Schmidt condition for the sharp transitions is still satisfied in $(2+1)$ dimensions, while the unitarity requirement fails in $(3+1)$ (or higher) spacetime dimensions. However, a remedy for this predicament is provided by the Bogoliubov transformations for smoothly varying accelerations. The corresponding coefficients $\Sbetahmn{1}{mn}$~(\ref{eq:smooth acceleration expansion alphas 1}) are given by Fourier transforms of the acceleration~$\mathbf{a}_{\mathrm{c}}(\tau)\,$. If the acceleration changes smoothly the rapid fall-off of the Fourier transform at infinity guarantees that the leading order of the sum
\begin{align}
    \sum\limits_{k_{\perp}}\,\sum\limits_{m,n}\,|\Sbetamn{mn}|^{2}\,
    \label{eq:unitarity smooth accelerations}
\end{align}
remains finite for all spacetime dimensions. Hence, unitarity is established.
\index{unitarity!bosons|)}

\subsubsection{Perturbative Expansion of the Transformed Bosonic Vacuum}

We return to the transformation of the vacuum state with the perturbative expansion of the normalization constant~$N_{\mathrm{vac}}\,$. We insert~(\ref{eq:bosonic V expansion}) into~(\ref{eq:boson vac trafo}) and require $\scpr{0}{0}=1$ to obtain
\begin{align}
    N_{\mathrm{vac}}    &=\,1\,-\,\tfrac{1}{4}\,\sum\limits_{p,q}\,|\betahmn{1}{pq}|^{2}\,h^{2}\,+\,O(h^{4})\,.
    \label{eq:boson vac normalization}
\end{align}
Consecutively we can express the transformed vacuum state as
\begin{align}
    \ket{0} &=\,
    \kethat{0}\,+\,h\,\tfrac{1}{2}\,\sum\limits_{p,q}\,\Vhmn{1}{pq}\,\adnhat{p}\,\adnhat{q}\,\kethat{0}
    \,+\,h^{2}\,\tfrac{1}{2}\,\sum\limits_{p,q}\,\Bigl(\Vhmn{2}{pq}\,\adnhat{p}\,\adnhat{q}
    \,-\,\tfrac{1}{2}\,|\betahmn{1}{pq}|^{2}\label{eq:transformed boson vac vector}\\[1mm]
    &\ +\,\tfrac{1}{4}\sum\limits_{p\pr\!,q\pr}\,\Vhmn{1}{pq}\,\Vhmn{1}{p\pr q\pr}\ \
    \adnhat{p}\,\adnhat{q}\,\adnhat{p\pr}\,\adnhat{q\pr}\Bigr)\,\kethat{0}\,+\,O(h^{3})\,,\nonumber
\end{align}
with $\Vhmn{1}{pq}$ and $\Vhmn{2}{pq}$ given by~(\ref{eq:bosonic V expansion 1}) and~(\ref{eq:bosonic V expansion 2}), respectively. To leading order the state remains unchanged, while the linear corrections add pairs of excitations to the superposition. To linear order the changes to the state are governed by the coefficients~$\betahmn{1}{mn}\,$. The density operator corresponding to~(\ref{eq:transformed boson vac vector}) is then simply\index{vacuum!bosons}
\begin{align}
    \ket{0}\!\bra{0} &=\,\kethat{0}\!\brahat{0}\,+\,h\,\tfrac{1}{2}\,\sum\limits_{p,q}\,
    \bigl(\Vhmn{1}{pq}\,\adnhat{p}\,\adnhat{q}\,\kethat{0}\!\brahat{0}\,+\,\mathrm{H.~c.}\Bigr)
    \label{eq:transformed boson vac density matrix}\\[1mm]
    &\ +\,h^{2}\,\tfrac{1}{2}\sum\limits_{p,q}\,
    \Bigl[\,\Bigl(\Vhmn{2}{pq}\adnhat{p}\,\adnhat{q}\,\kethat{0}\!\brahat{0}\,+\,\mathrm{H.~c.}\Bigr)\,
    -\,|\betahmn{1}{pq}|^{2}\,\kethat{0}\!\brahat{0}\nonumber\\[1mm]
    &\ +\,\tfrac{1}{4}\,\sum\limits_{p\pr\!,q\pr}\,\Bigl(\Vhmn{1}{pq}\,\Vhmn{1}{p\pr q\pr}\ \
    \adnhat{p}\,\adnhat{q}\,\adnhat{p\pr}\,\adnhat{q\pr}\,\kethat{0}\!\brahat{0}\,+\,\mathrm{H.~c.}\,\Bigr)
    \nonumber\\[1mm]
    &\ +\,\tfrac{1}{2}\,\sum\limits_{p\pr\!,q\pr}\,\Vhmn{1}{pq}\,\Vhmnstar{1}{p\pr q\pr}\,
    \adnhat{p}\,\adnhat{q}\,\kethat{0}\!\brahat{0}\,\anhat{p\pr}\,\anhat{q\pr}\,\Bigr]\,+\,O(h^{3})\,,
    \nonumber
\end{align}
where ``$\mathrm{H.~c.}$" denotes the Hermitean conjugate, $(\mathcal{O}+\mathrm{H.~c.})=(\mathcal{O}+\mathcal{O}^{\dagger})$.
Noting that $\Vmn{pq}$ is symmetric and the diagonal leading order terms vanish, $\Vhmn{1}{nn}=0\,$, the right hand side of Eq.~(\ref{eq:transformed boson vac density matrix}) can be quickly seen to be normalized, i.e., $\tr(\ket{0}\!\bra{0})=1+O(h^{3})\,$.

\subsection{Transformation of Bosonic Particle States}\label{sec:Transformation of Bosonic Particle States}

To obtain the out-region decomposition of any other Fock states\index{Fock!states}\index{state!Fock} we express the in-region creation operators in terms of their Bogoliubov transformation to the out-region operators, i.e.,
\begin{align}
    \adn{m}  &=\,\sum\limits_{n}\bigl(\alphamnstar{nm}\,\adnhat{n}\,+\,\betamn{nm}\,\anhat{n}\,\bigr)\,.
    \label{eq:in region creation operator mode k}
\end{align}
Consecutively, we apply the operators to the vacuum state in the decomposition~(\ref{eq:transformed boson vac vector}) and expand the Bogoliubov coefficients as in~(\ref{eq:generic travel scenario expansion}). To illustrate the procedure we consider Fock states with a single excitation in a particular mode and such with an excitation each in two different modes.

\subsubsection{Bosonic Single Particle States}

For the single excitation in an in-region mode labelled by~$k$ we apply the creation operator~$\adn{k}$ to the vacuum to study the transformation of the state $\adn{k}\ket{0}=\ket{1_{k}}\,$. The power expansion in~$h$ gives
\begin{align}
    \ket{1_{k}} &=\,\Ghnstar{0}{k}\,\kethatk{1}{k}\,+\,h\,\Bigl(
    \sum\limits_{m}\,\alphahmnstar{1}{mk}\ \adnhat{m}\,+\,\tfrac{1}{2}\,\Ghnstar{0}{k}\,
    \sum\limits_{p,q}\,\Vhmn{1}{pq}\ \adnhat{p}\,\adnhat{q}\,\adnhat{k}\Bigr)\,\kethat{0}
    \label{eq:transformed boson 1k vector}\\[1mm]
    &\ +\,h^{2}\,\Bigl[\sum\limits_{m}\,\Bigl(\alphahmnstar{2}{mk}\,\adnhat{m}\,+\,
    \sum\limits_{p}\,\betahmn{1}{pk}\,\Vhmn{1}{pm}\,\adnhat{m}\Bigr)\,
    +\,\tfrac{1}{2}\,
    \Ghnstar{0}{k}\,\sum\limits_{p,q}\,\Bigl(
    \sum\limits_{m}\,\Ghn{0}{k}\,\alphahmnstar{1}{mk}\,\Vhmn{1}{pq}\ \adnhat{p}\,\adnhat{q}\,\adnhat{m}
    \nonumber\\[1mm]
    &\ +\,
    \Vhmn{2}{pq}\ \adnhat{p}\,\adnhat{q}\,\adnhat{k}
    \,-\,\tfrac{1}{2}\,|\betahmn{1}{pq}|^{2}\,\adnhat{k}
    \,+\,\tfrac{1}{4}\,\sum\limits_{p\pr\!,q\pr}\,\Vhmn{1}{pq}\,\Vhmn{1}{p\pr q\pr}\
    \adnhat{p}\,\adnhat{q}\,\adnhat{p\pr}\,\adnhat{q\pr}\,\adnhat{k}\Bigr)\Bigr]\,\kethat{0}
    \,+\,O(h^{3})\,,\nonumber
\end{align}
where $\Vhmn{1}{pq}$ and $\Vhmn{2}{pq}$ are given by~(\ref{eq:bosonic V expansion 1}) and~(\ref{eq:bosonic V expansion 2}), as previously. In addition to the creation of particle pairs the linear order terms now feature the coefficients $\alphahmn{1}{mk}$ shifting the excitation of mode~$k$ to other modes. The density operator for the state~(\ref{eq:transformed boson 1k vector}) is given by
\begin{align}
    \ket{1_{k}}\!\bra{1_{k}} &=\,\kethatk{1}{k}\!\brahatk{1}{k}\,+\,h\,\Bigl[
    \sum\limits_{m}\,\Bigl(\Ghn{0}{k}\,\alphahmnstar{1}{mk}\ \adnhat{m}\,\kethat{0}\!\brahatk{1}{k}
    \,+\,\mathrm{H.~c.}\Bigr)
    \label{eq:transformed boson 1k density matrix}\\[1mm]
    &\hspace*{-8mm}+\,\tfrac{1}{2}\,\sum\limits_{p,q}\,\Bigl(\Vhmn{1}{pq}\ \adnhat{p}\,\adnhat{q}\,
    \kethatk{1}{k}\!\brahatk{1}{k}\,+\,\mathrm{H.~c.}\Bigr)\Bigr]\,+\,h^{2}\,\Bigl[
    \sum\limits_{m,m\pr}\,\alphahmnstar{1}{mk}\,\alphahmn{1}{m\pr k}\,
    \adnhat{m}\,\kethat{0}\!\brahat{0}\,\anhat{m\pr}
    \nonumber\\[1mm]
    &\hspace*{-8mm}+\,\sum\limits_{m}\,\Bigl\{
    \Ghn{0}{k}\,\alphahmnstar{2}{mk}\,\adnhat{m}\,\kethat{0}\!\brahatk{1}{k}\,+\,
    \tfrac{1}{2}\,\sum\limits_{p,q}\,\Ghnstar{0}{k}\,\alphahmn{1}{mk}\,\Vhmn{1}{pq}\,\adnhat{p}\,\adnhat{q}\,
    \kethatk{1}{k}\!\brahat{0}\,\anhat{m}\nonumber\\[1mm]
    &\hspace*{-8mm}+\,\Ghn{0}{k}\,\sum\limits_{p}\,\betahmn{1}{pk}\,\Vhmn{1}{pm}\,\adnhat{m}\,
    \kethat{0}\!\brahatk{1}{k}\,+\,\mathrm{H.~c.}\Bigr\}\,+\,\tfrac{1}{2}\,
    \sum\limits_{p,q}\,\Bigl\{\Bigl(
    \sum\limits_{m}\,\Ghn{0}{k}\,\alphahmnstar{1}{mk}\,\Vhmn{1}{pq}\ \adnhat{p}\,\adnhat{q}\,\adnhat{m}
    \nonumber\\[1mm]
    &\hspace*{-8mm}+\,
    \Vhmn{2}{pq}\ \adnhat{p}\,\adnhat{q}\,\adnhat{k}
    \,-\,\tfrac{1}{2}\,|\betahmn{1}{pq}|^{2}\,\adnhat{k}
    \,+\,\tfrac{1}{4}\,\sum\limits_{p\pr\!,q\pr}\,\Vhmn{1}{pq}\,\Vhmn{1}{p\pr q\pr}\
    \adnhat{p}\,\adnhat{q}\,\adnhat{p\pr}\,\adnhat{q\pr}\,\adnhat{k}\Bigr)\,\kethat{0}\!\brahatk{1}{k}
    \nonumber\\[1mm]
    &\hspace*{-8mm}+\,
    \tfrac{1}{4}\,\sum\limits_{p\pr\!,q\pr}\,\Vhmn{1}{pq}\,\Vhmnstar{1}{p\pr q\pr}\,
    \adnhat{p}\,\adnhat{q}\,\kethatk{1}{k}\!\brahatk{1}{k}\,\anhat{p\pr}\,\anhat{q\pr}
    \,+\,\mathrm{H.~c.}\Bigr\}\Bigr]\,+\,O(h^{3})\,,\nonumber
\end{align}
The normalization of~(\ref{eq:transformed boson 1k density matrix}) can be verified using the Bogoliubov identity~(\ref{eq:generic travel scenario bogo identities bosons alphas 2}) and the trace
\begin{align}
    \hspace*{-1cm}\tr\Bigl(\adnhat{p}\,\adnhat{q}\,\kethatk{1}{k}\!\brahatk{1}{k}\,
    \anhat{p\pr}\,\anhat{q\pr}\Bigr)    &=\,
    \Bigl(\delta_{pp\pr}\delta_{qq\pr}\,+\,\delta_{pq\pr}\delta_{qp\pr}\Bigr)
    \Bigl(2\delta_{pk}(1-\delta_{qk})
    \label{eq:trace second order term}\\[1mm]
    &\ +\,2\delta_{qk}(1-\delta_{pk})\,+\,
    (1-\delta_{pk})(1-\delta_{qk})\,+\,6\delta_{pk}\delta_{qk}\Bigr)\,,\nonumber
\end{align}
where~(\ref{eq:bosonic creation factors}) is taken into account. As before with~(\ref{eq:transformed boson vac vector}) the state remains pure, as required by the unitarity of the transformation, if no modes are traced over. We shall consider tracing over subsets of the modes in Chapters~\ref{Chapter 6 Motion Generates Entanglement} and~\ref{Chapter 7 Degradation of Entanglement between Moving Cavities} to study entanglement of the remaining modes.

\subsubsection{Bosonic Particle Pair}

For the state $\ket{1_{k}}\ket{1_{k\pr}}$ we apply the creation operator~$\adn{k\pr}$ to~(\ref{eq:transformed boson 1k vector}). For simplicity of notation in this illustration we keep terms up to linear order in~$h$ and obtain
\begin{align}
    \ket{1_{k}}\ket{1_{k\pr}}   &=\,\Ghnstar{0}{k}\,\Ghnstar{0}{k\pr}\,\kethatk{1}{k}\kethatk{1}{k\pr}\,
    +\,h\,\Bigl[\tfrac{1}{2}\,\bigl(\Ghnstar{0}{k}\,\betahmn{1}{kk\pr}\,+\,
    \Ghnstar{0}{k\pr}\,\betahmn{1}{k\pr k}\bigr)\,\kethat{0}
    \label{eq:transformed boson 1k 1kpr vector}\\[1mm]
    &\ +\,\sum\limits_{m}\,\Ghnstar{0}{k}\,\alphahmnstar{1}{mk\pr}\,\adnhat{m}\,\kethatk{1}{k}\,
    +\,\sum\limits_{m}\,\Ghnstar{0}{k\pr}\,\alphahmnstar{1}{mk}\,\adnhat{m}\,\kethatk{1}{k\pr}\,
    \nonumber\\[1mm]
    &\ -\,\tfrac{1}{2}\,\Ghnstar{0}{k}\,\Ghnstar{0}{k\pr}\,\sum\limits_{p,q}\,
    \Ghnstar{0}{q}\,\betahmnstar{1}{pq}\,\adnhat{p}\,\adnhat{q}\,\kethatk{1}{k}\kethatk{1}{k\pr}\Bigr]
    \,+\,O(h^{2})\,.\nonumber
\end{align}
The corresponding density operator is given by the projector on $\ket{1_{k}}\ket{1_{k\pr}}\,$. As can be inferred from~(\ref{eq:transformed boson vac density matrix}) and~(\ref{eq:transformed boson 1k density matrix}) the density operator decomposition becomes more and more involved when additional excitations are added. For more complicated states it thus becomes cumbersome to study the Bogoliubov transformation in this fashion.

\section{Transformation of Bosonic Gaussian States}\label{sec:Transformation of Bosonic Gaussian States}

\subsection{Symplectic Representation of Non-uniform Motion}\label{sec:Symplectic Representation of Non-uniform Motion}

A computationally much simpler way of handling more complicated states is the \emph{symplectic representation} of the Bogoliubov transformation in phase space as explained in Section~\ref{sec:symplectic operations}. The symplectic transformation~$S$\index{symplectic!transformation}\index{transformation!symplectic} for an arbitrary travel scenario can be decomposed into blocks~$\mathcal{M}_{mn}$ as given in~(\ref{eq:Gaussian n-mode Bogo transformation}), see Ref.~\cite[(\ref{Paper:FriisFuentes2013})]{FriisFuentes2013}. For fixed~$m$ and~$n$ the $2\times2$ matrix~$\mathcal{M}_{mn}$ is given by~(\ref{eq:M Bogo matrix}), which we may expand in a power series in~$h\,$,
\begin{align}
    \Mmn{ij}    &=\,\Mhmn{0}{ij}\,+\,\Mhmn{1}{ij}\,h\,+\,\Mhmn{2}{ij}\,h^{2}\,+\,O(h^{3})\,,
    \label{eq:M bogo matrix expanded}
\end{align}
where the non-vanishing leading order coefficient matrices in the expansion are given by
\begin{align}
    \Mhmn{0}{ii}    &=\,\begin{pmatrix}\ \cos(\omega_{i}\tilde{\tau}) &   \sin(\omega_{i}\tilde{\tau})\,\\
    -\sin(\omega_{i}\tilde{\tau})   &   \cos(\omega_{i}\tilde{\tau})\,\end{pmatrix}\,,
    \label{eq:M bogo matrix expansion coefficients 0}
\end{align}
with~$\omega_{i}$ given by~(\ref{eq:inertial modes frequencies}). The coefficient matrices of~$h^{n}$ take the form
\begin{align}
    \Mhmn{n}{ij}\,=\,
        \begin{pmatrix}
            \ \operatorname{Re}(\alphahmn{n}{ij}\,-\,\betahmn{n}{ij})     &
            \operatorname{Im}(\alphahmn{n}{ij}\,+\,\betahmn{n}{ij})\, \\[1.5mm]
            -\operatorname{Im}(\alphahmn{n}{ij}\,-\,\betahmn{n}{ij})    &
            \operatorname{Re}(\alphahmn{n}{ij}\,+\,\betahmn{n}{ij})\, \\
        \end{pmatrix}\,,
    \label{eq:M bogo matrix expansion coefficients p}
\end{align}
where $\alphahmn{n}{ij}$ and $\betahmn{n}{ij}$ are as in~(\ref{eq:generic travel scenario expansion}).

\subsection{Transformed Covariance Matrix Example}\label{sec:transformed covariance matrix example}

Specializing to \emph{Gaussian states}\index{state!Gaussian}\index{Gaussian!state} we are interested in the effect of the non-uniform motion described in Chapter~\ref{Chapter 4 Constructing Non Uniformly Moving Cavities} on the covariance matrix\index{covariance!matrix}, which encodes all the information about the entanglement of the state. The symplectic transformation~$S$ for a given travel scenario takes the in-region covariance matrix~$\Gamma$ to
\begin{align}
    \widehat{\Gamma}    &=\,S\,\Gamma\,S^{T}\,.
    \label{eq:out region covariance matrix general}
\end{align}
We examine a particular example for an initial state more closely. A fully separable state, represented by the covariance matrix~$\Gamma=\bigoplus_{n}\Gammak{n}$ of an arbitrary number of modes with individual covariance matrices~$\Gammak{n}$ $(n=1,2,\ldots)\,$. The transformed covariance matrix $\widehat{\Gamma}$ decomposes into the diagonal blocks~$\Gammahatk{m}$ for the individual modes, and off-diagonal blocks $\Chatmn{mn}$ encoding the correlations between modes~$m$ and~$n\,$. In terms of the matrices~$\Mmn{mn}$ from~(\ref{eq:M Bogo matrix}) these $2\times2$ matrices read
\begin{subequations}
\label{eq:transformed cov matrix blocks general}
\begin{align}
    \Gammahatk{m}   &=\,\sum\limits_{i}\,\Mmn{mi}\,\Gammak{i}\,\Mmntrans{mi}\,,
    \label{eq:transformed cov matrix blocks general gamma}\\[1mm]
    \Chatmn{mn} &=\,\sum\limits_{i}\,\Mmn{mi}\,\Gammak{i}\,\Mmntrans{ni}\,.
    \label{eq:transformed cov matrix blocks general C}
\end{align}
\end{subequations}

\subsubsection{Transformed Single-mode Squeezed States}\label{page:single mode squeezed Bogo transformation}

To examine our example more closely we select particular initial states~$\Gammak{n}\,$, i.e., we assume that each single mode can be \emph{squeezed} with squeezing parameters~$s_{n}\in\mathbb{R}\,$, such that\index{state!single-mode squeezed}\index{squeezed state!single-mode} the matrices~$\Gammak{n}$ are given by~(\ref{eq:single mode squeezed covariance matrix})
\begin{align}
    \Gammak{n}(s_{n})    &=\,
    \begin{pmatrix}e^{2s_{n}}   &   0   \\  0   &   e^{-2s_{n}}\end{pmatrix}\,.
    \label{eq:single mode squeezed states}
\end{align}
Employing the perturbative expansion of~(\ref{eq:M bogo matrix expanded}) we get~$\Gammahatk{m}$ from~(\ref{eq:transformed cov matrix blocks general gamma}) as a power series in~$h\,$,
\begin{align}
    \Gammahatk{m}   &=\,\Gammahatkh{m}{0}\,+\,\Gammahatkh{m}{1}\,h\,+\,\Gammahatkh{m}{2}\,h^{2}\,+\,O(h^{3})\,.
    \label{eq:transformed single mode squeezed cov matrix gamma}
\end{align}
The symmetric coefficient matrices in the expansion are expressed in components order by order, that is
\begin{subequations}
\label{eq:transformed single mode squeezed cov matrix gamma 0}
\begin{align}
    (\Gammahatkh{m}{0})_{11}    &=\,\cosh(2s_{m})\,+\,\sinh(2s_{m})\,\cos(2\omega_{m}\tilde{\tau})\,,
    \label{eq:transformed single mode squeezed cov matrix 0 gamma 11}\\[1mm]
    (\Gammahatkh{m}{0})_{22}    &=\,\cosh(2s_{m})\,-\,\sinh(2s_{m})\,\cos(2\omega_{m}\tilde{\tau})\,,
    \label{eq:transformed single mode squeezed cov matrix 0 gamma 22}\\[1mm]
    (\Gammahatkh{m}{0})_{12}    &=\,-\,\sinh(2s_{m})\,\sin(2\omega_{m}\tilde{\tau})\,.
    \label{eq:transformed single mode squeezed cov matrix 0 gamma 12}
\end{align}
\end{subequations}
The coefficients of~$h$ in the expansion of~$\Gammahatk{m}$ in~(\ref{eq:transformed single mode squeezed cov matrix gamma}) vanish identically but the second order coefficients are non-zero and given by
\begin{subequations}
\label{eq:transformed single mode squeezed cov matrix gamma 2}
\begin{align}
(\Gammahatkh{m}{2})_{11}    &=\,2\cosh(2s_{m})\operatorname{Re}
    \bigl(\Ghn{0}{m}\,\left[\alphahmnstar{2}{mm}-\betahmn{2}{mm}\right]\bigr)
    \,+\,\sinh(2s_{m})\operatorname{Re}
    \bigl(\Ghn{0}{m}\,\left[\alphahmn{2}{mm}-\betahmnstar{2}{mm}\right]\bigr)
    \nonumber\\[1.5mm]
    &\ \ +\,\sum\limits_{n}\,\Bigl[\cosh(2s_{m})\,\Bigl(|\alphahmn{1}{mn}|^{2}\,-\,
    2\operatorname{Re}(\alphahmn{1}{mn}\betahmn{1}{mn})\,+\,|\betahmn{1}{mn}|^{2}\Bigr)
    \label{eq:transformed single mode squeezed cov matrix 2 gamma 11}\\[0.5mm]
    &\ \hspace*{9mm}+\,\sinh(2s_{m})\operatorname{Re}\Bigl((\alphahmn{1}{mn})^{2}\,-\,
    2\alphahmn{1}{mn}\betahmnstar{1}{mn}\,+\,(\betahmn{1}{mn})^{2}\Bigr)\Bigr]\,,
    \nonumber\\[2.5mm]
(\Gammahatkh{m}{2})_{22}    &=\,2\cosh(2s_{m})\operatorname{Re}
    \bigl(\Ghn{0}{m}\,\left[\alphahmnstar{2}{mm}+\betahmn{2}{mm}\right]\bigr)
    \,-\,\sinh(2s_{m})\operatorname{Re}
    \bigl(\Ghn{0}{m}\,\left[\alphahmn{2}{mm}+\betahmnstar{2}{mm}\right]\bigr)
    \nonumber\\[1.5mm]
    &\ \ +\,\sum\limits_{n}\,\Bigl[\cosh(2s_{m})\,\Bigl(|\alphahmn{1}{mn}|^{2}\,+\,
    2\operatorname{Re}(\alphahmn{1}{mn}\betahmn{1}{mn})\,+\,|\betahmn{1}{mn}|^{2}\Bigr)
    \label{eq:transformed single mode squeezed cov matrix 2 gamma 22}\\[0.5mm]
    &\ \hspace*{9mm}-\,\sinh(2s_{m})\operatorname{Re}\Bigl((\alphahmn{1}{mn})^{2}\,+\,
    2\alphahmn{1}{mn}\betahmnstar{1}{mn}\,+\,(\betahmn{1}{mn})^{2}\Bigr)\Bigr]\,,
    \nonumber\\[2.5mm]
(\Gammahatkh{m}{2})_{12}    &=\,2\cosh(2s_{m})\operatorname{Im}\bigl(\Ghn{0}{m}\betahmn{2}{mm}\bigr)\,-\,
    2\sinh(2s_{m})\operatorname{Im}\bigl(\Ghn{0}{m}\alphahmn{2}{mm}\bigr)
    \label{eq:transformed single mode squeezed cov matrix 2 gamma 12}\\[1.5mm]
    &\ \ +\,\sum\limits_{n}\,\Bigl[2\cosh(2s_{m})\operatorname{Im}\bigl(\alphahmn{1}{mn}\betahmn{1}{mn}\bigr)
    \,-\,\sinh(2s_{m})\operatorname{Im}\bigl((\alphahmn{1}{mn})^{2}+(\betahmn{1}{mn})^{2}\bigr)\Bigr]\,.
    \nonumber
\end{align}
\end{subequations}
Similarly, the off-diagonal blocks~$\Chatmn{mn}$ are expanded as
\begin{align}
    \Chatmn{mn}   &=\,\Chatmnh{mn}{1}\,h\,+\,\Chatmnh{mn}{2}\,h^{2}\,+\,O(h^{3})\,,
    \label{eq:transformed single mode squeezed cov matrix C}
\end{align}
where the lowest order in the expansion is linear in~$h\,$. The corresponding components of these coefficients are given by
\begin{subequations}
\label{eq:transformed single mode squeezed cov matrix C 1}
\begin{align}
(\Chatmnh{mn}{1})_{11}    &=\,
    \cosh(2s_{m})\operatorname{Re}
    \bigl(\Ghn{0}{m}\left[\alphahmnstar{1}{nm}-\betahmn{1}{nm}\right]\bigr)\,+\,
    \cosh(2s_{n})\operatorname{Re}
    \bigl(\Ghn{0}{n}\left[\alphahmnstar{1}{mn}-\betahmn{1}{mn}\right]\bigr)
    \nonumber\\[1.5mm]
    &\ +\,
    \sinh(2s_{m})\operatorname{Re}
    \bigl(\Ghn{0}{m}\left[\alphahmn{1}{nm}-\betahmnstar{1}{nm}\right]\bigr)\,+\,
    \sinh(2s_{n})\operatorname{Re}
    \bigl(\Ghn{0}{n}\left[\alphahmn{1}{mn}-\betahmnstar{1}{mn}\right]\bigr)\,,
    \label{eq:transformed single mode squeezed cov matrix 1 C 11}\\[1.5mm]
(\Chatmnh{mn}{1})_{22}    &=\,
    \cosh(2s_{m})\operatorname{Re}
    \bigl(\Ghn{0}{m}\left[\alphahmnstar{1}{nm}+\betahmn{1}{nm}\right]\bigr)\,+\,
    \cosh(2s_{n})\operatorname{Re}
    \bigl(\Ghn{0}{n}\left[\alphahmnstar{1}{mn}+\betahmn{1}{mn}\right]\bigr)
    \nonumber\\[1.5mm]
    &\ -\,
    \sinh(2s_{m})\operatorname{Re}
    \bigl(\Ghn{0}{m}\left[\alphahmn{1}{nm}+\betahmnstar{1}{nm}\right]\bigr)\,-\,
    \sinh(2s_{n})\operatorname{Re}
    \bigl(\Ghn{0}{n}\left[\alphahmn{1}{mn}+\betahmnstar{1}{mn}\right]\bigr)\,,
    \label{eq:transformed single mode squeezed cov matrix 1 C 22}\\[1.5mm]
(\Chatmnh{mn}{1})_{12}    &=\,
    \cosh(2s_{m})\operatorname{Im}
    \bigl(\Ghn{0}{m}\left[\betahmn{1}{nm}+\alphahmnstar{1}{nm}\right]\bigr)\,+\,
    \cosh(2s_{n})\operatorname{Im}
    \bigl(\Ghn{0}{n}\left[\betahmn{1}{mn}-\alphahmnstar{1}{mn}\right]\bigr)
    \nonumber\\[1.5mm]
    &\ -\,
    \sinh(2s_{m})\operatorname{Im}
    \bigl(\Ghn{0}{m}\left[\alphahmn{1}{nm}+\betahmnstar{1}{nm}\right]\bigr)\,-\,
    \sinh(2s_{n})\operatorname{Im}
    \bigl(\Ghn{0}{n}\left[\alphahmn{1}{mn}-\betahmnstar{1}{mn}\right]\bigr)\,,
    \label{eq:transformed single mode squeezed cov matrix 1 C 12}\\[1.5mm]
(\Chatmnh{mn}{1})_{21}    &=\,
    \cosh(2s_{m})\operatorname{Im}
    \bigl(\Ghn{0}{m}\left[\betahmn{1}{nm}-\alphahmnstar{1}{nm}\right]\bigr)\,+\,
    \cosh(2s_{n})\operatorname{Im}
    \bigl(\Ghn{0}{n}\left[\betahmn{1}{mn}+\alphahmnstar{1}{mn}\right]\bigr)
    \nonumber\\[1.5mm]
    &\ -\,
    \sinh(2s_{m})\operatorname{Im}
    \bigl(\Ghn{0}{m}\left[\alphahmn{1}{nm}-\betahmnstar{1}{nm}\right]\bigr)\,-\,
    \sinh(2s_{n})\operatorname{Im}
    \bigl(\Ghn{0}{n}\left[\alphahmn{1}{mn}+\betahmnstar{1}{mn}\right]\bigr)\,.
    \label{eq:transformed single mode squeezed cov matrix 1 C 21}
\end{align}
\end{subequations}
We shall study the entanglement of the transformed single mode squeezed states in Chapter~\ref{Chapter 6 Motion Generates Entanglement}. For now, let us briefly return to the Fock space treatment, this time for the fermions.

\section{Fermionic State Transformation}\label{sec:Fermionic State Transformation}

\subsection{Fermionic Vacuum Transformation}\label{sec:fermionic Vacuum Transformation}

We pursue the construction of the Bogoliubov transformation on the \emph{fermionic Fock space}\index{Fock!space, fermionic} (see Section~\ref{sec:fermionic Fock space}) in a completely analogous fashion as the previous bosonic case in Section~\ref{sec:Bosonic State Transformation} by starting from the
fermionic vacuum\index{vacuum!fermions} state~$\fket{0}\,$. Following Ref.~[(\ref{Paper:FriisLeeBruschiLouko2012})]\cite{FriisLeeBruschiLouko2012}, a similar argument as that made for bosons on page~\pageref{eq:boson vac trafo} allows us to make the ansatz
\begin{align}
    \fket{0} &=\,\tilde{N}_{\text{vac}}\,\exp(\mathcal{W})\,\fkethat{0}\,=\,
    \tilde{N}_{\text{vac}}\,\exp\bigl(\tfrac{1}{2}\sum\limits_{\substack{p\geq0\\ q<0}}\,\fVmn{pq}\,\bdnhat{p}\,\cdnhat{q}\bigr)\,
    \fkethat{0}\,,
    \label{eq:fermion vac trafo}
\end{align}
where $\fVmn{pq}\in\mathbb{C}$ and $\tilde{N}_{\text{vac}}$ is a normalization constant. We then examine the property~$\bn{n}\fket{0}=0$ by inserting the Bogoliubov transformation of~$\bn{n}\,$,
\begin{align}
    \bn{n}  &=\,\sum\limits_{m\geq0}\,\Amn{mn}\,\bnhat{m}\,+\,\sum\limits_{m<0}\,\Amn{mn}\,\cdnhat{m}\,.
    \label{eq:Dirac bn Bogo}
\end{align}
Now turning again to the Hadamard Lemma\index{Hadamard Lemma} of~(\ref{eq:Hadamard lemma}) the commutators
\begin{subequations}
\label{eq:derivation of fermionic V commutators}
\begin{align}
    \comm{\bnhat{m}}{\mathcal{W}} &=\,\sum\limits_{q<0}\,\fVmn{mq}\,\cdnhat{q}\,,
    \label{eq:derivation of fermionic V commutators nonzero}\\[1mm]
    \comm{\cdnhat{m}}{\mathcal{W}} &=\,\comm{\comm{\bnhat{m}}{\mathcal{W}}}{\mathcal{W}}\,=\,0\,,
    \label{eq:derivation of fermionic V commutators zero}
\end{align}
\end{subequations}
provide the criterion (for $n\geq0,q<0$)
\begin{subequations}
\label{eq:fermionc V criteria}
\begin{align}
    \sum\limits_{m\geq0}\,\Amn{mn}\,\fVmn{mq}   &=\,-\,\Amn{qn}\,,
    \label{eq:fermionic V conditions 1}
\end{align}
while the same procedure using $\cn{n}\fket{0}=0$ gives (for $n<0,p\geq0$)
\begin{align}
    \sum\limits_{m<0}\,\Amnstar{mn}\,\fVmn{pm}   &=\,\Amnstar{pn}\,.
    \label{eq:fermionic V conditions 2}
\end{align}
\end{subequations}
If one of the blocks of~$A=(\Amn{mn})$ where the indices are either both non-negative or both negative is invertible then the conditions~(\ref{eq:fermionic V conditions 1}) or~(\ref{eq:fermionic V conditions 2}), respectively, uniquely determine the matrix $\mathcal{V}\,$. If both blocks are invertible~(\ref{eq:fermionic V conditions 1}) and~(\ref{eq:fermionic V conditions 2}) are equivalent by virtue of the unitarity of~$A\,$, i.e., $A^{\dagger}A=\mathds{1}\,$.

\subsubsection{Unitarity of the Transformation \textemdash\ Fermions}\label{page:unitarity fermions}

\index{unitarity!fermions|(}
To ensure that the fermionic in-region and out-region vacua can indeed be unitarily related a closer examination of the Hilbert-Schmidt condition is in order. For the fermionic case it manifests as the condition that the blocks of $A=(\Amn{mn})$ that relate positive and negative frequency solutions are \emph{Hilbert-Schmidt}\index{Hilbert-Schmidt!condition}, see Ref.~\cite{Labonte1974}. In complete analogy to~(\ref{eq:unitarity condition bosons leading order}) the leading order presents the requirement
\begin{align}
    \sum\limits_{\substack{p\geq0\\ q<0}}|\Ahmn{1}{pq}|^{2} &<\,\infty\,.
\label{eq:fermionic unitarity leading order}
\end{align}
We proceed, as before, with the sharp transitions between the Minkowski and Rindler solutions in $(1+1)$ dimensions, and we consider first the case of zero mass. The leading order coefficients for $M=0$ are given by~(\ref{eq:Dirac Mink to Rindler bogos massless linear})
\begin{align}
    \sum\limits_{\substack{p\geq0\\ q<0}}\,|\MRAhmn{1}{pq}(M=0)|^{2} &=\,
    \frac{1}{96\pi^{4}}\bigl(28\,\zeta(3)\,-\,31\,\zeta(5)\bigr)\,<\,\infty\,,
    \label{eq:unitarity conditions fermions zero mass}
\end{align}
where $\zeta(z)$ is the \emph{Riemann zeta function}\index{Riemann!zeta function}. For the massive case we use an analysis as in~(\ref{eq:unitarity conditions bosons nonzero mass}), where we consider an estimate in terms of a \emph{Riemann sum}\index{Riemann!sum}~\cite[(\ref{Paper:FriisLeeLouko2013})]{FriisLeeLouko2013}. Substituting~$x=m\pi/M$ and~$y=n\pi/M$ we write
\begin{align}
    M^{2}\,\sum\limits_{\substack{p\geq0\\ q<0}}\,|\MRAhmn{1}{pq}(M)|^{2} &\rightarrow\,
    \frac{8}{\pi^{2}}\,\int_{\substack{x>0\\ y>0}}\,
    \frac{\bigl(\sqrt{1+x^{2}}+x-\sqrt{1+y^{2}}-y\bigr)^{2}}
    {\bigl(\sqrt{1+x^{2}}+x+\sqrt{1+y^{2}}+y\bigr)^{6}}\label{eq:unitarity conditions fermions nonzero mass}\\[1mm]
    &\hspace*{-3cm} \times\frac{\bigl[(\sqrt{1+x^{2}}+x)(\sqrt{1+y^{2}}+y)-1\bigr]^{2}}{\bigl[(\sqrt{1+x^{2}}+x)(\sqrt{1+y^{2}}+y)+1\bigr]^{6}}\,
     \times\,\frac{x^{2}y^{2}(\sqrt{1+x^{2}}+x)^{4}(\sqrt{1+y^{2}}+y)^{4}}{(1+x^{2})(1+y^{2})}\,dx\,dy\,,
    \nonumber
\end{align}
where $\MRAhmn{1}{pq}$ are the coefficients of~(\ref{eq:Dirac Mink to Rindler bogos linear}). The integral in~(\ref{eq:unitarity conditions fermions nonzero mass}), where we have set $x=|p|/\mu$ and $y=|q|/\mu\,$, can be evaluated by the substitution $x=(u-u^{-1})/2$ and $y=(v-v^{-1})/2\,$, which reveals
\begin{align}
    M^{2}\,\sum\limits_{\substack{p\geq0\\ q<0}}\,|\MRAhmn{1}{pq}(M)|^{2} &\rightarrow\,
    \frac{7}{45\pi^{2}}\,-\,\frac{1}{64}\,<\,\infty\,.
    \label{eq:unitarity conditions fermions nonzero mass result}
\end{align}
We can thus conclude that the Hilbert-Schmidt condition is satisfied for the $(1+1)$ dimensional Dirac cavity field for all~$M\,$. The addition of extra spatial dimensions suffers from the same limitations as the bosonic case [see Eq.~(\ref{eq:extra dimensions Riemann sum estimations})] such that the unitarity of the Bogoliubov transformation for sharp transitions holds only in $(1+1)$ and $(2+1)$ dimensions, but fails in $(3+1)$ dimensions and beyond. For smooth accelerations the rapid fall-off of the Fourier transform in Eq.~(\ref{eq:smooth acceleration expansion fermions 1}) again guarantees the unitarity for all cases. With this in mind we return to the transformation of the vacuum state.
\index{unitarity!fermions|)}

\subsubsection{Perturbative Expansion of the Transformed Fermionic Vacuum}

We now perform the perturbative expansion of Eq.~(\ref{eq:fermionic V conditions 1}) with the coefficients of~(\ref{eq:generic travel scenario expansion A}), which yields
\begin{align}
    \fVmn{pq}   &=\,\fVhmn{1}{pq}\,h\,+\,\fVhmn{2}{pq}\,h^{2}\,+\,O(h^{3})\,,
    \label{eq:fermionic V perturbative expansion}
\end{align}
where the expansion coefficients are given by
\begin{subequations}
\label{eq:fermionic V perturbatively order}
\begin{align}
    \fVhmn{1}{pq}   &=\,-\,\Ghnstar{0}{p}\,\Ahmn{1}{qp}\,,
    \label{eq:fermionic V perturbatively order 1}\\[1mm]
    \fVhmn{2}{pq}   &=\,-\,\Ghnstar{0}{p}\,\Ahmn{2}{qp}\,-\,\Ghnstar{0}{p}\,\Ghn{0}{q}\,\sum\limits_{m\geq0}\,\Ahmn{1}{mp}\,\Ahmnstar{1}{mq}\,.
    \label{eq:fermionic V perturbatively order 2}
\end{align}
\end{subequations}
With this the normalization constant in~(\ref{eq:fermion vac trafo}) is immediately obtained as
\begin{align}
    \tilde{N}_{\text{vac}}    &=\,1\,-\,\tfrac{1}{2}\,\sum\limits_{\substack{p\geq0\\ q<0}}\,|\Ahmn{1}{pq}|^{2}\,h^{2}\,+\,O(h^{4})\,.
    \label{eq:fermion vac normalization}
\end{align}
In the following we will assume that the first (second) index of $\fVmn{pq}$ is always non-negative (negative) unless otherwise stated. The vacuum state can then be straightforwardly expanded in terms of powers of~$h$\,,
\begin{align}
    \fket{0} &=\,
    \fkethat{0}\,+\,h\,\sum\limits_{p,q}\,\fVhmn{1}{pq}\,\bdnhat{p}\,\cdnhat{q}\,\fkethat{0}
    \,+\,h^{2}\,\sum\limits_{p,q}\,\Bigl(\fVhmn{2}{pq}\,\bdnhat{p}\,\cdnhat{q}
    \,-\,\tfrac{1}{2}\,|\Ahmn{1}{pq}|^{2}\label{eq:transformed fermion vac vector}\\[1mm]
    &\ +\,\tfrac{1}{2}\sum\limits_{p\pr\!,q\pr}\,\fVhmn{1}{pq}\,\fVhmn{1}{p\pr q\pr}\ \
    \bdnhat{p}\,\cdnhat{q}\,\bdnhat{p\pr}\,\cdnhat{q\pr}\Bigr)\,\fkethat{0}\,+\,O(h^{3})\,,\nonumber
\end{align}
where we keep in mind the fermionic anticommutation relations~(\ref{eq:anticomm vanishing}) that imply that no second particle or antiparticle can be added to the same mode \textemdash\ the \emph{Pauli exclusion principle}\index{Pauli exclusion principle}. To progress further it is convenient to introduce an additional label in the fermionic Fock states of Section~\ref{sec:fermionic Fock space}. We distinguish excitations of positive (negative) frequency modes by a superscript sign $+$ ($-$) on the double-lined ket notation, i.e.,
\begin{subequations}
\label{eq:ket notation pos and neg frequencies}
\begin{align}
    \bdn{p}\,\fket{0}   &=\,\fketp{1_{p}}\,,
    \label{eq:ket notation pos freq}\\
    \cdn{q}\,\fket{0}   &=\,\fketm{1_{q}}\,,
    \label{eq:ket notation neg freq}
\end{align}
\end{subequations}
and similarly for the co-vectors. With this notation at hand we may rewrite Eq.~(\ref{eq:transformed fermion vac vector}),
\begin{align}
    \hspace*{-2mm}\fket{0} &=\,
    \fkethat{0}\,+\,h\,\sum\limits_{p,q}\,\fVhmn{1}{pq}\,\fkethatkp{1}{p}\!\fkethatkm{1}{q}
    \,+\,h^{2}\,\sum\limits_{p,q}\,\Bigl[\fVhmn{2}{pq}\,\fkethatkp{1}{p}\!\fkethatkm{1}{q}
    \,-\,\tfrac{1}{2}\,|\fVhmn{1}{pq}|^{2}\,\fkethat{0}\nonumber\\[1mm]
    &\hspace*{-5mm} +\,\tfrac{1}{2}\sum\limits_{p\pr\!,q\pr}\,\fVhmn{1}{pq}\,\fVhmn{1}{p\pr q\pr}\,
    (1-\delta_{pp\pr})(1-\delta_{qq\pr})\,
    \fkethatkp{1}{p}\!\fkethatkm{1}{q}\!\fkethatkp{1}{p\pr\!}\!\fkethatkm{1}{q\pr\!}
    \Bigr]\,+\,O(h^{3})\,,
    \label{eq:transformed fermion vac vector new notation}
\end{align}
where we have suppressed the symbol for the anti-symmetrized tensor product (see Section~\ref{sec:fermionic Fock space}). To conclude this section we form the density operator for the transformed vacuum state\index{vacuum!fermions}
\begin{align}
    \fket{0}\!\fbra{0}  &=
    \fkethat{0}\!\fbrahat{0}+h\sum\limits_{p,q}
    \Bigl[\fVhmn{1}{pq}\fkethatkp{1}{p}\!\fkethatkm{1}{q}\!\fbrahat{0}+\mathrm{H.~c.}\Bigr]
    -h^{2}\sum\limits_{p,q}\Bigl[|\fVhmn{1}{pq}|^{2}\fkethat{0}\!\fbrahat{0}
    \nonumber\\[1mm]
    &\hspace*{-1.5cm} -\,\tfrac{1}{2}\,\sum\limits_{p\pr\!,q\pr}\,
    \Bigl(\fVhmn{1}{pq}\,\fVhmn{1}{p\pr q\pr}\,(1-\delta_{pp\pr})(1-\delta_{qq\pr})\,
     \fkethatkp{1}{p}\!\fkethatkm{1}{q}\!\fkethatkp{1}{p\pr\!}\!\fkethatkm{1}{q\pr\!}\!\fbrahat{0}
     \,+\,\mathrm{H.~c.}\Bigr)\nonumber\\[1mm]
    &\hspace*{-1.5cm} -
    \Bigl(\fVhmn{2}{pq}\fkethatkp{1}{p}\!\fkethatkm{1}{q}\!\fbrahat{0}+\mathrm{H.~c.}\Bigr)
    -\sum\limits_{p\pr\!,q\pr}\,
    \fVhmn{1}{pq}\fVhmnstar{1}{p\pr q\pr}
    \fkethatkp{1}{p}\!\fkethatkm{1}{q}\!\fbrahatkm{1}{q\pr\!}\!\fbrahatkp{1}{p\pr\!}\Bigr]+O(h^{3}).
    \label{eq:transformed fermionic vac density matrix}
\end{align}

\subsection{Transformation of Fermionic Particle \& Anti-Particle States}
\label{sec:Transformation of Fermionic Particle and Anti-Particle States}

We continue with the \emph{fermionic Fock states}\index{Fock!states}\index{state!Fock} with particle or antiparticle content, by applying the respective creation operators~$\bdn{n}$ $(n\geq0)$ and~$\cdn{n}$ $(n<0)$ with Bogoliubov decompositions
\begin{subequations}
\label{eq:Dirac creation ops bogo}
\begin{align}
    \bdn{n}  &=\,\sum\limits_{m\geq0}\,\Amn{mn}^{*}\,\bdnhat{m}\,+\,
    \sum\limits_{m<0}\,\Amn{mn}^{*}\,\cnhat{m}\,,
    \label{eq:Dirac bdn Bogo}\\[1mm]
    \cdn{n}  &=\,\sum\limits_{m\geq0}\,\Amn{mn}\,\bnhat{m}\,+\,
    \sum\limits_{m<0}\,\Amn{mn}\,\cdnhat{m}\,.
    \label{eq:Dirac cdn Bogo}
\end{align}
\end{subequations}

\vspace*{2mm}
\subsubsection{Fermionic Single Particle States}

For the single fermion state $\fketp{1_{\kappa}}\,$, an excitation in a mode labelled by~$\kappa\geq0\,$, we apply the operator $\bdn{\kappa}$ of~(\ref{eq:Dirac bdn Bogo}), with coefficients expanded as in ~(\ref{eq:generic travel scenario expansion A}) to the vacuum state~(\ref{eq:transformed fermion vac vector new notation}) and we get
\begin{align}
\fketp{1_{\kappa\!}}   &=\,
    \Ghnstar{0}{\kappa}\fkethatkp{1}{\kappa\!}\,+\,h\Bigl[\sum\limits_{m\geq0}
    \Ahmnstar{1}{m\kappa}\fkethatkp{1}{m\!}+
    \Ghnstar{0}{\kappa}\sum\limits_{p,q}\fVhmn{1}{pq}(1-\delta_{p\kappa})
    \fkethatkp{1}{\kappa\!}\!\fkethatkp{1}{p}\!\fkethatkm{1}{q}\Bigr]\nonumber\\[1mm]
&\hspace*{-1.0cm}+\,h^{2}\,\Bigl\{\sum\limits_{m\geq0}\,
    \Ahmnstar{2}{m\kappa}\fkethatkp{1}{m\!}\,+\,\sum\limits_{p,q}\,\Bigl[\sum\limits_{m\geq0}\,
    \Ahmnstar{1}{m\kappa}\,\fVhmn{1}{pq}\,(1-\delta_{mp})
    \fkethatkp{1}{m\!}\!\fkethatkp{1}{p}\!\fkethatkm{1}{q}\nonumber\\[1mm]
&\hspace*{-1.0cm}
    -\,\Ahmnstar{1}{q\kappa}\,\fVhmn{1}{pq}\,\fkethatkp{1}{p}
    \,+\,\Ghnstar{0}{\kappa}\,\fVhmn{2}{pq}\,(1-\delta_{p\kappa})
    \fkethatkp{1}{\kappa\!}\!\fkethatkp{1}{p}\!\fkethatkm{1}{q}
    \,-\,\tfrac{1}{2}\,\Ghnstar{0}{\kappa}\,|\fVhmn{1}{pq}|^{2}\,\fkethatkp{1}{\kappa\!}
    \nonumber\\[2.5mm]
&\hspace*{-1.0cm}-\,\tfrac{1}{2}\,\Ghnstar{0}{\kappa}\,\sum\limits_{p\pr\!,q\pr}\,
    \fVhmn{1}{pq}\,\fVhmn{1}{p\pr q\pr}\,
    (1-\delta_{pp\pr})(1-\delta_{qq\pr})(1-\delta_{\kappa p})(1-\delta_{\kappa p\pr})
    \nonumber\\
&\hspace*{2cm} \times\,\fkethatkp{1}{\kappa\!}\!\fkethatkp{1}{p}\!\fkethatkp{1}{p\pr\!}
    \!\fkethatkm{1}{q}\!\fkethatkm{1}{q\pr\!}\Bigr]\Bigr\}\,+\,O(h^{3})\,.
    \label{eq:fermion single particle pure bogo}
\end{align}
To leading order the state remains unchanged apart from the phase that is picked up during time evolution. At linear order in~$h$ Eq.~(\ref{eq:fermion single particle pure bogo}) illustrates the role of the $\alpha$ and $\beta$-type coefficients. The block of $\Ahmn{1}{mn}$ where both indices are non-negative shifts the available excitation into a superposition of excitations in positive frequency modes. On the other hand, the block of $\Ahmn{1}{mn}$ that mixes negative and non-negative indices is responsible for terms in the superposition with additional pairs of particles and antiparticles. The appropriate expression, to second order in~$h\,$, for the density operator of the pure state $\fketp{1_{\kappa\!}}$ is given by
\begin{align}
    \fketp{1_{\kappa\!}}\!\fbrap{1_{\kappa\!}}  &=\,\fkethatkp{1}{\kappa\!}\!\fbrahatkp{1}{\kappa\!}\,+\,h\,
    \Bigl[\sum\limits_{p,q}\,\fVhmn{1}{pq}(1-\delta_{p\kappa})
    \fkethatkp{1}{\kappa\!}\!\fkethatkp{1}{p}\!\fkethatkm{1}{q}\!\fbrahatkp{1}{\kappa\!}\nonumber\\[1mm]
&\hspace*{-1.8cm}+\,\Ghn{0}{\kappa}\sum\limits_{m\geq0}\,\Ahmnstar{1}{m\kappa}
    \fkethatkp{1}{m\!}\!\fbrahatkp{1}{\kappa\!}
    \,+\,\mathrm{H.~c.}\Bigr]\,+\,h^{2}\,\Bigl\{\sum\limits_{m,m\pr\geq0}
    \Ahmnstar{1}{m\kappa}\Ahmn{1}{m\pr\!\kappa}\fkethatkp{1}{m\!}\!\fbrahatkp{1}{m\pr\!}\nonumber\\[1mm]
&\hspace*{-1.8cm}+\sum\limits_{\substack{p,q\\ m\geq0}}\!
    \Bigl(\Ghnstar{0}{\kappa}\!\Ahmn{1}{m\kappa}\fVhmn{1}{pq}(1-\delta_{p\kappa})
    \fkethatkp{1}{\kappa\!}\!\fkethatkp{1}{p}\!\fkethatkm{1}{q}\!\fbrahatkp{1}{m}\,+\,\mathrm{H.~c.}\Bigl)
    \,+\!\sum\limits_{p,q,p\pr\!,q\pr}\!\fVhmn{1}{pq}\,\fVhmnstar{1}{p\pr q\pr}\nonumber\\[1mm]
&\hspace*{-1.8cm}\times(1-\delta_{p\kappa})(1-\delta_{p\pr\!\kappa})
    \fkethatkp{1}{\kappa\!}\!\fkethatkp{1}{p}\!\fkethatkm{1}{q}\!
    \fbrahatkm{1}{q\pr\!}\!\fbrahatkp{1}{p\pr\!}\!\fbrahatkp{1}{\kappa\!}
    \,+\,
    \Bigl[\Ghn{0}{\kappa}\sum\limits_{m\geq0}\,
    \Ahmnstar{2}{m\kappa}
    \nonumber\\[1mm]
&\hspace*{-1.8cm}\times\fkethatkp{1}{m\!}\!\fbrahatkp{1}{\kappa\!}\,
    +\,\sum\limits_{p,q}\,\Bigl(\Ghn{0}{\kappa}\sum\limits_{m\geq0}\,
    \Ahmnstar{1}{m\kappa}\,\fVhmn{1}{pq}\,(1-\delta_{mp})
    \fkethatkp{1}{m\!}\!\fkethatkp{1}{p}\!\fkethatkm{1}{q}\!\fbrahatkp{1}{\kappa\!}\nonumber\\[1mm]
&\hspace*{-1.8cm}
    -\,\Ghn{0}{\kappa}\Ahmnstar{1}{q\kappa}\,\fVhmn{1}{pq}\,\fkethatkp{1}{p}\!\fbrahatkp{1}{\kappa\!}
    \,+\,\fVhmn{2}{pq}\,(1-\delta_{p\kappa})
    \fkethatkp{1}{\kappa\!}\!\fkethatkp{1}{p}\!\fkethatkm{1}{q}\!\fbrahatkp{1}{\kappa\!}
    \nonumber\\[3.5mm]
&\hspace*{-1.8cm}-\,\tfrac{1}{2}\,|\fVhmn{1}{pq}|^{2}\,\fkethatkp{1}{\kappa\!}\!\fbrahatkp{1}{\kappa\!}
    -\,\tfrac{1}{2}\,\sum\limits_{p\pr\!,q\pr}\,
    \fVhmn{1}{pq}\,\fVhmn{1}{p\pr q\pr}\,
    (1-\delta_{pp\pr})(1-\delta_{qq\pr})(1-\delta_{\kappa p})(1-\delta_{\kappa p\pr})
    \nonumber\\
&\hspace*{-1.8cm}\times\,\fkethatkp{1}{\kappa\!}\!\fkethatkp{1}{p}\!\fkethatkp{1}{p\pr\!}
    \!\fkethatkm{1}{q}\!\fkethatkm{1}{q\pr\!}\!\fbrahatkp{1}{\kappa\!}\Bigr)\,+\,\mathrm{H.~c.}\Bigr]\Bigr\}
    \,+\,O(h^{3})\,.
    \label{eq:fermion single particle density operator bogo}
\end{align}

\subsubsection{Fermionic Single Anti-Particle States}

For the single anti-fermion state $\fketm{1_{\kappa\pr\!}}$, an excitation in a mode labelled by~$\kappa\pr<0\,$, we apply the operator $\cdn{\kappa\pr}$ of~(\ref{eq:Dirac cdn Bogo}), with coefficients expanded as in ~(\ref{eq:generic travel scenario expansion A}), to the vacuum state~(\ref{eq:transformed fermion vac vector new notation}) to get
\begin{align}
\fketm{1_{\kappa\pr\!}}   &=
    \Ghn{0}{\kappa\pr}\fkethatkm{1}{\kappa\pr\!}+h\Bigl[\sum\limits_{m<0}
    \Ahmn{1}{m\kappa\pr}\fkethatkm{1}{m\!}+\Ghn{0}{\kappa\pr}\sum\limits_{p,q}
    \fVhmn{1}{pq}(1-\delta_{q\kappa\pr})
    \fkethatkp{1}{p}\!\fkethatkm{1}{q}\!\fkethatkm{1}{\kappa\pr\!}\Bigr]\nonumber\\[1mm]
&\hspace*{-1.0cm}+\,h^{2}\,\Bigl\{\sum\limits_{m<0}\,
    \Ahmn{2}{m\kappa\pr}\fkethatkm{1}{m\!}\,+\,\sum\limits_{p,q}\,\Bigl[\sum\limits_{m<0}\,
    \Ahmn{1}{m\kappa\pr}\,\fVhmn{1}{pq}\,(1-\delta_{mq})
    \fkethatkp{1}{p}\!\fkethatkm{1}{q}\!\fkethatkm{1}{m\!}\nonumber\\[1mm]
&\hspace*{-1.0cm}
    +\,\Ahmn{1}{p\kappa\pr}\,\fVhmn{1}{pq}\,\fkethatkm{1}{q}
    \,+\,\Ghn{0}{\kappa\pr}\,\fVhmn{2}{pq}\,(1-\delta_{q\kappa\pr})
    \fkethatkp{1}{p}\!\fkethatkm{1}{q}\!\fkethatkm{1}{\kappa\pr\!}
    \,-\,\tfrac{1}{2}\,\Ghn{0}{\kappa\pr}\,|\fVhmn{1}{pq}|^{2}\,\fkethatkm{1}{\kappa\pr\!}
    \nonumber\\[2.5mm]
&\hspace*{-1.0cm}-\,\tfrac{1}{2}\,\Ghn{0}{\kappa\pr}\,\sum\limits_{p\pr\!,q\pr}\,
    \fVhmn{1}{pq}\,\fVhmn{1}{p\pr q\pr}\,
    (1-\delta_{pp\pr})(1-\delta_{qq\pr})(1-\delta_{\kappa\pr q})(1-\delta_{\kappa\pr q\pr})
    \nonumber\\
&\hspace*{2cm} \times\,\fkethatkp{1}{p}\!\fkethatkp{1}{p\pr\!}\!\fkethatkm{1}{q}\!
    \fkethatkm{1}{q\pr\!}\!\fkethatkm{1}{\kappa\pr\!}\Bigr]\Bigr\}\,+\,O(h^{3})\,.
    \label{eq:fermion single antiparticle pure bogo}
\end{align}
In complete analogy to~(\ref{eq:fermion single particle density operator bogo}) we express the density operator for the pure state~$\fketm{1_{\kappa\pr\!}}$ as
\begin{align}
    \fketm{1_{\kappa\pr\!}}\!\fbram{1_{\kappa\pr\!}}  &=\fkethatkm{1}{\kappa\pr\!}\!\fbrahatkm{1}{\kappa\pr\!}+h
    \Bigl[\sum\limits_{p,q}\fVhmn{1}{pq}(1-\delta_{q\kappa\pr})
    \fkethatkp{1}{p\!}\!\fkethatkm{1}{q\!}\!\fkethatkm{1}{\kappa\pr\!}
    \!\fbrahatkm{1}{\kappa\pr\!}\nonumber\\[1mm]
&\hspace*{-1.8cm}+\,\Ghnstar{0}{\kappa\pr}\sum\limits_{m<0}
    \Ahmn{1}{m\kappa\pr}\fkethatkm{1}{m\!}\!\fbrahatkm{1}{\kappa\pr\!}
    \,+\,\mathrm{H.~c.}\Bigr]\,+\,h^{2}\,\Bigl\{\sum\limits_{m,m\pr<0}
    \Ahmnstar{1}{m\kappa\pr}\Ahmn{1}{m\pr\!\kappa\pr}\fkethatkm{1}{m\!}\!\fbrahatkm{1}{m\pr\!}
    \nonumber\\[1mm]
&\hspace*{-1.8cm}+\,\sum\limits_{\substack{p,q\\ m<0}}\Bigl(\Ghn{0}{\kappa\pr}\Ahmnstar{1}{m\pr\!\kappa\pr}
    \fVhmn{1}{pq}(1-\delta_{mq})
    \fkethatkp{1}{p}\!\fkethatkm{1}{q}\!\fkethatkm{1}{\kappa\pr\!}\!\fbrahatkm{1}{m}\,+\,\mathrm{H.~c.}\Bigl)
    \,+\!\sum\limits_{p,q,p\pr\!,q\pr}\!\fVhmn{1}{pq}\,\fVhmnstar{1}{p\pr q\pr}\nonumber\\[1mm]
&\hspace*{-1.8cm}\times(1-\delta_{q\kappa\pr})(1-\delta_{q\pr\!\kappa\pr})
    \fkethatkp{1}{p}\!\fkethatkm{1}{q}\!\fkethatkm{1}{\kappa\pr\!}\!
    \fbrahatkm{1}{\kappa\pr\!}\!\fbrahatkm{1}{q\pr\!}\!\fbrahatkp{1}{p\pr\!}
    \,+\,
    \Bigl[\Ghnstar{0}{\kappa\pr}\sum\limits_{m<0}\,
    \Ahmn{2}{m\kappa\pr}
    \nonumber\\[1mm]
&\hspace*{-1.8cm}\times\fkethatkm{1}{m\!}\!\fbrahatkm{1}{\kappa\pr\!}\,
    +\,\sum\limits_{p,q}\,\Bigl(\Ghnstar{0}{\kappa\pr}\sum\limits_{m<0}\,
    \Ahmn{1}{m\kappa\pr}\,\fVhmn{1}{pq}\,(1-\delta_{mq})
    \fkethatkp{1}{p}\!\fkethatkm{1}{q}\!\fkethatkm{1}{m\!}\!\fbrahatkm{1}{\kappa\pr\!}\nonumber\\[1.5mm]
&\hspace*{-1.8cm}
    +\,\Ghnstar{0}{\kappa\pr}\Ahmn{1}{p\kappa\pr}\,\fVhmn{1}{pq}\,\fkethatkm{1}{q}\!
    \fbrahatkm{1}{\kappa\pr\!}
    \,+\,\fVhmn{2}{pq}\,(1-\delta_{q\kappa\pr})
    \fkethatkp{1}{p}\!\fkethatkm{1}{q}\!\fkethatkm{1}{\kappa\pr\!}\!\fbrahatkm{1}{\kappa\pr\!}
    \nonumber\\[3.5mm]
&\hspace*{-1.8cm}-\,\tfrac{1}{2}\,|\fVhmn{1}{pq}|^{2}\,\fkethatkm{1}{\kappa\pr\!}\!\fbrahatkm{1}{\kappa\pr\!}
    -\,\tfrac{1}{2}\,\sum\limits_{p\pr\!,q\pr}\,
    \fVhmn{1}{pq}\,\fVhmn{1}{p\pr q\pr}\,
    (1-\delta_{pp\pr})(1-\delta_{qq\pr})(1-\delta_{\kappa\pr q})(1-\delta_{\kappa\pr q\pr})
    \nonumber\\
&\hspace*{-1.8cm}\times\,\fkethatkp{1}{p}\!\fkethatkp{1}{p\pr\!}\!\fkethatkm{1}{q}\!\fkethatkm{1}{q\pr\!}\!
    \fkethatkm{1}{\kappa\pr\!}\!\fbrahatkm{1}{\kappa\pr\!}\Bigr)\,+\,\mathrm{H.~c.}\Bigr]\Bigr\}
    \,+\,O(h^{3})\,.
    \label{eq:fermion single antiparticle density operator bogo}
\end{align}

\subsubsection{Fermionic Particle--Anti-Particle Pair}

As a last example we consider the leading order correction in the transformation of a pair of one particle in mode~$\kappa\geq0$ and one antiparticle in mode~$\kappa\pr<0\,$. By applying either~$\bdn{\kappa}$ to the state of Eq.~(\ref{eq:fermion single antiparticle pure bogo}), or, equivalently, applying~$\cdn{\kappa\pr}$ to~(\ref{eq:fermion single particle pure bogo}) we obtain
\begin{align}
    \fketp{1_{\kappa\!}}\!\fketm{1_{\kappa\pr\!}}   &=\,
    \Ghnstar{0}{\kappa}\Ghn{0}{\kappa\pr}\,\fkethatkp{1}{\kappa\!}\!\fkethatkm{1}{\kappa\pr\!}
    \,+\,h\,\Bigl[\Ghn{0}{\kappa\pr\!}\,\Ahmnstar{1}{\kappa\pr\kappa}\,\fkethat{0}
    \label{eq:fermion particle antiparticle pair pure bogo}\\[2mm]
    &\hspace*{-1cm} +\,\Ghn{0}{\kappa\pr}\,\sum\limits_{m\geq0}\,\Ahmnstar{1}{m\kappa}\,
    \fkethatkp{1}{m\!}\!\fkethatkm{1}{\kappa\pr\!}
    \,+\,\Ghnstar{0}{\kappa}\,\sum\limits_{n<0}\,\Ahmn{1}{n\kappa\pr}\,
    \fkethatkp{1}{\kappa\!}\!\fkethatkm{1}{n}
    \nonumber\\[1mm]
    &\hspace*{-1cm} +\,\Ghnstar{0}{\kappa}\Ghn{0}{\kappa\pr}\,\sum\limits_{p,q}\,\fVhmn{1}{pq}\,
    (1-\delta_{p\kappa})(1-\delta_{q\kappa\pr})
    \fkethatkp{1}{\kappa\!}\!\fkethatkp{1}{p}\!\fkethatkm{1}{q}\!\fkethatkm{1}{\kappa\pr\!}\Bigr]
    \,+\,O(h^{2})\,.
    \nonumber
\end{align}
To linear order in~$h$ the density operator for the out-region Fock space is
\begin{align}
    \fketp{1_{\kappa\!}}\!\fketm{1_{\kappa\pr\!\!}}\!\fbram{1_{\kappa\pr\!\!}}\!\fbrap{1_{\kappa\!}}   &=\,
    \fkethatkp{1}{\kappa\!}\!\fkethatkm{1}{\kappa\pr\!\!}\!\fbrahatkm{1}{\kappa\pr\!\!}\!\fbrahatkp{1}{\kappa\!}
    \,+\,h\,\Bigl[\Ghn{0}{\kappa\!}\Ahmnstar{1}{\kappa\pr\kappa}
    \fkethat{0}\!\!\fbrahatkm{1}{\kappa\pr\!\!}\!\fbrahatkp{1}{\kappa\!}\nonumber\\[2.0mm]
    &\hspace*{-4.35cm} +\,\Ghn{0}{\kappa}\sum\limits_{m\geq0}\Ahmnstar{1}{m\kappa}
    \fkethatkp{1}{m\!}\!\fkethatkm{1}{\kappa\pr\!\!}\!\fbrahatkm{1}{\kappa\pr\!\!}\!\fbrahatkp{1}{\kappa\!}
    \,+\,\Ghnstar{0}{\kappa\pr}\,\sum\limits_{n<0}\,\Ahmn{1}{n\kappa\pr}\,
    \fkethatkp{1}{\kappa\!}\!\fkethatkm{1}{n}\!\fbrahatkm{1}{\kappa\pr\!\!}\!\fbrahatkp{1}{\kappa\!}
    \nonumber\\[1mm]
    &\hspace*{-4.35cm} +\,\sum\limits_{p,q}\,\fVhmn{1}{pq}\,
    (1-\delta_{p\kappa})(1-\delta_{q\kappa\pr})
    \fkethatkp{1}{\kappa\!}\!\fkethatkp{1}{p}\!\fkethatkm{1}{q}\!\fkethatkm{1}{\kappa\pr\!}\!
    \fbrahatkm{1}{\kappa\pr\!\!}\!\fbrahatkp{1}{\kappa\!}
    \,+\,\mathrm{H.~c.}\Bigr]+O(h^{2})\,.
    \label{eq:fermion particle antiparticle pair pure bogo}
\end{align}
We find that the linear Bogoliubov coefficients create coherence between the initial state and those states where either one of the excitations is shifted or an additional particle-antiparticle pair is created.\\

The states we have considered in this chapter illustrate how the Bogoliubov transformations manifest on the Fock space (or phase space) of all modes of a single cavity. It has been verified that when terms proportional to~$h^{2}$ are kept the transformed states are normalized, Hermitean and have a non-negative spectrum to second order in the perturbative expansion (see also the discussion in Section~\ref{sec:Bosonic Vacuum}). The transformations are hence \emph{implemented unitarily} for smoothly changing accelerations in any dimension, or for sharply varying accelerations in up to $(2+1)$ dimensions. In this sense the transformations are \emph{global unitaries} on the Fock space of the cavity modes. We shall see in Chapter~\ref{Chapter 6 Motion Generates Entanglement} how the transformations affect the entanglement between chosen sets of these modes. 
    \newpage\ \\
    \chapter{Motion Generates Entanglement}\label{Chapter 6 Motion Generates Entanglement}

In this chapter we discuss the structure of the entanglement that is generated between the modes of quantum fields that are confined to cavities in non-uniform motion. The motivation for this analysis lies in the prospect of employing the specific structure of the created quantum correlations in the verification of the \emph{genuine quantumness} of particle creation phenomena in quantum field theory. In other words, particle creation phenomena and related transformations in the Fock space that occur due to the motion of the cavity boundaries are not arbitrary. They have a rich structure that may allow to unambiguously identify the source of the created particles as a quantum field theory effect, see, e.g., Ref.~\cite{WilsonDynCasNature2012}. 
In addition, via monogamy arguments (see pp.~\pageref{page:monogamy of entanglement}) the insights into the entanglement generation within a single cavity will be useful in determining the source of entanglement degradation effects when several entangled cavities are considered in Chapter~\ref{Chapter 7 Degradation of Entanglement between Moving Cavities}.\\

This far we have established the Bogoliubov transformations for cavity modes of quantum fields in a cavity that undergoes a change in motion from an inertial in-region to an inertial out-region (see Chapter~\ref{Chapter 4 Constructing Non Uniformly Moving Cavities}). Consecutively, we have shown in Chapter~\ref{Chapter 5 State Transformation by Non-Uniform Motion} how initial quantum states in their Fock space or phase space representations are transformed to the out-region. In this chapter we are going to draw from material published in Refs.~\cite[(\ref{Paper:FriisBruschiLoukoFuentes2012}-\ref{Paper:FriisFuentes2013})]
{FriisBruschiLoukoFuentes2012,FriisFuentes2013,FriisHuberFuentesBruschi2012}: Based on the transformed states presented in Chapter~\ref{Chapter 5 State Transformation by Non-Uniform Motion} we are going to study the reduced states of two modes and the \emph{entanglement} generated therein. As previously the treatment of bosonic Fock states and Gaussian states is separated into Sections~\ref{sec:Entanglement Generation in Bosonic Fock States} and~\ref{sec:Entanglement Generation in Bosonic Gaussian States}, respectively, before we turn the attention to fermionic states in Section~\ref{sec:Entanglement Generation in Fermionic States}. The chapter is concluded in Section~\ref{sec:Generation of Genuine Multipartite Entanglement} by an analysis of the structure of multipartite entanglement within the non-uniformly moving cavity
~\cite[(\ref{Paper:FriisHuberFuentesBruschi2012})]{FriisHuberFuentesBruschi2012}.

\section{Entanglement Generation in Bosonic Fock States}\label{sec:Entanglement Generation in Bosonic Fock States}

\subsection{Bosonic Vacuum}\label{sec:Bosonic Vacuum}\index{vacuum!bosons|(}

A convenient starting point for the analysis of the entanglement in the transformed states is the vacuum state. The out-region decomposition of the in-region vacuum is given by the density matrix~(\ref{eq:transformed boson vac density matrix}). We are here interested in the bipartite entanglement between a chosen pair of modes labelled by~$k$ and $k\pr\,$, respectively. To obtain the reduced out-region state for these two modes we trace out all other modes from~(\ref{eq:transformed boson vac density matrix}). In contrast to the fermionic case (see Section~\ref{sec:the fermionic ambiguity}) this can be done in a straightforward way since mapping bosonic modes to a tensor product space
is free of ambiguities. For the purpose of partial traces we can assume a tensor product of mode subspaces and write the resulting reduced state again as a density operator on a Fock space with an appropriately reduced number of modes. Taking into account the structure of the Bogoliubov coefficients~(\ref{eq:generic travel scenario expansion}) some lengthy but straightforward algebra reveals
\begin{align}
    {}_{\mathrm{vac}}\rho_{\raisebox{-1pt}{\tiny{$kk\pr$}}}   &:=\,\tr_{\lnot k,k\pr}\bigl(\ket{0}\!\bra{0}\bigr) \,=\,\kethat{0}\!\brahat{0}\,-\,h\,
    \Bigl[\Ghn{0}{k\pr}\betahmn{1}{kk\pr}\kethat{0}\!\brahatk{1}{k\pr}\!\brahatk{1}{k}\,+\,\mathrm{H.~c.}\Bigr]
    \label{eq:bosonic vac bogo traced k and kpr}\\[1.5mm]
    &\hspace*{-1.0cm} +\,h^{2}\,\Bigl[\bigl(-2f^{\beta}_{k\lnot k\pr}\,-\,2f^{\beta}_{k\pr}\bigr)\kethat{0}\!\brahat{0}\,+\,
    2f^{\beta}_{k\lnot k\pr}\kethatk{1}{k}\!\brahatk{1}{k}\,+\,2f^{\beta}_{k\pr\lnot k}\kethatk{1}{k\pr}\!\brahatk{1}{k\pr}\nonumber\\[1.5mm]
    &\hspace*{-1.0cm} +\,|\betahmn{1}{kk\pr}|^{2}\kethatk{1}{k}\!\kethatk{1}{k\pr}\!\brahatk{1}{k\pr}\!\brahatk{1}{k}\,+\,\Bigl(
    \Vhmnstar{2}{kk\pr}\kethat{0}\!\brahatk{1}{k\pr}\!\brahatk{1}{k}\,+\,
    \tfrac{1}{\sqrt{2}}\Vhmnstar{2}{kk}\kethat{0}\!\brahatk{2}{k}\nonumber\\[1.5mm]
    &\hspace*{-1.0cm} +\,
    \tfrac{1}{\sqrt{2}}\Vhmnstar{2}{k\pr\!k\pr}\kethat{0}\!\brahatk{2}{k\pr}\,+\,
    (\Vhmnstar{1}{kk\pr})^{2}\kethat{0}\!\brahatk{2}{k\pr}\!\brahatk{2}{k}\,+\,\mathrm{H.~c.}\Bigr)\Bigr]\,+\,O(h^{3})\,,
    \nonumber
\end{align}
where $\tr_{\lnot k,k\pr}$ denotes the trace over all modes except $k$ and $k\pr$, $\Vhmn{1}{pq}$ and $\Vhmn{2}{pq}$ are given by~(\ref{eq:bosonic V expansion 1}) and~(\ref{eq:bosonic V expansion 2}), respectively, and we have introduced the abbreviation
\begin{align}
    f^{\beta}_{m\lnot\hspace*{1pt} n} &=\,\tfrac{1}{2}\,\sum\limits_{i\neq n}\,|\betahmn{1}{mi}|^{2}\,.
    \label{eq:f beta m not n}
\end{align}
Cutting off the power expansion at second order effectively truncates each of the two modes in~(\ref{eq:bosonic vac bogo traced k and kpr}) to a three dimensional system. Hence, the reduced state can be represented on the tensor product space of two qutrits by the matrix
\begin{align}
    {}_{\mathrm{vac}}\rho_{\raisebox{-1pt}{\tiny{$kk\pr$}}}    &=
    \nonumber\\[1mm]
    &\hspace*{-1cm}\nonumber
\begin{scriptsize}
\begin{pmatrix}
1-2h^{2}\bigl(f^{\beta}_{k\lnot k\pr}+f^{\beta}_{k\pr}\bigr)    &  0  &
    \hspace*{-1.5mm}\tfrac{h^{2}}{\sqrt{2}}V^{(2)}_{k\pr\!k\pr}  &  0  &
    \hspace*{-4mm}h\,G^{(0)*}_{k\pr}\beta^{(1)*}_{kk\pr}+h^{2}V^{(2)}_{kk\pr}  &  0  &
    \tfrac{h^{2}}{\sqrt{2}}V^{(2)}_{kk}  &  0  &  h^{2}(V^{(1)}_{kk\pr})^{2} \\
0 &  \hspace*{-1.5mm}2h^{2}f^{\beta}_{k\pr\lnot k}  &  0  &  0  &  0  &  0  &  0  &  0  &  0 \\
\tfrac{h^{2}}{\sqrt{2}}V^{(2)*}_{k\pr\!k\pr} &  0  &  0  &  0  &  0  &  0  &  0  &  0  &  0 \\
0 &  0  &  0  &  \hspace*{-1.5mm}2h^{2}f^{\beta}_{k\lnot k\pr}  &  0  &  0  &  0  &  0  &  0 \\
h\,G^{(0)}_{k\pr}\beta^{(1)}_{kk\pr}+h^{2}V^{(2)*}_{kk\pr} &  0  &  0  &  0  &
    h^{2}|\beta^{(1)}_{kk\pr}|^{2}  &  0  &  0  &  0  &  0 \\
0 &  0  &  0  &  0  &  0  &  0  &  0  &  0  &  0 \\
\tfrac{h^{2}}{\sqrt{2}}V^{(2)*}_{kk} &  0  &  0  &  0  &  0  &  0  &  0  &  0  &  0 \\
0 &  0  &  0  &  0  &  0  &  0  &  0  &  0  &  0 \\
h^{2}(V^{(1)*}_{kk\pr})^{2} &  0  &  0  &  0  &  0  &  0  &  0  &  0  &  0
\end{pmatrix},\end{scriptsize}\\[1mm]
&   \label{eq:bosonic vac bogo traced k and kpr matrix}
\end{align}
where we have neglected terms of order~$h^{3}\,$. The Hilbert-Schmidt condition~(\ref{eq:unitarity condition bosons leading order}) ensures that the quantities $f^{\beta}_{m\lnot\hspace*{1pt}n}$ are finite and, consequently, the state in Eq.~(\ref{eq:bosonic vac bogo traced k and kpr}) is a well defined density operator. One can easily see that it is normalized and Hermitean. Working out the eigenvalues is somewhat more complicated and needs a more detailed examination.

\subsubsection{Perturbative Diagonalization}\label{page:pert diagonalization}

To determine the eigenvalues of the density operator perturbatively one has to be careful. In typical perturbation theory approaches it is assumed that a given matrix is perturbed by a term that is linear in the expansion parameter. However, in our case the density operator is expanded as
\begin{align}
     \rho   &=\,\rhoh{0}\,+\,h\,\rhoh{1}\,+\,h^{2}\,\rhoh{2}\,+\,O(h^{3})\,,
     \label{eq:density matrix expansion}
\end{align}
and we wish to obtain the leading order corrections to the eigenvalues~$\lambdah{0}_{i}$ of the unperturbed matrix~$\rhoh{0}\,$. To begin one may approximate the perturbative corrections to $\rhoh{0}$ by only the linear term~$h\rhoh{1}\,$. In that case corrections to the eigenvalues of $\rhoh{0}$ that are linear in~$h$ can be computed using the standard procedure: first the unperturbed matrix is diagonalized, i.e.,
\begin{align}
    \rhoh{0}\,\ket{\lambdah{0}_{i}} &=\,\lambdah{0}_{i}\,\ket{\lambdah{0}_{i}}\,.
    \label{eq:unperturbed diagonalization}
\end{align}
Subsequently, any degeneracies in the unperturbed eigenvalues need to be taken into account. For any non-degenerate eigenvalues~$\lambdah{0}_{i}$ the corrections are computed as the expectation values of the leading order perturbation~$\rhoh{1}$ in the corresponding unperturbed eigenstate~$\ket{\lambdah{0}_{i}}$, i.e.,
\begin{align}
    \lambdah{1}_{i} &=\,\bra{\lambdah{0}_{i}}\,\rhoh{1}\,\ket{\lambdah{0}_{i}}\,,
    \label{eq:leading order corrections to eig values general}
\end{align}
and one arrives at $\lambda_{i}=\lambdah{0}_{i}+h\lambdah{1}_{i}+O(h^{2})$. For any degenerate eigenvalues $\lambdah{0}_{i_{1}}=\lambdah{0}_{i_{2}}=\ldots=\lambdah{0}_{i_{n}}$ the appropriate leading order corrections are given by the eigenvalues of the matrix $\Lambda_{i}$ with components
\begin{align}
    (\Lambda_{i})_{mn}  &=\,\bra{\lambdah{0}_{i_{m}}}\,\rhoh{1}\,\ket{\lambdah{0}_{i_{n}}}\,.
    \label{eq:degenerate eig values leading corrections}
\end{align}
Similarly, if the linear corrections to the density matrix vanish, i.e., $\rhoh{1}=0\,$, one may perform the same procedure for $\rhoh{2}$ to obtain leading order corrections that are quadratic in~$h\,$. We shall use this procedure throughout Chapters~\ref{Chapter 6 Motion Generates Entanglement} and~\ref{Chapter 7 Degradation of Entanglement between Moving Cavities} to compute the eigenvalues of partially transposed density operators.\\

Nonetheless, for finding the corrections to the eigenvalues of the density matrix itself this strategy is not successful. The density matrix of Eq.~(\ref{eq:bosonic vac bogo traced k and kpr}) has non-zero linear corrections~$\rhoh{1}$, but no linear corrections to the unperturbed eigenvalues $\{\lambdah{0}_{1}=1,\lambdah{0}_{2}=0,\lambdah{0}_{3}=0,\ldots\}\,$, i.e., $\lambdah{1}_{i}=0\,\forall i\neq1\,$. The leading order corrections~$\lambdah{2}$ must therefore be found by diagonalizing~(\ref{eq:bosonic vac bogo traced k and kpr}) by hand. In other words, to determine the diagonal form~$\rho_{\mathrm{diag}}=\diag\{\lambda_{i}|i=1,2,\ldots\}$ of the density matrix we make the ansatz
\begin{align}
    U   &=\,\Uh{0}\,+\,h\,\Uh{1}\,+\,h^{2}\,\Uh{2}\,+\,O(h^{3})\,
    \label{eq:ansatz for diagonalizing unitary}
\end{align}
for the \emph{diagonalizing unitary}, such that $\rho_{\mathrm{diag}}=U\rho\,U^{\dagger}\,$. Next, we switch to a more compact notation for the density matrix~(\ref{eq:bosonic vac bogo traced k and kpr matrix}) by splitting the matrix into the subspaces corresponding to the unperturbed eigenvalues~$\lambdah{0}_{1}=1$ and $\lambdah{0}_{i\neq1}=0\,$, i.e., we write
\begin{align}
    {}_{\mathrm{vac}}\rho_{\raisebox{-1pt}{\tiny{$kk\pr$}}}    &=
    \begin{pmatrix}
    \ 1   &   0\    \\  \ 0   &   0\
    \end{pmatrix}\,+\,
    h\,\begin{pmatrix}
    \,0   &   \vhsymb{1}{\dagger}\,   \\  \,\vhsymb{1}{\!}   &   0\,
    \end{pmatrix}\,+\,
    h^{2}\,\begin{pmatrix}
    \,\rhoh{2}_{11}   &   \vhsymb{2}{\dagger}   \\  \,\vhsymb{2}{\!}   &   \rhohsymb{2}{\prime}
    \end{pmatrix}\,+\,O(h^{3})\,,
    \label{eq:diagonalization compact notation density matrix}
\end{align}
where $\rhoh{2}_{11}=-2(f^{\beta}_{k\lnot k\pr}+f^{\beta}_{k\pr})$ and the components of the vectors~$\vhsymb{1}{\!\!}$, and $\vhsymb{2}{\!\!}$, and the $6\times6$ matrix $\rhohsymb{2}{\prime}$ can be read off directly from ${}_{\mathrm{vac}}\rho_{\raisebox{-1pt}{\tiny{$kk\pr$}}}$ in Eq.~(\ref{eq:bosonic vac bogo traced k and kpr matrix}). A straightforward computation provides the expansion of the diagonalizing unitary
\begin{align}
    U    &=
    \begin{pmatrix}
    \ 1   &   0\    \\  \ 0   &   \idN{6}\
    \end{pmatrix}\,+\,
    h\,\begin{pmatrix}
    \,0   &   \vhsymb{1}{\dagger}\,   \\  -\vhsymb{1}{\!}   &   0\,
    \end{pmatrix}\,+\,
    h^{2}\,\begin{pmatrix}
    -\tfrac{1}{2}\vhsymb{1}{\dagger}\vhsymb{1}{\!}   &   \vhsymb{2}{\dagger}   \\
    -\vhsymb{2}{\!}   &   -\tfrac{1}{2}\vhsymb{1}{\!}\vhsymb{1}{\dagger}
    \end{pmatrix}\,+\,O(h^{3})\,,
    \label{eq:diagonalizing unitary expansion compact}
\end{align}
and the second order corrections to the eigenvalues of~(\ref{eq:bosonic vac bogo traced k and kpr}). Including second order terms the non-zero eigenvalues turn out to be
\begin{subequations}
\label{eq:nonzero eig values transformed vac bosons}
\begin{align}
    \lambda_{1} &=\,1-\,h^{2}\,\bigl(2\,f^{\beta}_{k\lnot k\pr}\,+\,2\,f^{\beta}_{k\pr\lnot k}\bigr)\,+\,O(h^{3})\,,
    \label{eq:nonzero eig values transformed vac bosons 1}\\
    \lambda_{2} &=\,h^{2}\,2\,f^{\beta}_{k\lnot k\pr}\,+\,O(h^{3})\,,
    \label{eq:nonzero eig values transformed vac bosons 2}\\[1mm]
    \lambda_{3} &=\,h^{2}\,2\,f^{\beta}_{k\pr\lnot k}\,+\,O(h^{3})\,.
    \label{eq:nonzero eig values transformed vac bosons 3}
\end{align}
\end{subequations}

\subsubsection{Entanglement Generation in the Bosonic Vacuum}

\begin{figure}[hb!]
\centering
(a)\includegraphics[width=0.74\textwidth]{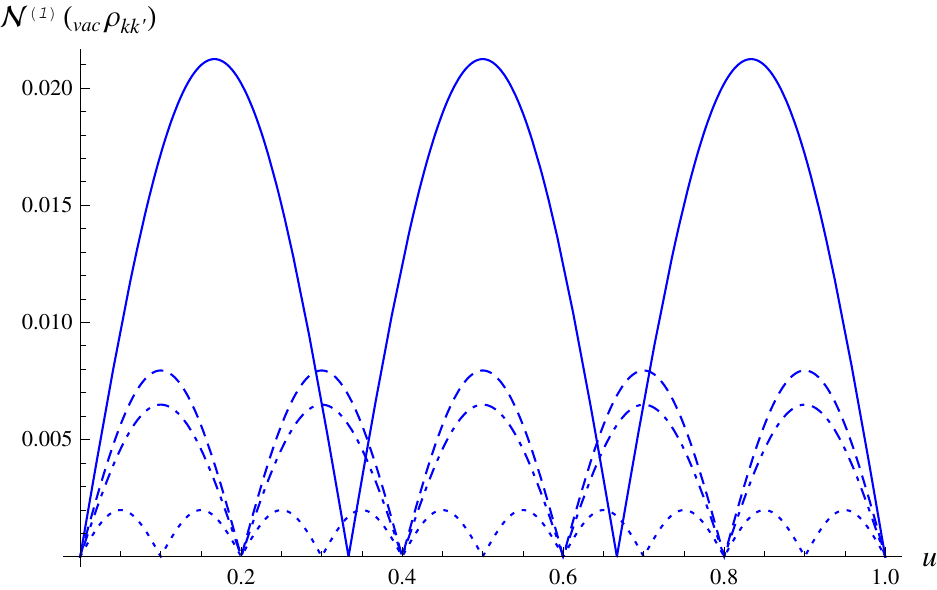}
(b)\includegraphics[width=0.74\textwidth]{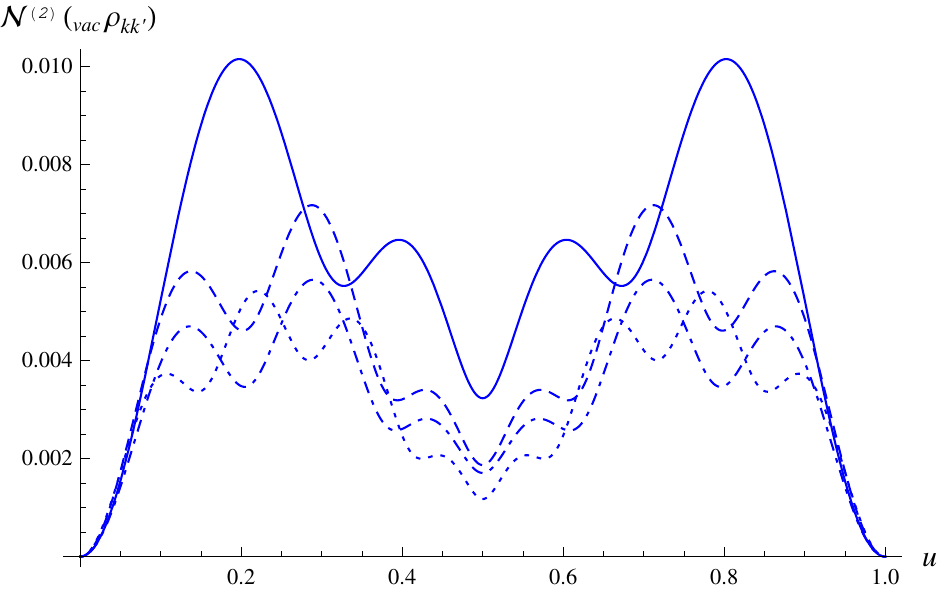}
\caption{
\textbf{Entanglement generation \textemdash\ bosonic vacuum:} The coefficients $\Negh{1}$ [see Eq.~(\ref{eq:linear negativity correction boson vac})] and $\Negh{2}$ [see Eq.~(\ref{eq:quadratic negativity correction boson vac})] of the negativity generated from the vacuum state are plotted in Fig.~\ref{fig:entanglement generation bosons vacuum}~(a) and Fig.~\ref{fig:entanglement generation bosons vacuum}~(b), respectively, for the basic building block travel scenario of Section~\ref{sec:basic building block}. For the $(1+1)$ dimensional massless scalar field used in this illustration the Bogoliubov coefficients are periodic in the dimensionless parameter $u:=h\tau/[4L\artanh(h/2)]\,$ [see Eq.~(\ref{eq:cavity centre proper freq})], where $\tau$ is the duration, as measured at the centre of the cavity, of the single segment of uniform acceleration. Curves are shown for the mode pairs $(k,k\pr)=(1,2)$ (solid), $(2,3)$ (dashed), $(3,4)$ (dotted), and $(1,4)$ (dotted-dashed) in Fig.~\ref{fig:entanglement generation bosons vacuum}~(a), and for $(1,3)$ (solid), $(2,4)$ (dashed), $(3,5)$ (dotted), and $(1,5)$ (dotted-dashed) in Fig.~\ref{fig:entanglement generation bosons vacuum}~(b).
\label{fig:entanglement generation bosons vacuum}}
\end{figure}

From Eqs.~(\ref{eq:nonzero eig values transformed vac bosons}) we clearly see that the eigenvalues of~(\ref{eq:bosonic vac bogo traced k and kpr}) are non-negative and well-defined, at least up to and including second order corrections. However, the partial trace leaves the transformed state \emph{mixed}, which can be quickly verified by computing the linear entropy~$S_{L}(\rho)$\index{entropy!linear} (see Definition~\ref{def:linear entropy}),
\begin{align}
    S_{L}\Bigl(\tr_{\lnot k,k\pr}\bigl(\ket{0}\!\bra{0}\bigr)\Bigr) &=\,
    h^{2}\,\bigl(2\,f^{\beta}_{k\lnot k\pr}\,+\,2\,f^{\beta}_{k\pr\lnot k}\bigr)\,+\,O(h^{3})\,.
    \label{eq:linear entropy transformed vac two modes}
\end{align}
To determine the entanglement between the modes~$k$ and~$k\pr$ we need to employ a measure that is computable for a mixed state of two qutrits. The \emph{negativity}\index{negativity} from Definition~\ref{def:negativity} provides such a tool, although one might miss bound entanglement. To calculate the negativity we determine the eigenvalues of the partially transposed density matrix. For the state~(\ref{eq:bosonic vac bogo traced k and kpr}) it is given by
\begin{align}
    ({}_{\mathrm{vac}}\rho_{\raisebox{-1pt}{\tiny{$kk\pr$}}})^{\,T\subtiny{-1}{-1}{k\pr}}    &=
    \nonumber\\[1mm]
    &\hspace*{-2cm}\nonumber
\begin{scriptsize}
\begin{pmatrix}
1-2h^{2}\bigl(f^{\beta}_{k\lnot k\pr}+f^{\beta}_{k\pr}\bigr)    &  \hspace*{-2mm}0  &
    \hspace*{-2mm}\tfrac{h^{2}}{\sqrt{2}}V^{(2)*}_{k\pr\!k\pr}  &  0  &
    \hspace*{-2mm}0  &  0  &
    \tfrac{h^{2}}{\sqrt{2}}V^{(2)}_{kk}  &  0  &  0 \\
0 &  \hspace*{-2mm}2h^{2}f^{\beta}_{k\pr\lnot k}  &  0  &
    \hspace*{-4mm}h\,G^{(0)*}_{k\pr}\beta^{(1)*}_{kk\pr}+h^{2}V^{(2)}_{kk\pr}  &  \hspace*{-2mm}0  &
    0  &  0  &  0  &  0 \\
\tfrac{h^{2}}{\sqrt{2}}V^{(2)}_{k\pr\!k\pr} &  \hspace*{-2mm}0  &  0  &  0  &  \hspace*{-2mm}0  &
    0  &  h^{2}(V^{(1)}_{kk\pr})^{2}  & 0  &  0 \\
0 &  \hspace*{-4mm}h\,G^{(0)}_{k\pr}\beta^{(1)}_{kk\pr}+h^{2}V^{(2)*}_{kk\pr}  &  0  &
    \hspace*{-1.5mm}2h^{2}f^{\beta}_{k\lnot k\pr}  &  \hspace*{-2mm}0  &  0  &  0  &  0  &  0 \\
0 &  \hspace*{-2mm}0  &  0  &  0  &
    \hspace*{-4mm}h^{2}|\beta^{(1)}_{kk\pr}|^{2}\hspace*{-1.5mm}  &  0  &  0  &  0  &  0 \\
0 &  \hspace*{-2mm}0  &  0  &  0  &  \hspace*{-2mm}0  &  0  &  0  &  0  &  0 \\
\tfrac{h^{2}}{\sqrt{2}}V^{(2)*}_{kk} &  \hspace*{-2mm}0  &
    \hspace*{-4mm}h^{2}(V^{(1)*}_{kk\pr})^{2}  &  0  & \hspace*{-2mm}0  &  0  &  0  &  0  &  0 \\
0 &  \hspace*{-2mm}0  &  0  &  0  &  \hspace*{-2mm}0  &  0  &  0  &  0  &  0 \\
0 &  \hspace*{-2mm}0  &  0  &  0  &  \hspace*{-2mm}0  &  0  &  0  &  0  &  0
\end{pmatrix}.\end{scriptsize}\\[1mm]
&   \label{eq:bosonic vac bogo traced k and kpr matrix par transposed}
\end{align}
We can proceed as laid out on page~\pageref{page:pert diagonalization} to determine the eigenvalues of the partial transpose\index{partial transposition}. If the modes~$k$ and~$k\pr$ have opposite parity, that is, if $(k+k\pr)$ is odd, the linear corrections to~(\ref{eq:bosonic vac bogo traced k and kpr matrix par transposed}) are non-zero , see Eq.~(\ref{eq:bosonic Bogo coeffs linear beta}). Keeping only the linear corrections it becomes evident that the eigenvalues of the matrix
\begin{align}
    h\,\begin{pmatrix}
    0  &  \Ghnstar{0}{k\pr}\betahmnstar{1}{kk\pr}    \\
    \Ghn{0}{k\pr}\betahmn{1}{kk\pr}    &   0
    \end{pmatrix}\,
    \label{eq:linear corr to part trans boson vac}
\end{align}
provide the corrections $\pm h|\beta^{(1)}_{kk\pr}|$ to the degenerate unperturbed eigenvalue~$0\,$. We thus find the linear contribution $\Negh{1}$ to the negativity
\begin{align}
    \mathcal{N}({}_{\mathrm{vac}}\rho_{\raisebox{-1pt}{\tiny{$kk\pr$}}})   &=\,
    h\,\Negh{1}({}_{\mathrm{vac}}\rho_{\raisebox{-1pt}{\tiny{$kk\pr$}}})\,+\,O(h^{2})\,=\,
    h\,|\betahmn{1}{kk\pr}|\,+\,O(h^{2})\,.
    \label{eq:linear negativity correction boson vac}
\end{align}
We see that, to linear order in~$h$, the entanglement as measured by the negativity~(\ref{eq:linear negativity correction boson vac}) is generated by the coherent excitation of two particles in the modes~$k$ and~$k\pr$. Now let us turn to the case where $(k+k\pr)$ is even, i.e., the two modes have the same parity. Then~$\betahmn{1}{kk\pr}=0$ and, consequently, also~$\Vhmn{1}{kk\pr}=0$, see Eq.~(\ref{eq:bosonic V expansion 1}). This leaves a non-zero $2\times2$ block of the second order corrections to the partially transposed state~(\ref{eq:bosonic vac bogo traced k and kpr matrix par transposed}) in the subspace of the degenerate eigenvalue~$0$, given by
\begin{align}
    h^{2}\,\begin{pmatrix}
    2\,f^{\beta}_{k\pr\lnot k}  &  \Vhmn{2}{kk\pr}    \\
    \Vhmnstar{2}{kk\pr} &   2\,f^{\beta}_{k\lnot k\pr}
    \end{pmatrix}\,.
    \label{eq:quadr corr to part trans boson vac}
\end{align}
There is only one possibly negative correction to the eigenvalue~$0$ and the leading order correction to the negativity is then simply found to be
\begin{align}
    \mathcal{N}({}_{\mathrm{vac}}\rho_{\raisebox{-1pt}{\tiny{$kk\pr$}}})   &=\,
    h^{2}\,\Negh{2}({}_{\mathrm{vac}}\rho_{\raisebox{-1pt}{\tiny{$kk\pr$}}})\,+\,O(h^{3})
    \label{eq:quadratic negativity correction boson vac}\\[1mm]
    &\ =\,h^{2}\max\Bigl\{0,
    \sqrt{\bigl(f^{\beta}_{k\lnot k\pr}\,-\,f^{\beta}_{k\pr\lnot k}\bigr)^{2}\,+\,|\Vhmn{2}{kk\pr}|^{2}}
    \,-\,\bigl(f^{\beta}_{k\lnot k\pr}\,+\,f^{\beta}_{k\pr\lnot k}\bigr)\Bigr\}\,+\,O(h^{3})\,.
    \nonumber
\end{align}
The entanglement is now generated by the coefficient~$\Vhmn{2}{kk\pr}\,$ from~(\ref{eq:bosonic V expansion 2}), which has contributions from $\betahmn{2}{kk\pr}$ \textemdash\ pairs of particles that are created directly in the modes~$k$ and~$k\pr$ \textemdash\ and products~$\betahmnstar{1}{mk}\alphahmn{1}{mk\pr}$ \textemdash\ a pair of particles is created in modes~$k$ and~$m$ and the excitation in~$m$ is subsequently shifted to~$k\pr$. These terms compete with the (anti)particle creation where only one constituent of the created pair is generated in~$k$ or~$k\pr$. An illustration of the corrections is shown in Fig.~\ref{fig:entanglement generation bosons vacuum}.
\vspace*{-2mm}
\index{vacuum!bosons|)}
\clearpage
\subsection{Bosonic Single Particle States}\label{sec:Bosonic Single Particle States}

\begin{figure}[ht!]
\centering
(a)\includegraphics[width=0.75\textwidth]{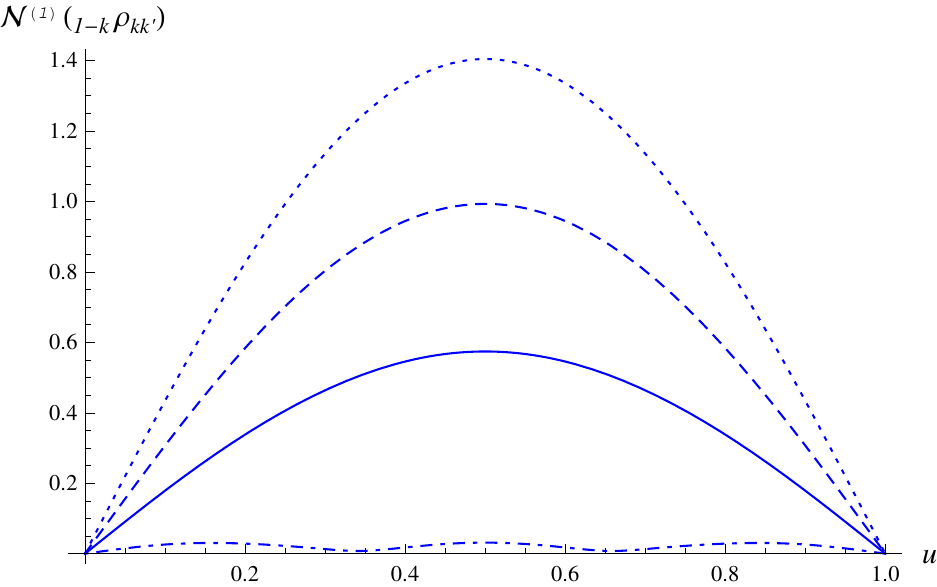}
(b)\includegraphics[width=0.75\textwidth]{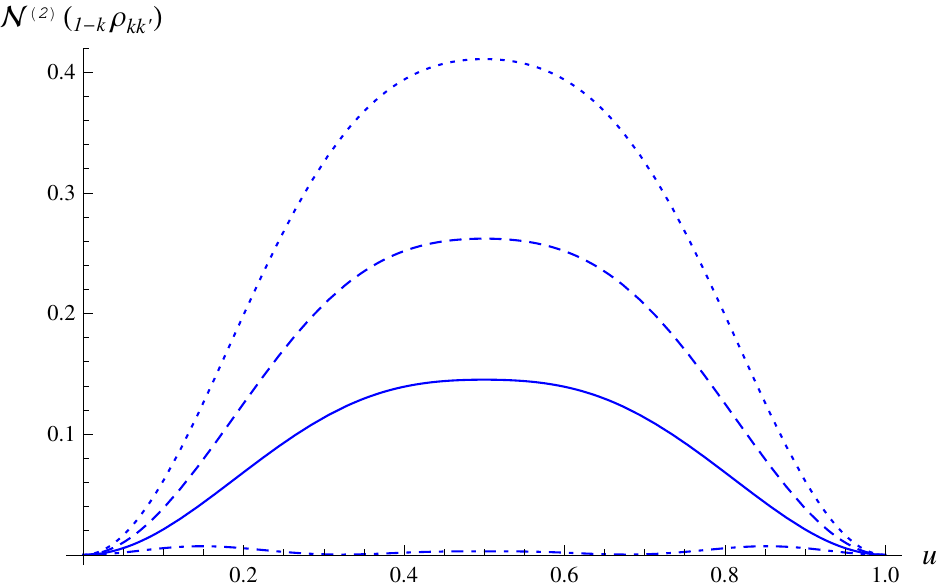}
\caption{
\textbf{Entanglement generation \textemdash\ bosonic single particle state $\ket{1_{k}}$:} The coefficients $\Negh{1}$ [see Eq.~(\ref{eq:linear negativity correction boson 1k})] and $\Negh{2}$ [see Eq.~(\ref{eq:quadratic negativity correction boson 1k})] of the negativity generated from the state~$\ket{1_{k}}$ are plotted in Fig.~\ref{fig:entanglement generation bosons 1k}~(a) and Fig.~\ref{fig:entanglement generation bosons 1k}~(b), respectively, for the basic building block travel scenario of Section~\ref{sec:basic building block}. For the $(1+1)$ dimensional massless scalar field used in this illustration the Bogoliubov coefficients are periodic in the dimensionless parameter $u:=h\tau/[4L\artanh(h/2)]\,$ [see Eq.~(\ref{eq:cavity centre proper freq})], where $\tau$ is the duration, as measured at the centre of the cavity, of the single segment of uniform acceleration. Curves are shown for the mode pairs $(k,k\pr)=(1,2)$ (solid), $(2,3)$ (dashed), $(3,4)$ (dotted), and $(1,4)$ (dotted-dashed) in Fig.~\ref{fig:entanglement generation bosons 1k}~(a), and for $(1,3)$ (solid), $(2,4)$ (dashed), $(3,5)$ (dotted), and $(1,5)$ (dotted-dashed) in Fig.~\ref{fig:entanglement generation bosons 1k}~(b).
\label{fig:entanglement generation bosons 1k}}
\end{figure}

For the bosonic single particle state~$\ket{1_{k}}$ we select the density operator from~(\ref{eq:transformed boson 1k density matrix}) and again trace over all modes except~$k$ and~$k\pr$ to obtain the reduced state
\begin{align}
    {}_{\operatorname{1-k}}\rho_{\raisebox{-1pt}{\tiny{$kk\pr$}}}   &:=\,
        \tr_{\lnot k,k\pr}\bigl(\ket{1_{k}}\!\bra{1_{k}}\bigr) \,=\,
        \kethatk{1}{k}\!\brahatk{1}{k}\,-\,h\,\Bigl[
        \sqrt{2}\,\Ghn{0}{k}\betahmn{1}{k\pr k}\kethatk{1}{k}\!\brahatk{1}{k\pr\!}\!\brahatk{2}{k}
        \label{eq:bosonic 1k bogo traced k and kpr}\\[2mm]
    &\hspace*{-1.0cm} -\,
        \Ghn{0}{k}\alphahmnstar{1}{k\pr k}\kethatk{1}{k\pr\!}\!\brahatk{1}{k}
        \,+\,\mathrm{H.~c.}\Bigr]\,+\,h^{2}\,\Bigl\{2\,f^{\alpha}_{k\lnot k\pr}\kethat{0}\!\brahat{0}\,+\,|\alphahmn{1}{kk\pr}|^{2}\kethatk{1}{k\pr\!}\!\brahatk{1}{k\pr\!}
        \nonumber\\[2mm]
    &\hspace*{-1.0cm} -\,2\,\bigl(f^{\alpha}_{k}+f^{\beta}_{k\pr\lnot k}+2f^{\beta}_{k}\bigr)
        \kethatk{1}{k}\!\brahatk{1}{k}+2\,f^{\beta}_{k\pr\lnot k}\kethatk{1}{k}\!\kethatk{1}{k\pr}\!
        \brahatk{1}{k\pr\!}\!\brahatk{1}{k}+4\,f^{\beta}_{k\lnot k\pr}\kethatk{2}{k}\!\brahatk{2}{k}
        \nonumber\\[2mm]
    &\hspace*{-1.0cm} +\,2\,|\betahmn{1}{kk\pr}|^{2}\kethatk{2}{k}\!\kethatk{1}{k\pr\!}\!
        \brahatk{1}{k\pr\!}\!\brahatk{2}{k}+\Bigl(\bigl[\Ghn{0}{k}\alphahmnstar{2}{k\pr k}-
        2\,\operatorname{Re}(\Ghn{0}{k}\Ghnstar{0}{k\pr}\sum\limits_{p}\betahmn{1}{pk}
        \betahmnstar{1}{pk\pr})\bigr]\kethatk{1}{k\pr\!}\!\brahatk{1}{k}\nonumber\\[0.5mm]
    &\hspace*{-1.0cm} -\,\Ghn{0}{k}\Ghn{0}{k\pr}\sum\limits_{p}\alphahmnstar{1}{pk}\betahmn{1}{pk\pr}
        \kethat{0}\!\brahatk{1}{k\pr\!}\!\brahatk{1}{k}-\sqrt{2}\,(\Ghn{0}{k})^{2}\sum\limits_{p\neq k\pr}
        \alphahmnstar{1}{pk}\betahmn{1}{pk}\kethat{0}\!\brahatk{2}{k}\nonumber\\[0.5mm]
    &\hspace*{-1.0cm} +\,\sqrt{2}\,\alphahmn{1}{kk\pr}\betahmn{1}{kk\pr}
        \kethatk{1}{k\pr\!}\!\brahatk{1}{k\pr\!}\!\brahatk{2}{k}+\sqrt{2}\,\alphahmn{1}{k\pr k}
        \betahmn{1}{k\pr k}\kethatk{1}{k}\!\brahatk{2}{k\pr\!}\!\brahatk{1}{k}+\sqrt{2}\,\Vhmnstar{2}{kk\pr}
        \kethatk{1}{k}\!\brahatk{1}{k\pr\!}\!\brahatk{2}{k}\nonumber\\[2mm]
    &\hspace*{-1.0cm} +\sqrt{\tfrac{3}{2}}\,\Vhmnstar{2}{kk}\kethatk{1}{k}\!\brahatk{3}{k}+
        \sqrt{3}\,(\Ghn{0}{k}\betahmn{1}{k\pr k})^{2}\kethatk{1}{k}\!\brahatk{2}{k\pr\!}\!\brahatk{3}{k}
        +\,\mathrm{H.~c.}\Bigr)\Bigr\}\,+\,O(h^{3})\,,\nonumber
\end{align}
where, in analogy to~(\ref{eq:f beta m not n}), we define the quantity
\begin{align}
    f^{\alpha}_{m\lnot\hspace*{1pt} n} &=\,\tfrac{1}{2}\,\sum\limits_{i\neq n}\,|\alphahmn{1}{mi}|^{2}\,.
    \label{eq:f alpha m not n}
\end{align}
As before, a quick computation provides the linear entropy~$S_{L}$\index{entropy!linear} (see Definition~\ref{def:linear entropy}),
\begin{align}
    S_{L}\bigl({}_{\operatorname{1-k}}\rho_{\raisebox{-1pt}{\tiny{$kk\pr$}}}\bigr) &=\,
    h^{2}\,\bigl(8\,f^{\beta}_{k\lnot k\pr}\,+\,4\,f^{\beta}_{k\pr\lnot k}
    \,+\,4\,f^{\alpha}_{k\lnot k\pr}\bigr)\,+\,O(h^{3})\,,
    \label{eq:linear entropy transformed 1k two modes}
\end{align}
which immediately reveals that the state~(\ref{eq:bosonic 1k bogo traced k and kpr}) is mixed, as expected. To evaluate the entanglement that is produced between the modes~$k$ and~$k\pr$ we again employ the negativity. If the modes~$k$ and~$k\pr$ have opposite parity the partial transpose\index{partial transposition} of~(\ref{eq:bosonic 1k bogo traced k and kpr}) features corrections linear in~$h\,$. In the subspace of the degenerate unperturbed eigenvalue~$0$ we find the linear perturbation
\begin{align}
    h\,\begin{pmatrix}
    0   &   \Ghn{0}{k}\alphahmnstar{1}{k\pr k}   &   0   \\
    \Ghnstar{0}{k}\alphahmn{1}{k\pr k}  &   0   &   -\sqrt{2}\,\Ghn{0}{k}\betahmn{1}{k\pr k}    \\
    0   &   -\sqrt{2}\,\Ghnstar{0}{k}\betahmnstar{1}{k\pr k}    &   0
    \end{pmatrix}\,.
    \label{eq:bosonic 1k traced par transp linear block}
\end{align}
To leading order this supplies one negative eigenvalue and we get the negativity
\begin{align}
    \mathcal{N}({}_{\operatorname{1-k}}\rho_{\raisebox{-1pt}{\tiny{$kk\pr$}}})   &=\,
    h\,\Negh{1}({}_{\operatorname{1-k}}\rho_{\raisebox{-1pt}{\tiny{$kk\pr$}}})\,+\,O(h^{2})\,=\,
    h\,\sqrt{|\alphahmn{1}{kk\pr}|^{2}\,+\,2\,|\beta^{(1)}_{kk\pr}|^{2}}\,+\,O(h^{2})\,.
    \label{eq:linear negativity correction boson 1k}
\end{align}
In contrast to~(\ref{eq:linear negativity correction boson vac}) the entanglement is now generated by both coherent excitations of particle pairs in the modes~$k$ and~$k\pr$, and the shift of excitations from mode~$k$ to mode~$k\pr$ by the coefficient $\alphahmn{1}{kk\pr}\,$. An illustration of these results is shown in Fig.~\ref{fig:entanglement generation bosons 1k}~(a).

When~$k$ and~$k\pr$ have the same parity the situation is slightly more complicated. As previously, all the corrections to the reduced density matrix that are linear in~$h$ vanish. We further ignore the row and column of the partial transpose corresponding to the subspace of the unperturbed eigenvalue~$1$ because a small perturbation cannot possibly change this eigenvalue enough to become negative. The non-zero corrections in the subspace of the unperturbed eigenvalue~$0$ decompose into two independent blocks. The first block, given by the matrix
\begin{align}
    h^{2}\,\begin{pmatrix}
    |\alphahmn{1}{kk\pr}|^{2}   &   \sqrt{2}\,\alphahmn{1}{kk\pr}\betahmn{1}{kk\pr}\\
    \sqrt{2}\,\alphahmnstar{1}{kk\pr}\betahmnstar{1}{kk\pr} &   2\,|\beta^{(1)}_{kk\pr}|^{2}
    \end{pmatrix}
    \label{eq:quadratic correction part transp boson 1k part 1}
\end{align}
provides one positive eigenvalue, while the other eigenvalue vanishes identically. The second block is represented by
\begin{align}
    \begin{scriptsize}
    h^{2}
    \begin{pmatrix}
     2\,f^{\alpha}_{k\lnot k\pr}  &
        \Ghn{0}{k}\alpha^{(2)*}_{k\pr k}-2\operatorname{Re}
        (\Ghn{0}{k}\Ghnstar{0}{k\pr} g^{\beta\beta}_{kk\pr}) &
        -\sqrt{2}\,(\Ghn{0}{k})^{2} g^{\alpha\beta\,*}_{kk\lnot k\pr}  \\[1mm]
     \Ghnstar{0}{k}\alpha^{(2)}_{k\pr k}-2\operatorname{Re}
        (\Ghn{0}{k}\Ghnstar{0}{k\pr}g^{\beta\beta\,*}_{kk\pr}) &
        2\,f^{\beta}_{k\pr\lnot k}  &
        \sqrt{2}\,V^{(2)*}_{kk\pr} \\[1mm]
    -\sqrt{2}\,(\Ghnstar{0}{k})^{2}g^{\alpha\beta}_{kk\lnot k\pr} &
        \sqrt{2}\,V^{(2)}_{kk\pr}    &
        4\,f^{\beta}_{k\lnot k\pr}
    \end{pmatrix}
    \label{eq:quadratic correction part transp boson 1k part 2}
    \end{scriptsize}
\end{align}
where we have simplified the notation with the abbreviations
\begin{align}
    g^{\beta\beta}_{kk\pr}  &=\,\sum\limits_{p}\betahmn{1}{pk}\betahmnstar{1}{pk\pr}\,,
    \hspace*{0.3cm}
    g^{\alpha\beta}_{kk\lnot k\pr}\,=\,\sum\limits_{p\neq k\pr}\alphahmn{1}{pk}\betahmnstar{1}{pk}\,.
    \label{eq:galphabeta}
\end{align}
The eigenvalues of~(\ref{eq:quadratic correction part transp boson 1k part 2}) are given by the solutions to a cubic equation. It is not difficult to see that the off-diagonal elements of this matrix are responsible for possible entanglement generation, competing with the noise that is introduced by the diagonal elements. If the quantities~$f^{\alpha}_{k\lnot k\pr}\,$, $f^{\beta}_{k\pr\lnot k}\,$, and $f^{\beta}_{k\lnot k\pr}$ were zero, while the off-diagonals are non-vanishing the matrix~(\ref{eq:quadratic correction part transp boson 1k part 2}) would supply at least one negative eigenvalue. Practically, the negative solutions to the cubic equation mentioned above are best evaluated numerically, and the modulus of the negative eigenvalue of the matrix~(\ref{eq:quadratic correction part transp boson 1k part 2}) provides the negativity
\vspace*{-2mm}
\begin{align}
    \mathcal{N}({}_{\operatorname{1-k}}\rho_{\raisebox{-1pt}{\tiny{$kk\pr$}}})   &=\,
    h^{2}\,\Negh{2}({}_{\operatorname{1-k}}\rho_{\raisebox{-1pt}{\tiny{$kk\pr$}}})\,+\,O(h^{3})
    \label{eq:quadratic negativity correction boson 1k}
\end{align}
illustrated in Fig.~\ref{fig:entanglement generation bosons 1k}~(b). Finally, it should be noted that the presence of a non-zero negativity allows us to unambiguously conclude that the transformation creates entanglement. Vanishing negativity, on the other hand, does not rule out the presence of entanglement in principle (see pp.~\pageref{page:NPT entanglement}) because neither are the states under consideration Gaussian, nor can the modes be truncated to qubits. However, in the explicit examples that we have analyzed, e.g., the basic building block (see Section~\ref{sec:basic building block}) in Fig.~\ref{fig:entanglement generation bosons vacuum} and Fig.~\ref{fig:entanglement generation bosons 1k}, the negativity vanishes only when the corresponding Bogoliubov coefficients also disappear and the state is left unchanged \textemdash\ separable. This indicates that no bound entanglement is produced in the situations considered.


\section{Entanglement Generation in Bosonic Gaussian States}\label{sec:Entanglement Generation in Bosonic Gaussian States}

The analysis of Section~\ref{sec:Entanglement Generation in Bosonic Fock States} has demonstrated that entanglement is generated from initially separable Fock states within the cavity. However, as excitations are added the calculations quickly become computationally demanding. Moreover, there seems to be no natural restriction to the choice of the initial states. Put bluntly, there does not seem to be ample motivation to study, for instance, the state $\ket{5_{k}}\!\ket{3_{k\pr}}$ rather than $\ket{17_{k}}\,$. A class of states that distinguishes itself from general bosonic states is the group of \emph{Gaussian states}, see Section~\ref{sec:gaussian states}, which we are going to restrict ourselves to in this section.

\subsection{Single-Mode Squeezed States}\label{sec:Single-Mode Squeezed States}

To study entanglement generation phenomena it is prudent to start with an initial state that is separable. For example, a state where all modes are uncorrelated but individually (single-mode) squeezed\index{squeezed state!single-mode}, see pp.~\pageref{page:single mode squeezed Bogo transformation}. Since we are particularly interested in the entanglement that is generated between the modes~$k$ and~$k\pr$ we allow for non-zero squeezing only for these modes, i.e., $s_{n}=0\ \forall n\neq k,k\pr\,$. The $4\times4$ covariance matrix for the modes~$k$ and~$k\pr$ is decomposed as
\begin{align}
    \widehat{\Gamma}    &=\,\begin{pmatrix}
    \Gammahatk{k}   &  \Chatmn{kk\pr} \\
    \Chatmn{kk\pr}^{\,T}    &   \Gammahatk{k\pr}
    \end{pmatrix}\,,
    \label{eq: cov matrix k and kp}
\end{align}
where the leading order coefficients in the series expansions of the components of the $2\times2$ matrices $\Gammahatk{k}\,,\Chatmn{kk\pr}\,$, and $\Gammahatk{k\pr}$ can be read off directly from Eqs.~(\ref{eq:transformed single mode squeezed cov matrix gamma 0}) and~(\ref{eq:transformed single mode squeezed cov matrix C 1}). Given these perturbative expressions we can proceed to evaluate the entanglement of this state. Taking Theorem~\ref{thm:PPT Gaussian} as a starting point we wish to obtain the smallest symplectic eigenvalue of the partial transpose\index{partial transposition} of the covariance matrix~(\ref{eq: cov matrix k and kp}). Since this essentially entails perturbatively determining the eigenvalues of the matrix
\begin{align}
    i\hspace*{1pt}\Omega\hspace*{1pt}\oversmile{T}_{\hspace*{-1pt}k\pr}\hspace*{1pt}
    \widehat{\Gamma}\hspace*{1pt}\hspace*{0.8pt}\oversmile{T}_{\hspace*{-1pt}k\pr}\,,
    \label{eq:i Omega partially transposed cov matr}
\end{align}
where $\Omega$ is the symplectic form~(\ref{eq:symplectic form}) and $\oversmile{T}_{\hspace*{-1pt} k\pr}=\mathds{1}\oplus\diag\{1,-1\}\,$ represents the partial transposition, we can turn to the procedure described on pp.~\pageref{page:pert diagonalization} to do so. The symplectic eigenvalues of the partial transpose of the unperturbed state are given by
\begin{align}
    \spectr(i\hspace*{1pt}\Omega\hspace*{1pt}\oversmile{T}_{\hspace*{-1pt}k\pr}\hspace*{1pt}
    \Gammahatkh{}{0}\hspace*{1pt}\hspace*{0.8pt}\oversmile{T}_{\hspace*{-1pt}k\pr})  &=\,
    \{-1,-1,+1,+1\}\,,
    \label{eq:unperturbed sympl eig values part transp}
\end{align}
as expected for a separable pure state, and we note that the eigenvalues are twice degenerate. Our aim is then to find the leading order negative correction to the eigenvalue $\numh{-}{0}=\numh{+}{0}=1$. At this stage we specialize to the case where $(k+k\pr)$ is odd, i.e., the modes have opposite parity, such that the leading order correction to $\Gammahatkh{}{0}$ is linear in~$h$. In the next step we have to diagonalize the subspace of the correction $i\hspace*{1pt}\Omega\hspace*{1pt}\oversmile{T}_{\hspace*{-1pt}k\pr}\hspace*{1pt}
\Gammahatkh{}{1}\hspace*{1pt}\hspace*{0.8pt}\oversmile{T}_{\hspace*{-1pt}k\pr}$ corresponding to the unperturbed eigenvalue~$\numh{\pm}{0}=1$. In other words, the eigenvalues of the matrix
\begin{align}
    \begin{pmatrix}
    \bra{\numh{+}{0}}i\hspace*{1pt}\Omega\hspace*{1pt}\oversmile{T}_{\hspace*{-1pt}k\pr}\hspace*{1pt}
    \Gammahatkh{}{1}\hspace*{1pt}\hspace*{0.8pt}\oversmile{T}_{\hspace*{-1pt}k\pr}\ket{\numh{+}{0}}   &
    \bra{\numh{+}{0}}i\hspace*{1pt}\Omega\hspace*{1pt}\oversmile{T}_{\hspace*{-1pt}k\pr}\hspace*{1pt}
    \Gammahatkh{}{1}\hspace*{1pt}\hspace*{0.8pt}\oversmile{T}_{\hspace*{-1pt}k\pr}\ket{\numh{-}{0}}   \\
    \bra{\numh{-}{0}}i\hspace*{1pt}\Omega\hspace*{1pt}\oversmile{T}_{\hspace*{-1pt}k\pr}\hspace*{1pt}
    \Gammahatkh{}{1}\hspace*{1pt}\hspace*{0.8pt}v{T}_{\hspace*{-1pt}k\pr}\ket{\numh{+}{0}}   &
    \bra{\numh{-}{0}}i\hspace*{1pt}\Omega\hspace*{1pt}\oversmile{T}_{\hspace*{-1pt}k\pr}\hspace*{1pt}
    \Gammahatkh{}{1}\hspace*{1pt}\hspace*{0.8pt}\oversmile{T}_{\hspace*{-1pt}k\pr}\ket{\numh{-}{0}}
    \end{pmatrix}\,
    \label{eq:subspace of pos eig value}
\end{align}
need to be determined, where the non-zero elements of~$\Gammahatkh{}{1}$ are given by~(\ref{eq:transformed single mode squeezed cov matrix C 1}) and $\ket{\numh{\pm}{0}\!}$ are the eigenstates of $i\hspace*{1pt}\Omega\hspace*{1pt}\oversmile{T}_{\hspace*{-1pt}k\pr}\hspace*{1pt}
\Gammahatkh{}{0}\hspace*{1pt}\hspace*{0.8pt}\oversmile{T}_{\hspace*{-1pt}k\pr}$ with eigenvalue~$\numh{\pm}{0}=1\,$. The eigenstates $\ket{\numh{\pm}{0}\!}$ are given by
\begin{subequations}
\label{eq:eigenstates pos unperturbed eigenvalue}
\begin{align}
    \ket{\numh{+}{0}\!} &=\,N(s_{k},\omega_{k}\tilde{\tau})\,
    \Bigl(\,
       \frac{i\hspace*{1pt}\cos(\omega_{k}\tilde{\tau})\,+\,\exp(s_{k})\sin(\omega_{k}\tilde{\tau})}
       {\exp(s_{k})\cos(\omega_{k}\tilde{\tau})\,-\,i\hspace*{1pt}\sin(\omega_{k}\tilde{\tau})}
       \,,\,1\,,\,0\,,\,0\,\Bigr)^{T}\,,\label{eq:eigenstates pos unperturbed eigenvalue 1}\\[1mm]
    \ket{\numh{-}{0}\!} &=\,N(s_{k\pr},\omega_{k\pr}\tilde{\tau})\,
    \Bigl(\,0\,,\,0\,,\,  \frac{i\hspace*{1pt}\cos(\omega_{k\pr}\tilde{\tau})\,-\,\exp(s_{k\pr})\sin(\omega_{k\pr}\tilde{\tau})}
       {\exp(s_{k\pr})\cos(\omega_{k\pr}\tilde{\tau})\,+\,i\hspace*{1pt}\sin(\omega_{k\pr}\tilde{\tau})}
       \,,\,1\,\Bigr)^{T}\,,
    \label{eq:eigenstates pos unperturbed eigenvalue 2}
\end{align}
\end{subequations}
where the normalization constants $N(s_{n},\omega_{n}\tilde{\tau})$ are given by
\begin{align}
    N(s_{n},\omega_{n}\tilde{\tau}) &=\,\Bigl(\frac{1\,+\,\exp(2s_{n})}
    {\exp(2s_{n})\cos^{2}(\omega_{n}\tilde{\tau})\,+\,\sin^{2}(\omega_{n}\tilde{\tau})}\Bigr)
    ^{\raisebox{0.0pt}{\scriptsize{$-\tfrac{1}{2}$}}}\,.
    \label{eq:eigenstates pos unperturbed eigenvalue normalization}
\end{align}
This allows us to obtain the perturbative expansion of the symplectic eigenvalues of the partially transposed covariance matrix to linear order in~$h\,$, i.e.,
\begin{align}
    \num{\pm}   &=\,1\,+\,h\,\numh{\pm}{1}\,+\,O(h^{2})\,.
    \label{eq:sympl eig val part transp expansion}
\end{align}
We find that $\numh{+}{1}=-\numh{-}{1}\geq0\,$, which further allows us to express the leading order correction to the negativity from Eq.~(\ref{eq:negativity Gaussian}) in the following form
\begin{align}
    \mathcal{N} &=\,h\,\Negh{1}\,+\,O(h^{2})\,=\,h\,\frac{|\numh{-}{1}|}{2}\,+\,O(h^{2})\,,
    \label{eq:Gaussian negativity power expansion}
\end{align}
where the leading order coefficient~$\Negh{1}$ is given by
\begin{align}
    \Negh{1}    &=\,
    \frac{1}{\sqrt{2}}\Bigl(\,|\alphahmn{1}{kk\pr}|^{2}\bigl[\cosh(2s_{k})\cosh(2s_{k\pr})-1\bigr]\,+\,
    |\betahmn{1}{kk\pr}|^{2}\bigl[\cosh(2s_{k})\cosh(2s_{k\pr})+1\bigr]
    \nonumber\\[1mm]
    &\hspace*{-0.7cm} -\,\operatorname{Re}\bigl[(\Ghnstar{0}{k}\alphahmn{1}{kk\pr})^{2}\,+\,
        (\Ghnstar{0}{k}\betahmn{1}{kk\pr})^{2}\bigr]\sinh(2s_{k})\sinh(2s_{k\pr})
        \,-\,2\,\operatorname{Re}\bigl[(\Ghnstar{0}{k}\alphahmn{1}{kk\pr})
        (\Ghn{0}{k}\betahmnstar{1}{kk\pr})\bigr]
        \nonumber\\[1mm]
    &\hspace*{-0.7cm} \times\cosh(2s_{k})\sinh(2s_{k\pr})
        \,+\,2\,\operatorname{Re}\bigl[(\Ghnstar{0}{k}\alphahmn{1}{kk\pr})
        (\Ghnstar{0}{k}\betahmn{1}{kk\pr})\bigr]\sinh(2s_{k})\cosh(2s_{k\pr})
    \,\Bigr)^{\raisebox{0.0pt}{\scriptsize{$\tfrac{1}{2}$}}}\,.
    \label{eq:single mode squeezed negativity corrections}
\end{align}
Alternatively, one may obtain the expression in~(\ref{eq:single mode squeezed negativity corrections}) by a different line of argument. To linear order in~$h$ the Bogoliubov transformation does not affect the purity of the initial state. In particular, the initially pure states we have chosen remain pure when terms proportional to~$h^{2}$ are neglected. Recall now that every pure two-mode Gaussian state is equivalent up to local symplectic transformations to a two-mode squeezed state~(\ref{eq:two mode squeezed covariance matrix}), see also Ref.~\cite{AdessoIlluminati2005b}. Since we work with small perturbations of the state, the corresponding two-mode squeezing parameter~$r$ can be assumed to satisfy $r\ll1$. We may therefore relate the local symplectic invariant $\det(\Chatmn{kk\pr})$ to~$r$ via the relation
\begin{align}
    \det(\Chatmn{kk\pr})    &=\,-\,\sinh^{2}(2r)\,=\,-\,4\,r^{2}\,+\,O(r^{4})\,,
    \label{eq:local sympl inv two mode squeezing}
\end{align}
where we have performed a power expansion assuming~$r\ll1$ in the last step. Since the squeezing parameter is also directly related to~$\num{-}\,$, i.e., $\num{-}=e^{-2|r|}\,$, the expression
\begin{figure}[ht!]
\centering
\includegraphics[width=0.75\textwidth]{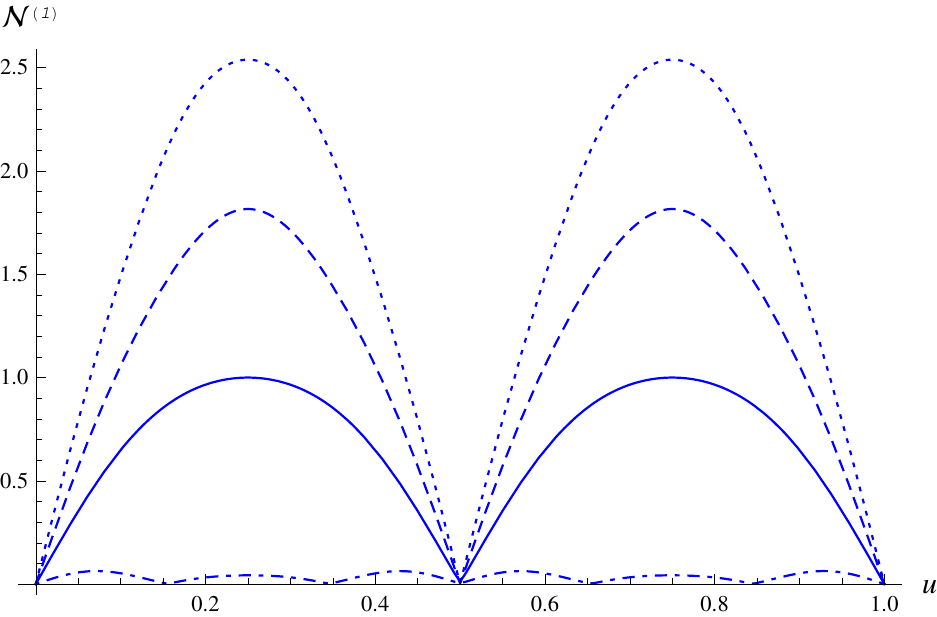}
\caption{
\textbf{Entanglement generation \textemdash\ symmetric single-mode squeezing:} The coefficient $\Negh{1}$ [see Eq.~(\ref{eq:single mode squeezed negativity corrections})] of the negativity generated from a symmetrically single-mode squeezed state is plotted for the basic building block travel scenario of Section~\ref{sec:basic building block} for squeezing parameters $s_{k}=s_{k\pr}=1\,$. For the $(1+1)$ dimensional massless scalar field in this illustration the Bogoliubov coefficients are periodic in the dimensionless parameter $u:=h\tau/[4L\artanh(h/2)]\,$ [see Eq.~(\ref{eq:cavity centre proper freq})], where $\tau$ is the duration, as measured at the centre of the cavity, of the single segment of uniform acceleration. Curves are shown for the mode pairs $(k,k\pr)=(1,2)$ (solid), $(2,3)$ (dashed), $(3,4)$ (dotted), and $(1,4)$ (dotted-dashed).
\label{fig:entanglement generation single mode squeezing}}
\end{figure}
in Eq.~(\ref{eq:single mode squeezed negativity corrections}) can be calculated from $\det(\Chatmn{kk\pr})$ in a straightforward manner. Moreover, we can conclude that, to leading order, the transformed state is locally equivalent to a two-mode squeezed state with squeezing parameter $r=h\Negh{1}\,$. The correction to the negativity is consistent with the expression obtained for symmetric single-mode squeezing, $s_{k}=s_{k\pr}\,$, in Ref.~\cite[(\ref{Paper:FriisFuentes2013})]{FriisFuentes2013}, which is illustrated in Fig.~\ref{fig:entanglement generation single mode squeezing}. In the limit of vanishing initial squeezing, i.e., for $s_{k}=s_{k\pr}=0\,$, we further recover the correct expression for the entanglement generated from the vacuum, see Eq.~(\ref{eq:linear negativity correction boson vac}).\\

As expected, and as can be inferred from a quick comparison of Figs.~\ref{fig:entanglement generation bosons vacuum} and~\ref{fig:entanglement generation single mode squeezing}, the presence of squeezing in the initial state can greatly enhance the entanglement production. However, the perturbative treatment restricts the validity of these considerations. We trust the perturbative corrections as long as the main features of the state, e.g., the \emph{mixedness}\index{mixedness}, are not significantly altered. More precisely, we quantify the mixedness by the \emph{linear entropy} $S_{L}$\index{entropy!linear} (see Definition~\ref{def:linear entropy}). For a Gaussian state corresponding to the covariance matrix~$\Gamma$ it is given by~(see, e.g., Ref.~\cite[p.~38]{Adesso:PhDThesis2007})
\begin{align}
    S_{L}(\Gamma)   &=\,1\,-\,1/\sqrt{\det(\Gamma)}\,.
    \label{eq:linear entropy of cov matrix}
\end{align}
For the symmetrically single-mode squeezed state $(s_{k}=s_{k\pr}=s)$ considered in Fig.~\ref{fig:entanglement generation single mode squeezing} we have
\begin{align}
    \det\bigl(\,\widehat{\Gamma}\,\bigr)    &=\,1\,+\,h^{2}\,\Bigl[
        4\bigl(f^{\beta}_{k\lnot k\pr}\,+\,f^{\beta}_{k\pr\lnot k}\bigr)[\cosh(2s)\,+\,1]\,+\,
        4\bigl(f^{\alpha}_{k\lnot k\pr} + f^{\alpha}_{k\pr\lnot k}\bigr)[\cosh(2s)\,-\,1] \nonumber \\[1mm]
        &\ \ \ -\,4\sinh(2s)\,\sum\limits_{n\neq k,k\pr}
        \operatorname{Re}\bigl(\alphahmn{1}{nk}\betahmnstar{1}{nk}\,+\,
        \alphahmn{1}{nk\pr}\betahmnstar{1}{nk\pr}\bigr)\Bigr]\,+\,O(h^{3})\,,
    \label{eq:sym squeezed determinant}
\end{align}
with $f^{\alpha}_{m\lnot n}$ and $f^{\beta}_{m\lnot n}$ as in Eqs.~(\ref{eq:f alpha m not n}) and~(\ref{eq:f beta m not n}), respectively. The dominant correction in~(\ref{eq:sym squeezed determinant}) is given by the terms proportional to $f^{\alpha}_{k\lnot k\pr}$ and $f^{\alpha}_{k\pr\lnot k}$ and it is then easy to see that the validity of the perturbative treatment is expressed in the condition
\begin{align}
    h^{2}\,\bigl(f^{\alpha}_{k\lnot k\pr} + f^{\alpha}_{k\pr\lnot k}\bigr)\,e^{2|s|}    &\ll\,1\,.
    \label{eq:validitiy of pert regime single mode squeezing}
\end{align}
Finally, a note on the choice of entanglement measure is in order. To linear order in~$h$ we have a symmetric two-mode Gaussian state for which the entanglement of formation~(\ref{eq:EoF for symmetric Gaussian states}) can be computed. However, this entanglement measure is based on the quantification of the mixedness in the reduced state of one mode arising from tracing out the other mode. But, as we have argued before, the mixedness does not change unless second order terms are included, i.e., if only terms linear in~$h$ are kept, the state remains pure. However, when terms proportional to~$h^{2}$ are kept, we are left with a non-symmetric state, $\det(\Gammahatk{k})\neq\det(\Gammahatk{k\pr})\,$, for which the entanglement of formation cannot be computed. Hence, the negativity is the most suitable measure for our purposes.

\subsection{Resonances of Entanglement Generation}\label{sec:Resonances of Entanglement Generation}

The analysis we have undertaken up to this point has established the entanglement generation from various initial states in terms of the Bogoliubov coefficients for generic travel scenarios, including smoothly varying accelerations, as described in Section~\ref{sec:Grafting Generic Cavity Trajectories}. As we have seen from the examples under scrutiny, for instance in Figs.~\ref{fig:entanglement generation bosons vacuum}, \ref{fig:entanglement generation bosons 1k} and~\ref{fig:entanglement generation single mode squeezing}, the choice of initial state influences the amount of generated entanglement. Now we shall inquire if it is possible to enhance the entanglement production simply by moving the cavity in a particular way \textemdash\ we want to find \emph{entanglement resonances}\index{entanglement!resonances} \textemdash\ possibly with an accompanying restriction of the initial states. The results we present here are based on the results of Section~\ref{sec:gaussian entanglement} and the insights gathered from Refs.~\cite{BruschiDraganLeeFuentesLouko2013} and~\cite{BruschiLoukoFaccioFuentes2013}.\\

Let us start with the resonance condition\index{resonance condition} of Eq.~(\ref{eq:resonance condition}). For initial states with a covariance matrix proportional to the identity, i.e., the vacuum or coherent states, a vanishing commutator $\comm{S}{S^{\,T}}$ indicates that the entanglement produced by the symplectic transformation~$S$ can be linearly increased with the number of repetitions of the transformation if the operation represented by~$S$ is restricted to contain no overall single-mode squeezing and if it is possible to perform the transformation successively in principle. Both of these conditions are met by the Bogoliubov transformation for non-uniform cavity motion when we select an arbitrary travel scenario between two inertial regions and terms are kept only up to linear order in~$h\,$. The latter condition ensures that the coefficients $\betahmn{2}{nn}$, which would introduce single-mode squeezing, can be neglected. Under these premises let us proceed by examining the mechanism of the \emph{resonance condition} perturbatively.\\

\begin{figure}[hb!]
\centering
\includegraphics[width=0.75\textwidth]{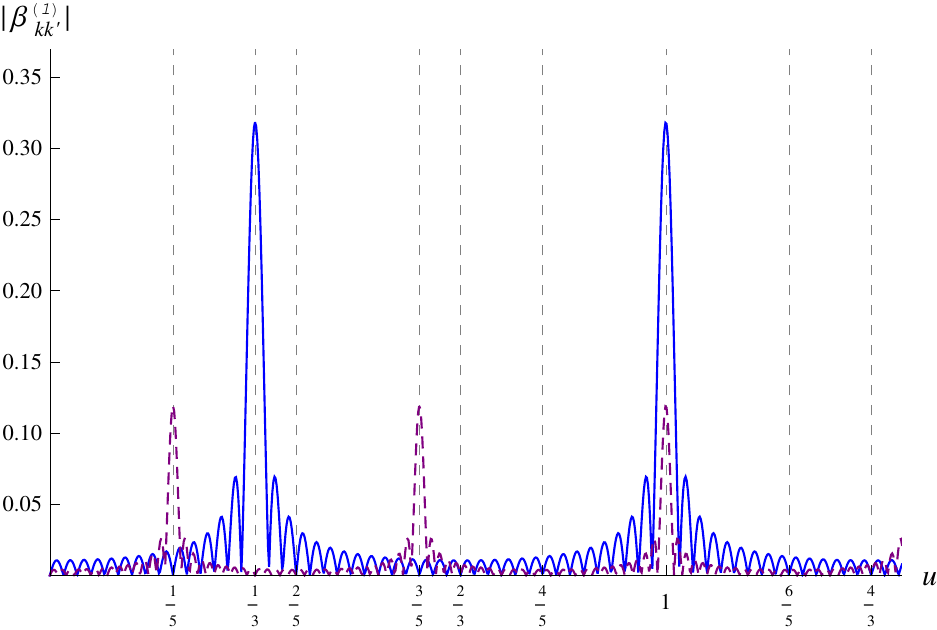}\\[1mm]
\caption{
\textbf{Entanglement resonances \textemdash\ $\beta$'s:} The coefficient~$|\betahmn{1}{kk\pr}|$ that generates entanglement from the vacuum [see Eq.~(\ref{eq:linear negativity correction boson vac})] is plotted against the dimensionless parameter $u:=h\tau/[4L\artanh(h/2)]$ for a massless $(1+1)$ dimensional scalar field. The travel scenario, which is illustrated in Fig.~\ref{fig:entanglement resonances alphas}~(b) for $N=2\,$, has $N$ segments of uniform proper acceleration $h/L$ and duration $\tau/2$ as measured at the centre of the cavity, separated by $(N-1)$ segments of inertial coasting of the same duration (to linear order in $h$). The curves are plotted for $N=15$,
$(k,k\pr)=(1,2)$ (blue, solid) and $(k,k\pr)=(2,3)$ (purple, dashed). The vertical dashed lines indicate the potential resonance times as given by Eq.~(\ref{eq:resonance duration}) for $(k,k\pr)=(1,2)$ and
$(k,k\pr)=(2,3)$, respectively.
\label{fig:entanglement resonances}}
\end{figure}

First we note that the linear order of corrections introduced by the Bogoliubov transformation for non-uniform cavity motion correlates modes only pairwise. Thus, neglecting second order corrections can be considered as a \emph{two-mode truncation}~\cite{BruschiDraganLeeFuentesLouko2013} and we can restrict the analysis to only two modes $k$ and~$k\pr\,$. Further assuming that the initial state is represented by $\Gammak{kk\pr}=\mathds{1}$ we write the transformed covariance matrix as
\begin{align}
    \Gammahatk{kk\pr}   &=\,S\,S^{T}\,=\,\mathds{1}\,+\,h\,\Gammahatkh{kk\pr}{1}\,+\,O(h^{2})\,.
    \label{eq:res derivation transformed state}
\end{align}
The entanglement is determined by the smallest symplectic eigenvalue of the partial transpose, in other words, the smallest positive entry of the diagonal matrix
\begin{align}
    iU\hspace*{1pt}\Omega\hspace*{1pt}\oversmile{T}_{\hspace*{-1pt}k\pr}\hspace*{1pt}
    \Gammahatk{kk\pr}\hspace*{1pt}\hspace*{0.8pt}\oversmile{T}_{\hspace*{-1pt}k\pr}U^{\dagger}\,,
\end{align}
where $U$ is the diagonalizing unitary. In particular, since we start with a separable state it is the correction term
\begin{align}
    iU\hspace*{1pt}\Omega\hspace*{1pt}\oversmile{T}_{\hspace*{-1pt}k\pr}\hspace*{1pt}
    \Gammahatkh{kk\pr}{1}\hspace*{1pt}\hspace*{0.8pt}\oversmile{T}_{\hspace*{-1pt}k\pr}U^{\dagger}
    &=\,\diag\{\pm\numh{-}{1},\pm\numh{+}{1}\}\,
    \label{eq:res derivation sympl eig val corr}
\end{align}
that generates the entanglement, specifically, the quantity~$\numh{-}{1}\,$, see Eq.~(\ref{eq:Gaussian negativity power expansion}). From Eq.~(\ref{eq:res derivation transformed state}) it then follows immediately that $N$-fold repetition of a transformation satisfying the resonance condition of Eq.~(\ref{eq:resonance condition}) will produce a state represented by the covariance matrix
\begin{align}
    (S)^{N}\,(S^{T})^{N}   &\,=\,(S\,S^{T})^{N}\,=\,\mathds{1}\,+\,h\,N\,\Gammahatkh{kk\pr}{1}\,+\,O(h^{2})\,.
    \label{eq:res derivation transformed state N times}
\end{align}
Consequently, the correction to the smallest symplectic eigenvalue after $N$ repetitions is given by~$N\numh{-}{1}\,$. In other words, the entanglement production grows linearly with the number of repetitions. Let us therefore investigate how the resonance condition can be satisfied by inserting the expansions of~(\ref{eq:M bogo matrix expanded})-(\ref{eq:M bogo matrix expansion coefficients p}) into~(\ref{eq:Gaussian n-mode Bogo transformation}), whilst restricting to the two modes~$k$ and~$k\pr\,$. To linear order the commutator of the resonance condition~(\ref{eq:resonance condition}) has two independent non-zero entries
\begin{subequations}
\label{eq:res cond perturbatively}
\begin{align}
    \operatorname{Re}(\Ghn{0}{k}-\Ghn{0}{k\pr})\operatorname{Re}(\betahmn{1}{kk\pr})\,+\,
    \operatorname{Im}(\Ghn{0}{k}+\Ghn{0}{k\pr})\operatorname{Im}(\betahmn{1}{kk\pr})    &=\,0\,,
    \label{eq:res cond perturbatively 1}\\[1mm]
    \operatorname{Re}(\Ghn{0}{k}-\Ghn{0}{k\pr})\operatorname{Im}(\betahmn{1}{kk\pr})\,-\,
    \operatorname{Im}(\Ghn{0}{k}+\Ghn{0}{k\pr})\operatorname{Re}(\betahmn{1}{kk\pr})    &=\,0\,.
    \label{eq:res cond perturbatively 2}
\end{align}
\end{subequations}
The two conditions in~(\ref{eq:res cond perturbatively}) can be conveniently combined into the single requirement
\begin{align}
    \bigl(\Ghnstar{0}{k}\,-\,\Ghn{0}{k\pr}\bigr)\,\betahmn{1}{kk\pr}  &=\,0\,.
    \label{eq:resonance condition perturbative simpel}
\end{align}
Further noting that $\betahmn{1}{kk\pr}$ needs to be non-zero to create entanglement at all [see Eq.~(\ref{eq:linear negativity correction boson vac})] one finds that the resonances are purely governed by the phases that are acquired during the free time evolution. For any mode pair~$k$ and~$k\pr$ the \emph{arbitrary} travel scenario that is to be repeated has to be timed appropriately to satisfy $(\Ghnstar{0}{k}-\Ghn{0}{k\pr})=0$, i.e., the duration of a single repetition as measured at the centre of the cavity has to take on one of the discrete values
\begin{align}
    \tau_{n}    &=\,\frac{2\hspace*{1pt}\pi\hspace*{1pt}n}{\omega_{k}\,+\,\omega_{k\pr}}\,,
    \label{eq:resonance duration}
\end{align}
with $n=1,2,3,\ldots$, for which $\betahmn{1}{kk\pr}\,$ takes on a non-zero value, see Ref.~\cite{BruschiDraganLeeFuentesLouko2013}. An illustration of the resonance peaks of the created entanglement for fixed mode pairs is shown in Fig.~\ref{fig:entanglement resonances}.

Subsequently, one may ask about possible resonances for states that are not described by a covariance matrix that is proportional to the identity. For instance, inspecting Eqs.~(\ref{eq:linear negativity correction boson 1k}) and~(\ref{eq:single mode squeezed negativity corrections}) it seems that a strong increase of $|\betahmn{1}{kk\pr}|$ [see Fig.~\ref{fig:entanglement resonances}~(a)] may also increase the entanglement that is produced from these states, but the role of the coefficients $\alphahmn{1}{kk\pr}$ requires separate inspection.
\begin{figure}[ht!]
\centering
(a)\includegraphics[width=0.8\textwidth]{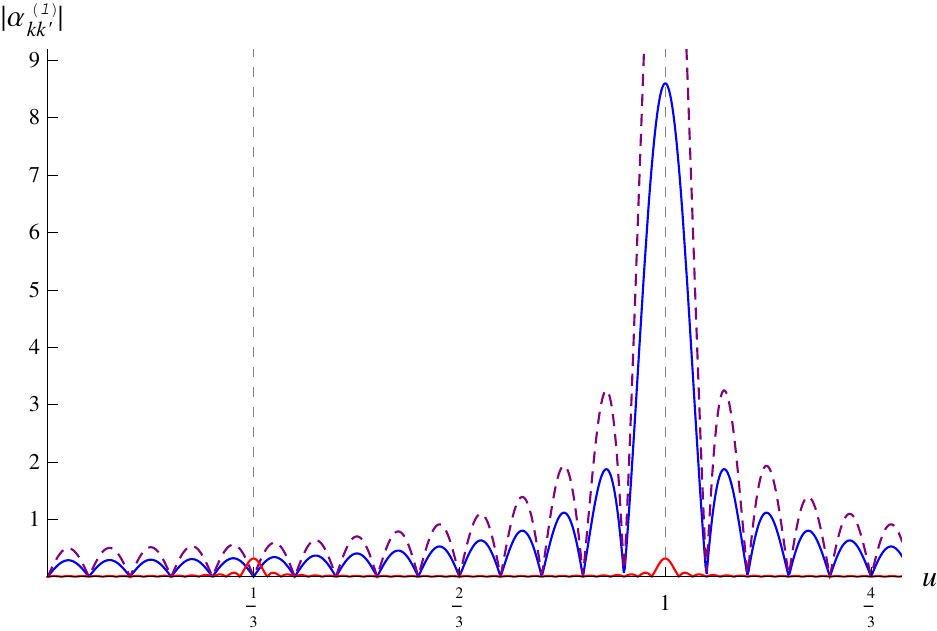}\\[4mm]
(b)\includegraphics[width=0.8\textwidth]{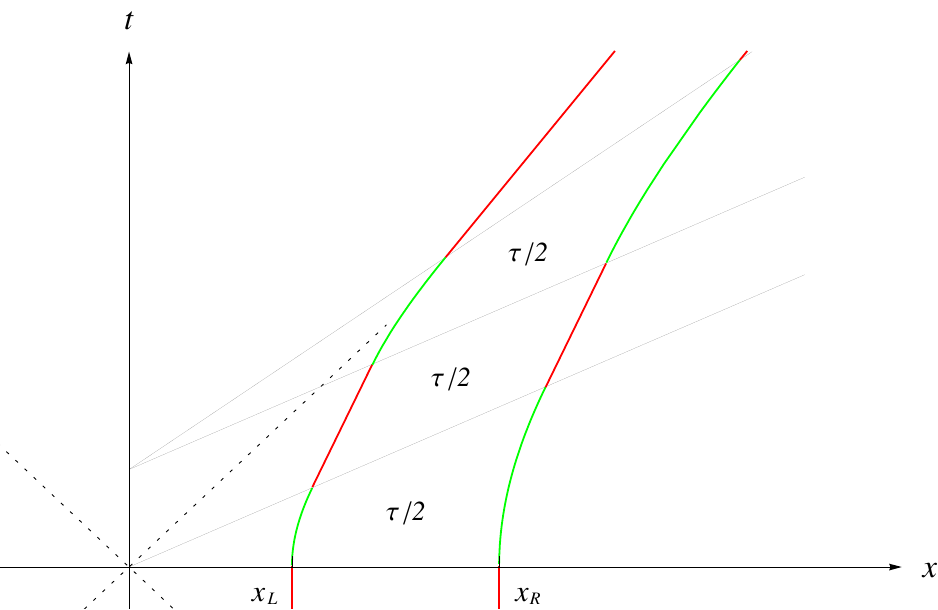}\\[4mm]
\caption{
\textbf{Entanglement resonances \textemdash\ $\alpha$'s:} The coefficient~$|\alphahmn{1}{kk\pr}|$ that contributes to the entanglement generation, e.g., for squeezed states [see Eqs.~(\ref{eq:linear negativity correction boson 1k}) and~(\ref{eq:single mode squeezed negativity corrections})] is plotted in Fig.~\ref{fig:entanglement resonances alphas}~(a) against the dimensionless parameter $u:=h\tau/[4L\artanh(h/2)]\,$ for a massless $(1+1)$ dimensional scalar field. The travel scenario, which is illustrated in Fig.~\ref{fig:entanglement resonances alphas}~(b) for $N=2\,$, has $N$ segments of uniform proper acceleration $h/L$ and duration $\tau/2$ as measured at the centre of the cavity, separated by $(N-1)$ segments of inertial coasting of the same duration (to linear order in $h$). The curves in Fig.~\ref{fig:entanglement resonances alphas}~(a) are plotted for $N=15$, $(k,k\pr)=(1,2)$ (blue, solid) and $(k,k\pr)=(2,3)$ (purple, dashed). For comparison the quantity $|\betahmn{1}{kk\pr}|$ (red, dashed) from Fig.~\ref{fig:entanglement resonances} is shown for $(k,k\pr)=(1,2)$.
\label{fig:entanglement resonances alphas}}
\end{figure}
Indeed, a graphical analysis [see Fig.~\ref{fig:entanglement resonances alphas}~(a)] shows that some resonances occur for both the coefficients $\alphahmn{1}{kk\pr}$ and $\betahmn{1}{kk\pr}$, which suggests that the entanglement production is significantly enhanced, growing (at most) linearly with the number of repetitions, also for squeezed states and single particle states.\\

\vspace*{-3mm}
Finally, following Ref.~\cite{BruschiLoukoFaccioFuentes2013} we turn to the case of smoothly varying acceleration discussed in Section~\ref{sec:smoothly varying accelerations}. Let us assume that the proper acceleration\index{proper!acceleration} at the centre of the cavity is a sinusoidal function
\begin{align}
    \mathbf{a}_{\mathrm{c}}(\tau)   &=\,\mathbf{a}_{0}\,\sin(\omega_{\mathrm{c}}\tau)\,,
    \label{eq:sinusoidal acceleration}
\end{align}
with an amplitude $\mathbf{a}_{0}$ that is much smaller than the inverse length of the cavity. The expressions for the leading order Bogoliubov coefficients from Eqs.~(\ref{eq:smooth acceleration expansion alphas 1}) and~(\ref{eq:smooth acceleration expansion betas 1}) are oscillatory for an arbitrary value of the oscillation frequency $\omega_{\mathrm{c}}\,$. However, for specific choices of~$\omega_{\mathrm{c}}$ two cases can be distinguished. If $\omega_{\mathrm{c}}=\omega_{k}+\omega_{k\pr}$ the integral in~(\ref{eq:smooth acceleration expansion betas 1}) for the coefficient $\Sbetamn{kk\pr}$ grows linearly with the overall time of acceleration. Such particle creation resonances are at the heart of the \emph{dynamical Casimir effect}\index{dynamical Casimir effect} (DCE). We refer the interested reader to the recent review Ref.~\cite{Dodonov2010} and references therein. In particular, the DCE has been investigated in a variety of media, such as Bose-Einstein condensates (see, e.g., Ref.~\cite{JaskulaPartridgeBonneauRuaudelBoironWestbrook2012}), or superconducting microwave circuits~\cite{JohanssonJohanssonWilsonNori2010,WilsonDynCasNature2012,LaehteenmaekiParaoanuHasselHakonen2013}.\\

\vspace*{-3mm}
On the other hand, for $\omega_{\mathrm{c}}=|\omega_{k}-\omega_{k\pr}|$ we obtain a linear growth of the coefficient $\Salphamn{kk\pr}$ with increasing overall time of the oscillation. Such a \emph{mode-mixing} resonance could in principle be exploited for desktop experiments at mechanical frequencies~\cite{BruschiLoukoFaccioFuentes2013}, as well as for entanglement generation, e.g., in simulations of cavity motion in microwave circuitry, see Refs.~\cite[(\ref{Paper:FriisLeeTruongSabinSolanoJohanssonFuentes2013})]
{FriisLeeTruongSabinSolanoJohanssonFuentes2013} and~\cite{SvenssonMScThesis2012}.\\

\vspace*{-3mm}
Note that in all resonance scenarios that we have discussed here we are still limited by the perturbative regime. The overall perturbation to any quantity of interest still needs to remain small if the approximations we have made are to hold. Nonetheless, the resonance formalism presents an elegant way of enhancing the corrections by orders of magnitude, possibly even to observable levels.

\section{Entanglement Generation in Fermionic States}\label{sec:Entanglement Generation in Fermionic States}

In this section we study the entanglement generation between the modes of a Dirac field that is confined to a non-uniformly moving cavity. Mirroring the analysis of the bosonic case in Section~\ref{sec:Entanglement Generation in Bosonic Fock States} we start from the vacuum state in Section~\ref{sec:entanglement from the fermionic vacuum} before we add particle content in Sections~\ref{sec:entanglement from the fermionic particle state} and~\ref{sec:entanglement from the fermionic particle-antiparticle pair}. In contrast to the bosonic situation two practical issues already arise at this stage of the analysis. First, in the selection of the two modes between which entanglement generation is studied we have the choice between distinct particle and antiparticle modes. Second, in the partial tracing to recover the reduced states of the two chosen modes the consistency conditions~(\ref{eq:consistency condition}) have to be respected, which requires tracing ``inside-out." With these considerations in mind we proceed with the vacuum state.

\subsection{Entanglement from the Fermionic Vacuum}\label{sec:entanglement from the fermionic vacuum}\index{vacuum!fermions|(}

\begin{figure}[hb!]
\centering
(a)\includegraphics[width=0.69\textwidth]{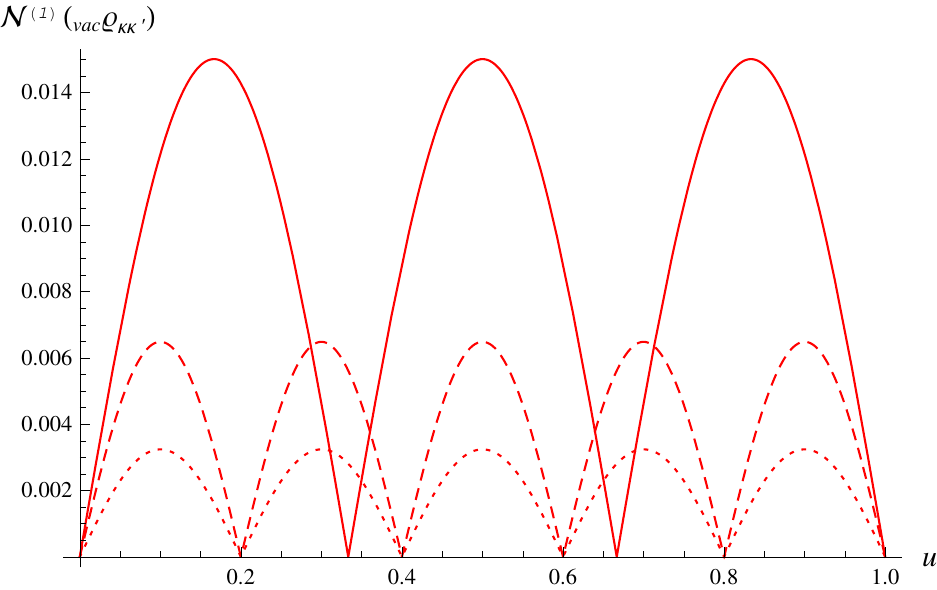}
(b)\includegraphics[width=0.69\textwidth]{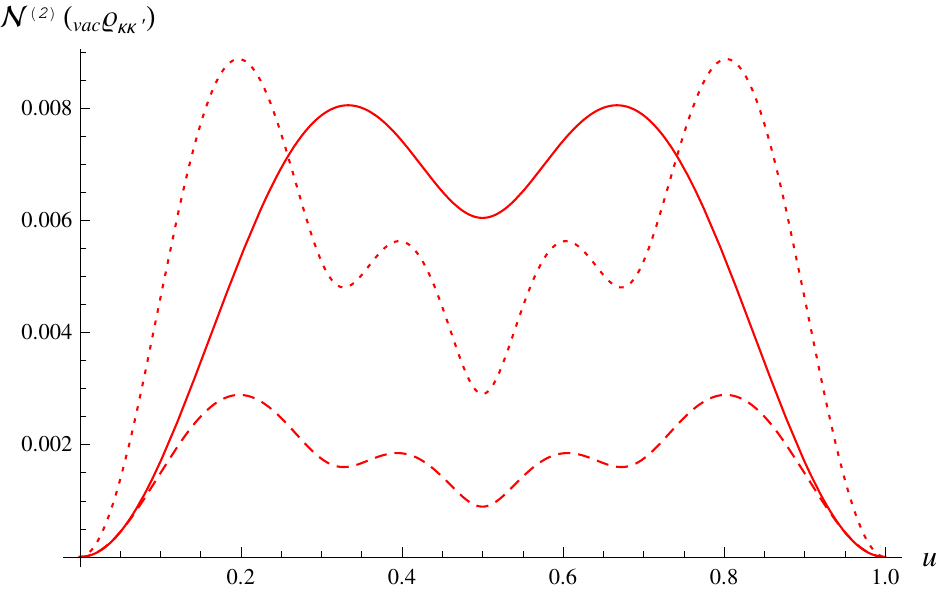}
\caption{
\textbf{Entanglement generation \textemdash\ fermionic vacuum:} The coefficients $\Negh{1}$ and $\Negh{2}$ [see Eq.~(\ref{eq:linear negativity correction fermion vac}) and~(\ref{eq:quadratic negativity correction fermion vac})] of the negativity generated from the fermionic vacuum are plotted in Fig.~\ref{fig:entanglement generation fermions vacuum}~(a) and Fig.~\ref{fig:entanglement generation fermions vacuum}~(b), respectively, for the basic building block travel scenario of Section~\ref{sec:basic building block}. The effects for the $(1+1)$ dimensional Dirac field used here are periodic in the dimensionless parameter $u:=h\tau/[4L\artanh(h/2)]\,$ [see Eq.~(\ref{eq:massless Rindler spinor frequencies})], where $\tau$ is the duration of the acceleration, as measured at the centre of the cavity. Curves are shown for the modes $(\kappa,\kappa\pr)=(0,-3),(2,-1)$ (solid), $(0,-5),(4,-1)$ (dashed), and $(3,-2),(1,-4)$ (dotted) in Fig.~\ref{fig:entanglement generation fermions vacuum}~(a), and for $(1,-1),(0,-2)$ (solid), $(1,-3),(2,-2)$ (dashed), and $(0,-4),(3,-1)$ (dotted) in Fig.~\ref{fig:entanglement generation fermions vacuum}~(b).
\label{fig:entanglement generation fermions vacuum}}
\end{figure}

A quick inspection of the off-diagonal elements of Eq.~(\ref{eq:transformed fermionic vac density matrix}) suggests to start by tracing over all modes except a particle mode labelled by~$\kappa$ and an antiparticle mode labelled by~$\kappa\pr\,$. Using~(\ref{eq:fermionic V perturbatively order}) and~(\ref{eq:generic travel scenario bogo identities fermions 1}) one quickly arrives at
\begin{align}
    {}_{\mathrm{vac}}\varrho_{\raisebox{-1pt}{\tiny{$\kappa\kappa\pr$}}}\,=\,\tr_{\lnot \kappa,\kappa\pr}\bigl(\fket{0}\!\fbra{0}\bigr)  &=\,
    \fkethat{0}\!\fbrahat{0}\,+\,h\,
    \Bigl[\Ghn{0}{\kappa\pr}\Ahmnstar{1}{\kappa\kappa\pr}
    \fkethatkp{1}{\kappa\!}\!\fkethatkm{1}{\kappa\pr}\!\fbrahat{0}+\mathrm{H.~c.}\Bigr]
    \nonumber\\[1mm]
    &\hspace*{-4.8cm}
    +\,h^{2}\,\Bigl[2\bar{f}^{A}_{\kappa\lnot\kappa\pr}\,\fkethatkp{1}{\kappa\!}\!\fbrahatkp{1}{\kappa\!}\,
    +\,2f^{A}_{\kappa\pr\lnot\kappa}\,\fkethatkm{1}{\kappa\pr\!}\!\fbrahatkm{1}{\kappa\pr\!}
    \,-\,2(\bar{f}^{A}_{\kappa\lnot\kappa\pr}+f^{A}_{\kappa\pr})\,\fkethat{0}\!\fbrahat{0}
    \label{eq:transformed fermionic vac partial trace kplus kprminus}\\[1mm]
    &\hspace*{-4.8cm} +\,|\Ahmn{1}{\kappa\kappa\pr}|^{2}\,
    \fkethatkp{1}{\kappa\!}\!\fkethatkm{1}{\kappa\pr\!}\!\fbrahatkm{1}{\kappa\pr\!}\!\fbrahatkp{1}{\kappa\!}
    \,+\,
    \Bigl(\fVhmn{2}{\kappa\kappa\pr}\fkethatkp{1}{\kappa\!}\!\fkethatkm{1}{\kappa\pr\!}\!\fbrahat{0}
    \,+\,\mathrm{H.~c.}\Bigr)\Bigr]\,+\,O(h^{3})\,,\nonumber
\end{align}
where we have defined the abbreviations
\begin{align}
    f^{A}_{m\lnot\hspace*{1pt} n} &=\,\tfrac{1}{2}\,\sum\limits_{\substack{i\geq0\\ i\neq n}}\,|\Ahmn{1}{mi}|^{2}\,,\ \ \ \
    \bar{f}^{A}_{m\lnot\hspace*{1pt} n}\,=\,\tfrac{1}{2}\,\sum\limits_{\substack{i<0\\ i\neq n}}\,|\Ahmn{1}{mi}|^{2}\,.
    \label{eq:f A and fbar A m not n}
\end{align}
For two modes we may unambiguously map the two fermionic modes to two qubits, see Section~\ref{sec:entanglement of fermionic modes} and Ref.~\cite[(\ref{Paper:FriisLeeBruschi2013})]{FriisLeeBruschi2013}, and compute entanglement measures such as the \emph{negativity}\index{negativity} (see Definition~\ref{def:negativity}) or the concurrence~(\ref{eq:wootters concurrence two qubits}) with respect to the tensor product of the two-qubit space. However, perturbative calculations of the concurrence present practical difficulties, see Ref.~\cite[(\ref{Paper:FriisLeeBruschiLouko2012})]{FriisLeeBruschiLouko2012} or Section~\ref{sec:fermionic bell states}. It is thus more convenient to compute the negativity instead, which allows for simple comparisons with our previous results for bosons. Hence, we continue by representing the partial transpose\index{partial transposition} of the two-qubit state associated to Eq.~(\ref{eq:transformed fermionic vac partial trace kplus kprminus}) as
\begin{align}
\begin{scriptsize}
    \begin{pmatrix}
    1-h^{2}\,2(\bar{f}^{A}_{\kappa\lnot\kappa\pr}+f^{A}_{\kappa\pr})  &   0   &   0   &   0   \\
    0   &   h^{2}\,2f^{A}_{\kappa\pr\lnot\kappa}   &   h\,G^{(0)}_{\kappa\pr}A^{(1)*}_{\kappa\kappa\pr}+h^{2}\,\mathcal{V}^{(2)*}_{\kappa\kappa\pr}   &   0   \\
    0   & \hspace*{-3mm}h\,G^{(0)*}_{\kappa\pr}A^{(1)}_{\kappa\kappa\pr}+h^{2}\,
    \mathcal{V}^{(2)}_{\kappa\kappa\pr}   &   h^{2}\,2\bar{f}^{A}_{\kappa\lnot\kappa\pr} &   0 \\
    0   &   0   &   0   &   h^{2}\,|A^{(1)}_{\kappa\kappa\pr}|^{2}
    \end{pmatrix}
    \end{scriptsize}\,.
    \label{eq:partial transpose fermion vacuum}
\end{align}
Specializing to the case where $(\kappa+\kappa\pr)$ is odd the linear corrections to the off-diagonal elements persist and we find the corrections to the degenerate unperturbed eigenvalues $\lambdah{0}_{1,2,3}=0$ to linear order as $\{\pm h|\Ahmn{1}{\kappa\kappa\pr}|,0\}$ using the procedure from page~\pageref{page:pert diagonalization}. We thus find the negativity that is generated from the fermionic vacuum to linear order in~$h\,$, i.e.,
\vspace*{-4mm}
\begin{align}
    \mathcal{N}({}_{\mathrm{vac}}\varrho_{\raisebox{-1pt}{\tiny{$\kappa\kappa\pr$}}})   &=\,
    h\,\Negh{1}({}_{\mathrm{vac}}\varrho_{\raisebox{-1pt}{\tiny{$\kappa\kappa\pr$}}})\,+\,O(h^{2})\,=\,
    h\,|\Ahmn{1}{\kappa\kappa\pr}|\,+\,O(h^{2})\,.
    \label{eq:linear negativity correction fermion vac}
\end{align}
Alternatively, we may select two modes with the same parity, $(\kappa+\kappa\pr)$ even, such that only corrections quadratic in~$h$ remain in~(\ref{eq:partial transpose fermion vacuum}). In this case the diagonalization of the $3\times3$ sub-block corresponding to the unperturbed eigenvalue~$0$ also has one possibly negative eigenvalue and the negativity
\vspace*{-1mm}
\begin{align}
    \begin{small}\begin{matrix}
    \mathcal{N}({}_{\mathrm{vac}}\varrho_{\raisebox{-1pt}{\tiny{$\kappa\kappa\pr$}}})   
    =h^{2}\max\Bigl\{0,
    \sqrt{\bigl(\bar{f}^{A}_{\kappa\lnot\kappa\pr}-f^{A}_{\kappa\pr\lnot \kappa}\bigr)^{2}+|\mathcal{V}^{(2)}_{\kappa\kappa\pr}|^{2}}
    -\bigl(\bar{f}^{A}_{\kappa\lnot\kappa\pr}+f^{A}_{\kappa\pr\lnot\kappa}\bigr)\Bigr\}+O(h^{3})
    \end{matrix}
    \end{small}
    \label{eq:quadratic negativity correction fermion vac}
\end{align}
is obtained, see Fig.~\ref{fig:entanglement generation fermions vacuum}. Note the similarity between the bosonic and fermionic case by comparing Eq.~(\ref{eq:linear negativity correction boson vac}) with~(\ref{eq:linear negativity correction fermion vac}), and Eq.~(\ref{eq:quadratic negativity correction boson vac}) with~(\ref{eq:quadratic negativity correction fermion vac}), respectively.
\vspace*{-1mm}
\index{vacuum!fermions|)}

\subsection{Entanglement from the Fermionic Particle State}\label{sec:entanglement from the fermionic particle state}

\begin{figure}[hb!]
\centering
(a)\includegraphics[width=0.73\textwidth]{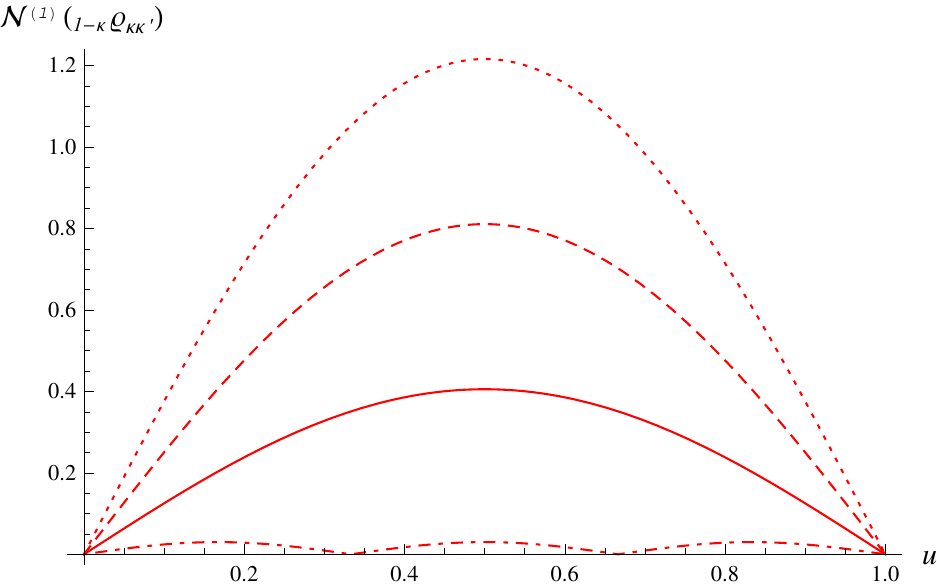}
(b)\includegraphics[width=0.73\textwidth]{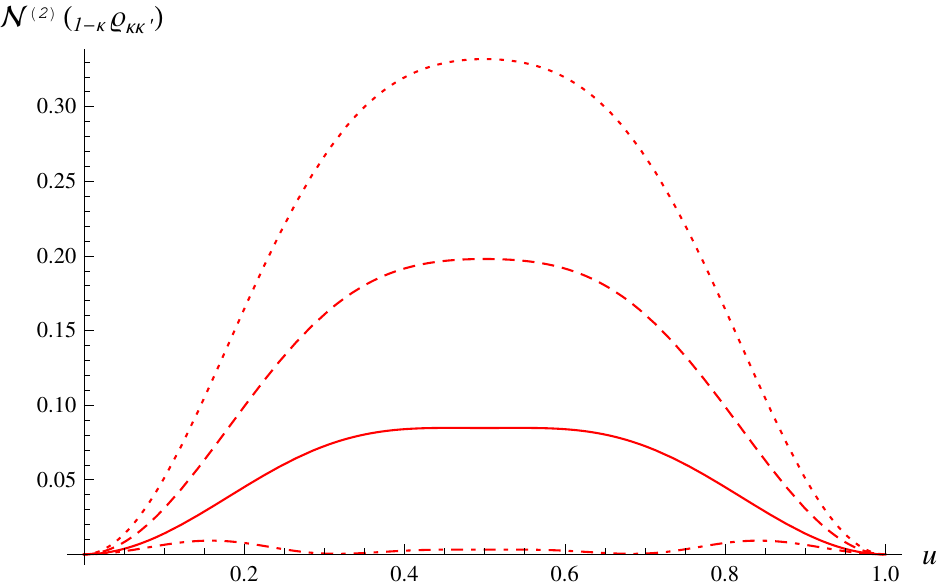}
\caption{
\textbf{Entanglement generation \textemdash\ fermionic particle state:} The coefficients $\Negh{1}$ and $\Negh{2}$ [see Eq.~(\ref{eq:linear negativity correction fermion 1k pos}) and~(\ref{eq:quadratic negativity correction fermion 1k pos})] of the negativity generated from $\fketp{1_{\kappa}}$ are plotted in Fig.~\ref{fig:entanglement generation fermions 1k pos}~(a) and Fig.~\ref{fig:entanglement generation fermions 1k pos}~(b), respectively, for the basic building block travel scenario of Section~\ref{sec:basic building block}. The effects for the $(1+1)$ dimensional Dirac field used here are periodic in the dimensionless parameter $u:=h\tau/[4L\artanh(h/2)]\,$ [see Eq.~(\ref{eq:massless Rindler spinor frequencies})], where $\tau$ is the duration of the acceleration, as measured at the centre of the cavity. Curves are shown for the modes $(\kappa,\kappa\pr)=(0,1)$ (solid), $(1,2)$ (dashed), $(2,3)$ (dotted), and $(0,3)$ (dotted-dashed) in Fig.~\ref{fig:entanglement generation fermions 1k pos}~(a), and for $(0,2)$ (solid), $(1,3)$ (dashed), $(2,4)$ (dotted), and $(0,4)$ (dotted-dashed) in Fig.~\ref{fig:entanglement generation fermions 1k pos}~(b).
\label{fig:entanglement generation fermions 1k pos}}
\end{figure}

Let us pursue the same strategy for the single fermion state as chosen before for the fermionic vacuum in Section~\ref{sec:entanglement from the fermionic vacuum}, i.e., we consider the density matrix of Eq.~(\ref{eq:fermion single particle density operator bogo}) and trace over (see Section~\ref{sec:partial trace ambiguity}) all modes except for~$\kappa\geq0$ and~$\kappa\pr<0$ to arrive at
\begin{align}
    \tr_{\lnot\kappa,\kappa\pr}\bigl(\fketp{1_{\kappa}}\!\fbrap{1_{\kappa}}\bigr) &=\,
    \fkethatkp{1}{\kappa}\!\fbrahatkp{1}{\kappa}\,+\,h^{2}\,\Bigl[\,2\,f^{A}_{\kappa\pr\lnot\kappa}\,
    \fkethatkp{1}{\kappa}\!\fkethatkm{1}{\kappa\pr\!}\!\fbrahatkm{1}{\kappa\pr\!}\fbrahatkp{1}{\kappa}
    \nonumber\\[1mm]
    &\ -\,2\,\bigl(f^{A}_{\kappa\pr\lnot\kappa}\,+\,f^{A}_{\kappa}\bigr)
    \fkethatkp{1}{\kappa}\!\fbrahatkp{1}{\kappa}\,+\,2\,f^{A}_{\kappa}\,\fkethat{0}\!\fbrahat{0}
    \label{eq:fermion 1k traced over k pos kp neg}\\[1mm]
    \mbox{\scriptsize{(for\ $\kappa\geq0,\,\kappa\pr<0$)}}\hspace*{5mm}&\ -\,
    \Bigl(\Ghnstar{0}{\kappa}\Ghn{0}{\kappa\pr\!}\sum\limits_{m\geq0}\Ahmn{1}{m\kappa}\Ahmnstar{1}{m\kappa\pr}
    \fkethatkp{1}{\kappa}\!\fkethatkm{1}{\kappa\pr\!}\!\fbrahat{0}\,+\,\mathrm{H.~c.}\Bigr)\Bigr]+O(h^{3})\,,
    \nonumber
\end{align}
where $f^{A}_{\kappa}$ and $f^{A}_{\kappa\pr\lnot\kappa}$ are as in Eq.~(\ref{eq:f A and fbar A m not n}). Mapping this state to two qubits it is quite straightforward to see that the subspace of degenerate eigenvalues of the partial transpose\index{partial transposition} of has no negative corrections to second order in~$h\,$. Thus, no entanglement is generated from the state $\fket{1_{\kappa}}$ between any chosen pair of modes with opposite sign of frequency. However, we may select two modes of positive frequency instead, such that $\kappa,\kappa\pr\geq0\,$. One then finds the reduced state
\begin{align}
    \tr_{\lnot\kappa,\kappa\pr}\bigl(\fketp{1_{\kappa}}\!\fbrap{1_{\kappa}}\bigr) &=\,
    \fkethatkp{1}{\kappa}\!\fbrahatkp{1}{\kappa}\,-\,h\,\Bigl[\fVhmn{1}{\kappa\kappa\pr}
    \fkethatkp{1}{\kappa\!}\!\fbrahatkp{1}{\kappa\pr\!}\,+\,\mathrm{H.~c.}\Bigr]
    \label{eq:fermion 1k traced over k pos kp pos}\\[1mm]
    &\hspace*{-2.5cm}+\,h^{2}\,\Bigl[\,2\,f^{A}_{\kappa\lnot\kappa\pr}\,\fkethat{0}\!\fbrahat{0}\,+\,
    |\Ahmn{1}{\kappa\kappa\pr}|^{2}\fkethatkp{1}{\kappa\pr\!}\!\fbrahatkp{1}{\kappa\pr\!}
    \,-\,\bigl(f^{A}_{\kappa}\,+\,\bar{f}^{A}_{\kappa\pr}\bigr)\fkethatkp{1}{\kappa}\!\fbrahatkp{1}{\kappa}
    \nonumber\\[1mm]
    &\hspace*{-2.5cm}+\,2\,\bar{f}^{A}_{\kappa\pr}\,
    \fkethatkp{1}{\kappa}\!\fkethatkp{1}{\kappa\pr\!}\!\fbrahatkp{1}{\kappa\pr\!}\fbrahatkp{1}{\kappa}
    \,-\,\Bigl(\fVhmn{2}{\kappa\kappa\pr}\,\fkethatkp{1}{\kappa\!}\!\fbrahatkp{1}{\kappa\pr\!}\,+\,
    \mathrm{H.~c.}\Bigr)\Bigr]\,+\,O(h^{3})\,,
    \nonumber\\[-1mm]
\mbox{\scriptsize{(for\ $\kappa,\kappa\pr\geq0$)}}\hspace*{15mm}&\
    \nonumber
\end{align}
where we have used the Bogoliubov identities~(\ref{eq:generic travel scenario bogo identities fermions 1}) and~(\ref{eq:generic travel scenario bogo identities fermions 2}), and the definition of the components $\fVmn{pq}\ (p\geq0,q<0)$ has been extended to indices of all sign combinations via their perturbative expansions in~(\ref{eq:fermionic V perturbatively order}), such that the indices $p$ and $q$ can take on both negative and non-negative values. For two fermionic modes we can consistently represent (see Section~\ref{sec:entanglement of fermionic modes}) the reduced state to second order in~$h$ as a two-qubit density matrix ${}_{\operatorname{1-\kappa}}\varrho_{\raisebox{-1pt}{\tiny{$\kappa\kappa\pr$}}}$ with partial transpose
\begin{align}
    \begin{pmatrix}
    2\,h^{2}\,f^{A}_{\kappa\lnot\kappa\pr}   &   0   &   0   &
        -\,h\,\fVhmnstar{1}{\kappa\kappa\pr}\,-\,h^{2}\,\fVhmnstar{2}{\kappa\kappa\pr}   \\
    0   &   h^{2}\,|\Ahmn{1}{\kappa\kappa\pr}|^{2} &  0   &   0   \\
    0   &   0   &   1\,-\,h^{2}\,(f^{A}_{\kappa}+\bar{f}^{A}_{\kappa\pr})   &   0   \\
    -\,h\,\fVhmn{1}{\kappa\kappa\pr}\,-\,h^{2}\,\fVhmn{2}{\kappa\kappa\pr}   &   0   &   0   &
        2\,h^{2}\,\bar{f}^{A}_{\kappa\pr}
    \end{pmatrix}\,.
    \label{eq:fermion 1k traced over k pos kp pos part transp}
\end{align}
If the modes $\kappa$ and $\kappa\pr$ have opposite parity, $(\kappa+\kappa\pr)$ is odd, we find the negativity to be
\begin{align}
    \mathcal{N}({}_{\operatorname{1-\kappa}}\varrho_{\raisebox{-1pt}{\tiny{$\kappa\kappa\pr$}}})   &=\,
    h\,\Negh{1}({}_{\operatorname{1-\kappa}}\varrho_{\raisebox{-1pt}{\tiny{$\kappa\kappa\pr$}}})\,+\,O(h^{2})\,=\,
    h\,|\Ahmn{1}{\kappa\kappa\pr}|\,+\,O(h^{2})\,,
    \label{eq:linear negativity correction fermion 1k pos}
\end{align}
formally the same expression as in Eq.~(\ref{eq:linear negativity correction fermion vac}), but with the appropriate non-negative
\newpage
\noindent
value for~$\kappa\pr$. Similarly, if the modes have the same parity, i.e., if $(\kappa+\kappa\pr)$ is even, we find
\begin{align}
    \mathcal{N}({}_{\operatorname{1-\kappa}}\varrho_{\raisebox{-1pt}{\tiny{$\kappa\kappa\pr$}}})   &=\,
    h^{2}\,\Negh{2}({}_{\operatorname{1-\kappa}}\varrho_{\raisebox{-1pt}{\tiny{$\kappa\kappa\pr$}}})\,+\,O(h^{3})
    \label{eq:quadratic negativity correction fermion 1k pos}\\[1mm]
    &\ =\,h^{2}\max\Bigl\{0,
    \sqrt{\bigl(f^{A}_{\kappa\lnot\kappa\pr}\,-\,\bar{f}^{A}_{\kappa\pr}\bigr)^{2}\,+\,
    |\fVhmn{2}{\kappa\kappa\pr}|^{2}}
    \,-\,\bigl(f^{A}_{\kappa\lnot\kappa\pr}\,+\,\bar{f}^{A}_{\kappa\pr}\bigr)\Bigr\}\,+\,O(h^{3})\,.
    \nonumber
\end{align}
An illustration of the entanglement generated from $\fketp{1_{\kappa}}$ is shown in Fig.~\ref{fig:entanglement generation fermions 1k pos}.

\subsubsection{Entanglement from the Fermionic Antiparticle State}

The analysis of Section~\ref{sec:entanglement from the fermionic particle state} can be repeated step by step if we start instead from a single-antiparticle state $\fketm{1_{\hat{\kappa}}}$ and trace over all modes except two modes $\hat{\kappa},\hat{\kappa}\pr<0\,$. However, since the positive and negative frequency solutions appear symmetrically in the spectrum (see Section~\ref{sec:Dirac Fields in Rigid Cavities}) the corresponding results can be directly obtained by setting $\kappa$ and $\kappa\pr$ in Eq.~(\ref{eq:linear negativity correction fermion 1k pos}) and~(\ref{eq:quadratic negativity correction fermion 1k pos}) to $|\hat{\kappa}+1|$ and $|\hat{\kappa}\pr+1|\,$. The same is true for the sample plots in Fig.~\ref{fig:entanglement generation fermions 1k pos}.

\subsection{Entanglement from the Fermionic Particle-Antiparticle Pair}\label{sec:entanglement from the fermionic particle-antiparticle pair}

\begin{figure}[hb!]
\centering
\includegraphics[width=0.73\textwidth]{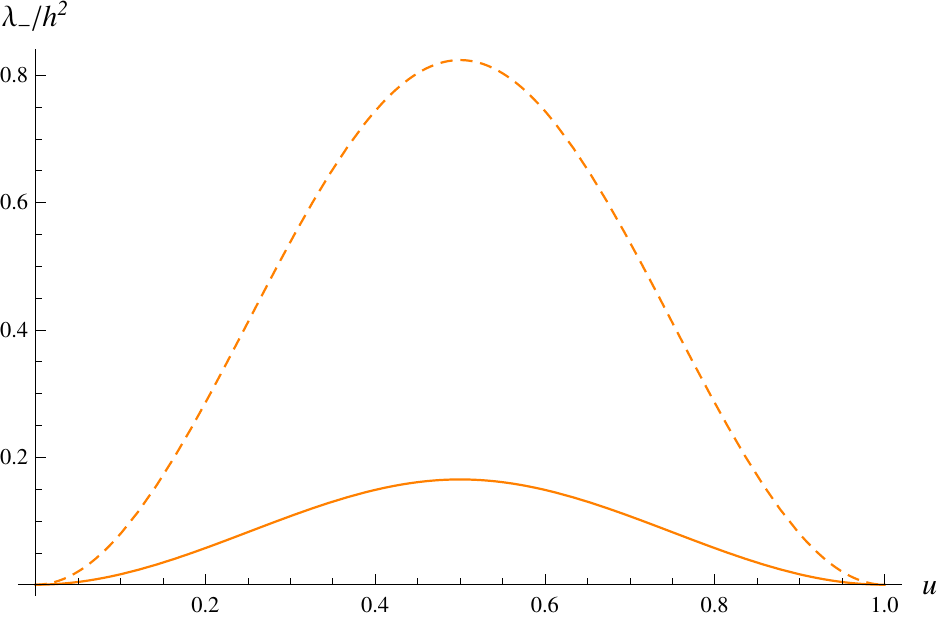}
\caption{
\textbf{Entanglement generation \textemdash\ fermionic particle-antiparticle pair:}
The second order coefficient of the only possibly negative eigenvalue $\lambda_{-}=h^{2}[f^{A}_{\kappa}+\bar{f}^{A}_{\kappa}]-h^{2}
\sqrt{[f^{A}_{\kappa}-\bar{f}^{A}_{\kappa\pr}]^{2}+|\fVhmn{2}{\kappa\kappa\pr}|^{2}}$
of the partial transpose of the two qubit representation for the state Eq.~(\ref{eq:fermion 1k 1kp traced over k pos kp neg}) is shown. Curves are plotted for the basic building block travel scenario of Section~\ref{sec:basic building block} for a $(1+1)$ dimensional Dirac field against the dimensionless parameter $u:=h\tau/[4L\artanh(h/2)]\,$ [see Eq.~(\ref{eq:massless Rindler spinor frequencies})], where $\tau$ is the duration of the acceleration, as measured at the centre of the cavity. Curves are shown for the modes $(\kappa,\kappa\pr)=(1,-1),(0,-2),(0,-4),(3,-1)$ (solid), and $(1,-3),(2,-2)$ (dashed). Since the correction is positive throughout no entanglement is generated.
\label{fig:entanglement generation fermions 1k 1kp}}
\end{figure}

For the initial state $\fketp{1_{\kappa}}\!\fketm{1_{\kappa\pr\!}}$ of a pair of one particle $(\kappa\geq0)$ and one antiparticle $(\kappa\pr<0)$ we expand Eq.~(\ref{eq:fermion particle antiparticle pair pure bogo}) to second order in~$h\,$. Keeping terms proportional to~$h^{2}$ and tracing over all modes except~$\kappa$ and~$\kappa\pr$ one arrives at the expression
\begin{align}
    \tr_{\lnot\kappa,\kappa\pr}\bigl(\fketp{1_{\kappa}}\!\fketm{1_{\kappa\pr\!}}\!
    \fbram{1_{\kappa\pr\!}}\!\fbrap{1_{\kappa}}\bigr) &=\,
    \fkethatkp{1}{\kappa}\!\fkethatkm{1}{\kappa\pr\!}\!\fbrahatkm{1}{\kappa\pr\!}\!\fbrahatkp{1}{\kappa}
    \label{eq:fermion 1k 1kp traced over k pos kp neg}\\[1.5mm]
    &\hspace*{-6cm}-\,h\,\Bigl[\fVhmn{1}{\kappa\kappa\pr}
    \fkethatkp{1}{\kappa\!}\!\fkethatkm{1}{\kappa\pr\!}\!\fbrahat{0}\,+\,\mathrm{H.~c.}\Bigr]
    \,+\,h^{2}\,\Bigl[\,|\Ahmn{1}{\kappa\kappa\pr}|^{2}\,\fkethat{0}\!\fbrahat{0}\,+\,
    2\,f^{A}_{\kappa}\,\fkethatkm{1}{\kappa\pr\!}\!\fbrahatkm{1}{\kappa\pr\!}\,
    \nonumber\\[1.5mm]
    &\hspace*{-6cm}+\,2\,\bar{f}^{A}_{\kappa\pr}\,\fkethatkp{1}{\kappa}\!\fbrahatkp{1}{\kappa}
    \,-\,\bigl(2\,^{A}_{\kappa}\,+\,2\,\bar{f}^{A}_{\kappa\pr}\,+\,|\Ahmn{1}{\kappa\kappa\pr}|^{2}\bigr)
    \fkethatkp{1}{\kappa}\!\fkethatkm{1}{\kappa\pr\!}\!\fbrahatkm{1}{\kappa\pr\!}\fbrahatkp{1}{\kappa}
    \nonumber\\[1.5mm]
    &\hspace*{-6cm}-\,\Bigl(\fVhmn{2}{\kappa\kappa\pr}\,
    \fkethatkp{1}{\kappa\!}\!\fkethatkm{1}{\kappa\pr\!}\!\fbrahat{0}\,+\,
    \mathrm{H.~c.}\Bigr)\Bigr]\,+\,O(h^{3})\,.
    \nonumber
\end{align}
Comparing with the case for the fermionic vacuum one immediately finds the negativity
\begin{align}
    \mathcal{N}({}_{\operatorname{1-\kappa,1-\kappa\pr}}\varrho_{\raisebox{-1pt}{\tiny{$\kappa\kappa\pr$}}})   &=\,
    h\,\Negh{1}({}_{\operatorname{1-\kappa,1-\kappa\pr}}\varrho_{\raisebox{-1pt}{\tiny{$\kappa\kappa\pr$}}})\,+\,O(h^{2})\,=\,
    h\,|\Ahmn{1}{\kappa\kappa\pr}|\,+\,O(h^{2})\,,
    \label{eq:linear negativity correction fermion 1k 1kp}
\end{align}
for modes with opposite parity, that is, if $(\kappa+\kappa\pr)$ is odd, while mode pairs with equal parity provide a correction to the negativity that is quadratic in~$h\,$, i.e.,
\begin{align}
    \mathcal{N}({}_{\operatorname{1-\kappa,1-\kappa\pr}}\varrho_{\raisebox{-1pt}{\tiny{$\kappa\kappa\pr$}}})   &=\,
    h^{2}\,\Negh{2}({}_{\operatorname{1-\kappa,1-\kappa\pr}}\varrho_{\raisebox{-1pt}{\tiny{$\kappa\kappa\pr$}}})\,+\,O(h^{3})
    \label{eq:quadratic negativity correction fermion 1k 1kp}\\[1mm]
    &\ =\,h^{2}\max\Bigl\{0,
    \sqrt{\bigl(f^{A}_{\kappa}\,-\,\bar{f}^{A}_{\kappa\pr}\bigr)^{2}\,+\,|\fVhmn{2}{\kappa\kappa\pr}|^{2}}
    \,-\,\bigl(f^{A}_{\kappa}\,+\,\bar{f}^{A}_{\kappa}\bigr)\Bigr\}\,+\,O(h^{3})\,.
    \nonumber
\end{align}
Formally, these expressions are remarkably similar to the results for the fermionic vacuum in Eq.~(\ref{eq:linear negativity correction fermion vac}) and Eq.~(\ref{eq:quadratic negativity correction fermion vac}). In fact, the linear corrections are exactly the same and one may consult Fig.~\ref{fig:entanglement generation fermions vacuum}~(a) for an illustration. The quadratic corrections, on the other hand, are slightly different. The sums $\bar{f}^{A}_{\kappa}$ and $f^{A}_{\kappa\pr}$ in Eq.~(\ref{eq:quadratic negativity correction fermion vac}) represent particle creation coefficients, while the quantities $f^{A}_{\kappa}$ and $\bar{f}^{A}_{\kappa\pr}$ in Eq.~(\ref{eq:quadratic negativity correction fermion 1k 1kp}) are responsible for shifting excitations from the mode $\kappa$\ $(\kappa\pr)$ to other positive (negative) frequency modes \textemdash\ mode mixing $\alpha$-type coefficients. This alteration makes for all the difference: even for a choice of modes with minimal energy for a $(1+1)$ dimensional Dirac field the entanglement generation by the particle creation coefficient $|\fVhmn{2}{\kappa\kappa\pr}|$ cannot compete with the much larger contributions by $f^{A}_{\kappa}$ and $\bar{f}^{A}_{\kappa\pr}$ that add noise to the reduced state. The partial transpose\index{partial transposition} of the two-qubit density matrix representing the state~(\ref{eq:fermion 1k 1kp traced over k pos kp neg}) has one possibly negative eigenvalue, but graphical analysis shows it remains non-negative, see Fig.~\ref{fig:entanglement generation fermions 1k 1kp}. For higher dimensions, higher mode numbers, or increased mass, the noise introduced by $f^{A}_{\kappa}$ and $\bar{f}^{A}_{\kappa\pr}$ will only increase. Consequently, no entanglement is created from the state $\fketp{1_{\kappa}}\fketm{1_{\kappa\pr\!}}$ for mode pairs $(\kappa,\kappa\pr)$ with equal parity.

\section{Generation of Genuine Multipartite Entanglement}
\label{sec:Generation of Genuine Multipartite Entanglement}

The investigation carried out up to this point has revealed that entanglement is created by the non-uniform motion between pairs of modes of the quantum fields. The amount of entanglement and, indeed, if any entanglement is created at all, depends on the choice of initial state and chosen modes. In particular, the fermionic systems suffer from limitations in the creation of entanglement due to the Pauli exclusion principle\index{Pauli exclusion principle}, while bosonic systems are more susceptible to the required particle creation and shifting of excitations. Conceptually, it is of further interest to learn how the quantum correlations connect more than two modes. We may ask if and how \emph{genuine multipartite entanglement}\index{entanglement!genuine multipartite} (GME) emerges from the Bogoliubov transformations. The analysis here is based on material published in Ref.~\cite[(\ref{Paper:FriisHuberFuentesBruschi2012})]{FriisHuberFuentesBruschi2012}. For simplicity we restrict our deliberations to the multipartite entanglement of the transformed vacuum states, the bosonic vacuum in Section~\ref{sec:Genuine Multipartite Entanglement from the Bosonic Vacuum}, and the fermionic counterpart in Section~\ref{sec:Genuine Multipartite Entanglement from the Fermionic Vacuum}. It is further useful to recall the discussion of Section~\ref{sec:multipartite entanglement} for the basic concepts and definitions.

\subsection{Genuine Multipartite Entanglement \textemdash\ Bosonic Vacuum}\label{sec:Genuine Multipartite Entanglement from the Bosonic Vacuum}

For the case of bosonic GME we return to the transformed vacuum state of Eq.~(\ref{eq:transformed boson vac density matrix}) and we reduce the state to three modes, $k$, $k\pr$, and $k\prpr$. At this stage we specialize to the case where not all three modes have the same parity. Without loss of generality we pick $(k+k\pr)$ and $(k\pr+k\prpr)$ to be odd, which implies that $(k+k\prpr)$ is even, such that the first order coefficient $\Vhmn{1}{kk\prpr}$ as well as the second order coefficients $\Vhmn{2}{kk\pr}$ and $\Vhmn{2}{k\pr\hspace*{-1pt}k\prpr}$ vanish. One then obtains the reduced state
\begin{align}
    {}_{\mathrm{vac}}\rho_{\raisebox{-1pt}{\tiny{$kk\pr\hspace*{-1pt}k\prpr$}}}   &:=
        \tr_{\lnot k,k\pr\hspace*{-1pt},k\prpr}\bigl(\ket{0}\!\bra{0}\bigr) =\kethat{0}\!\brahat{0}-h
        \Bigl[\Vhmn{1}{kk\pr}\kethatk{1}{k}\!\kethatk{1}{k\pr\!}\!\brahat{0}+
        \Vhmn{1}{k\pr\hspace*{-1pt}k\prpr}\kethatk{1}{k\pr\!}\!\kethatk{1}{k\prpr\!}\!\brahat{0}
        \nonumber\\[1.5mm]
    &\hspace*{-1.2cm}+\mathrm{H.~c.}\Bigr] +h^{2}\Bigl[
        2f^{\beta}_{k\lnot k\pr}\kethatk{1}{k}\!\brahatk{1}{k}
        +2f^{\beta}_{k\pr\lnot k,k\prpr}\kethatk{1}{k\pr\!}\!\brahatk{1}{k\pr\!}
        +2f^{\beta}_{k\prpr\lnot k\pr}\kethatk{1}{k\prpr\!}\!\brahatk{1}{k\prpr\!}
        -2\bigl(f^{\beta}_{k\lnot k\pr}
        \nonumber\\[1.5mm]
    &\hspace*{-1.2cm}
        +f^{\beta}_{k\pr\lnot k\prpr}+f^{\beta}_{k\prpr}\bigr)\kethat{0}\!\brahat{0}
        +|\betahmn{1}{kk\pr}|^{2}\kethatk{1}{k}\!\kethatk{1}{k\pr\!}\!\brahatk{1}{k\pr\!}\!\brahatk{1}{k}
        +|\betahmn{1}{k\pr \hspace*{-1pt}k\prpr}|^{2}
        \kethatk{1}{k\pr\!}\!\kethatk{1}{k\prpr\!}\!\brahatk{1}{k\prpr\!}\!\brahatk{1}{k\pr\!}
        \nonumber\\[1.5mm]
    &\hspace*{-1.2cm}
        +\Bigl(\Vhmn{2}{kk\prpr}\kethatk{1}{k}\!\kethatk{1}{k\prpr\!}\!\brahat{0}
        +\tfrac{1}{\sqrt{2}}\Vhmn{2}{kk}\kethatk{2}{k}\!\brahat{0}
        +\tfrac{1}{\sqrt{2}}\Vhmn{2}{k\pr\hspace*{-1pt}k\pr}\kethatk{2}{k\pr\!}\!\brahat{0}
        +\tfrac{1}{\sqrt{2}}\Vhmn{2}{k\prpr\hspace*{-1pt}k\prpr}\kethatk{2}{k\prpr\!}\!\brahat{0}
        \nonumber\\[1.5mm]
    &\hspace*{-1.2cm}
        +(\Vhmn{1}{kk\pr})^{2}\kethatk{2}{k}\!\kethatk{2}{k\pr\!}\!\kethat{0}
        +(\Vhmn{1}{k\pr\hspace*{-1pt}k\prpr})^{2}\kethatk{2}{k\pr\!}\!\kethatk{2}{k\prpr\!}\!\kethat{0}
        +\sqrt{2}\,\Vhmn{1}{kk\pr}\Vhmn{1}{k\prpr\hspace*{-1pt}k\pr}
        \kethatk{1}{k}\!\kethatk{2}{k\pr\!}\!\kethatk{1}{k\prpr\!}\!\brahat{0}
        \nonumber\\[1.5mm]
    &\hspace*{-1.2cm}
        +\sum\limits_{p\neq k\pr}\Vhmn{1}{pk}\Vhmn{1}{pk\prpr}\kethatk{1}{k}\!\brahatk{1}{k\prpr\!}
        \,+\,\mathrm{H.~c.}\Bigr)\Bigr]\,+\,O(h^{3})\,,
\label{eq:bosonic vac bogo traced k kpr and kprpr}
\end{align}
where $\Vhmn{1}{mn}$ and $\Vhmn{2}{mn}$ are given by~(\ref{eq:bosonic V expansion 1}) and~(\ref{eq:bosonic V expansion 2}), respectively. In the face of the complicated decomposition of the reduced state~(\ref{eq:bosonic vac bogo traced k kpr and kprpr}) a simple method for the detection of GME is invaluable. It is particularly convenient to invoke the \emph{GME witness inequalities} of Theorem~\ref{thm:GME witness theorem}. Keeping terms proportional to~$h^{2}$ in the Taylor-Maclaurin expansion effectively truncates the problem at hand to a three qutrit system, i.e., each mode is mapped to a three dimensional Hilbert space. For this situation we use the techniques from \cite{GabrielHiesmayrHuber2010,MaChenChenSpenglerGabrielHuber2011,WuKampermannBruszKloecklHuber2012} to construct the particular witness inequality~\cite{FriisHuberFuentesBruschi2012}
\begin{align}
    2\,\Bigl(\,\left|\bra{\tilde{0}}
    {}_{\mathrm{vac}}\rho_{\raisebox{-1pt}{\tiny{$kk\pr\hspace*{-1pt} k\prpr$}}}
    \ket{\!\tilde{1}_{k}\!}\ket{\!\tilde{2}_{k\pr}\!}\ket{\!\tilde{1}_{k\prpr}\!}\right|
    &-\,\sqrt{\bra{\!\tilde{1}_{k}\!}
        {}_{\mathrm{vac}}\rho_{\raisebox{-1pt}{\tiny{$kk\pr\hspace*{-1pt} k\prpr$}}}
        \ket{\!\tilde{1}_{k}\!}
        \bra{\!\tilde{1}_{k\prpr}\!}\bra{\!\tilde{2}_{k\pr}\!}
        {}_{\mathrm{vac}}\rho_{\raisebox{-1pt}{\tiny{$kk\pr\hspace*{-1pt} k\prpr$}}}
        \ket{\!\tilde{2}_{k\pr}\!}\ket{\!\tilde{1}_{k\prpr}\!}}
    \nonumber\\[1mm]
    &\hspace*{-3cm}-\,\sqrt{\bra{\!\tilde{1}_{k\prpr}\!}
        {}_{\mathrm{vac}}\rho_{\raisebox{-1pt}{\tiny{$kk\pr\hspace*{-1pt} k\prpr$}}}
        \ket{\!\tilde{1}_{k\prpr}\!}
        \bra{\!\tilde{2}_{k\pr}\!}\bra{\!\tilde{1}_{k}\!}
        {}_{\mathrm{vac}}\rho_{\raisebox{-1pt}{\tiny{$kk\pr\hspace*{-1pt} k\prpr$}}}
        \ket{\!\tilde{1}_{k}\!}\ket{\!\tilde{2}_{k\pr}\!}}
    \label{eq:BosonsB GME witness complete}\\[1mm]
    &\hspace*{-3cm}-\,\sqrt{\bra{\!\tilde{2}_{k\pr}\!}
        {}_{\mathrm{vac}}\rho_{\raisebox{-1pt}{\tiny{$kk\pr\hspace*{-1pt} k\prpr$}}}
        \ket{\!\tilde{2}_{k\pr}\!}
        \bra{\!\tilde{1}_{k\prpr}\!}\bra{\!\tilde{1}_{k}\!}
        {}_{\mathrm{vac}}\rho_{\raisebox{-1pt}{\tiny{$kk\pr\hspace*{-1pt} k\prpr$}}}
        \ket{\!\tilde{1}_{k}\!}\ket{\!\tilde{1}_{k\prpr}\!}}
    \,\Bigr)\,\leq0\,.\nonumber
\end{align}
\newpage
As we have discussed in Section~\ref{sec:multipartite entanglement} such inequalities are always satisfied by any bi-separable states and their violation therefore unambiguously detects GME. Moreover, the violation of this type of inequality can be regarded as a lower bound to actual measures of GME, see Refs.~\cite{WuKampermannBruszKloecklHuber2012}. Performing the perturbative expansion of Eq.~(\ref{eq:BosonsB GME witness complete}) we find the simple inequality
\begin{align}
    2\,\bigl|\brahat{0}{}_{\mathrm{vac}}\rho_{\raisebox{-1pt}{\tiny{$kk\pr\hspace*{-1pt}k\prpr$}}}
    \kethatk{1}{k}\!\kethatk{2}{k\pr\!}\!\kethatk{1}{k\prpr\!}\bigr|
    \,-\,O(h^{3})   &=\,
    2\sqrt{2}\,h^{2}\,|\,\betahmn{1}{kk\pr}|\,|\,\betahmn{1}{k\pr\hspace*{-1pt}k\prpr}|
    \,-\,O(h^{3}) &\leq0\,.
    \label{eq:BosonsB GME witness}
\end{align}
We find that the inequality indeed is generally violated, showing that GME is created between the three chosen modes by coherently exciting pairs of particles in $(k,k\pr)$ and $(k\pr,k\prpr)$. Moreover, a quick glance at Fig.~\ref{fig:entanglement resonances} reveals that joint \emph{entanglement resonances}\index{entanglement!resonances} can occur. For instance, the individual coefficients $|\,\betahmn{1}{kk\pr}|$ and $|\,\betahmn{1}{k\pr\hspace*{-1pt}k\prpr}|$ for a massless scalar field in $(1+1)$ dimensions increase linearly with the number of repetitions of some basic travel scenario when the basic travel time (as measured at the centre of the cavity) is $\tau=2nL/(k+k\pr)$ or $\tau=2mL/(k\pr+k\prpr)$, $n,m\in\mathds{N}_{\!+}$, respectively, see Eq.~(\ref{eq:resonance duration}). Both of these resonances coincide if $n=p(k+k\pr)$ and $m=p(k\pr+k\prpr)$, such that $\tau=2pL$, $p\in\mathds{N}_{\!+}$, see Fig.~\ref{fig:entanglement resonances}.\\

At this resonance time, which happens to be independent of the chosen modes, the lower bound on the GME increases quadratically with the number $N$ of repetitions of the basic travel scenario. Simultaneously, the terms $f^{\beta}_{k\lnot k\pr}$, $f^{\beta}_{k\pr\lnot k,k\prpr}\,$, and $f^{\beta}_{k\prpr\lnot k\pr}\,$, which introduce mixedness into the reduced state, scale quadratically at the mode-independent resonance. Nonetheless, the validity of the
perturbative approach is guaranteed because all second order terms are at most proportional to
$N^{2}h^{2}\ll N h$, which in turn is required to be much smaller than~$1$ if the perturbative approach is to be justified.\\
\vspace*{3mm}

\subsection{Genuine Multipartite Entanglement \textemdash\ Fermionic Vacuum}\label{sec:Genuine Multipartite Entanglement from the Fermionic Vacuum}

For the fermionic counterpart of the situation studied in Section~\ref{sec:Genuine Multipartite Entanglement from the Bosonic Vacuum} we may also select three modes $\kappa$, $\kappa\pr$, and $\kappa\prpr$, that do not all have the same parity. In addition, we now have the choice between positive and negative frequency modes. The analysis of Section~\ref{sec:entanglement from the fermionic vacuum} taught us that entanglement is generated from the vacuum between modes of opposite frequency sign. We thus choose two positive frequency modes, $\kappa\geq0$ and $\kappa\pr\geq0$, of the same parity, $(\kappa+\kappa\pr)$ is even, while the third mode $\kappa\prpr<0$ is selected from the negative frequencies such that it has opposite parity to the particle modes, i.e., $(\kappa+\kappa\prpr)$ and $(\kappa\pr+\kappa\prpr)$ are odd. Tracing out all other modes from the transformed vacuum of Eq.~(\ref{eq:transformed fermionic vac density matrix}) we arrive at
\begin{align}
    {}_{\mathrm{vac}}\varrho_{\raisebox{-1pt}{\tiny{$\kappa\kappa\pr\hspace*{-1pt}\kappa\prpr$}}}\,=\,
        \tr_{\lnot \kappa,\kappa\pr\hspace*{-1pt}\kappa\prpr}\bigl(\fket{0}\!\fbra{0}\bigr)  &=\,
        \fkethat{0}\!\fbrahat{0}\,+\,h\,\Bigl[
        \fVhmn{1}{\kappa\kappa\prpr}\,\fkethatkp{1}{\kappa\!}\!\fkethatkm{1}{\kappa\prpr\!}\!\fbrahat{0}
        \label{eq:transformed fermionic vac partial trace kplus kprplus kprpr minus}\\[1.5mm]
    &\hspace{-5.3cm}+\,\fVhmn{1}{\kappa\pr\hspace*{-1pt}\kappa\prpr}\,
        \fkethatkp{1}{\kappa\pr\!}\!\fkethatkm{1}{\kappa\prpr\!}\!\fbrahat{0}
        \,+\,\mathrm{H.~c.}\Bigr]\,+\,h^{2}\,\Bigl[
        -\,2\,\bigl(\bar{f}^{A}_{\kappa\lnot\kappa\prpr}+\bar{f}^{A}_{\kappa\pr\lnot\kappa\prpr}+
        f^{A}_{\kappa\prpr}\bigr)\,\fkethat{0}\!\fbrahat{0}
        \nonumber\\[1.5mm]
    &\hspace*{-5.3cm}
        +\,2\,\bar{f}^{A}_{\kappa\lnot\kappa\prpr}\,\fkethatkp{1}{\kappa\!}\!\fbrahatkp{1}{\kappa\!}
        \,+\,2\,\bar{f}^{A}_{\kappa\pr\lnot\kappa\prpr}\,\fkethatkp{1}{\kappa\pr\!}\!\fbrahatkp{1}{\kappa\pr\!}
        \,+\,2\,f^{A}_{\kappa\prpr\lnot\kappa,\kappa\pr}\,
        \fkethatkm{1}{\kappa\prpr\!}\!\fbrahatkm{1}{\kappa\prpr\!}
        \nonumber\\[1.5mm]
    &\hspace*{-5.3cm} +\,|\Ahmn{1}{\kappa\kappa\prpr}|^{2}\,
        \fkethatkp{1}{\kappa\!}\!\fkethatkm{1}{\kappa\prpr\!}\!
        \fbrahatkm{1}{\kappa\prpr\!}\!\fbrahatkp{1}{\kappa\!}
        +\,|\Ahmn{1}{\kappa\pr\hspace*{-1pt}\kappa\prpr}|^{2}\,
        \fkethatkp{1}{\kappa\pr\!}\!\fkethatkm{1}{\kappa\prpr\!}\!
        \fbrahatkm{1}{\kappa\prpr\!}\!\fbrahatkp{1}{\kappa\pr\!}
        \nonumber\\[1.5mm]
    &\hspace*{-5.3cm}+
        \Bigl(\fVhmn{1}{\kappa\kappa\prpr}\fVhmnstar{1}{\kappa\pr\hspace*{-1pt}\kappa\prpr}
        \fkethatkp{1}{\kappa\!}\!\fkethatkm{1}{\kappa\prpr\!}\!
        \fbrahatkm{1}{\kappa\prpr\!}\!\fbrahatkp{1}{\kappa\pr\!}
        +\!\sum\limits_{\substack{q<0\\q\neq\kappa\prpr}}\!\fVhmn{1}{\kappa q}
        \fVhmnstar{1}{\kappa\pr\hspace*{-1pt}q}
        \fkethatkp{1}{\kappa\!}\!\fbrahatkp{1}{\kappa\pr\!}
        +\mathrm{H.~c.}\Bigr)\Bigr]+O(h^{3}).\nonumber
\end{align}
At this stage an impasse is reached. As we have argued in Section~\ref{sec:fermionic entanglement beyond 2 modes}, three fermionic modes cannot in general be consistently mapped to three qubits without changing the entanglement properties. However, the specific structure of the Bogoliubov transformations at hand removes some of the otherwise possible elements in the state of Eq.~(\ref{eq:transformed fermionic vac partial trace kplus kprplus kprpr minus}) as compared to Eq.~(\ref{eq:general three-mode fermion state supseselection}). As it happens, this difference is already enough to allow us to write down a consistent three-qubit density matrix representation of~(\ref{eq:transformed fermionic vac partial trace kplus kprplus kprpr minus}) keeping only terms up to order~$h^{2}$, i.e.,

\begin{align}
    \begin{scriptsize}
    \begin{pmatrix}
        1-2h^{2}\bigl(\bar{f}^{A}_{\kappa\lnot\kappa\prpr}+\bar{f}^{A}_{\kappa\pr\lnot\kappa\prpr}+
            f^{A}_{\kappa\prpr}\bigr)   &   \hspace*{-5mm}0   &   \hspace*{-5mm}0   &   \hspace*{-4mm}h\mathcal{V}^{(1)*}_{\kappa\pr\hspace*{-1pt}\kappa\prpr}   &   \hspace*{-4mm}0   &   \hspace*{-4mm}h\mathcal{V}^{(1)*}_{\kappa\kappa\prpr}    &   0   &   0  \\[2.5mm]
        0   &   \hspace*{-5mm}2h^{2}f^{A}_{\kappa\prpr\lnot\kappa,\kappa\pr} &  \hspace*{-5mm}0  &
            \hspace*{-4mm}0  &  \hspace*{-4mm}0  & \hspace*{-4mm}0  &  0  &  0 \\[2.5mm]
        0   &   \hspace*{-5mm}0   &   \hspace*{-5mm}2h^{2}\bar{f}^{A}_{\kappa\pr\lnot\kappa\prpr} &
            \hspace*{-4mm}0 & \hspace*{-4mm}h^{2}\!\!\!
            \sum\limits_{\substack{q<0\\q\neq\kappa\prpr}}\!\mathcal{V}^{(1)*}_{\kappa q}  \mathcal{V}^{(1)}_{\kappa\pr\hspace*{-1pt}q} & \hspace*{-4mm}0 & 0 & 0 \\
        h\mathcal{V}^{(1)}_{\kappa\pr\hspace*{-1pt}\kappa\prpr} & \hspace*{-5mm}0 & \hspace*{-5mm}0 &
            \hspace*{-4mm}h^{2}|A^{(1)}_{\kappa\pr\hspace*{-1pt}\kappa\prpr\!}|^{2} & \hspace*{-4mm}0 &
            \hspace*{-4mm}h^{2}\mathcal{V}^{(1)*}_{\kappa\kappa\prpr}
            \mathcal{V}^{(1)}_{\kappa\pr\hspace*{-1pt}\kappa\prpr} & 0 & 0\\[2.5mm]
        0 & \hspace*{-5mm}0 & \hspace*{-5mm}h^{2}\!\!\!\sum\limits_{\substack{q<0\\q\neq\kappa\prpr}}\!
            \mathcal{V}^{(1)}_{\kappa q}\mathcal{V}^{(1)*}_{\kappa\pr\hspace*{-1pt}q} & \hspace*{-4mm}0 & \hspace*{-4mm}2h^{2}\bar{f}^{A}_{\kappa\lnot\kappa\prpr} & \hspace*{-4mm}0 & 0 & 0 \\
        h\mathcal{V}^{(1)}_{\kappa\kappa\prpr} & \hspace*{-5mm}0 & \hspace*{-5mm}0 &
            \hspace*{-4mm}h^{2}\mathcal{V}^{(1)}_{\kappa\kappa\prpr}
            \mathcal{V}^{(1)*}_{\kappa\pr\hspace*{-1pt}\kappa\prpr} & \hspace*{-4mm}0 & \hspace*{-4mm}h^{2}|A^{(1)}_{\kappa\kappa\prpr}|^{2} & 0 & 0\\[2.5mm]
        0 & \hspace*{-5mm}0 & \hspace*{-5mm}0 & \hspace*{-4mm}0 & \hspace*{-4mm}0 & \hspace*{-4mm}0 & 0 & 0 \\[2.5mm]
        0 & \hspace*{-5mm}0 & \hspace*{-5mm}0 & \hspace*{-4mm}0 & \hspace*{-4mm}0 & \hspace*{-4mm}0 & 0 & 0
    \end{pmatrix}.
    \end{scriptsize}\nonumber\\
    \label{eq:three mode ferm vac matrix rep}
\end{align}

In other words, all reductions of the fermionic three-mode state~(\ref{eq:transformed fermionic vac partial trace kplus kprplus kprpr minus}) are equivalent to the corresponding partial traces of the three-qubit density matrix~(\ref{eq:three mode ferm vac matrix rep}). This allows us to employ a witness inequality for GME. However, since we are dealing with fermions restricted by the Pauli exclusion principle\index{Pauli exclusion principle} the particular witness used in Eq.~(\ref{eq:BosonsB GME witness complete}) will be of no use. Instead we use the techniques described in Ref.~\cite{HuberMintertGabrielHiesmayr2010} to construct a witness. For convenience let us map the three-qubit witness for \emph{genuine tripartite entanglement}\index{entanglement!genuine tripartite} back to the three fermionic modes and write it as
\begin{align}
   &\left|\fbrahat{0}
    \frhovackkprkprpr
    \fkethatkp{1}{\kappa\!}\!\fkethatkm{1}{\kappa\prpr\!}\right|\,+\,
    \left|\fbrahat{0}
    \frhovackkprkprpr
    \fkethatkp{1}{\kappa\pr\!}\!\fkethatkm{1}{\kappa\prpr\!}\right|
    -\sqrt{\fbrahat{0}\frhovackkprkprpr\fkethat{0}}
    \nonumber\\[2.5mm]
   &\ \times\sqrt{\fbrahatkmup{1}{\kappa\prpr\!}\!\fbrahatkpup{1}{\kappa\!}
    \frhovackkprkprpr\fkethatkp{1}{\kappa\!}\!\fkethatkm{1}{\kappa\prpr\!}
    +\fbrahatkmup{1}{\kappa\prpr\!}\!\fbrahatkpup{1}{\kappa\pr\!}
    \frhovackkprkprpr\fkethatkp{1}{\kappa\pr\!}\!\fkethatkm{1}{\kappa\prpr\!}}
    \nonumber\\[2.5mm]
    &\ -\sqrt{\fbrahatkmup{1}{\kappa\prpr\!}\frhovackkprkprpr\fkethatkm{1}{\kappa\prpr\!}
    \fbrahatkpup{1}{\kappa\pr\!}\frhovackkprkprpr\fkethatkp{1}{\kappa\pr\!}}
    -\sqrt{\fbrahatkmup{1}{\kappa\prpr\!}\frhovackkprkprpr\fkethatkm{1}{\kappa\prpr\!}}
    \nonumber\\[2.5mm]
    &\ \times\sqrt{\fbrahatkpup{1}{\kappa\!}\frhovackkprkprpr\fkethatkp{1}{\kappa\!}}\,\leq 0\,.
    \label{eq:FermionsB witness complete}
\end{align}
As previously, the inequality is satisfied for all bi-separable pure states and its validity is extended to mixed states by the virtue of the convexity of the absolute value, see Eq.~(\ref{eq:convex function of rho}), and the concavity of the square roots of the density matrix elements, see Eq.~(\ref{eq:concave function of rho}). This means that a positive value for the right hand side of~(\ref{eq:FermionsB witness complete}) unambiguously detects GME. We insert the perturbative expansion of the transformed vacuum state of Eq.~(\ref{eq:transformed fermionic vac partial trace kplus kprplus kprpr minus}) to reduce the witness inequality to
\begin{align}
    |\Ahmn{1}{\kappa\kappa\prpr}|\,+\,|\Ahmn{1}{\kappa\pr\hspace*{-1pt}\kappa\prpr}|\,-\,
    \sqrt{|\Ahmn{1}{\kappa\kappa\prpr}|^{2}\,+\,|\Ahmn{1}{\kappa\pr\hspace*{-1pt}\kappa\prpr}|^{2}}\,+\,O(h) &\leq\,0\,.
    \label{eq:ferm vac witness expansion}
\end{align}
One can then use the triangle inequality to see that Eq.~(\ref{eq:ferm vac witness expansion}) can be violated whenever $\Ahmn{1}{\kappa\kappa\prpr}$ and $\Ahmn{1}{\kappa\pr\hspace*{-1pt}\kappa\prpr}$ are both nonzero. Hence, we find that even though no mode can be occupied by more than one excitation it is nonetheless the combination of the coefficients that create bipartite entanglement between $\kappa$ and $\kappa\prpr$, as well as $\kappa\pr$ and $\kappa\prpr$, respectively, that are responsible for the generation of genuine tripartite entanglement from the fermionic vacuum.\\

\vspace*{-3.5mm}
We can thus conclude this chapter noting that both bipartite and genuine multipartite entanglement are created from a variety of initial states. The Bogoliubov transformations correlate modes depending on their relative parity and energy levels. Most importantly, entanglement can be created between specific modes by selecting appropriate travel scenarios, see Section~\ref{sec:Resonances of Entanglement Generation}, which may be used to verify the \emph{quantumness} of the created radiation. In other words, the entanglement that is produced may serve as a clear indicator of the origin of the produced radiation being a quantum field theory effect.\\

\vspace*{-3.5mm}
The transformations induced by the motion of the cavity may further be interpreted as \emph{quantum gates}\index{quantum gates} \textemdash\ weak two-mode squeezing~\cite{BruschiDraganLeeFuentesLouko2013} or beam-splitting gates~\cite{BruschiLoukoFaccioFuentes2013}, or even as gates generating GME states~\cite[(\ref{Paper:FriisHuberFuentesBruschi2012})]{FriisHuberFuentesBruschi2012}. This, in turn, is of conceptual interest and can be considered to be a first step towards the possible future implementation of quantum information processing on the basis of relativistic motion, possibly complemented by alternative approaches~\cite{MartinMartinezAasenKempf2013}.\\

All modes that are being traced over, e.g., which we do not have access to due to limited measurement possibilities, add to the mixedness of the reduced state since information about their correlations with the modes under scrutiny is lost. It is exactly this issue that leads to the entanglement degradation that will be discussed in the final Chapter~\ref{Chapter 7 Degradation of Entanglement between Moving Cavities}.


%
%
%
%

    \chapter{Degradation of Entanglement between Moving Cavities}
\label{Chapter 7 Degradation of Entanglement between Moving Cavities}

The previous chapters have analyzed the entanglement generation between the modes of quantum fields that are confined to non-uniformly moving cavities. We have argued that, indeed, the radiation produced due to the transitions between orbits of different Killing vector fields (see Section~\ref{sec:relativistically rigid cavity}) is entangled for most initial states and chosen pairs of modes. Such effects may be of interest to identify effects of quantum field theory by distinguishing the produced particles from uncorrelated background noise, see, e.g., Refs.~\cite{WilsonDynCasNature2012} and~\cite[(\ref{Paper:BruschiFriisFuentesWeinfurtner2013})]{BruschiFriisFuentesWeinfurtner2013}. In principle, the entanglement that is being produced is distillable and could be utilized for quantum information tasks. The transformations effectively act as weak entangling gates on pairs of modes~\cite{BruschiDraganLeeFuentesLouko2013,FriisHuberFuentesBruschi2012}, but the perturbative approach limits the practical applications for this scheme of entanglement generation as a resource. However, as we shall see in this chapter, the motion of the cavities may influence quantum information processing tasks in a different way.\\

\begin{figure}[ht!]
\centering
\includegraphics[width=0.75\textwidth]{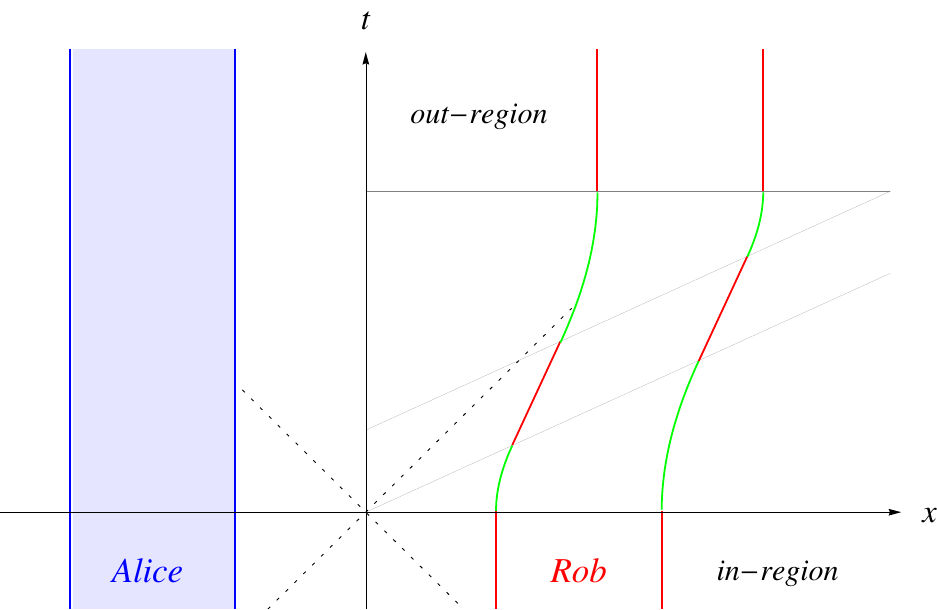}
\caption{
\textbf{Quantum communication between two cavities:} Alice and Rob each control a cavity containing a quantum field. They share an initially entangled state between one mode in each cavity to be used in a quantum communication task, e.g., quantum teleportation. Rob's cavity is undergoing non-uniform motion, which entangles all modes in the spectrum of his cavity. Consequently the entanglement shared between Alice and Rob is degraded by the motion.
\label{fig:two cavities setup}}
\end{figure}

\vspace*{-2mm}
Let us now consider \emph{two cavities}, controlled by the observers \emph{Alice} and \emph{Rob}, respectively, see Fig.~\ref{fig:two cavities setup}. Alice and Rob wish to use entanglement between their cavities as a resource for quantum communication tasks, for instance, for quantum teleportation (see Sections~\ref{sec:quantum teleportation}, \ref{sec:fidelity of teleportation bosons} and~\ref{sec:nonlocality and teleportation fermions}). For practical reasons the entanglement shared between the cavities will be restricted to certain finite sets of modes that are controlled by the observers. For the sake of the argument let us consider entanglement between one mode in each cavity only. The entanglement generation inside individual cavities then entangles the selected modes with all other modes in the respective spectra. Since Alice and Rob do not have access to the entire spectrum of their cavities, information is lost and the resource entanglement between their initial modes is \emph{degraded}. This process may be interpreted as decoherence, and is indeed a consequence of the monogamy of entanglement (see page~\pageref{page:monogamy of entanglement}).\\

In the following we shall make this phenomenological description more precise. In Section~\ref{sec:Entanglement between Two Bosonic Cavities} we analyze the entanglement degradation between two types of entangled initial states for the scalar field \textemdash\ \emph{Bell-states} (based on results presented in Ref.~\cite{BruschiFuentesLouko2012}) and \emph{two-mode squeezed states} (based on Ref.~\cite[(\ref{Paper:FriisLeeTruongSabinSolanoJohanssonFuentes2013})]{FriisLeeTruongSabinSolanoJohanssonFuentes2013}), accompanied by an application to the continuous variable \emph{teleportation protocol} in Section~\ref{sec:fidelity of teleportation bosons}. We finalize the investigation of the bosonic case with a brief look at a \emph{simulation in superconducting circuits} in Section~\ref{sec:Simulations in Superconducting Circuits}. At last we perform the corresponding analysis for the Dirac field as investigated in Ref.~\cite[(\ref{Paper:FriisLeeBruschiLouko2012})]{FriisLeeBruschiLouko2012}, including effects on teleportation and Bell inequality violation, in Section~\ref{sec:Entanglement between Two Fermionic Cavities}.

\section{Entanglement between Two Bosonic Cavities}\label{sec:Entanglement between Two Bosonic Cavities}

\subsection{Bosonic Bell States}\label{sec:Bosonic Bell States}

For the scalar field we start with the \emph{Bell states}\index{Bell!states}\index{state!Bell} $\phi^{+}$ and $\phi^{-}$ from Eq.~(\ref{eq:Bell states phi plus minus}) between two selected modes, $k$ in Alice's cavity and $k\pr$ in Rob's cavity, i.e.,
\begin{align}
    \ket{\phi^{\pm}}_{\hspace*{-0.5pt}\raisebox{0.0pt}{\tiny{$A\hspace*{-0.5pt}R$}}}    &=\,\tfrac{1}{\sqrt{2}}\Bigl(\ket{0}_{\hspace*{-0.5pt}\raisebox{0.0pt}{\tiny{$A$}}} \otimes\ket{0}_{\hspace*{-0.5pt}\raisebox{0.0pt}{\tiny{$R$}}}\pm
    \ket{1_{k}}_{\hspace*{-0.5pt}\raisebox{0.0pt}{\tiny{$A$}}}\otimes
    \ket{1_{k\pr\!}}_{\hspace*{-0.5pt}\raisebox{0.0pt}{\tiny{$R$}}}\Bigr)\,,
    \label{eq:Bell states phi plus minus Alice Rob}
\end{align}
where we have written the tensor product explicitly to point out that particles in the two cavities are now distinguishable by their association to either Alice's or Rob's cavity. The corresponding in-region density operator is then given by
\begin{align}
    \rho^{\pm}_{\hspace*{-0.5pt}\raisebox{0.0pt}{\tiny{$A\hspace*{-0.5pt}R$}}} &=\,\tfrac{1}{2}\Bigl(\ket{0}\!\bra{0}\otimes\ket{0}\!\bra{0}\,\pm\,\ket{0}\!\bra{1_{k}}\otimes\ket{0}\!\bra{1_{k\pr\!}}
    \label{eq:scalar field in region bell state density matrix}\\[1mm]
    &\ \ \pm\,\ket{1_{k}}\!\bra{0}\otimes\ket{1_{k\pr\!}}\!\bra{0}\,+\,\ket{1_{k}}\!\bra{1_{k}}\otimes\ket{1_{k\pr\!}}\!\bra{1_{k\pr\!}}\Bigr)\,,
    \nonumber
\end{align}
and we have dropped the label for Alice's and Rob's cavity and assume that the position in the tensor product is sufficient for this distinction. Rob now undergoes non-uniform motion as described in Chapter~\ref{Chapter 4 Constructing Non Uniformly Moving Cavities}, which means that we have to transform the right hand sides of the tensor products in Eq.~(\ref{eq:scalar field in region bell state density matrix}) to the out-region. The expressions for the transformed versions of the projectors $\ket{0}\!\bra{0}$ and $\ket{1_{k\pr\!}}\!\bra{1_{k\pr\!}}$ were already obtained in Chapter~\ref{Chapter 5 State Transformation by Non-Uniform Motion} and are given by Eq.~(\ref{eq:transformed boson vac density matrix}) and, with appropriate relabelling, by Eq.~(\ref{eq:transformed boson 1k density matrix}). For the off-diagonal matrix element $\ket{0}\!\bra{1_{k\pr\!}}$ we combine Eq.~(\ref{eq:transformed boson vac vector}) with the Hermitean conjugate of Eq.~(\ref{eq:transformed boson 1k vector}) with the relabelling $k\rightarrow k\pr$. Subsequently we trace over all of Rob's modes except $k\pr$ and we obtain the transformed matrix element
\begin{align}
    \tr_{\lnot k\pr}\bigl(\ket{0}\!\bra{1_{k\pr\!}}\bigr) &=\Ghn{0}{k\pr}\,\kethat{0}\!\brahatk{1}{k\pr\!}+h^{2}\Ghn{0}{k\pr}\Bigl[
    (\Ghnstar{0}{k\pr}\alphahmn{2}{k\pr k\pr}-4f^{\beta}_{k\pr})\kethat{0}\!\brahatk{1}{k\pr\!}-g_{k\pr k\pr}^{\alpha\beta}\kethatk{1}{k\pr\!}\!\brahat{0}
    \nonumber\\[1mm]
    &\hspace*{-2.7cm} +\,2\sqrt{2}\,f^{\beta}_{k\pr}\kethatk{1}{k\pr\!}\!\brahatk{2}{k\pr\!}\,+\,
    \sqrt{2}\,\Vhmn{2}{k\pr k\pr}\kethatk{2}{k\pr\!}\!\brahatk{1}{k\pr\!}\,+\,
    \sqrt{\tfrac{3}{2}}\,\Vhmnstar{2}{k\pr k\pr}\kethat{0}\!\brahatk{3}{k\pr\!}\,\Bigr]\,+\,O(h^{3})\,,
    \label{eq:bosonic Bell off diag partial trace}
\end{align}
where we have used the abbreviations $g^{\alpha\beta}_{k\pr k\pr}$ and $f^{\beta}_{k\pr}$ from Eqs.~(\ref{eq:galphabeta}) and Eq.~(\ref{eq:f beta m not n}), respectively, and $\Vhmn{2}{k\pr k\pr}$ is given by Eq.~(\ref{eq:bosonic V expansion 2}). The partial traces of the transformed diagonal elements $\ket{0}\!\bra{0}$ and $\ket{1_{k\pr\!}}\!\bra{1_{k\pr\!}}$ are most easily obtained from Eq.~(\ref{eq:bosonic vac bogo traced k and kpr}) and, again with appropriate relabelling, from Eq.~(\ref{eq:bosonic 1k bogo traced k and kpr}), respectively. We obtain
\begin{subequations}
\label{bosonic Bell diagonal partial traces}
\begin{align}
    \tr_{\lnot k\pr}\bigl(\ket{0}\!\bra{0}\bigr) &=\,\kethat{0}\!\brahat{0}\,+\,h^{2}\Bigl[
    -2f^{\beta}_{k\pr}\kethat{0}\!\brahat{0}\,+\,2f^{\beta}_{k\pr}\kethatk{1}{k\pr\!}\!\brahatk{1}{k\pr\!}
    \label{eq:bosonic Bell diagonal 0 partial trace}\\[1mm]
    &\ +\,\tfrac{1}{\sqrt{2}}\Bigl(\,\Vhmnstar{2}{k\pr k\pr}\kethat{0}\!\brahatk{2}{k\pr\!}\,+\,\mathrm{H.~c.}\Bigr)\,\Bigr]\,+\,O(h^{3})\,,
    \nonumber\\[2mm]
    \tr_{\lnot k\pr}\bigl(\ket{1_{k\pr\!}}\!\bra{1_{k\pr\!}}\bigr) &=\,\kethatk{1}{k\pr\!}\!\brahatk{1}{k\pr\!}\,+\,h^{2}\Bigl[
    2f^{\alpha}_{k\pr}\kethat{0}\!\brahat{0}\,-\,2\bigl(f^{\alpha}_{k\pr}+2f^{\beta}_{k\pr}\bigr)\kethatk{1}{k\pr\!}\!\brahatk{1}{k\pr\!}
    \label{eq:bosonic Bell diagonal 1 partial trace}\\[1mm]
    &\ +\,4f^{\beta}_{k\pr}\kethatk{2}{k\pr\!}\!\brahatk{2}{k\pr\!}\,
    -\,\Bigl(\,\sqrt{2}(\Ghn{0}{k\pr})^{2}g^{\alpha\beta*}_{k\pr k\pr}\kethat{0}\!\brahatk{2}{k\pr\!}\,+\,\mathrm{H.~c.}\Bigr)\,\Bigr]\,+\,O(h^{3})\,,
    \nonumber
\end{align}
\end{subequations}
where we have also used the shorthand $f^{\alpha}_{k\pr}$ from Eq.~(\ref{eq:f alpha m not n}). Inserting Eqs.~(\ref{eq:bosonic Bell off diag partial trace}) and~(\ref{bosonic Bell diagonal partial traces}) into $\tr_{\lnot k,k\pr}(\rho^{\pm}_{AR})$ we find that the perturbative expansion truncates the transformed state to a $2\times4$ dimensional system. Since we are now dealing with a mixed state we can quantify the entanglement of the transformed state via the \emph{negativity}\index{negativity} (see Definition~\ref{def:negativity}). As can be easily seen from Eq.~(\ref{eq:scalar field in region bell state density matrix}) the unperturbed partial transposition\index{partial transposition} has three positive eigenvalues $\lambda_{+}=\tfrac{1}{2}$ and one negative eigenvalue $\lambda_{-}=-\tfrac{1}{2}$, while all other eigenvalues vanish, regardless of the sign in $\ket{\phi^{\pm}}$ of Eq.~(\ref{eq:Bell states phi plus minus Alice Rob}). The subspace of the vanishing unperturbed eigenvalues contains only one positive correction $4h^{2}f^{\beta}_{k\pr}$. The positive unperturbed eigenvalues cannot be turned into negative eigenvalues by the small perturbative corrections. Hence, the only correction to the negativity stems from the leading order perturbation to the unperturbed, non-degenerate eigenvalue $\lambda_{-}=-\tfrac{1}{2}$. Following the prescription detailed on page~\pageref{page:pert diagonalization} we quickly get the corrected value of the negativity as reported in Ref.~\cite{BruschiFuentesLouko2012}
\begin{align}
    \mathcal{N}(\rho^{\pm}_{\hspace*{-0.5pt}\raisebox{0.0pt}{\tiny{$A\hspace*{-0.5pt}R$}}})    &=\,
    \Negh{0}\,-\,h^{2}\Negh{2}\,+\,O(h^{3})\,=\,
    \tfrac{1}{2}\,-\,h^{2}\,(2f^{\beta}_{k\pr}\,+\,f^{\alpha}_{k\pr})\,+\,O(h^{3})\,,
    \label{eq:bosons Bell state negativity}
\end{align}
where we have used Eq.~(\ref{eq:generic travel scenario bogo identities bosons alphas 2}). The entanglement is thus degraded by information loss due to the generation of particle pairs, where one constituent is created in Rob's mode~$k\pr$, as well as information loss due to the possibility of shifting excitations from the mode~$k\pr$ to other energy levels in Rob's cavity. The coefficients $f^{\alpha}_{k\pr}$ and $f^{\beta}_{k\pr}$ that are degrading the entanglement are illustrated in Fig.~\ref{fig:entanglement degradation bosons alphas} and Fig.~\ref{fig:entanglement degradation bosons betas}, respectively.

\subsection{Two-Mode Squeezed States}\label{sec:two mode squeezed states}

\begin{figure}[hb!]
\centering
\includegraphics[width=0.73\textwidth]{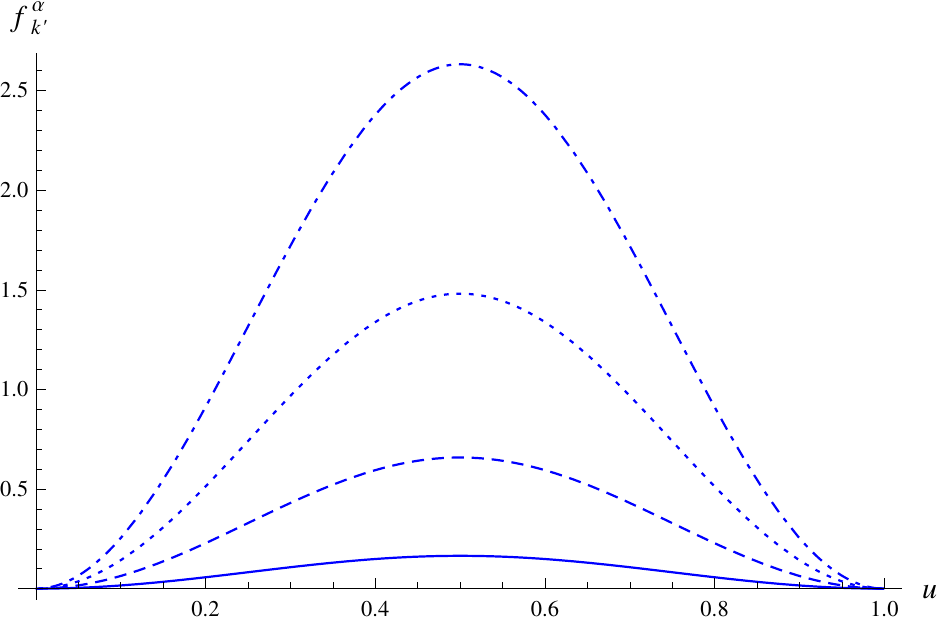}
\caption{
\textbf{Entanglement degradation \textemdash\ $f^{\alpha}_{k\pr}$:} The coefficient $f^{\alpha}_{k\pr}$ that is degrading the entanglement between Alice and Rob as quantified by the negativity [see Eq.~(\ref{eq:bosons Bell state negativity}) and Eq.~(\ref{eq:bosons TMS state negativity})] is plotted for the basic building block travel scenario of Section~\ref{sec:basic building block} for a $(1+1)$ dimensional massless scalar field. The effects are periodic in the dimensionless parameter $u:=h\tau/[4L\artanh(h/2)]\,$ [see Eq.~(\ref{eq:cavity centre proper freq})], where $\tau$ is the duration of the uniform acceleration, as measured at the centre of the cavity. Curves are shown for the modes $k\pr=1$ (solid), $k\pr=2$ (dashed), $k\pr=3$ (dotted), and $k\pr=4$ (dotted-dashed).
\label{fig:entanglement degradation bosons alphas}}
\end{figure}

\begin{figure}[ht!]
\centering
\includegraphics[width=0.75\textwidth]{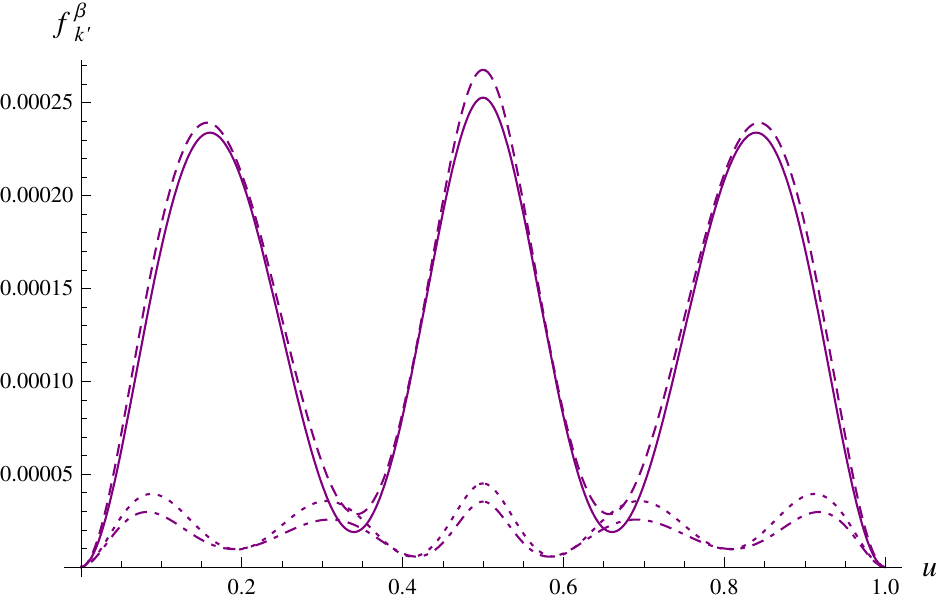}
\caption{
\textbf{Entanglement degradation \textemdash\ $f^{\beta}_{k\pr}$:} The coefficient $f^{\beta}_{k\pr}$ that is degrading the entanglement between Alice and Rob as quantified by the negativity [see Eq.~(\ref{eq:bosons Bell state negativity}) and Eq.~(\ref{eq:bosons TMS state negativity})] is plotted for the basic building block travel scenario of Section~\ref{sec:basic building block} for a $(1+1)$ dimensional massless scalar field. The effects are periodic in the dimensionless parameter $u:=h\tau/[4L\artanh(h/2)]\,$ [see Eq.~(\ref{eq:cavity centre proper freq})], where $\tau$ is the duration of the uniform acceleration, as measured at the centre of the cavity. Curves are shown for the modes $k\pr=1$ (solid), $k\pr=2$ (dashed), $k\pr=3$ (dotted), and $k\pr=4$ (dotted-dashed).
\label{fig:entanglement degradation bosons betas}}
\end{figure}

Although the Bell state that we have analyzed in Section~\ref{sec:Bosonic Bell States} is a simple example for an entangled two-mode state, more general entangled states \textemdash\ \emph{two-mode squeezed states}\index{squeezed state!two-mode}\index{state!two-mode squeezed} (see Section~\ref{sec:two-mode squeezed states}), with superpositions of various particle numbers are allowed. For convenience we shall switch again from the Fock space treatment to the \emph{phase space} and work with the covariance matrix only. As we have explained in Section~\ref{sec:entanglement in bosonic quantum fields} the covariance matrix encodes all relevant information about the entanglement between modes of Gaussian states. Let us assume now that Alice and Rob are sharing a two-mode squeezed state between their modes $k$ (Alice) and $k\pr$ (Rob), represented by the covariance matrix $\Gamma_{\raisebox{0pt}{\tiny{$\mathrm{TMS}$}}}(r)$ from Eq.~(\ref{eq:two mode squeezed covariance matrix}) that is decomposed into $2\times2$ blocks, i.e.,
\begin{align}
    \Gamma_{\raisebox{0pt}{\tiny{$\mathrm{TMS}$}}}(r)   &=\,
    \begin{pmatrix}
    \Gammak{k}  &   \Cmn{kk\pr} \\
    \Cmn{kk\pr}^{T} &   \Gammak{k\pr}
    \end{pmatrix}\,,
    \label{eq:two modes squeezed state decomposition}
\end{align}
where $\Gammak{k}=\Gammak{k\pr}=\cosh(2r)\idN{2}\,$, $\Cmn{kk\pr}=\sinh(2r)\sigma_{3}\,$, and $\sigma_{3}$ is the third Pauli matrix\index{Pauli matrices} from Eq.~(\ref{eq:Pauli matrices}). Now we employ the formalism of Section~\ref{sec:symplectic operations} to transform the covariance matrix to the out-region after Rob has undergone non-uniform motion. The $2\times2$ blocks of the out-region covariance matrix $\widehat{\Gamma}_{\raisebox{0pt}{\tiny{$\mathrm{TMS}$}}}(r)$ are given by $\Gammahatk{k}=\Gammak{k}\,$,
\begin{align}
    \Gammahatk{k\pr}    &=\,\cosh(2r)\mathcal{M}_{k\pr k\pr}\mathcal{M}_{k\pr k\pr}^{T}\,+\,
    \sum\limits_{n\neq k\pr}\mathcal{M}_{m\pr n}\mathcal{M}_{k\pr n}^{T}\,,
    \label{eq:gamma hat kpr}
\end{align}
and $\Chatmn{kk\pr}=\Cmn{kk\pr}\mathcal{M}_{k\pr k\pr}^{T}\,$, where the matrices $\mathcal{M}_{mn}$ are decomposed into the corresponding Bogoliubov coefficients according to Eq.~(\ref{eq:M Bogo matrix}). We then proceed with the perturbative expansion of these blocks by inserting the expansions from~(\ref{eq:generic travel scenario expansion alphas}) and~(\ref{eq:generic travel scenario expansion betas}). We find
\begin{subequations}
\label{eq:transformed TMS state expansion}
\begin{align}
    \Gammahatk{k\pr}    &=\,\Gammahatkh{k\pr}{0}\,+\,h^{2}\,\Gammahatkh{k\pr}{2}\,+\,O(h^{3})\,,
    \label{eq:transformed TMS state expansion Gammahatkp}\\[1mm]
    \Chatmn{kk\pr}    &=\,\Chatmnh{kk\pr}{0}\,+\,h^{2}\,\Chatmnh{kk\pr}{2}\,+\,O(h^{3})\,,
    \label{eq:transformed TMS state expansion Chatkkp}
\end{align}
\end{subequations}
where the leading order is given by $\Gammahatkh{k\pr}{0}=\Gammak{k}$ and
\begin{align}
    \Chatmnh{kk\pr}{0}  &=\,\sinh(2r)\,
    \begin{pmatrix}
    \cos(\omega_{k\pr}\tilde{\tau}) &   -\sin(\omega_{k\pr}\tilde{\tau})    \\
    -\sin(\omega_{k\pr}\tilde{\tau})    &   -\cos(\omega_{k\pr}\tilde{\tau})
    \end{pmatrix}\,,
    \label{eq:transformed TMS state expansion Chatkkp h0}
\end{align}
with $\omega_{k\pr}$ from Eq.~(\ref{eq:inertial modes frequencies}) and $\tilde{\tau}$ is the proper time at the centre of the cavity (see Section~\ref{sec:relativistically rigid cavity}). Note that we have not included the time evolution of the mode~$k$ in Alice's cavity here explicitly but this may be achieved by replacing $\omega_{k\pr}\tilde{\tau}$ by $(\omega_{k\pr}\tilde{\tau}+\omega_{k}\tilde{\tau}_{\raisebox{0.0pt}{\tiny{A}}})$, where $\omega_{k}$ is the frequency of the mode~$k$ and $\tilde{\tau}_{\raisebox{0.0pt}{\tiny{A}}}$ is Alice's proper time\index{proper!time}. The coefficients of the corrections that are quadratic in $h$ are
\begin{subequations}
\label{eq:transformed TMS state expansion second order}
\begin{align}
    \Gammahatkh{k\pr}{2}    &=\,2\cosh(2r)\,
    \begin{pmatrix}
    f^{\beta}_{k\pr}-f^{\alpha}_{k\pr}-\operatorname{Re}\bigl(\Ghn{0}{k\pr}\betahmn{2}{k\pr k\pr}\bigr)  &
    \operatorname{Im}\bigl(\Ghn{0}{k\pr}\betahmn{2}{k\pr k\pr}\bigr)  \\
    \operatorname{Im}\bigl(\Ghn{0}{k\pr}\betahmn{2}{k\pr k\pr}\bigr)  &
    f^{\beta}_{k\pr}-f^{\alpha}_{k\pr}+\operatorname{Re}\bigl(\Ghn{0}{k\pr}\betahmn{2}{k\pr k\pr}\bigr)
    \end{pmatrix}
    \label{eq:transformed TMS state expansion Gammahatkp h2}\\[1mm]
    &\hspace*{3mm}
    \,+\,2\,\begin{pmatrix}
    f^{\alpha}_{k\pr}+f^{\beta}_{k\pr}+
    \operatorname{Re}\bigl[(\Ghn{0}{k\pr})^{2}g^{\alpha\beta*}_{k\pr k\pr}\bigr]    &
    -\operatorname{Im}\bigl[(\Ghn{0}{k\pr})^{2}g^{\alpha\beta*}_{k\pr k\pr}\bigr]   \\
    -\operatorname{Im}\bigl[(\Ghn{0}{k\pr})^{2}g^{\alpha\beta*}_{k\pr k\pr}\bigr]   &
    f^{\alpha}_{k\pr}+f^{\beta}_{k\pr}-
    \operatorname{Re}\bigl[(\Ghn{0}{k\pr})^{2}g^{\alpha\beta*}_{k\pr k\pr}\bigr]
    \end{pmatrix}\,,
    \nonumber\\[2mm]
    \Chatmnh{kk\pr}{2}    &=\,\sinh(2r)\,
    \begin{pmatrix}
    \operatorname{Re}\bigl(\alphahmn{2}{k\pr k\pr}\,-\,\betahmn{2}{k\pr k\pr}\bigr)  &
    -\operatorname{Im}\bigl(\alphahmn{2}{k\pr k\pr}\,-\,\betahmn{2}{k\pr k\pr}\bigr)  \\
    -\operatorname{Im}\bigl(\alphahmn{2}{k\pr k\pr}\,+\,\betahmn{2}{k\pr k\pr}\bigr)  &
    -\operatorname{Re}\bigl(\alphahmn{2}{k\pr k\pr}\,+\,\betahmn{2}{k\pr k\pr}\bigr)
    \end{pmatrix}\,,
\end{align}
\end{subequations}
where $f^{\alpha}_{k\pr}$ and $f^{\beta}_{k\pr}$ are given by Eqs.~(\ref{eq:f alpha m not n}) and~(\ref{eq:f beta m not n}), respectively, while $g^{\alpha\beta}_{k\pr k\pr}$ is as in Eq.~(\ref{eq:galphabeta}), and we have used the Bogoliubov identity of Eq.~(\ref{eq:generic travel scenario bogo identities bosons alphas 2}). As expected from our deliberations in Chapter~\ref{Chapter 6 Motion Generates Entanglement} the reduced state of the two modes~$k$ and~$k\pr$ is \emph{mixed}, as can be seen from the determinant of~$\widehat{\Gamma}_{\raisebox{0pt}{\tiny{$\mathrm{TMS}$}}}(r)$ [see Eq.~(\ref{eq:linear entropy of cov matrix})], which is found to be
\begin{align}
    \det\bigl(\widehat{\Gamma}_{\raisebox{0pt}{\tiny{$\mathrm{TMS}$}}}\bigr)    &=\,
    1\,+\,4h^{2}\,\Bigl(\bigl[\cosh(2r)+1\bigr]f^{\beta}_{k\pr}\,+\,\bigl[\cosh(2r)-1\bigr]f^{\alpha}_{k\pr}\Bigr)
    \,+\,O(h^{3})\,.
    \label{eq:transformed TMS state determinant}
\end{align}
The perturbative nature of the calculation demands that the corrections do not drastically change the state, in particular, the mixedness. From Eq.~(\ref{eq:transformed TMS state determinant}) it can be seen that this imposes the restriction $e^{2|r|}h^{2}\ll1$. Let us now proceed by evaluating the entanglement of the transformed state. Since the transformed state is a mixed Gaussian two-mode state, but it is not symmetric, we again employ the \emph{negativity} (see Definition~\ref{def:negativity}). More specifically, we use Eq.~(\ref{eq:negativity Gaussian}), which means we have to determine the perturbative corrections to the \emph{smallest symplectic eigenvalue} of the partial transpose\index{partial transposition}. The eigenvalues of the unperturbed matrix $i\hspace*{0.5pt}\Omega\hspace*{0.5pt}\oversmile{T}_{\hspace*{-1pt} k\pr}\hspace*{0.5pt}\GammaTMShath{0}\hspace*{0.5pt}\oversmile{T}_{\hspace*{-1pt} k\pr}$ are found to be $\{\pm e^{2r},\pm e^{-2r}\}$. When $\operatorname{sgn}(r)=\pm1$ the smallest positive eigenvalue is given by $e^{\mp2r}$. The corresponding eigenvectors are
\begin{align}
    \ket{\numh{-}{0}}   &=\,\tfrac{1}{2}\bigl(\,\mp i \Ghnstar{0}{k\pr},\ \mp\Ghnstar{0}{k\pr},\ i,\ 1\,\bigr)^{T}\,.
    \label{eq:par transp TMS cov matrix eig vectors}
\end{align}
Since the eigenvalues are non-degenerate and the leading order corrections to the covariance matrix are quadratic in $h$ we expect an expansion of the form
\begin{align}
    \num{-} &=\,\numh{-}{0}\,+\,h^{2}\,\numh{-}{2}\,+\,O(h^{3})\,,
    \label{eq:num TMS state expansion}
\end{align}
\vspace*{-2mm}
and we can compute the correction to the smallest symplectic eigenvalue of the partial transpose as the expectation value~\cite[(\ref{Paper:FriisLeeTruongSabinSolanoJohanssonFuentes2013})]{FriisLeeTruongSabinSolanoJohanssonFuentes2013}
\begin{align}
    \numh{-}{2} &=\,\bra{\numh{-}{0}}\,i\hspace*{0.5pt}\Omega\hspace*{0.5pt}\oversmile{T}_{\hspace*{-1pt} k\pr}\hspace*{0.5pt}\GammaTMShath{2}\hspace*{0.5pt}\oversmile{T}_{\hspace*{-1pt} k\pr}\,\ket{\numh{-}{0}}\,=\,
    (1\,-\,e^{-2|r|})f^{\alpha}_{k\pr}\,+\,(1\,+\,e^{-2|r|})f^{\beta}_{k\pr}\,,
    \label{eq:numinus TMS state correction}
\end{align}
where we have again used the identity~(\ref{eq:generic travel scenario bogo identities bosons alphas 2}) in the last step. The result depends on the value of the squeezing parameter~$r$ but its validity is limited by the perturbative approach. In particular the small corrections cannot remove the non-degeneracy of the symplectic eigenvalues of the partial transpose, such that $h^{2}\ll \sinh(2|r|)$. With this in mind we finally obtain the corrected negativity
\begin{align}
    \mathcal{N}(\widehat{\Gamma}_{\raisebox{0pt}{\tiny{$\mathrm{TMS}$}}})    &=\,
    \Negh{0}\,-\,h^{2}\Negh{2}\,+\,O(h^{3})\label{eq:bosons TMS state negativity}\\[1mm]
    &\ =\,
    \tfrac{1}{2}\bigl(e^{2|r|}-1\bigr)\,-\,h^{2}\,
    e^{2|r|}\Bigl(\tfrac{1}{2}\bigl[e^{2|r|}-1\bigr]\bigl(f^{\beta}_{k\pr}\,+\,f^{\alpha}_{k\pr}\bigr)\,
    +\,f^{\beta}_{k\pr}\Bigr)\,+\,O(h^{3})\,.
    \nonumber
\end{align}
Illustrations of the functions $f^{\alpha}_{k\pr}$ and $f^{\beta}_{k\pr}$ that are responsible for the entanglement degradation are shown in Fig.~\ref{fig:entanglement degradation bosons alphas} and Fig.~\ref{fig:entanglement degradation bosons betas}, respectively.
\vspace*{-2mm}

\subsection{Fidelity of Teleportation}\label{sec:fidelity of teleportation bosons}

\begin{figure}[ht!]
\centering
\includegraphics[width=0.50\textwidth]{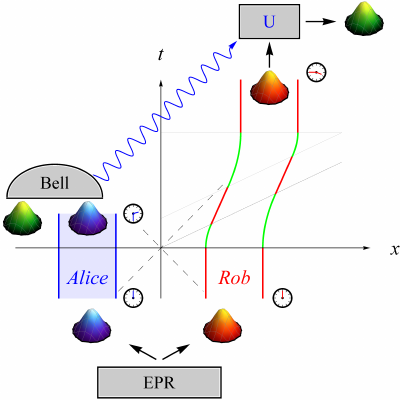}
\caption{
\textbf{Quantum teleportation between two cavities:} Alice and Rob wish to use the initially shared entanglement that is supplied by the EPR source to teleport an unknown coherent state. After the initial state has been prepared Alice performs a Bell measurement on the (unknown) state that is to be teleported and her mode~$k$ of the entangled resource state. Subsequently, she sends the measurement outcomes to Rob via a classical channel. Meanwhile, Rob undergoes a finite period of non-uniform motion, after which he receives the classical information necessary to retrieve the unknown input state by performing a local unitary~$U$. By measuring their respective proper times and applying corresponding local rotations Alice and Rob can optimize their teleportation scheme.
\label{fig:teleportation scheme}}
\end{figure}

The entanglement between Alice's mode~$k$ and Rob's mode~$k\pr$ is degraded due to the non-uniform motion. For practical reasons we have used the negativity to quantify the loss of correlations even though this measure does not have a direct operational interpretation. Now we wish to place this result in the context of a practical application \textemdash\ the \emph{teleportation} protocol (see pp.~\pageref{page:Gaussian teleportation}). We wish to analyze the influence of the entanglement degradation on the teleportation scheme that is illustrated in Fig.~\ref{fig:teleportation scheme}. The fidelity of the teleportation protocol for Gaussian states is given by Eq.~(\ref{eq:Gaussian teleportation fidelity}) (see Ref.~\cite{MariVitali2008}). We insert the perturbative expansions of the transformed covariance matrix elements from Eqs.~(\ref{eq:transformed TMS state expansion})-(\ref{eq:transformed TMS state expansion second order}) into Eq.~(\ref{eq:Gaussian teleportation fidelity}) to obtain the expression
\begin{align}
    \mathcal{F}(\widehat{\Gamma}_{\raisebox{0pt}{\tiny{$\mathrm{TMS}$}}})    &=\,
    \Fidh{0}\,-\,h^{2}\,\Fidh{2}\,+\,O(h^{3})\,,
    \label{eq:Gaussian teleportation fidelity expansion}
\end{align}
where the coefficients are found to be
\begin{subequations}
\label{eq:Gaussian tele fid expansion coefficients}
\begin{align}
    \Fidh{0}    &=\,\bigl(1\,+\,\cosh(2r)\,-\,\cos(\omega_{k\pr}\tilde{\tau}+
        \omega_{k}\tilde{\tau}_{\raisebox{0.0pt}{\tiny{A}}})\sinh(2r)\bigr)^{-1}\,,
    \label{eq:Gaussian tele fid expansion coefficients h0}\\[1mm]
    \Fidh{2}    &=\,\bigl(\Fidh{0}\bigr)^{2}\bigl(1+e^{-2r}\bigr)
        \bigl[f^{\beta}_{k\pr}\,+\,f^{\alpha}_{k\pr}\,\tanh(2r)\bigr]\,.
    \label{eq:Gaussian tele fid expansion coefficients h2}
\end{align}
\end{subequations}
We have specifically included the time evolution of Alice's mode in Eq.~(\ref{eq:Gaussian tele fid expansion coefficients h2}) and one should note that the phases accumulated by both modes~$k$ and~$k\pr$ affect the un-
\newpage
\noindent
-perturbed teleportation fidelity~$\Fidh{0}$. However, the effect of the time evolution can be easily corrected \textemdash\ Alice and Rob can simply keep track of their respective proper times\index{proper!time} and apply local rotations to remove the phases. These operations can be performed independently by the two observers and they do not require any knowledge about the other's state of motion. Since these corrections can be implemented by local unitaries the amount of entanglement that is shared is not altered. Incidentally, the local rotations also remove the phase dependence from the correction term in Eq.~(\ref{eq:Gaussian tele fid expansion coefficients h2}) and we arrive at the \emph{optimal teleportation fidelity}
\begin{align}
    \mathcal{F}_{\mathrm{opt}}(\widehat{\Gamma}_{\raisebox{0pt}{\tiny{$\mathrm{TMS}$}}})    &=\,
    \Fidh{0}\,-\,h^{2}\,\Fidh{2}\,+\,O(h^{3})\,,
    \label{eq:Gaussian opt teleportation fidelity expansion}
\end{align}
where the coefficients are given by~\cite[(\ref{Paper:FriisLeeTruongSabinSolanoJohanssonFuentes2013})]{FriisLeeTruongSabinSolanoJohanssonFuentes2013}
\begin{subequations}
\label{eq:Gaussian opt tele fid expansion coefficients}
\begin{align}
    \Fidopth{0}    &=\,\bigl(1\,+\,e^{-2r}\bigr)^{-1}\,,
    \label{eq:Gaussian opt tele fid expansion coefficients h0}\\[1mm]
    \Fidopth{2}    &=\,\Fidopth{0}\bigl[f^{\beta}_{k\pr}\,+\,f^{\alpha}_{k\pr}\,\tanh(2r)\bigr]\,.
    \label{eq:Gaussian opt tele fid expansion coefficients h2}
\end{align}
\end{subequations}
Using Eq.~(\ref{eq:num TMS state expansion}) and~(\ref{eq:numinus TMS state correction}) it can be immediately seen that the upper bound of~(\ref{eq:bounds on Gaussian fidelity}) is achieved. The correction $\Fidopth{2}$ in Eq.~(\ref{eq:Gaussian opt tele fid expansion coefficients h2}) thus isolates the degrading effect of Rob's non-uniform motion.

\subsection{Simulations in Superconducting Circuits}\label{sec:Simulations in Superconducting Circuits}

With the practical application to the teleportation protocol in mind we now want to gain insight about the numerical values of the relative size of the perturbative corrections. So far we have considered the dimensionless expansion parameter~$h$ in units where the speed of light is set to unity, $c=1$. Inserting the speed of light explicitly the perturbative parameter is
\begin{align}
    h   &:=\,\frac{\mathbf{a}_{\mathrm{c}}L}{c^{2}}\,,
    \label{eq:perturbative parameter with c}
\end{align}
where~$L$ is the length of the cavity and $\mathbf{a}_{\mathrm{c}}$ is the proper acceleration\index{proper!acceleration} at its centre. Assuming that the cavity size is well below the length scale of one meter it becomes clear that the perturbative approach can easily accommodate accelerations of $10^{17}\hspace*{0.3pt}\mathrm{m}\hspace*{0.3pt}\mathrm{s}^{-2}$. In other words, the accelerations must reach extremely large values to produce observable effects for arbitrary setups. However, selecting particular initial states and exploiting the effects of transverse momenta (see Eq.~(\ref{eq:extra dimensions mass}) and Fig.~\ref{fig:KG alphas}) the overall corrections may yet reach observable levels~\cite{BruschiLoukoFaccioFuentes2013}. We shall explore a different route here by studying a setup that simulates the mechanical motion of the cavity walls.\\

\begin{figure}[h]
\centering
\includegraphics[width=0.75\textwidth]{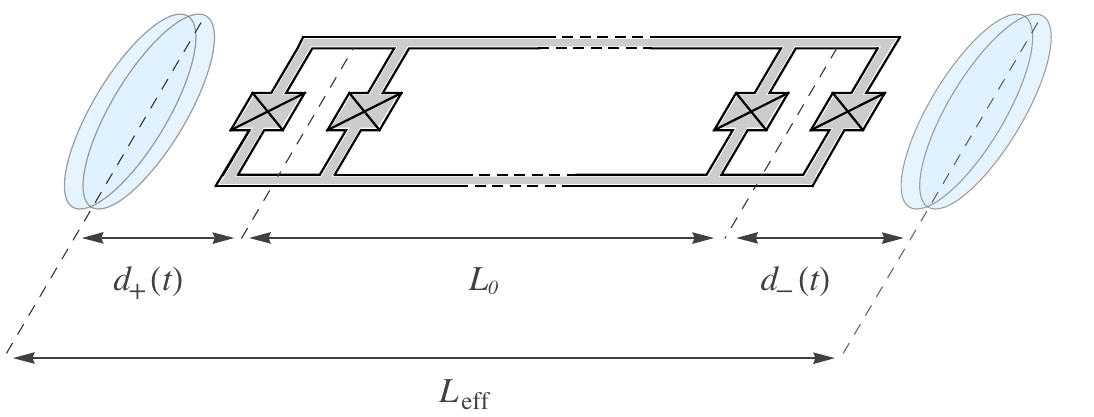}
\caption{
\textbf{Superconducting cavity simulation:} A one-dimensional transmission line for microwave radiation is interrupted by two SQUIDs at a distance $L_{0}$ with respect to each other. The SQUIDs are threaded by time-dependent magnetic fluxes that create boundary conditions \textemdash\ effective mirrors \textemdash\ at distances $d_{\pm}(t)$ away from the SQUIDs. This creates a cavity of effective length $L_{\mathrm{eff}}$ that can simulate non-uniform motion by changing the fluxes in time such that the length $L_{\mathrm{eff}}=d_{+}+L_{0}+d_{-}=\mathrm{const.}$, as measured by a potentially co-moving observer, remains constant.
\label{fig:cavity simulation}}
\end{figure}

Following Ref.~\cite[(\ref{Paper:FriisLeeTruongSabinSolanoJohanssonFuentes2013})]{FriisLeeTruongSabinSolanoJohanssonFuentes2013} we envisage a one-dimensional transmission line for electromagnetic radiation in the microwave domain that is terminated by two \emph{superconducting quantum interference devices} (SQUIDs)\index{SQUID}\index{superconducting quantum interference device|see{SQUID}}. Similar setups, e.g., with an open transmission line terminated by a single SQUID, have been extensively used to study the related \emph{dynamical Casimir effect}\index{dynamical Casimir effect}, see, for instance, Refs.~\cite{JohanssonJohanssonWilsonNori2010,WilsonDynCasNature2012,LaehteenmaekiParaoanuHasselHakonen2013}.
The role of the SQUID, which consists of a superconducting circuit with two parallel \emph{Josephson junctions}\index{Josephson junction}, is to generate the boundary conditions for the electromagnetic field in the transmission line (see Ref.~\cite{JohanssonJohanssonWilsonNori2010} for details). Each SQUID is threaded by a magnetic flux, which can be externally tuned at will, that determines the boundary condition. In particular, the parameters can be tuned to mimic perfectly reflecting mirrors whose distances to the SQUIDs depend on the chosen magnetic fluxes, see Fig.~\ref{fig:cavity simulation}. Two such SQUIDs thus constitute a cavity for the electromagnetic radiation and the position of the ``walls", i.e., the boundary conditions, can be varied by adjusting the magnetic fluxes. Indeed, the cavity setup we propose has already been implemented in a laboratory, see Ref.~\cite{SvenssonMScThesis2012}.\\

To emulate the motion of a relativistically rigid cavity as described in Section~\ref{sec:relativistically rigid cavity} the fluxes of the two SQUIDs need to be changed in a particular fashion. Let us imagine an observer undergoing a chosen travel scenario (see Section~\ref{sec:Grafting Generic Cavity Trajectories} for examples). If the magnetic fluxes are selected such that the positions of the boundary conditions remain at a fixed distance $L_{\mathrm{eff}}$ with respect to this observer the effective cavity can be thought of as rigid and undergoing the same travel scenario. From the point of view of the laboratory no piece of the equipment is in motion, and the magnetic fluxes are not changing symmetrically. The imaginary observer, on the other hand, would see the distance $L_{\mathrm{0}}$ between the SQUIDs vary in time \textemdash\ the cavity would be contracting and expanding according to the relative velocity of the observer with respect to the laboratory.
Finally, let us insert typical values for the parameters to estimate the relative correction. We consider a cavity of length $L=1.2\hspace*{0.3pt}\mathrm{cm}$ that is undergoing accelerations of up to $\mathbf{a}_{\mathrm{c}}=3\times10^{17}\mathrm{ms}^{-2}$, while the effective speed of light in the transmission line is around $1.2\times10^{8}\mathrm{ms}^{-1}$, similar to the setting in Ref.~\cite{WilsonDynCasNature2012}. This combines to an estimate of $h^{2}\approx0.06$ for the expansion parameter. As a resource we select a two-mode squeezed state with squeezing parameter $r=\tfrac{1}{2}$, which is well within the limits of current technology~\cite{EichlerEtal2011,FlurinRochMalletDevoretHuard2012,MenzelEtal2012}. For the mode $k\pr=3$, corresponding to an (angular) frequency of $2\pi\times15$GHz, the approximate value of the function $f^{\alpha}_{k\pr}$ is $1.5$, while the contribution from $f^{\beta}_{k\pr}$ can be neglected, see Fig.~\ref{fig:entanglement degradation bosons betas}. Hence, the relative correction to the optimal teleportation fidelity of Eq.~(\ref{eq:Gaussian opt teleportation fidelity expansion}) is given by
\begin{align}
    \frac{h^{2}|\Fidopth{2}|}{\Fidopth{0}}   &=\,h^{2}\bigl[f^{\beta}_{k\pr}\,+\,f^{\alpha}_{k\pr}\,\tanh(2r)\bigr]\,
    \approx\,0.04\,,
    \label{eq:teleportation fid rel correction}
\end{align}
that is, a $4\%$ relative correction, which is both reasonably small to fit the perturbative regime, but also large enough to be detected in possible future experiments. We thus conclude that the experimental verification or simulation of the effects of non-uniform motion on entanglement can in principle be achieved in current laboratories.

\newpage
\section{Entanglement between Two Fermionic Cavities}\label{sec:Entanglement between Two Fermionic Cavities}

To complete our analysis, let us consider the fermionic counterpart of the situation studied in Section~\ref{sec:Entanglement between Two Bosonic Cavities}. We copy the previous scenario, i.e., an entangled state shared between Alice's and Rob's cavity and we let Rob undergo non-uniform motion, see Fig.~\ref{fig:two cavities setup}. However, this time Dirac fields are confined to the cavities in question. In Section~\ref{sec:fermionic bell states} we consider the entanglement degradation of an initially maximally entangled state as reported in Ref.~\cite[(\ref{Paper:FriisLeeBruschiLouko2012})]{FriisLeeBruschiLouko2012}, before we briefly analyze the consequences for practical applications, \emph{Bell inequalities} and \emph{teleportation}, in Section~\ref{sec:nonlocality and teleportation fermions}.

\subsection{Fermionic Bell States}\label{sec:fermionic bell states}

For the Dirac field we can consider a maximally entangled \emph{Bell state}\index{Bell!states}\index{state!Bell} [see Eqs.~(\ref{eq:Bell states})] between Alice's mode~$\kappa$ and Rob's mode~$\kappa\pr\,$, given by
\begin{align}
    \fket{\phi^{\pm}}_{\hspace*{-0.5pt}\raisebox{0.0pt}{\tiny{$A\hspace*{-0.5pt}R$}}}    &=\,\tfrac{1}{\sqrt{2}}\Bigl(\fket{0}_{\hspace*{-0.5pt}\raisebox{0.0pt}{\tiny{$A$}}} \otimes\fket{0}_{\hspace*{-0.5pt}\raisebox{0.0pt}{\tiny{$R$}}}\pm
    \fket{1_{\kappa\!}}_{\hspace*{-0.5pt}\raisebox{0.0pt}{\tiny{$A$}}}\otimes
    \fketp{1_{\kappa\pr\!}}_{\hspace*{-0.5pt}\raisebox{0.0pt}{\tiny{$R$}}}\Bigr)\,,
    \label{eq:fermion Bell states phi plus minus Alice Rob}
\end{align}
where we have assumed that $\kappa\pr\geq0$ is a positive frequency mode, while the frequency of the mode~$\kappa$ is inconsequential for our present analysis and we therefore have not specifically indicated it in Eq.~(\ref{eq:fermion Bell states phi plus minus Alice Rob}). Since the positive and negative frequency modes appear symmetrically in the spectrum we can be content to study the case of $\kappa\pr\geq0$, but the interested reader may find the expressions for the case $\kappa\pr<0$ in Ref.~\cite[(\ref{Paper:FriisLeeBruschiLouko2012})]{FriisLeeBruschiLouko2012}. Assuming that the fermions can be distinguished by their appearance in either Alice's or Rob's cavity we can assume a tensor product between the Fock spaces of different cavities. The density operator that corresponds to the state in Eq.~(\ref{eq:fermion Bell states phi plus minus Alice Rob}) is given by
\begin{align}
    \varrho^{\pm}_{\hspace*{-0.5pt}\raisebox{0.0pt}{\tiny{$A\hspace*{-0.5pt}R$}}} &=\,\tfrac{1}{2}\Bigl(\fket{0}\!\fbra{0}\otimes\fket{0}\!\fbra{0}\,\pm\,
    \fket{0}\!\fbra{1_{\kappa\!}}\otimes\fket{0}\!\fbrap{1_{\kappa\pr\!}}
    \label{eq:Dirac field in region bell state density matrix}\\[1mm]
    &\ \ \pm\,\fket{1_{\kappa\!}}\!\fbra{0}\otimes\fketp{1_{\kappa\pr\!}}\!\fbra{0}\,+\,
    \fket{1_{\kappa\!}}\!\fbra{1_{\kappa\!}}\otimes\fketp{1_{\kappa\pr\!}}\!\fbrap{1_{\kappa\pr\!}}\Bigr)\,,
    \nonumber
\end{align}
where we have dropped the labels for Alice and Rob. Subsequently, we transform the matrix elements on the right hand side of the tensor product to the out-region to take into account Rob's motion. The corresponding transformed versions of $\fket{0}$ and $\fket{1_{\kappa\pr\!}}$ are given by~(\ref{eq:transformed fermion vac vector}) and, with appropriate relabelling~(\ref{eq:fermion single particle pure bogo}), respectively. We then trace out all modes except~$\kappa$ and~$\kappa\pr$ from the relevant matrix elements, i.e.,
\begin{subequations}
\label{eq:fermion bell state partial traces}
\begin{align}
    \hspace*{-3mm}\tr_{\lnot\kappa\pr}\Bigl(\fket{0}\!\fbra{0}\Bigr)    &=
    (1-2 h^{2}\bar{f}^{A}_{\kappa\pr})\fkethat{0}\!\fbrahat{0}+2h^{2}\bar{f}^{A}_{\kappa\pr}
    \fkethatkp{\!1}{\kappa\pr\!}\!\fbrahatkp{\!1}{\kappa\pr\!}+O(h^{3}),
    \label{eq:fermion bell state partial traces 00}\\[1.8mm]
    \hspace*{-3mm}\tr_{\lnot\kappa\pr}\Bigl(\fketp{\!1_{\kappa\pr\!}}\!\fbrap{\!1_{\kappa\pr\!}}\Bigr)    &=
    (1-2h^{2}f^{A}_{\kappa\pr})\fkethatkp{\!1}{\kappa\pr\!}\!\fbrahatkp{\!1}{\kappa\pr\!}
    +2h^{2}f^{A}_{\kappa\pr}\fkethat{0}\!\fbrahat{0}+O(h^{3}),
    \label{eq:fermion bell state partial traces 11}\\[1.8mm]
    \hspace*{-3mm}\tr_{\lnot\kappa\pr}\Bigl(\fket{0}\!\fbrap{\!1_{\kappa\pr\!}}\Bigr)    &=\,
    \Ghn{0}{\kappa\pr}\fket{0}\!\fbrahatkp{1}{{\kappa\pr\!}}\,+\,h^{2}\,\Ahmn{2}{\kappa\pr\!\kappa\pr}\,
    \fket{0}\!\fbrahatkp{1}{{\kappa\pr\!}}\,+\,O(h^{3})\,,
    \label{eq:fermion bell state partial traces 01}
\end{align}
\end{subequations}
where the functions $f^{A}_{\kappa\pr}$ and $\bar{f}^{A}_{\kappa\pr}$ are as in Eq.~(\ref{eq:f A and fbar A m not n}). For the situation we are dealing with here the two fermionic modes can be mapped to two qubits without problems (see Section~\ref{sec:Entanglement in Fermionic Quantum Fields}). We represent the transformed state of the modes~$\kappa$ and $\kappa\pr$ by the two-qubit density matrix
\begin{align}
\begin{pmatrix}
\tfrac{1}{2}-h^{2}\bar{f}^{A}_{\kappa\pr} &   0   &   0   &   \pm\tfrac{1}{2}\Ghn{0}{\kappa\pr}\pm\tfrac{1}{2} h^{2}\Ahmn{2}{\kappa\pr\!\kappa\pr}\\
0   &   h^{2}\bar{f}^{A}_{\kappa\pr} &   0   & 0   \\
0   &   0   &   h^{2}f^{A}_{\kappa\pr} &   0   \\
\pm\tfrac{1}{2}\Ghnstar{0}{\kappa\pr}\pm \tfrac{1}{2}h^{2}\Ahmnstar{2}{\kappa\pr\!\kappa\pr}    &   0   &   0   &   \tfrac{1}{2}-h^{2}f^{A}_{\kappa\pr}
\label{eq:fermion Bell state density matrix}
\end{pmatrix}\,,
\end{align}
where we have neglected terms of $O(h^{3})$. Next, we can compute the \emph{negativity}\index{negativity} for this state. The partial transposition\index{partial transposition} shifts the off-diagonals towards the centre along the anti-diagonal and for the unperturbed state one immediately finds three positive eigenvalues $\lambdah{0}_{+}=\tfrac{1}{2}$ and one negative eigenvalue $\lambdah{0}_{-}=-\tfrac{1}{2}$. Since there are no corrections linear in~$h$ we can find the leading order correction to the negative eigenvalue as the expectation value of the perturbations of $\varrho^{\pm}_{\hspace*{-0.5pt}\raisebox{0.0pt}{\tiny{$A\hspace*{-0.5pt}R$}}}$ in the eigenvector $\ket{\lambdah{0}_{-}}$ corresponding to the negative unperturbed eigenvalue. That eigenvector is given by
\begin{align}
\ket{\lambdah{0}_{-}}   &=\,\tfrac{1}{\sqrt{2}}\bigl(\,0,\ 1,\,\mp\Ghn{0}{\kappa\pr},\ 0\,\bigr)^{T}\,,
\end{align}
and, using the Bogoliubov identity~(\ref{eq:generic travel scenario bogo identities fermions 2}), we find the negativity
\begin{align}
    \mathcal{N}(\varrho^{\pm}_{\hspace*{-0.5pt}\raisebox{0.0pt}{\tiny{$A\hspace*{-0.5pt}R$}}})  &=\,
    \Negh{0}\,-\,h^{2}\,\Negh{2}\,+\,O(h^{3})\,=\,
    \tfrac{1}{2}\,-\,h^{2}\bigl(f^{A}_{\kappa\pr}+\bar{f}^{A}_{\kappa\pr}\bigr)\,+\,O(h^{3})\,.
    \label{eq:fermion bell state negativity}
\end{align}
The quantities $f^{A}_{\kappa\pr}$ and $\bar{f}^{A}_{\kappa\pr}$ are illustrated in Fig.~\ref{fig:entanglement degradation fermions}.

\subsubsection{Perturbative Expressions for the Concurrence}

With the results of the previous sections at hand it is not surprising that the entanglement in Eq.~(\ref{eq:fermion bell state negativity}) is degraded due to Rob's motion. Nonetheless, we also wish to supply a quantitative description of the entanglement loss that relates to practical applications. One such measure is the \emph{entanglement of formation}, which, for two qubits, is fully determined by the \emph{concurrence}\index{concurrence}, see Eq.~(\ref{eq:wootters concurrence two qubits}). However, as we have hinted at in Section~\ref{sec:entanglement from the fermionic vacuum}, computing the concurrence in a perturbative approach proves to be somewhat impractical, as we shall demonstrate here. For the calculation we need to determine the eigenvalues of the matrix $\varrho^{\pm}_{\hspace*{-0.5pt}\raisebox{0.0pt}{\tiny{$A\hspace*{-0.5pt}R$}}}(\sigma_{2}\otimes\sigma_{2})
\varrho^{\pm\,*}_{\hspace*{-0.5pt}\raisebox{0.0pt}{\tiny{$A\hspace*{-0.5pt}R$}}}(\sigma_{2}\otimes\sigma_{2})$, where $\sigma_{2}$ is the second Pauli matrix\index{Pauli matrices} from Eq.~(\ref{eq:Pauli matrices}) and $\varrho^{\pm}_{\hspace*{-0.5pt}\raisebox{0.0pt}{\tiny{$A\hspace*{-0.5pt}R$}}}$ is taken from~(\ref{eq:fermion Bell state density matrix}), which, to second order in~$h$, can be written as
\begin{align}
\begin{pmatrix}
\tfrac{1}{2}-h^{2}\bigl(f^{A}_{\kappa\pr}+\bar{f}^{A}_{\kappa\pr}\bigr) &   0   &   0   &
\pm\tfrac{1}{2}\Ghn{0}{\kappa\pr}\pm h^{2}\bigl(\tfrac{1}{2}\Ahmn{2}{\kappa\pr\!\kappa\pr}-\Ghn{0}{\kappa\pr}\bar{f}^{A}_{\kappa\pr}\bigr)\\
0   &   0   &   0   &   0   \\
0   &   0   &   0   &   0   \\
\pm\tfrac{1}{2}\Ghnstar{0}{\kappa\pr}\pm h^{2}\bigl(\tfrac{1}{2}\Ahmnstar{2}{\kappa\pr\!\kappa\pr}-\Ghnstar{0}{\kappa\pr}f^{A}_{\kappa\pr}\bigr)    &   0   &   0   &
\tfrac{1}{2}-h^{2}\bigl(f^{A}_{\kappa\pr}+\bar{f}^{A}_{\kappa\pr}\bigr)
\label{eq:fermion Bell state matrix concurrence}
\end{pmatrix}\,.
\end{align}
The unperturbed matrix has the eigenvalues $\lambdah{0}_{1}=1$ and $\lambdah{0}_{2,3,4}=0$. Applying the techniques described on pp.~\pageref{page:pert diagonalization} to determine the corrections to these eigenvalues, and with the help of Eq.~(\ref{eq:generic travel scenario bogo identities fermions 2}) we find that none of the degenerate eigenvalues are perturbed when terms proportional to~$h^{2}$ are included. The non-degenerate eigenvalue, on the other hand, is corrected such that
\begin{align}
    \lambda_{1} &=\,\lambdah{0}_{1}\,+\,h^{2}\,\lambdah{2}_{1}\,+\,O(h^{3})\,=\,1\,-\,2\,h^{2}\,\bigl(f^{A}_{\kappa\pr}+\bar{f}^{A}_{\kappa\pr}\bigr)\,+\,O(h^{3})\,.
    \label{eq:fermion bell state concurrence eigenvalue}
\end{align}
When we now wish to evaluate the concurrence from Eq.~(\ref{eq:wootters concurrence two qubits}) we have to take the square roots of the perturbed eigenvalues and we encounter an issue. The expansion $\sqrt{\lambda_{1}}=1-h^{2}\bigl(f^{A}_{\kappa\pr}+\bar{f}^{A}_{\kappa\pr}\bigr)+O(h^{3})$ is easily determined, but all other square roots vanish to leading order. However, without further computations we cannot exclude the possibility that the eigenvalues $\lambda_{2,3,4}$ receive fourth order corrections when terms proportional to~$h^{4}$ are kept throughout the calculation. These corrections could contribute to the second order corrections of the concurrence. The present calculation thus only allows us to specify an \emph{upper bound} on the degraded concurrence. But, with the aid of the inequality~(\ref{eq:concurrence negativity bounds}) and the negativity from Eq.~(\ref{eq:fermion bell state negativity}) we can supply also a \emph{lower bound}, such that the perturbed concurrence is bounded by
\begin{align}
    1\,-\,2\,h^{2}\,\bigl(f^{A}_{\kappa\pr}+\bar{f}^{A}_{\kappa\pr}\bigr)\,\leq
    C(\varrho^{\pm}_{\hspace*{-0.5pt}\raisebox{0.0pt}{\tiny{$A\hspace*{-0.5pt}R$}}})\,+\,O(h^{3})    &\leq\,
    1\,-\,h^{2}\,\bigl(f^{A}_{\kappa\pr}+\bar{f}^{A}_{\kappa\pr}\bigr)\,.
    \label{eq:fermion bell state concurrence bounds}
\end{align}

\subsection{Non-Locality \& Fidelity of Teleportation}\label{sec:nonlocality and teleportation fermions}

\begin{figure}[hb!]
\centering
(a)\includegraphics[width=0.72\textwidth]{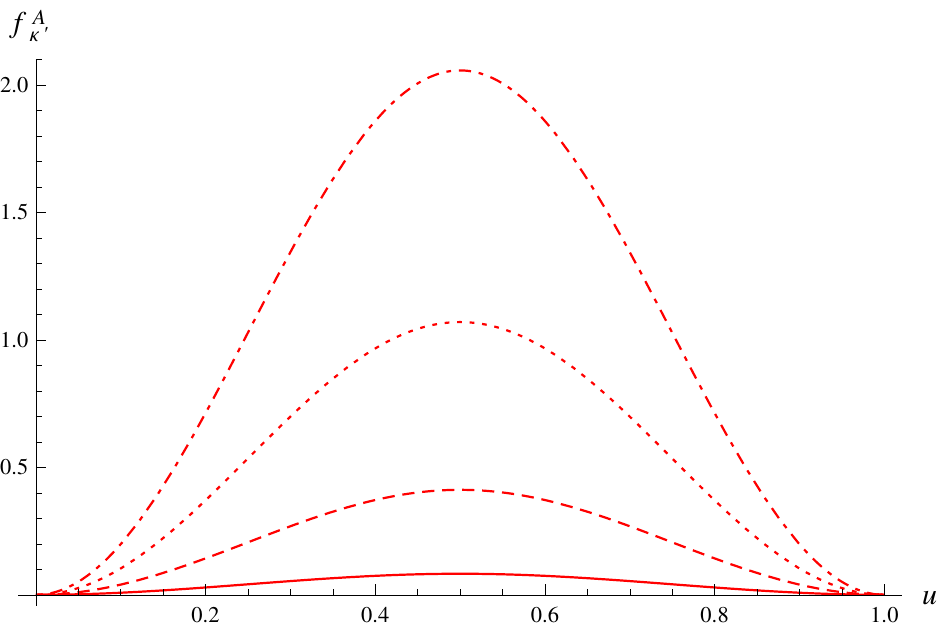}
(b)\includegraphics[width=0.72\textwidth]{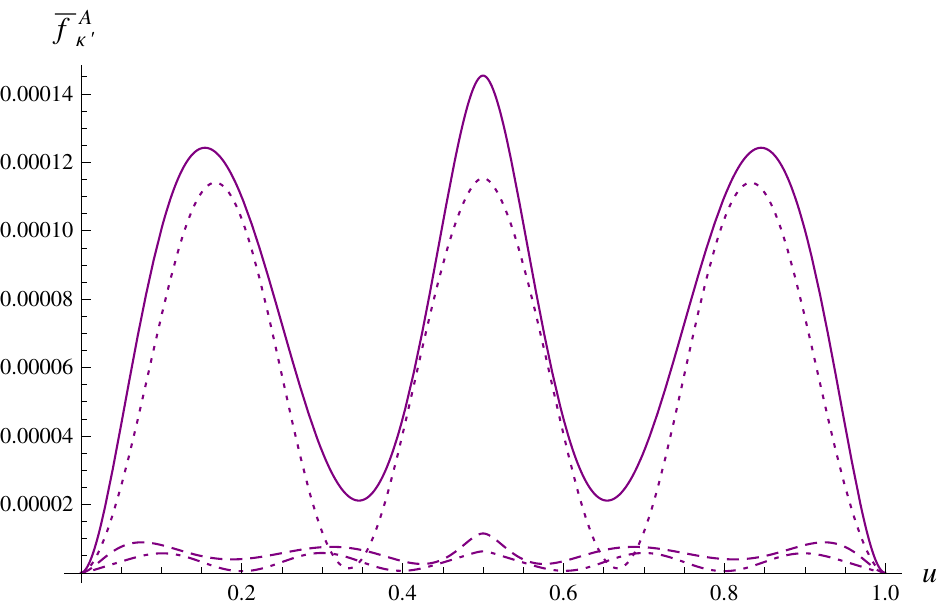}
\caption{
\textbf{Entanglement degradation \textemdash\ $f^{A}_{\kappa\pr}$ and $\bar{f}^{A}_{\kappa\pr}$:} The quantities $f^{A}_{\kappa\pr}$ and $\bar{f}^{A}_{\kappa\pr}$ that are degrading the entanglement between Alice and Rob [see Eqs.~(\ref{eq:fermion bell state negativity}), (\ref{eq:max CHSH violation fermion Bell state}) and (\ref{eq:max teleportation fidelity fermion Bell state})] are shown in Fig.~\ref{fig:entanglement degradation fermions}~(a) and Fig.~\ref{fig:entanglement degradation fermions}~(b), respectively, for the basic building block travel scenario of Section~\ref{sec:basic building block} for a $(1+1)$ dimensional massless Dirac field. The horizontal axis shows the dimensionless parameter $u:=h\tau/[4L\artanh(h/2)]\,$ [see Eq.~(\ref{eq:cavity centre proper freq})], where $\tau$ is the duration of the uniform acceleration at the centre of the cavity. Curves are shown for the modes $\kappa\pr=0$ (solid), $\kappa\pr=1$ (dashed), $\kappa\pr=2$ (dotted), and $\kappa\pr=3$ (dotted-dashed).
\label{fig:entanglement degradation fermions}\vspace*{-2mm}}
\end{figure}

Since the perturbative evaluation of the concurrence proved to be rather intricate, let us turn to more accessible means of supplying an operational picture for the entanglement degradation in fermionic systems. In Sections~\ref{sec:bell inequalities and nonlocality} and~\ref{sec:quantum teleportation} we have seen that the \emph{correlation matrix}\index{correlation matrix} $t[\rho]$ (see Theorem~\ref{thm:chsh criterion 2} or Ref.~\cite{HorodeckiRPM1995}) of a two-qubit state can be used to determine the maximally possible violation of the \emph{CHSH inequality}\index{CHSH!inequality} as well as the optimal \emph{teleportation fidelity}\index{teleportation!fidelity}. We thus determine the matrix
$M_{\varrho^{\pm}_{\hspace*{-0.5pt}\raisebox{0.0pt}{\tiny{$A\hspace*{-0.5pt}R$}}}}=t[\varrho^{\pm}_{\hspace*{-0.5pt}\raisebox{0.0pt}{\tiny{$A\hspace*{-0.5pt}R$}}}]^{T}
t[\varrho^{\pm}_{\hspace*{-0.5pt}\raisebox{0.0pt}{\tiny{$A\hspace*{-0.5pt}R$}}}]$ for the two-qubit density matrix in Eq.~(\ref{eq:fermion Bell state density matrix}) and we obtain
\begin{align}
\begin{pmatrix}
1\,-\,2\,h^{2}\,\bigl(f^{A}_{\kappa\pr}+\bar{f}^{A}_{\kappa\pr}\bigr)   &   0   &   0   \\
0   &   1\,-\,2\,h^{2}\,\bigl(f^{A}_{\kappa\pr}+\bar{f}^{A}_{\kappa\pr}\bigr)   &   0   \\
0   &   0   &   1\,-\,4\,h^{2}\,\bigl(f^{A}_{\kappa\pr}+\bar{f}^{A}_{\kappa\pr}\bigr)
\end{pmatrix}\,+\,O(h^{3})\,,
\end{align}
where we have again used Eq.~(\ref{eq:generic travel scenario bogo identities fermions 2}). Since the matrix is already diagonal we can straightforwardly find the maximally possible violation of the CHSH inequality from Theorem~\ref{thm:chsh criterion 2} as
\begin{align}
    \expval{\mathcal{O}^{\mathrm{max}}\subtiny{0}{0}{\mathrm{CHSH}}}_{\varrho^{\pm}_{\hspace*{-0.5pt}\raisebox{0.0pt}{\tiny{$A\hspace*{-0.5pt}R$}}}}
    &=\,2\sqrt{2}\,\bigl(1\,-\,h^{2}\,\bigl[f^{A}_{\kappa\pr}+\bar{f}^{A}_{\kappa\pr}\bigr]\bigr)\,+\,O(h^{3})\,,
    \label{eq:max CHSH violation fermion Bell state}
\end{align}
while the maximal teleportation fidelity, optimized over Rob's local rotations, is found to be
\vspace*{-4mm}
\begin{align}
    \mathcal{F}_{\mathrm{max}}(\varrho^{\pm}_{\hspace*{-0.5pt}\raisebox{0.0pt}{\tiny{$A\hspace*{-0.5pt}R$}}})
    &=\,1\,-\,\tfrac{2}{3}\,h^{2}\,\bigl(f^{A}_{\kappa\pr}+\bar{f}^{A}_{\kappa\pr}\bigr)\,+\,O(h^{3})\,.
    \label{eq:max teleportation fidelity fermion Bell state}
\end{align}
An illustration of the functions $f^{A}_{\kappa\pr}$ and $\bar{f}^{A}_{\kappa\pr}$ is shown in Fig.~\ref{fig:entanglement degradation fermions}. The quantities of Eqs.~(\ref{eq:max CHSH violation fermion Bell state}) and~(\ref{eq:max teleportation fidelity fermion Bell state}) provide clear operational meaning for the entanglement degradation effects of the fermionic modes, and may hopefully allow for simulations of these effects in analogue materials, see, e.g., Refs.~\cite{BoadaCeliLatorreLewenstein2011,ZhangWangZhu2012,Iorio2012}.
\vspace*{-2mm}

    \pdfbookmark[-1]{Conclusions}{Conclusions}
    \chapter*{Conclusions}
\addcontentsline{toc}{chapter}{Conclusions}
\vspace*{2mm}
In this thesis we have presented the model of \emph{relativistically rigid cavities} in the context of relativistic quantum information (RQI), which was first introduced in Ref.~\cite{BruschiFuentesLouko2012}. We have discussed the geometric aspects of the rigid cavity in Minkowski spacetime and we have analyzed the confinement of bosonic scalar fields as well as fermionic Dirac fields to the cavity when it is undergoing non-uniform motion. The quantum fields can be massless or have non-zero mass and are confined to the cavity by boundary conditions that enforce that either the mode functions or the spatial probability current vanish at the cavity walls. The motion can consist of individual segments of inertial motion and uniform acceleration that are related by sharp transitions, or the proper acceleration can vary smoothly. For $(1+1)$ and $(2+1)$ dimensions both options can be implemented unitarily on the Fock spaces of the bosonic and fermionic field operators, respectively. However, in $(3+1)$ dimensions unitarity fails for non-smooth transitions~\cite[(\ref{Paper:FriisLeeLouko2013})]{FriisLeeLouko2013}.\\

The main focus of the analysis was aimed at the investigation of the role of the cavity as a system for the storage and manipulation of quantum information. We have shown how the Bogoliubov transformations that are induced by the non-uniform motion create entanglement between previously unentangled modes of the quantum fields inside the cavities. Quantum correlations are created for various initial states, including bosonic Fock states~\cite[(\ref{Paper:FriisBruschiLoukoFuentes2012})]{FriisBruschiLoukoFuentes2012} and squeezed states~\cite[(\ref{Paper:FriisFuentes2013})]{FriisFuentes2013}, as well as for different fermionic Fock states. Moreover, we have reviewed how the production of entanglement can be resonantly enhanced, see Refs.~\cite{BruschiDraganLeeFuentesLouko2013} and~\cite[(\ref{Paper:BruschiFriisFuentesWeinfurtner2013})]{BruschiFriisFuentesWeinfurtner2013} and even generate genuine multipartite entanglement~\cite[(\ref{Paper:FriisHuberFuentesBruschi2012})]{FriisHuberFuentesBruschi2012}. These entanglement generation effects may be of interest for the identification of the quantumness of particle creation phenomena similar to the \emph{dynamical Casimir effect} via the specific signature of the created quantum correlations. This may allow to assign observed radiation unambiguously to the effects of non-uniform motion. Moreover, the entanglement generation is conceptually interesting since it suggests that the motion of the cavity may be interpreted as (weak) quantum gates~\cite{BruschiDraganLeeFuentesLouko2013,BruschiLoukoFaccioFuentes2013}\index{quantum gates}. Certainly, this opens avenues for further investigation around the central motive: \emph{``Can quantum information processing tasks or quantum computation be performed by simply moving quantum systems in spacetime?"}\\

Finally, we have turned our attention to \emph{entanglement degradation} effects when quantum communication tasks, for instance, \emph{teleportation} between two different cavities, are considered. If the observers do not have access to all of the modes in the spectrum \textemdash\  typically only a finite number of modes can be addressed \textemdash\ the motion of the individual cavities degrades the initially shared entanglement. This is the case because the particle creation and shifting of excitations within one cavity entangles the modes in the spectrum with each other. Subsequently, some of the entangled modes are traced over, which leads to a loss of information that can be viewed as \emph{decoherence}. We have studied such situations for cavities containing scalar fields~\cite{BruschiFuentesLouko2012} as well as Dirac fields~\cite[(\ref{Paper:FriisLeeBruschiLouko2012})]{FriisLeeBruschiLouko2012}. For the special case of Gaussian two-mode squeezed states of the bosonic fields we have investigated the effects on the \emph{quantum teleportation} protocol~\cite[(\ref{Paper:FriisLeeTruongSabinSolanoJohanssonFuentes2013})]{FriisLeeTruongSabinSolanoJohanssonFuentes2013}, and we have presented a setup where the mechanical motion of the cavity mirrors may be simulated in superconducting circuits.\\

\vspace*{-2mm}
The significance of this direction of our research lies in the basic need to establish assessments of the \emph{robustness} of quantum communication procedures against the effects of relativistic motion. Our treatment has significantly advanced the previous toy models in RQI addressing such questions, taking them from the realm of thought experiments with global modes, and eternal uniform accelerations towards practical settings that may be emulated with current technology, see Refs.~\cite{JohanssonJohanssonWilsonNori2010,WilsonDynCasNature2012,SvenssonMScThesis2012,LaehteenmaekiParaoanuHasselHakonen2013} and~\cite[(\ref{Paper:FriisLeeTruongSabinSolanoJohanssonFuentes2013})]{FriisLeeTruongSabinSolanoJohanssonFuentes2013}. However, the analysis presented here covers only one specific type of quantum system used for the manipulation of quantum information, and it will hence be of interest for future investigations to study relativistic effects on other tools for quantum communication. In addition, a whole zoo of other relativistic effects, besides those described here, may emerge from further research in this direction.\\

\vspace*{-2mm}
We have also come across issues relating to the practical treatment of fermionic modes for the purpose of quantum information processing~\cite[(\ref{Paper:FriisLeeBruschi2013})]{FriisLeeBruschi2013}. Although computations can be carried out in a meaningful way for the situations we have considered here, we showed that this is not the case in general. Briefly summarized, \emph{fermionic modes are not qubits}, which calls for a reevaluation of standard techniques in quantum information for fermionic modes. In particular, the quantification of fermionic mode entanglement remains an open question for theoretical research that might possibly also inspire experimental tests.\\

We conclude that the effects of the non-uniform motion, although small compared to common day-to-day experience, may be large enough for experimental observation in modern cutting-edge laboratories, for instance using superconducting technology that was recently employed for the confirmation of the dynamical Casimir effect~\cite{WilsonDynCasNature2012}\index{dynamical Casimir effect}. The rapidly progressing technological advancement, e.g., in the control and manipulation of individual quantum systems~\cite{RaimondBruneHaroche2001}, suggests that even previously negligible effects may become relevant in the near future. Quantum communication is already operating at length scales where relativity plays a role~\cite{Ma-Zeilinger-tele2012,RideoutEtal2012,BruschiSabinWhiteBaccettiOiFuentes2013}, and so it seems prudent to study relativistic effects appearing in such tasks. Moreover, relativistic effects may provide novel ways to estimate kinematical parameters and spacetime properties~\cite{AhmadiBruschiSabinAdessoFuentes2013,DownesMilburnCaves2012}.\\

In combination, recently established, as well as well-known theoretical and experimental techniques, and yet-to-be-made discoveries in the overlap of relativity and quantum information science will form the core for the next generation of quantum technologies.\\

\newpage\ \\ 

    \pdfbookmark[-1]{References}{References}
	\phantomsection

    \newpage\ \\ \newpage

    \pdfbookmark[-1]{List of Figures}{List of Figures}
    \phantomsection
    \addcontentsline{toc}{chapter}{List of Figures}
    \listoffigures

    \newpage\ \\ \newpage
    \phantomsection
    \pdfbookmark[-1]{Index}{Index}
    \addcontentsline{toc}{chapter}{Index}
    \printindex


\end{document}